\documentclass{article}
\usepackage[T1]{fontenc}
\usepackage[latin1]{inputenc}
\usepackage{graphics}

\makeatletter

\newcommand{\LyX}{L\kern-.1667em\lower.25em\hbox{Y}\kern-.125emX\spacefactor1000}

\makeatother

\begin{document}

\title{One Particle (improper) states and Scattering States in Nelson's Massless Model}

\author{Alessandro Pizzo\thanks{
Pizzo@sissa.it
}\\
S.I.S.S.A. -I.S.A.S., via Beirut 2-4 ~~34014 Trieste, Italia\\
}

\date{October 25, 2000 \\
}

\maketitle
\begin{abstract}
In the one nucleon sector of Nelson's massless model with an ultraviolet cut-off
and no infrared regularization, one particle improper states are constructed
and a scattering theory is developed both for the massless scalar field and
for the non-relativistic particle. One particle improper states are obtained
constructively by iterating an analytic perturbation of isolated eigenvalues.
Scattering states are constructed by exploiting some properties of (non-relativistic)
locality of the model and starting from ``minimal asymptotic nucleon states''.
They represent asymptotic configurations where a cloud of (soft) mesons always
appears even if their energy may be arbitrarily small.\\
\\
\\
\\
\\
\\
\\
\\
\\
\\
\\
\\
\\
\\
\\
\\
\\
\\
\\
\\
\\
\\

\end{abstract}
\textbf{\Large Introduction}{\large }\\
{\large }\\
{\large }\\
{\large The model I will examine describes the covariant interaction (under
spatial translation) between mesons and a non-relativistic nucleon. The rigorous
study of the model was started by Nelson {[}1{]}, in order to remove the ultraviolet
cut-off in the interaction, in the case of massive mesons.}\\
{\large Then, it was used as a toy model to give a consistent explanation of
interaction between radiation and matter, in particular of a single charged
particle with radiation field.}\\
{\large The most important results on the subject are contained in {[}2{]} .
They regard both the case of massive mesons and the massless case. In particular
they show the existence of a subspace of one particle states for the hamiltonian
(it means that the hamiltonian acts on them as a function of the total impulse)
in the massive case and the absence of such a subspace in the other case (if
there is no infrared regularization in the interaction term). Recently {[}3{]}
the model was studied in order to prove asymptotic completeness, adding some
more conditions (confinement of the non-relativistic particle).}\\
{\large In this paper I study the interaction of a non-relativistic nucleon
with massless mesons, without infrared regularization. The aim is to prove that
a description of the asymptotic dynamics exists once the ultraviolet cut-off
is fixed in the interaction and for sufficiently small values of the coupling
constant; the result is a scattering theory in a physical relevant subspace
with an explicit interpretation with respect to the asymptotic dynamical variables.}\\
{\large }\\
{\large The interaction without infrared regularization involves additional
difficulties with respect to previous treated versions:}\\
{\large }\\
{\large - the absence of a ground state for the hamiltonians at fixed total
momentum;}\\
{\large }\\
{\large - the massless dispersion of smearing test functions in L.S.Z. meson
field operators. These operators have to define a free meson field in the asymptotic
limit.}\\
{\large }\\
{\large My approach to scattering is different from the traditional one in two
aspects:}\\
{\large }\\
{\large - the first one consists in using some mechanism of the scattering in
quantum field theory (Haag-Ruelle formulation), by exploiting the locality properties
at fixed time of the model;}\\
{\large }\\
{\large - the second one is technical and regards the determination of the limit,
for \( \sigma \rightarrow 0 \), of the ground states of the hamiltonians with
an infrared cut-off \( \sigma  \), at fixed total momentum and properly transformed.
I will use an iterative procedure (different from the operatorial renormalization
group {[}4{]} by Bach, Froehlich and Segal) which gives a strong convergence
with an error estimable in terms of the infrared cut-off} {\large that we have
to remove}.\\
\\
{\large Finally, I observe that some of these ideas and techniques can be used
in the study of scattering in the one-electron sector of non relativistic quantum
electrodynamics.}\\
\\
\\
\textbf{\large Definition of the model.} {\large }\\
{\large }\\
{\large The system consists of a non-relativistic spinless quantum particle
of mass \( m \), which is coupled to the massless boson field. The non-relativistic
particle is described by position and momentum variables with usual canonic
commutation rules (c.c.r.) }\\
{\large }\\
{\large \( \left[ x_{i},p_{j}\right] =i\delta _{i,j} \)} \textbf{\large }{\large (}\textbf{\large \( \hbar =1 \)}{\large )
;}\\
{\large }\\
{\large the meson field is described by} {\large \( \mathbf{A}\left( 0,\mathbf{y}\right) =\frac{1}{\sqrt{2\pi }^{3}}\cdot \int \left( a^{\dagger }\left( \mathbf{k}\right) e^{-i\mathbf{k}\cdot \mathbf{y}}+a\left( \mathbf{k}\right) e^{i\mathbf{k}\cdot \mathbf{y}}\right) \frac{d^{3}k}{\sqrt{2\left| \mathbf{k}\right| }} \)
\( \left( c=1\right)  \),} {\large where \( a^{\dagger }\left( \mathbf{k}\right) ,a\left( \mathbf{k}\right)  \)
are creation and destruction operator valued distributions which satisfy the
c.c.r. }\\
{\large }\\
{\large \( \left[ a\left( \mathbf{k}\right) ,a\left( \mathbf{q}\right) \right] =\left[ a^{\dagger }\left( \mathbf{k}\right) ,a^{\dagger }\left( \mathbf{q}\right) \right] =0 \)
}\\
{\large \( \left[ a\left( \mathbf{k}\right) ,a^{\dagger }\left( \mathbf{q}\right) \right] =\delta ^{3}\left( \mathbf{k}-\mathbf{q}\right)  \)}
\textbf{\large }\\
\textbf{\large }\\
\textbf{\large }\\
{\large The spatial translations are implemented by the total momentum}\\
{\large }\\
{\large \( \mathbf{P}=\mathbf{p}+\mathbf{P}^{e.m.}=\mathbf{p}+\int \mathbf{k}a^{\dagger }\left( \mathbf{k}\right) a\left( \mathbf{k}\right) d^{3}k \)}
{\large ;}\\
{\large }\\
{\large the time evolution is given by the covariant hamiltonian (\( \left[ H,\mathbf{P}\right] =0 \))}\\
{\large }\\
{\large \( H=\frac{\mathbf{p}^{2}}{2m}+g\int _{0}^{\kappa }\left( a\left( \mathbf{k}\right) e^{i\mathbf{k}\cdot \mathbf{x}}+a^{\dagger }\left( \mathbf{k}\right) e^{-i\mathbf{k}\cdot \mathbf{x}}\right) \frac{d^{3}\mathbf{k}}{\sqrt{2}\left| \mathbf{k}\right| ^{\frac{1}{2}}}+H^{mes} \)}\\
{\large }\\
{\large where \( \kappa  \) is the ultraviolet} \emph{\large cut-off} {\large and
\( g \) is the coupling constant.}\\
{\large }\\
{\large The Hilbert space of the system is}{\large ~H}{\large \( =L^{2}\left( R^{3}\right) \otimes F \),
where \( F \) is the Fock} {\large space with respect to the operator valued
distributions \( \left\{ a^{\dagger }\left( \mathbf{k}\right) ,a\left( \mathbf{k}\right) \right\}  \).
An element of}{\large  ~H} {\large is a sequence \( \left\{ \psi ^{n}\right\}  \)
of functions on \( R^{3n+1} \) with \( \left\Vert \psi \right\Vert <\infty  \),
where}\\
{\large \( \left\Vert \psi \right\Vert ^{2}=\sum ^{\infty }_{n=0}\int \overline{\psi ^{n}\left( \mathbf{x},\mathbf{k}_{1},..\mathbf{k}_{n}\right) }\psi ^{n}\left( \mathbf{x},\mathbf{k}_{1},..\mathbf{k}_{n}\right) d^{3}k_{1}...d^{3}k_{n}d^{3}x \)
and each \( \psi ^{n}\left( \mathbf{x},\mathbf{k}_{1},..\mathbf{k}_{n}\right)  \)
is symmetric in \( \mathbf{k}_{1},...\mathbf{k}_{n} \). The \( n=0 \) component
corresponds to the vacuum subspace tensorized with the non-relativistic particle
space \( L^{2}\left( R^{3}\right)  \).}\\
\emph{\large }\\
\emph{\large Standard results about \( H \)and \( P \):}{\large }\\
{\large }\\
{\large i) The operators \( \mathbf{P}=\mathbf{p}\otimes 1+1\otimes \int \mathbf{k}a^{\dagger }\left( \mathbf{k}\right) a\left( \mathbf{k}\right) d^{3}k \)
 are essentially self-adjoint (e.s.a.) in \( D\equiv \bigvee _{n\in \aleph }h\otimes \psi ^{n} \),
which is the set of finite linear combinations of vectors of wave function \( h\left( \mathbf{x}\right) \psi ^{n}\left( \mathbf{k}_{1},...\mathbf{k}_{n}\right)  \),
where \( h\left( \mathbf{x}\right) \in S\left( R^{3}\right)  \) and where \( \psi _{}^{n}\left( \mathbf{k}_{1},...\mathbf{k}_{n}\right) \in S\left( R^{3n}\right)  \)
\( simm. \) ,\( n\in \aleph  \). Since \( \mathbf{p},\int \mathbf{k}a^{\dagger }\left( \mathbf{k}\right) a\left( \mathbf{k}\right) d^{3}k \)
are e.s.a. respectively in \( S\left( R^{3}\right)  \) and in \( \bigvee _{n\in \aleph }\psi ^{n} \),
the result follows for \( \mathbf{P} \) operators. }\\
{\large }\\
{\large ii) The interaction term in the hamiltonian is a Kato's small perturbation
with respect to \( H_{0}\equiv \frac{\mathbf{p}^{2}}{2m}+H^{mes} \); therefore
the hamiltonian \( H \) is e.s.a. in \( D\equiv \bigvee _{n\in \aleph }h\otimes \psi ^{n} \)and
\( D\left( H\right) \equiv D\left( H_{0}\right)  \).}\\
{\large }\\
{\large iii) The groups \( e^{i\mathbf{a}\cdot \mathbf{P}} \)e \( e^{i\tau H} \)
( \( \tau ,a^{i}\in R \) ) commute.}\\
{\large }\\
{\large iv) The joined spectral decomposition of the space ~}{\large H,} {\large with
respect to \( \mathbf{P} \) operators, is written~~}{\large H}{\large \( =\oplus \int  \)}{\large H}{\large \( _{\mathbf{P}}d^{3}P \),
where~~}{\large  H}{\large \( _{\mathbf{P}} \) is isomorphic to \( F \).}
\\
{\large In fact, to the improper eigenvectors (of the \( \mathbf{P} \) operators)
\( \psi _{\mathbf{P}}^{n} \)} {\large of wave function }\\
{\large }\\
{\large 
\[
\psi _{\mathbf{P}}^{n}\left( \mathbf{x},\mathbf{k}_{1},..\mathbf{k}_{n}\right) =\left( 2\pi \right) ^{\frac{-3}{2}}e^{i\left( \mathbf{P}-\mathbf{k}_{1}-...-\mathbf{k}_{n}\right) \cdot \mathbf{x}}\psi _{\mathbf{P}}^{n}\left( \mathbf{k}_{1},..\mathbf{k}_{n}\right) \]
}\\
{\large we can relate a natural scalar product: }\\
{\large 
\[
\left( \phi _{\mathbf{P}}^{n},\psi _{\mathbf{P}}^{m}\right) =\delta _{n,m}\int \overline{\phi _{\mathbf{P}}^{n}\left( \mathbf{k}_{1},..\mathbf{k}_{n}\right) }\psi _{\mathbf{P}}^{m}\left( \mathbf{k}_{1},..\mathbf{k}_{n}\right) d^{3}k_{1}...d^{3}k_{n}\]
}\marginpar{
{\large (0.1)}{\large \par}
}{\large }\\
{\large }\\
{\large The vectorial space \( \left\{ \overline{\bigvee _{n\in \aleph }\psi _{\mathbf{P}}^{n}}\, \right\}  \)
is obtained as closure of the finite linear combinations of the \( \psi _{\mathbf{P}}^{n} \),
in the norm which derives from the scalar product (0.1). Starting from this
space we uniquely define the linear application}\\
{\large 
\[
I_{\mathbf{P}}:\overline{\bigvee _{n\in \aleph }\psi _{\mathbf{P}}^{n}}\rightarrow F^{b}\]
}  {\large }\\
{\large by the prescription: }\\
{\large }\\
{\large \( I_{\mathbf{P}}\left( \psi _{\mathbf{P}}^{n}\left( \mathbf{x},\mathbf{k}_{1},..\mathbf{k}_{n}\right) \right) =\frac{1}{\sqrt{n!}}\int b^{\dagger }\left( \mathbf{k}_{1}\right) ....b^{\dagger }\left( \mathbf{k}_{n}\right) \psi _{\mathbf{P}}^{n}\left( \mathbf{k}_{1},..\mathbf{k}_{n}\right) d^{3}k_{1}...d^{3}k_{n}\psi _{0} \)}
{\large }\\
{\large }\\
{\large where \( b\left( \mathbf{k}\right) ,b^{\dagger }\left( \mathbf{k}\right)  \)
formally correspond to \( a\left( \mathbf{k}\right) e^{i\mathbf{k}\cdot \mathbf{x}},a^{\dagger }\left( \mathbf{k}\right) e^{-i\mathbf{k}\cdot \mathbf{x}} \).
They are destruction and creation operator valued distributions in the Fock
space \( F^{b}\cong F \) . The norm given by (0.1) for \( \psi _{\mathbf{P}}^{n} \)
is equal to \( \left\Vert I_{\mathbf{P}}\left( \psi _{\mathbf{P}}^{n}\right) \right\Vert _{F} \)
( \( \left\Vert \right\Vert _{F} \) is the Fock norm).}\\
{\large }\\
\textbf{\large Main results.}{\large }\\
{\large }\\
{\large The results will be presented in two parts. }\\
{\large }\\
{\large The first one regards the determination of the ground states \( \psi _{\mathbf{P}}^{\sigma _{j}} \)
of the hamiltonians}\\
{\large }\\
{\large \( H_{\mathbf{P},\sigma _{j}}=\frac{\left( \mathbf{P}^{mes}\right) ^{2}}{2m}-\frac{\mathbf{P}\cdot \mathbf{P}^{mes}}{m}+\frac{\mathbf{P}^{2}}{2m}+g\int _{\sigma _{j}}^{\kappa }\left( b\left( \mathbf{k}\right) +b^{\dagger }\left( \mathbf{k}\right) \right) \frac{d^{3}\mathbf{k}}{\sqrt{2\left| \mathbf{k}\right| }}+H^{mes} \)}{\large }\\
{\large }\\
{\large at fixed total momentum \( \mathbf{P} \),} {\large restricted to the
set} \emph{\large \( \Sigma \equiv \left\{ \mathbf{P}:\, \left| \mathbf{P}\right| \leq \sqrt{m}\right\}  \),}
{\large where} {\large ~\( \sigma _{j}=\epsilon ^{\frac{j+1}{2}},\, 0<\epsilon <\left( \frac{1}{4}\right) ^{16}\, and\, j\in \aleph  \),
is the infrared cut-off.}\\
{\large }\\
{\large }\\
{\large The main results are:}\\
{\large }\\
{\large 1)} \textbf{\large Theorem} {\large }\textbf{\large 1.5}{\large }\\
{\large }\\
\( H_{\mathbf{P},\epsilon \sqrt{\epsilon }}\mid _{F_{\epsilon \sqrt{\epsilon }}^{+}} \)
has unique ground state \textbf{\( \psi _{\mathbf{P}}^{\epsilon \sqrt{\epsilon }} \)}
with gap bigger than \( \frac{\epsilon \sqrt{\epsilon }}{2} \); \textbf{\( \psi _{\mathbf{P}}^{\epsilon \sqrt{\epsilon }} \)}(unnormalized))
is\\
\\

\[
\psi _{\mathbf{P}}^{\epsilon \sqrt{\epsilon }}=-\frac{1}{2\pi i}\oint \frac{1}{H_{\mathbf{P},\epsilon \sqrt{\epsilon }}-E}dE\: \psi _{\mathbf{P}}^{\epsilon }\]
where the integral is calculated around the point \( E_{\mathbf{P}}^{\epsilon } \)
, \( E\in \mathcal{C} \) and s.t. \( \left| E-E_{\mathbf{P}}^{\epsilon }\right| =\frac{11\epsilon \sqrt{\epsilon }}{20} \)
\\
{\large }\\
{\large In its generalization to generic \( j \), this result and} \textbf{\large Corollary
1.6} {\large allow to construct the sequence of ground states \( \left\{ \psi _{\mathbf{P}}^{\epsilon ^{\frac{j+1}{2}}}\right\}  \)
for sufficiently small \( g \). }\\
{\large }\\
{\large 2) In} \textbf{\large theorem 2.3} {\large and} \textbf{\large 2.3bis.}
{\large the strong convergence of the sequence \( \left\{ \phi _{\mathbf{P}}^{\epsilon ^{\frac{j+1}{2}}}\right\}  \)
is proved for small coupling constant \( g \), with an error} {\large \( \left\Vert \phi _{\mathbf{P}}-\phi ^{\epsilon ^{\frac{j+1}{2}}}_{\mathbf{P}}\right\Vert \leq \epsilon ^{\frac{j+1}{16}} \),
where \( \phi _{\mathbf{P}}=s-lim_{j\rightarrow \infty }\, \phi ^{\epsilon ^{\frac{j+1}{2}}}_{\mathbf{P}} \)}{\large .
The vectors \( \left\{ \phi _{\mathbf{P}}^{\epsilon ^{\frac{j+1}{2}}}\right\}  \)
are the ground states of the transformed hamiltonians}\\
{\large }\\
\( H^{w}_{\mathbf{P},\epsilon ^{\frac{j+1}{2}}}=W_{\epsilon ^{\frac{j+1}{2}}}\left( \nabla E^{\epsilon ^{\frac{j+1}{2}}}\left( \mathbf{P}\right) \right) H_{\mathbf{P},\epsilon ^{\frac{j+1}{2}}}W^{\dagger }_{\epsilon ^{\frac{j+1}{2}}}\left( \nabla E^{\epsilon ^{\frac{j+1}{2}}}\left( \mathbf{P}\right) \right)  \)
{\large }\\
{\large (where} \( W_{\epsilon ^{\frac{j+1}{2}}}\left( \nabla E^{\epsilon ^{\frac{j+1}{2}}}\left( \mathbf{P}\right) \right) =e^{-g\int _{\epsilon ^{\frac{j+1}{2}}}^{\kappa }\frac{b\left( \mathbf{k}\right) -b^{\dagger }\left( \mathbf{k}\right) }{\left| \mathbf{k}\right| \left( 1-\widehat{\mathbf{k}}\cdot \nabla E^{\epsilon ^{\frac{j+1}{2}}}\left( \mathbf{P}\right) \right) }\frac{d^{3}k}{\sqrt{2\left| \mathbf{k}\right| }}} \){\large )}\\
\textbf{\large }\\
\textbf{\large 
\[
\Leftrightarrow \]
}\\
\textbf{\large }\\
{\large In the second part, I will treat the} {\large scattering. I will assume
a technical hypothesis that is not proved in the spectral analysis:}\footnote{
In this respect, I cite the result T.Chen communicated me: the function \( E\left( \mathbf{P}\right)  \)
has continuous \( 2^{nd} \) derivatives.
}{\large }\\
{\large there exists a positive constant \( m_{r} \) such that}\\
{\large }\\
{\large hypothesis} \textbf{\large B1 ~~~~~~~~~~~~\( \frac{\partial E^{\sigma }\left( \mathbf{P}\right) }{\partial \left| \mathbf{P}\right| }\geq \frac{\left| \mathbf{P}\right| }{m_{r}} \)~~~~~e
~~~~\( \frac{\partial ^{2}E^{\sigma }\left( \mathbf{P}\right) }{\partial ^{2}\left| \mathbf{P}\right| }\geq \frac{1}{m_{r}} \)}{\large ~~~~~~~~~~~~~\( \forall \sigma  \)}\\
{\large }\\
{\large }\\
{\large The main result is the construction of the generic} \emph{\large minimal
asymptotic nucleon state} {\large \( \psi ^{out\left( in\right) }_{G} \) (\( G \)
is the wave function, in variables} \textbf{\large \( \mathbf{P} \)}{\large ,
of the one-particle state from which the construction of the minimal asymptotic
nucleon state starts). It is defined by the strong convergence of the approximating
vector \( \psi _{G}\left( t\right)  \). This subject is discussed in the paragraph
4.2. The result is proved in} \textbf{\large theorem 4.1}{\large . }\\
{\large }\\
{\large ~~~~~~~~~~~~~~~~~~~\( \exists \, C,\rho >0 \)~~~~~~~~~~~s.t.~~~~~~~~~~~~~~~~\( \left\Vert \psi _{G}\left( t\right) -\psi ^{out}_{G}\right\Vert \leq \frac{C}{t^{\rho }} \)
}\\
{\large }\\
{\large From this result, the construction of} \emph{\large s}{\large cattering
subspaces}{\large  ~H}{\large \( ^{out\left( in\right) } \) and the definition
of asymptotic dynamical variables easily follow:}\\
{\large }\\
\textbf{\large theorem 5.2}{\large }\\
{\large }\\
{\large The functions of nucleon mean velocity, continuous and of compact support,
have asymptotically strong limits in}{\large  ~H}{\large \( ^{out} \); in particular:
}\\
{\large }\\
{\large \( s-lim_{t\rightarrow +\infty }e^{iHt}f\left( \frac{\mathbf{x}}{t}\right) e^{-iHt}\psi _{G,\varphi }^{out}=\psi _{G\cdot \widehat{f},\varphi }^{out} \)}
{\large ~~~~~~where~~\( \widehat{f}\left( \mathbf{P}\right) \equiv lim_{\sigma \rightarrow 0}f\left( \nabla E^{\sigma }\left( \mathbf{P}\right) \right)  \)~}\\
{\large }\\
\textbf{\large corollary 5.3}{\large }\\
{\large }\\
{\large In the spaces}{\large  ~H}{\large \( ^{out} \), the asymptotic meson
algebra \( \mathcal{A}^{out} \) is defined as the norm closure of the {*}algebra
generated by the set of Weyl operators \( \left\{ W^{out}\left( \mu \right) :\widetilde{\mu }\left( \mathbf{k}\right) \in C^{\infty }_{0}\left( R^{3}\setminus 0\right) \right\}  \):
}\\
{\large 
\[
W^{out}\left( \mu \right) =s-\lim _{t\rightarrow +\infty }e^{iHt}e^{-iH^{mes}t}e^{i\left( a\left( \mu \right) +a^{\dagger }\left( \mu \right) \right) }e^{iH^{mes}t}e^{-iHt}\]
} {\large }\\
{\large }\\
{\large \par}

\part{Spectral Analysis.}

{\large The iterative procedure aims at constructing the sequence \( \left\{ \psi ^{\sigma _{j}}_{\mathbf{P}}\right\}  \)
of ground states of the hamiltonians \( H_{\mathbf{P},\sigma _{j}} \)}\\
{\small }\\
{\large \( H_{\mathbf{P},\sigma _{j}}=\frac{\left( \mathbf{P}^{mes}\right) ^{2}}{2m}-\frac{\mathbf{P}\cdot \mathbf{P}^{mes}}{m}+\frac{\mathbf{P}^{2}}{2m}+g\int _{\sigma _{j}}^{\kappa }\left( b\left( \mathbf{k}\right) +b^{\dagger }\left( \mathbf{k}\right) \right) \frac{d^{3}\mathbf{k}}{\sqrt{2\left| \mathbf{k}\right| }}+H^{mes} \)}\\
{\large }\\
{\large w}{\large here \( \sigma _{j}=\epsilon ^{\frac{j+1}{2}}\quad 0<\epsilon <\left( \frac{1}{4}\right) ^{16}\quad j\in \aleph  \)
.}\\
{\large }\\
{\large The procedure starts from the comparisons between the resolvents of
the hamiltonians \( H_{\mathbf{P},\sigma _{j}} \)and \( H_{\mathbf{P},\sigma _{j+1}} \).
It recursively uses the Kato's theorem on analytic perturbation of isolated
eigenvalues (of self-adjoint operators) to relate the ground states \( \psi ^{\sigma _{j}}_{\mathbf{P}} \)
and \( \psi ^{\sigma _{j+1}}_{\mathbf{P}} \); at each step two pieces of information
are required:}\\
{\large }\\
{\large 1) a lower bound for the} {\large gap} {\large (with respect to the
ground eigenvalue) of the hamiltonian \( H_{\mathbf{P},\sigma _{j}} \) restricted
to the subspace}\\
{\large \( F_{\sigma _{j+1}}^{+}\equiv \left\{ \overline{\bigvee _{n\in \aleph }\psi ^{n}\left( \mathbf{k}_{1},...\mathbf{k}_{n}\right) }:n\in \aleph ,\: \psi ^{n}\left( \mathbf{k}_{1},...\mathbf{k}_{n}\right) \in L^{2}\left( R^{3n}\right) \, simm.\: ,\, \left| \mathbf{k}_{i}\right| \geq \sigma _{j+1}\: 1\leq i\leq n\right\}  \);}\\
{\large }\\
{\large 2) an estimate of the difference} {\large \( \Delta H_{\mathbf{P}}\mid ^{\sigma _{j}}_{\sigma _{j+1}}\equiv H_{\mathbf{P},\sigma _{j+1}}\mid _{F_{\sigma _{j+1}}^{+}}-H_{\mathbf{P},\sigma _{j}}\mid _{F_{\sigma _{j+1}}^{+}} \)
between two subsequent infrared cut-off hamiltonians; this is small with respect}
to {\large \( H_{\mathbf{P},\sigma _{j}}\mid _{F_{\sigma _{j+1}}^{+}} \) in
a generalized sense, which means that it is possible to expand the} {\large spectral
projection of \( H_{\mathbf{P},\sigma _{j+1}}\mid _{F_{\sigma _{j+1}}^{+}} \)on
the ground eigenvalue in a perturbative series in terms of the resolvent of}
{\large \( H_{\mathbf{P},\sigma _{j}}\mid _{F_{\sigma _{j+1}}^{+}} \) and of
the difference} {\large \( \Delta H_{\mathbf{P}}\mid ^{\sigma _{j}}_{\sigma _{j+1}} \).}\\
{\large }\\
{\large }\\
{\large The convergence of the sequence} {\large \( \left\{ \psi ^{\sigma _{j}}_{\mathbf{P}}\right\}  \)}
{\large is a problem of perturbation of an eigenvalue in the continuum, precisely
of the ground eigenvalue of the hamiltonian \( H_{\mathbf{P}}^{0}= \)}\\
{\large \( =\frac{\left( \mathbf{P}^{mes}\right) ^{2}}{2m}-\frac{\mathbf{P}\cdot \mathbf{P}^{mes}}{m}+\frac{\mathbf{P}^{2}}{2m}+H^{mes} \)
. If the exponent of \( \left| \mathbf{k}\right|  \) in the interaction term
of the hamiltonian were larger than} \( -\frac{1}{2} \), {\large the norm estimates
about resolvents would be sufficient not only to construct the sequence} {\large \( \left\{ \psi ^{\sigma _{j}}_{\mathbf{P}}\right\}  \)}
{\large but also to gain the convergence.}\\
{\large The physical case} \( -\frac{1}{2} \) {\large is a limit case which
requires inequivalent representations of the variables \( \left\{ b\left( \mathbf{k}\right) ,b^{\dagger }\left( \mathbf{k}\right) \right\}  \)
at different} \textbf{\large \( \mathbf{P} \)}{\large , and strong estimates
of the series expansion of the difference between two subsequent ground eigenvectors
in order to have convergence.}  {\large }\\
{\large In the case \( \mathbf{P}=0 \) the right representation is explicitly
known by symmetry. The only problem is to improve some estimates of some terms
in the iterative procedure, in order to gain the convergence;} {\large in the
case \( \mathbf{P}\neq 0 \),} {\large the representation problem is solved
by transforming, step by step, the hamiltonian in a canonic form} {\large .
Such canonic form is analogous, as regards the convergence problem, to the case
\( \mathbf{P}=0 \). In the limit we obtain the representation of \( \left\{ b\left( \mathbf{k}\right) ,b^{\dagger }\left( \mathbf{k}\right) \right\}  \)
given by the non-Fock coherent transformation: 
\[
\mathcal{W}\left( \mathbf{P}\right) =e^{-g\int _{0}^{\kappa }\frac{b\left( \mathbf{k}\right) -b^{\dagger }\left( \mathbf{k}\right) }{\left| \mathbf{k}\right| \left( 1-\widehat{\mathbf{k}}\cdot \nabla E\left( \mathbf{P}\right) \right) }\frac{d^{3}k}{\sqrt{2\left| \mathbf{k}\right| }}}\qquad \]
}\\
{\large Since such representations of \( \left\{ b\left( \mathbf{k}\right) ,b^{\dagger }\left( \mathbf{k}\right) \right\}  \)
are inequivalent for different \( \mathbf{P} \), we have to face the problem
of the existence of the vector \( \int \phi _{\mathbf{P}}d^{3}P \) where \( \phi _{\mathbf{P}} \)
is the limit in Fock space of the ground states of the transformed hamiltonians
\( \mathcal{W}_{\sigma }\left( \mathbf{P}\right) H_{\mathbf{P},\sigma }\mathcal{W}_{\sigma }^{\dagger }\left( \mathbf{P}\right)  \).
The sufficient conditions to define \( \int \phi _{\mathbf{P}}d^{3}P \) are
explored in chapter 3.}\\
{\large \par}

\section{Construction of the sequence \protect\( \left\{ \psi ^{\sigma _{j}}_{\mathbf{P}}\right\} \protect \).\\
}

{\large In the present chapter I only construct the sequence. In order to do
it, I introduce some preliminary lemmas (1.1, 1.2, 1.3, 1.4). The are necessary
to perform the first step of the iterative procedure contained in theorem 1.5.
In the results of theorem 1.5 there are the hypotheses to repeat the same procedure
with a smaller infrared cut-off. Finally in corollary 1.6 the sequence of ground
states \( \left\{ \psi _{\mathbf{P}}^{\epsilon ^{\frac{j+1}{2}}},\: j\in \aleph \right\}  \)}
{\large is constructed. }\\
{\large The lemma 1.4 is crucial to the prove theorem 1.5. Starting from the
perturbative series of the resolvents of the hamiltonians, it allows to state
that the norm difference between the ground states of \( H_{\mathbf{P},\epsilon ^{\frac{j+2}{2}}} \)
and of \( H_{\mathbf{P},\epsilon ^{\frac{j+1}{2}}} \)is of order \( 1 \).}
{\large }\\
{\large }\\
{\large }\\
\emph{\large The initial hypotheses are:}\\
 \emph{\large }\\
\emph{\large - at the first step, the infrared cut-off is ~~ \( \epsilon <\kappa  \)
,~~\( 0<\epsilon <\left( \frac{1}{4}\right) ^{16} \);}\\
{\large -} \emph{\large the mass \( m \) satisfies} {\large \( m>25\cdot 4^{20} \);}\\
{\large -}\emph{\large the coupling constant \( g>0 \) and the ultraviolet
cut-off \( \kappa  \) satisfy the relation }\\
{\large ~~\( 2\pi g^{2}\kappa \leq \frac{1}{4} \);}\\
{\large -} \emph{\large the momentum \( \mathbf{P} \) are restricted to the
set \( \Sigma \equiv \left\{ \mathbf{P}:\, \left| \mathbf{P}\right| \leq \sqrt{m}\right\}  \)}
{\large .}\\
{\large }\\
{\large }\\
{\large }\\
{\large We synthesize the content of the lemmas:}\\
{\large }\\
{\large - lemma 1.1 is a simple application of Kato's theorem to the hamiltonian
with infrared cut-off \( \epsilon  \) in order to fix a coupling constant \( \overline{g_{\epsilon }} \)
such that a unique ground state} \textbf{\large \( \psi _{\mathbf{P}}^{\epsilon } \)}
{\large exists of energy \( E_{\mathbf{P}}^{\epsilon } \) where the} \emph{\large gap}
{\large is bigger than \( \frac{\epsilon }{2} \), for} \textbf{\large \( \mathbf{P} \)}
{\large in the set \( \Sigma \equiv \left\{ \mathbf{P}:\, \left| \mathbf{P}\right| \leq \sqrt{m}\right\}  \);}
{\large }\\
{\large }\\
{\large - in the lemma 1.2 I study the operator \( H_{\mathbf{P},\epsilon } \)
restricted to the subspace \( F_{\epsilon \sqrt{\epsilon }}^{+} \). Under the
initial assumptions,} \textbf{\large \( \psi _{\mathbf{P}}^{\epsilon }\otimes \psi _{0} \)}
{\large is the unique ground state of \( H_{\mathbf{P},\epsilon }\mid _{F_{\epsilon \sqrt{\epsilon }}^{+}} \)}
{\large of energy \( E_{\mathbf{P}}^{\epsilon } \) and its gap is bigger than}
\textbf{\large \( \frac{3}{5}\epsilon \sqrt{\epsilon } \)}{\large ;}{\large \(  \)}
{\large }\\
{\large }\\
{\large - lemma 1.3 proves that the ground energy is increasing in the infrared
cut-off: \( E_{\mathbf{P}}^{\epsilon }\geq E_{\mathbf{P}}^{\epsilon \sqrt{\epsilon }} \).}\\
{\large }\\
{\large }\\
{\large }\\
\textbf{\large Lemma 1.1}{\large }\\
{\large }\\
{\large Given \( \epsilon  \) and for \( \mathbf{P}\in \Sigma \equiv \left\{ \mathbf{P}:\, \left| \mathbf{P}\right| \leq \sqrt{m}\right\}  \),
there is a value for the coupling constant \( \overline{g_{\epsilon }}\equiv \overline{g_{\epsilon }}\left( m,\kappa \right)  \)
such that the ground state \( \psi _{\mathbf{P}}^{\epsilon } \) of \( H_{\mathbf{P},\epsilon }\mid _{F_{\epsilon }^{+}} \)
exists unique and the corresponding eigenvalue \( E_{\mathbf{P}}^{\epsilon } \)
is isolated. Its gap is bigger than \( \frac{\epsilon }{2} \).}\\
{\large }\\
{\large Proof.}\\
{\large }\\
{\large If \( \mathbf{P}\in \Sigma \equiv \left\{ \mathbf{P}:\, \left| \mathbf{P}\right| \leq \sqrt{m}\right\}  \)
and \( \varphi \in D^{b}\bigcap F^{+}_{\epsilon } \) (} {\large \( D^{b}=\bigvee _{n\in \aleph }\psi ^{n} \)
the dense set in \( F^{+} \) which is generated by the finite linear combinations
of vectors of a finite number of mesons (\( b,b^{\dagger } \)) and of (symmetric)
wave function \( \psi ^{n}\left( \mathbf{k}_{1},..,\mathbf{k}_{n}\right) \in S\left( R^{3n}\right) ,n\in \aleph  \)}{\large )
we have:}\\
{\large }\\
{\large \( \left( \varphi ,\left( H_{\mathbf{P}}^{0}-\frac{\mathbf{P}^{2}}{2m}\right) \varphi \right) =\left( \varphi ,\left( \frac{\left( \mathbf{P}^{mes}\right) ^{2}}{2m}-\frac{\mathbf{P}\cdot \mathbf{P}^{mes}}{m}+H^{mes}\right) \varphi \right) \geq  \)}\\
{\large \( \geq \left( \varphi ,\left( -\frac{\mathbf{P}\cdot \mathbf{P}^{mes}}{m}+H^{mes}\right) \varphi \right) =\left( \varphi ,\, \int _{\epsilon }^{+\infty }\left( \left| \mathbf{k}\right| -\mathbf{k}\cdot \frac{\mathbf{P}}{m}\right) b^{\dagger }\left( \mathbf{k}\right) b\left( \mathbf{k}\right) d^{3}k\varphi \right)  \);}\\
{\large }\\
{\large therefore, if \( \mathbf{P}\in \Sigma  \), the vacuum vector \( \psi _{0} \)
is the ground eigenvector of \( H_{\mathbf{P}}^{0} \), of energy \( \frac{\mathbf{P}^{2}}{2m} \).
}\\
{\large }\\
{\large Let us consider the perturbation}\\
{\large }\\
{\large 
\[
H_{\mathbf{P},\epsilon }^{I}\equiv g\int _{\epsilon }^{\kappa }\frac{1}{\sqrt{2}\left| \mathbf{k}\right| ^{\frac{1}{2}}}\left( b\left( \mathbf{k}\right) +b^{\dagger }\left( \mathbf{k}\right) \right) d^{3}k\]
}\\
{\large and an integration circle \( \gamma  \) in the complex plane of radius
\( \frac{3}{4}\epsilon  \) and centered in \( E^{0}_{\mathbf{P}}=\frac{\mathbf{P}^{2}}{2m} \).
\( E^{0}_{\mathbf{P}}=\frac{\mathbf{P}^{2}}{2m} \) is the eigenvalue of the
ground state \( \psi _{0} \) of \( H_{\mathbf{P}}^{0}\mid _{F_{\epsilon }^{+}} \)when
\( \mathbf{P}\in \Sigma  \). By Kato's theorem , for \( g \) sufficiently
small and uniform in \( \mathbf{P}\in \Sigma  \), \( g\leq \overline{g_{\epsilon }} \),
there is a unique ground state \( \psi _{\mathbf{P}}^{\epsilon } \) of \( H_{\mathbf{P},\epsilon }\mid _{F_{\epsilon }^{+}} \),
of energy \( E_{\mathbf{P}}^{\epsilon }<\frac{1}{4}\epsilon +\frac{\mathbf{P}^{2}}{2m} \);
moreover no other point of the spectrum of \( H_{\mathbf{P},\epsilon }\mid _{F_{\epsilon }^{+}\ominus \psi _{\mathbf{P}}^{\epsilon }} \)
is inside \( \gamma  \) . }\\
{\large Therefore the \( \inf spec\left( H_{\mathbf{P},\epsilon }\mid _{F^{+}_{\epsilon }}\right)  \)is
\( E_{\mathbf{P}}^{\epsilon } \) and the related} {\large gap is} {\large bigger
than \( \frac{1}{2}\epsilon  \).}\\
{\large }\\
\textbf{\emph{\large Remark}}\\
{\large }\\
{\large The ultraviolet} \emph{\large cut-off} {\large \( \kappa  \) and the
mass \( m \), with the initial constraints, are fixed.} {\large The value of
\( g \) will be constrained several times during the procedure; at each time
I will call \( \overline{g} \) the maximum value such that the constraint under
examination is satisfied and the previous constraints too.}\textbf{\large }\\
\textbf{\large }\\
\textbf{\large }\\
\textbf{\large Lemma 1.2}\\
\textbf{\large }\\
{\large If} \textbf{\large \( \psi _{\mathbf{P}}^{\epsilon } \)} {\large is
the ground state of \( H_{\mathbf{P},\epsilon }\mid _{F_{\epsilon }^{+}} \)
with gap bigger than \( \frac{1}{2}\epsilon  \), then} \textbf{\large \( \psi _{\mathbf{P}}^{\epsilon } \)}
{\large is the ground state of \( H_{\mathbf{P},\epsilon }\mid _{F_{\epsilon \sqrt{\epsilon }}^{+}} \)}
{\large (with the same eigenvalue) and its gap is bigger than} \textbf{\large \( \frac{3}{5}\epsilon \sqrt{\epsilon } \).}\\
{\large }\\
{\large Proof. }\\
{\large }\\
{\large The proof is in two steps: }\\
{\large }\\
\textbf{\large a)} {\large at first, I analyze the hamiltonian \( H_{\mathbf{P},\epsilon }\mid _{F_{\epsilon }^{+}} \)
plus terms of the difference }\\
{\large \( H_{\mathbf{P},\epsilon }\mid _{F_{\epsilon \sqrt{\epsilon }}^{+}}-H_{\mathbf{P},\epsilon }\mid _{F_{\epsilon }^{+}} \)
in which the meson field modes of frequency between \( \epsilon \sqrt{\epsilon } \)
and \( \epsilon  \) do not interact with frequencies bigger than \( \epsilon  \);}
\textbf{\large }\\
\textbf{\large }\\
\textbf{\large b)} {\large then I will consider the interaction with frequencies
bigger than \( \epsilon  \) too.}\\
{\large }\\
{\large 
\[
\longleftrightarrow \]
}\\
{\large }\\
\textbf{\large a)} {\large }\\
{\large }\\
{\large We decompose \( F_{\epsilon \sqrt{\epsilon }}^{+} \) as \( F_{\epsilon }^{+}\otimes F^{\epsilon }_{\epsilon \sqrt{\epsilon }} \),
where \( F^{\epsilon }_{\epsilon \sqrt{\epsilon }} \) is the tensorial sub-product
defined as follows}\\
{\large 
\[
F^{\epsilon }_{\epsilon \sqrt{\epsilon }}\equiv \left\{ \overline{\bigvee \psi ^{n}\left( \mathbf{k}_{1},...\mathbf{k}_{n}\right) }:n\in \aleph ,\: \psi ^{n}\left( \mathbf{k}_{1},...\mathbf{k}_{n}\right) \in L^{2}\left( R^{3n}\right) \, simm.\: ,\, \epsilon \geq \left| \mathbf{k}_{i}\right| \geq \epsilon \sqrt{\epsilon }\: ,\! 1\leq i\leq n\right\} \]
}\\
{\large I introduce the intermediate hamiltonian \( \widehat{H}_{\mathbf{P},\epsilon }\mid _{F_{\epsilon \sqrt{\epsilon }}^{+}} \)
:}\\
{\large }\\
{\large \( \widehat{H}_{\mathbf{P},\epsilon }\mid _{F_{\epsilon \sqrt{\epsilon }}^{+}}\equiv \frac{1}{2m}\left( \mathbf{P}^{mes}\mid ^{+\infty }_{\epsilon }\right) ^{2}\otimes 1-\frac{\mathbf{P}\cdot \mathbf{P}^{mes}\mid ^{+\infty }_{\epsilon }}{m}\otimes 1+\frac{\mathbf{P}^{2}}{2m}\otimes 1+\int _{\epsilon }^{+\infty }\left| \mathbf{k}\right| b^{\dagger }\left( \mathbf{k}\right) b\left( \mathbf{k}\right) d^{3}k\otimes 1+ \)}\\
{\large }\\
{\large \( +g\int _{\epsilon }^{\kappa }\left( b\left( \mathbf{k}\right) +b^{\dagger }\left( \mathbf{k}\right) \right) \frac{d^{3}\mathbf{k}}{\sqrt{2}\left| \mathbf{k}\right| ^{\frac{1}{2}}}\otimes 1+1\otimes \left\{ \frac{1}{2m}\left( \mathbf{P}^{mes}\mid ^{\epsilon }_{\epsilon \sqrt{\epsilon }}\right) ^{2}+\frac{3}{4}\int _{\epsilon \sqrt{\epsilon }}^{\epsilon }\left| \mathbf{k}\right| b^{\dagger }\left( \mathbf{k}\right) b\left( \mathbf{k}\right) d^{3}k\right\}  \)}\\
{\large }\\
{\large that I denote as \( H_{1}\otimes 1+1\otimes H_{2} \).}\\
{\large }\\
{\large I observe that \( \forall \varphi _{1}\in F_{\epsilon }^{+} \)and \( \forall \varphi _{2}\in F^{\epsilon }_{\epsilon \sqrt{\epsilon }} \),
\( \varphi _{1} \)and \( \varphi _{2} \) normalized vectors,  we have}\\
{\large }\\
{\large 
\[
\left( \varphi _{1}\otimes \varphi _{2},\widehat{H}_{\mathbf{P},\epsilon }\mid _{F_{\epsilon \sqrt{\epsilon }}^{+}}\varphi _{1}\otimes \varphi _{2}\right) =\left( \varphi _{1},H_{1}\varphi _{1}\right) +\left( \varphi _{2},H_{2}\varphi _{2}\right) \geq \left( \varphi _{1},H_{1}\varphi _{1}\right) \geq E_{\mathbf{P}}^{\epsilon }\]
} \marginpar{
{\large (1)}{\large \par}
}{\large }\\
{\large since }{\large \par}

\begin{itemize}
\item {\large \( H_{2}=\frac{1}{2m}\left( \mathbf{P}^{mes}\right) ^{2}\mid ^{\epsilon }_{\epsilon \sqrt{\epsilon }}+\frac{3}{4}\int _{\epsilon \sqrt{\epsilon }}^{\epsilon }\left| \mathbf{k}\right| b^{\dagger }\left( \mathbf{k}\right) b\left( \mathbf{k}\right) d^{3}k \)
is a positive operator}{\large \par}
\item {\large \( \left( \varphi _{1},H_{1}\varphi _{1}\right) =\left( \varphi _{1},H_{\mathbf{P},\epsilon }\mid _{F_{\epsilon }^{+}}\varphi _{1}\right)  \)
\( \quad  \)if \( \varphi _{1}\in F_{\epsilon }^{+} \)}\\
{\large \par}
\end{itemize}
{\large Starting from the joined spectral decomposition of \( H_{1}\otimes 1 \)
and \( 1\otimes H_{2} \) in \( F_{\epsilon }^{+}\otimes F^{\epsilon }_{\epsilon \sqrt{\epsilon }} \)}
{\large we conclude that: }\\
{\large }\\
\textbf{\emph{\large 1)}} {\large \( E_{\mathbf{P}}^{\epsilon } \) is the ground
energy of \( \widehat{H}_{\mathbf{P},\epsilon }\mid _{F_{\epsilon \sqrt{\epsilon }}^{+}} \),
otherwise the condition (1) is not valid;}\\
{\large }\\
\textbf{\emph{\large 2)}} {\large the gap} {\large of \( E_{\mathbf{P}}^{\epsilon } \)
(corresponding to the eigenvector \( \psi _{\mathbf{P}}^{\epsilon }\otimes \psi _{0} \),
\( \psi _{0} \) is the Fock vacuum) is bigger than \( \min \left\{ \frac{3}{4}\epsilon \sqrt{\epsilon },\frac{\epsilon }{2}\right\}  \);
by the hypothesis \( \epsilon <\left( \frac{1}{4}\right) ^{16} \), then the
gap is bigger than \( \frac{3}{4}\epsilon \sqrt{\epsilon } \) .}\\
{\large }\\
{\large }\\
\textbf{\large b)} {\large }\\
{\large }\\
{\large \( H_{\mathbf{P},\epsilon }\mid _{F_{\epsilon \sqrt{\epsilon }}^{+}}=\widehat{H}_{\mathbf{P},\epsilon }\mid _{F_{\epsilon \sqrt{\epsilon }}^{+}}+\frac{\mathbf{P}^{mes}\mid ^{\epsilon }_{\epsilon \sqrt{\epsilon }}}{m}\left( \mathbf{P}^{mes}\mid ^{+\infty }_{\epsilon }-\mathbf{P}\right) +\frac{1}{4}\int _{\epsilon \sqrt{\epsilon }}^{\epsilon }\left| \mathbf{k}\right| b^{\dagger }\left( \mathbf{k}\right) b\left( \mathbf{k}\right) d^{3}k \)}\\
{\large }\\
{\large I define: }\\
{\large }\\
{\large \( \Delta \widehat{H}_{\mathbf{P},\epsilon }\mid _{F_{\epsilon \sqrt{\epsilon }}^{+}}\equiv \frac{\mathbf{P}^{mes}\mid ^{\epsilon }_{\epsilon \sqrt{\epsilon }}}{m}\left( \mathbf{P}^{mes}\mid ^{+\infty }_{\epsilon }-\mathbf{P}\right) +\frac{1}{4}\int _{\epsilon \sqrt{\epsilon }}^{\epsilon }\left| \mathbf{k}\right| b^{\dagger }\left( \mathbf{k}\right) b\left( \mathbf{k}\right) d^{3}k= \)}\\
{\large }\\
{\large \( =\frac{\mathbf{P}^{mes}\mid ^{\epsilon }_{\epsilon \sqrt{\epsilon }}}{m}\left( \mathbf{P}^{mes}\mid ^{+\infty }_{\epsilon }-\mathbf{P}\right) +\frac{1}{5}\int _{\epsilon \sqrt{\epsilon }}^{\epsilon }\left| \mathbf{k}\right| b^{\dagger }\left( \mathbf{k}\right) b\left( \mathbf{k}\right) d^{3}k+\frac{1}{20}\int _{\epsilon \sqrt{\epsilon }}^{\epsilon }\left| \mathbf{k}\right| b^{\dagger }\left( \mathbf{k}\right) b\left( \mathbf{k}\right) d^{3}k \)}
{\large }\\
{\large }\\
{\large Let us observe that from the decomposition of \( F_{\epsilon \sqrt{\epsilon }}^{+}=F_{\epsilon }^{+}\otimes F^{\epsilon }_{\epsilon \sqrt{\epsilon }} \)
in the two orthogonal subspaces \( F_{1}\oplus F_{2} \), where \( F_{1}\equiv \left\{ \varphi \otimes \psi _{0}:\: \varphi \in F^{+}_{\epsilon }\right\}  \)
and \( F_{2}\equiv \left\{ \varphi \otimes \psi :\: \varphi \in F^{+}_{\epsilon },\: \psi \in F^{\epsilon }_{\epsilon \sqrt{\epsilon }}\: \psi \bot \psi _{0}\right\}  \),
we have:}\\
{\large }\\
{\large \( H_{\mathbf{P},\epsilon }\mid _{F_{\epsilon \sqrt{\epsilon }}^{+}}:\: F_{1}\rightarrow F_{1} \)
}\\
{\large \( H_{\mathbf{P},\epsilon }\mid _{F_{\epsilon \sqrt{\epsilon }}^{+}}:\: F_{2}\rightarrow F_{2} \).}\\
{\large }\\
{\large \par}

\emph{\large Spectrum of \( H_{\mathbf{P},\epsilon }\mid _{F_{1}} \)}{\large .}\\
{\large }\\
{\large Since \( H_{\mathbf{P},\epsilon }\mid _{F_{1}}\equiv \widehat{H}_{\mathbf{P},\epsilon }\mid _{F_{1}} \)the
final results} \textbf{\large 1)} {\large and} \textbf{\large 2)} {\large of}
\textbf{\large a)} {\large are valid}\textbf{\large .} {\large }\\
{\large }\\
{\large }\\
 {\large }{\large \par}

\emph{\large Spectrum of \( H_{\mathbf{P},\epsilon }\mid _{F_{2}} \).}{\large }\\
{\large }\\
{\large In order to verify that \( \inf spec\left( H_{\mathbf{P},\epsilon }\mid _{F_{2}}\right) \geq E_{\mathbf{P}}^{\epsilon }+\frac{3}{5}\epsilon \sqrt{\epsilon } \)~,
I will prove the inequality}\\
 {\large 
\[
\inf spec\left( H_{\mathbf{P}\epsilon }\mid _{F_{2}}-\frac{1}{20}\int _{\epsilon \sqrt{\epsilon }}^{\epsilon }\left| \mathbf{k}\right| b^{\dagger }\left( \mathbf{k}\right) b\left( \mathbf{k}\right) d^{3}k\mid _{F_{2}}\right) \geq E_{\mathbf{P}}^{\epsilon }+\frac{3}{5}\epsilon \sqrt{\epsilon }\]
which implies the previous one and that will be useful in the lemma 1.4. }\\
{\large }\\
{\large Given the joined spectral decomposition of the operators: }\\
{\large }\\
{\large - \( H_{2}=\frac{1}{2m}\left( \mathbf{P}^{mes}\right) ^{2}\mid ^{\epsilon }_{\epsilon \sqrt{\epsilon }}+\frac{3}{4}H^{mes}\mid ^{\epsilon }_{\epsilon \sqrt{\epsilon }} \),
}\\
{\large - \( P^{mes^{l}}\mid _{\epsilon \sqrt{\epsilon }}^{\epsilon }\; l=1,2,3 \)}
{\large }\\
{\large - \( \frac{1}{5}H^{mes}\mid ^{\epsilon }_{\epsilon \sqrt{\epsilon }} \)
}\\
{\large }\\
{\large it is easy to verify that if two vectors \( \varphi _{j} \) and \( \varphi _{j'} \)
in \( F^{\epsilon }_{\epsilon \sqrt{\epsilon }} \) have disjoint spectral supports,
we have}\\
{\large }\\
{\large 
\[
\left( \varphi \otimes \varphi _{j'},H_{\mathbf{P},\epsilon }\mid _{F_{2}}\varphi \otimes \varphi _{j}\right) =\left( \varphi \otimes \varphi _{j'},\: \left( \widehat{H}_{\mathbf{P},\epsilon }\mid _{F_{2}}+\Delta \widehat{H}_{\mathbf{P},\epsilon }\mid _{F_{2}}\right) \varphi \otimes \varphi _{j}\right) =0\]
}\\
{\large }\\
{\large Therefore, it is possible to restrict the analysis to the mean value
of }\\
{\large \( H_{\mathbf{P},\epsilon }\mid _{F_{2}}-\frac{1}{20}\int _{\epsilon \sqrt{\epsilon }}^{\epsilon }\left| \mathbf{k}\right| b^{\dagger }\left( \mathbf{k}\right) b\left( \mathbf{k}\right) d^{3}k\mid _{F_{2}} \)
applied to normalized vectors like \( \varphi \otimes \varphi _{j} \), where
\( \varphi \in F^{+}_{\epsilon } \) is in the domain of \( \widehat{H}_{\mathbf{P},\epsilon } \),
\( \varphi _{j}\in F^{\epsilon }_{\epsilon \sqrt{\epsilon }} \) is in the domain
of \( H^{mes}\mid ^{\epsilon }_{\epsilon \sqrt{\epsilon }} \):}\\
{\large }\\
{\large \( \left( \varphi \otimes \varphi _{j},\: H_{\mathbf{P},\epsilon }\mid _{F_{2}}-\frac{1}{20}\int _{\epsilon \sqrt{\epsilon }}^{\epsilon }\left| \mathbf{k}\right| b^{\dagger }\left( \mathbf{k}\right) b\left( \mathbf{k}\right) d^{3}k\mid _{F_{2}}\varphi \otimes \varphi _{j}\right) = \)
}\\
{\large }\\
{\large \( =\left\{ \left( \varphi \otimes \varphi _{j},\widehat{H}_{\mathbf{P},\epsilon }\mid _{F_{2}}\varphi \otimes \varphi _{j}\right) +2\left( \varphi \otimes \varphi _{j},\frac{\mathbf{P}^{mes}\mid ^{\epsilon }_{\epsilon \sqrt{\epsilon }}}{2m}\left( \mathbf{P}^{mes}\mid ^{+\infty }_{\epsilon }-\mathbf{P}\right) \varphi \otimes \varphi _{j}\right) +\right.  \)}\\
{\large }\\
{\large \( \left. +\left( \varphi \otimes \varphi _{j},\frac{1}{5}H^{mes}\mid ^{\epsilon }_{\epsilon \sqrt{\epsilon }}\varphi \otimes \varphi _{j}\right) \right\}  \)}\\
{\large }\\
{\large }\\
{\large I study the quantity written above as a function of two independent
variables:}\\
{\large \par}

\begin{itemize}
\item {\large \( c_{j}\equiv \left( \varphi _{j},\int _{\epsilon \sqrt{\epsilon }}^{\epsilon }\left| \mathbf{k}\right| b^{\dagger }\left( \mathbf{k}\right) b\left( \mathbf{k}\right) d^{3}k\varphi _{j}\right)  \)} 
\item {\large \( E_{_{\mathbf{P},}\varphi \otimes \varphi _{j}}\equiv \left( \varphi \otimes \varphi _{j},\widehat{H}_{\mathbf{P},\epsilon }\mid _{F_{\epsilon \sqrt{\epsilon }}^{+}}\varphi \otimes \varphi _{j}\right)  \)} 
\end{itemize}
{\large with the constraints} {\large \( c_{j}>0 \)~~~~~ and~~~~\( E_{\mathbf{P},\varphi \otimes \varphi _{j}}\geq E_{\mathbf{P}}^{\epsilon }+\frac{3}{4}\epsilon \sqrt{\epsilon } \).}
{\large }\\
{\large }\\
{\large Let us note that}\\
{\large }\\
 {\large }\textbf{\large \( \begin{array}{ccc}
 & \widehat{H}_{\mathbf{P},\epsilon }\mid _{F_{\epsilon \sqrt{\epsilon }}^{+}}-\frac{1}{2m}\left( \mathbf{P}^{mes}\mid ^{+\infty }_{\epsilon }-\mathbf{P}\right) ^{2}-\frac{1}{2m}\left( \mathbf{P}^{mes}\mid ^{\epsilon }_{\epsilon \sqrt{\epsilon }}\right) ^{2}+2\pi g^{2}\kappa \geq 0 & \\
\left\{ \right.  &  & \\
 & \left[ \mathbf{P}^{mes}\mid ^{\epsilon }_{\epsilon \sqrt{\epsilon }},\mathbf{P}^{mes}\mid ^{+\infty }_{\epsilon }\right] =0\qquad \qquad \qquad \qquad \quad \quad \qquad \qquad  & 
\end{array} \)} {\large }\\
{\large }\\
{\large }\\
{\large from which it follows that}\\
{\large }\\
{\large \( \left( \varphi \otimes \varphi _{j},\widehat{H}_{\mathbf{P},\epsilon }\mid _{F_{2}}\varphi \otimes \varphi _{j}\right) +\left( \varphi \otimes \varphi _{j},\frac{\mathbf{P}^{mes}\mid ^{\epsilon }_{\epsilon \sqrt{\epsilon }}}{m}\left( \mathbf{P}^{mes}\mid ^{+\infty }_{\epsilon }-\mathbf{P}\right) \varphi \otimes \varphi _{j}\right) +2\pi g^{2}\kappa \geq 0 \)
}\\
{\large }\\
{\large Since }\\
{\large }\\
{\large \( H_{\mathbf{P},\epsilon }\mid _{F_{\epsilon \sqrt{\epsilon }}^{+}}=\widehat{H}_{\mathbf{P},\epsilon }\mid _{F_{\epsilon \sqrt{\epsilon }}^{+}}+\frac{\mathbf{P}^{mes}\mid ^{\epsilon }_{\epsilon \sqrt{\epsilon }}}{m}\left( \mathbf{P}^{mes}\mid ^{+\infty }_{\epsilon }-\mathbf{P}\right) +\frac{1}{4}\int _{\epsilon \sqrt{\epsilon }}^{\epsilon }\left| \mathbf{k}\right| b^{\dagger }\left( \mathbf{k}\right) b\left( \mathbf{k}\right) d^{3}k \)}\\
{\large }\\
{\large for \( c_{j}\geq 5 \)}\\
{\large }\\
{\large \( \left( \varphi \otimes \varphi _{j},\: \left\{ \right. H_{\mathbf{P},\epsilon }\mid _{F_{2}}-\frac{1}{20}\int _{\epsilon \sqrt{\epsilon }}^{\epsilon }\left| \mathbf{k}\right| b^{\dagger }\left( \mathbf{k}\right) b\left( \mathbf{k}\right) d^{3}k\mid _{F_{2}}\left. \right\} \varphi \otimes \varphi _{j}\right) = \)}
\textbf{\large }\\
{\large }\\
{\large \( =\left( \varphi \otimes \varphi _{j},\: \left\{ \right. \widehat{H}_{\mathbf{P},\epsilon }\mid _{F_{\epsilon \sqrt{\epsilon }}^{+}}+\frac{\mathbf{P}^{mes}\mid ^{\epsilon }_{\epsilon \sqrt{\epsilon }}}{m}\left( \mathbf{P}^{mes}\mid ^{+\infty }_{\epsilon }-\mathbf{P}\right) +\frac{1}{5}\int _{\epsilon \sqrt{\epsilon }}^{\epsilon }\left| \mathbf{k}\right| b^{\dagger }\left( \mathbf{k}\right) b\left( \mathbf{k}\right) d^{3}k\left. \right\} \varphi \otimes \varphi _{j}\right) \geq  \)}\\
{\large }\\
{\large \( \geq \frac{1}{5}c_{j}-2\pi g^{2}\kappa \geq 1-2\pi g^{2}\kappa  \)}\\
{\large }\\
{\large }\\
{\large In order to study the case \( 0<c_{j}<5 \), since \( \widehat{H}_{\mathbf{P},\epsilon }+2\pi g^{2}\kappa -\frac{\left( P^{mes^{l}}\mid ^{+\infty }_{\epsilon }-P^{l}\right) ^{2}}{2m} \)
is a positive operator,} {\large I observe that}\\
{\large }\\
{\large }\\
{\large \( \left| \left( \varphi \otimes \varphi _{j},\frac{\mathbf{P}^{mes}\mid ^{\epsilon }_{\epsilon \sqrt{\epsilon }}}{m}\left( \mathbf{P}^{mes}\mid ^{+\infty }_{\epsilon }-\mathbf{P}\right) \varphi \otimes \varphi _{j}\right) \right| =\left| \left( \varphi ,\left( \mathbf{P}^{mes}\mid ^{+\infty }_{\epsilon }-\mathbf{P}\right) \varphi \right) \cdot \left( \varphi _{j},\frac{\mathbf{P}^{mes}\mid ^{\epsilon }_{\epsilon \sqrt{\epsilon }}}{m}\varphi _{j}\right) \right| \leq  \)}\\
{\large }\\
{\large \( \leq \sum ^{3}_{l=1}\frac{1}{m}\left| \left( \varphi ,\left( P^{mes^{l}}\mid ^{+\infty }_{\epsilon }-P^{l}\right) \varphi \right) \right| \left| \left( \varphi _{j},\int _{\epsilon \sqrt{\epsilon }}^{\epsilon }\left| \mathbf{k}\right| b^{\dagger }\left( \mathbf{k}\right) b\left( \mathbf{k}\right) d^{3}k\varphi _{j}\right) \right| \leq  \)}\\
{\large }\\
{\large \( \leq \left( E_{\mathbf{P},\varphi \otimes \varphi _{j}}+2\pi g^{2}\kappa \right) ^{\frac{1}{2}}\cdot 3\sqrt{\frac{2}{m}}\left( \varphi _{j},\int _{\epsilon \sqrt{\epsilon }}^{\epsilon }\left| \mathbf{k}\right| b^{\dagger }\left( \mathbf{k}\right) b\left( \mathbf{k}\right) d^{3}k\varphi _{j}\right) \leq  \)
}\\
{\large }\\
{\large \( \leq \left( E_{\mathbf{P},\varphi \otimes \varphi _{j}}+2\pi g^{2}\kappa \right) ^{\frac{1}{2}}\cdot 3c_{j}\sqrt{\frac{2}{m}} \)}\\
{\large }\\
{\large from which}\textbf{\large }\\
\textbf{\large }\\
\textbf{\large \( \left( \varphi \otimes \varphi _{j},\, \left( H_{\mathbf{P},\epsilon }\mid _{F_{2}}-\frac{1}{20}\int _{\epsilon \sqrt{\epsilon }}^{\epsilon }\left| \mathbf{k}\right| b^{\dagger }\left( \mathbf{k}\right) b\left( \mathbf{k}\right) d^{3}k\mid _{F_{2}}\right) \varphi \otimes \varphi _{j}\right) = \)
}\\
\textbf{\large }\\
{\large \( =\left\{ \left( \varphi \otimes \varphi _{j},\widehat{H}_{\mathbf{P},\epsilon }\mid _{F_{2}}\varphi \otimes \varphi _{j}\right) +\left( \varphi \otimes \varphi _{j},\frac{\mathbf{P}^{mes}\mid ^{\epsilon }_{\epsilon \sqrt{\epsilon }}}{m}\left( \mathbf{P}^{mes}\mid ^{+\infty }_{\epsilon }-\mathbf{P}\right) \varphi \otimes \varphi _{j}\right) +\right.  \)}\\
{\large }\\
{\large \( \left. +\left( \varphi \otimes \varphi _{j},\frac{1}{5}H^{mes}\mid ^{\epsilon }_{\epsilon \sqrt{\epsilon }}\varphi \otimes \varphi _{j}\right) \right\} \geq E_{\mathbf{P},\varphi \otimes \varphi _{j}}-3c_{j}\sqrt{\frac{2}{m}}\left( E_{\mathbf{P},\varphi \otimes \varphi _{j}}+2\pi g^{2}\kappa \right) ^{\frac{1}{2}}+\frac{1}{5}c_{j} \)}\\
\textbf{\large }\\
\textbf{\large }\\
\textbf{\large }\\
{\large I define the function:}\textbf{\large }\\
{\large }\\
{\large }\\
{\large \( f\left( E_{\mathbf{P},\varphi \otimes \varphi _{j}},c_{j}\right) \equiv E_{\mathbf{P},\varphi \otimes \varphi _{j}}-3c_{j}\sqrt{\frac{2}{m}}\left( E_{\mathbf{P},\varphi \otimes \varphi _{j}}+2\pi g^{2}\kappa \right) ^{\frac{1}{2}}+\frac{1}{5}c_{j}-\left\{ \frac{4}{5}\left( E_{\mathbf{P},\varphi \otimes \varphi _{j}}-E_{\mathbf{P}}^{\epsilon }\right) +E_{\mathbf{P}}^{\epsilon }\right\} = \)
}\\
{\large \( =\frac{1}{5}E_{\mathbf{P},\varphi \otimes \varphi _{j}}-3c_{j}\sqrt{\frac{2}{m}}\left( E_{\mathbf{P},\varphi \otimes \varphi _{j}}+2\pi g^{2}\kappa \right) ^{\frac{1}{2}}+\frac{1}{5}c_{j}-\frac{1}{5}E_{\mathbf{P}}^{\epsilon } \)}\\
{\large }\\
{\large }\\
\textbf{\emph{\large Analysis of}} \textbf{\large \( f\left( E_{\mathbf{P},\varphi \otimes \varphi _{j}},c_{j}\right)  \)}{\large .}\\
{\large }\\
{\large First of all we observe that the positivity of} \textbf{\large \( f\left( E_{\mathbf{P},\varphi \otimes \varphi _{j}},c_{j}\right)  \)}
{\large implies:}\\
{\large }\\
{\large \( \left( \varphi \otimes \varphi _{j},\, \left( H_{\mathbf{P},\epsilon }\mid _{F_{2}}-\frac{1}{20}\int _{\epsilon \sqrt{\epsilon }}^{\epsilon }\left| \mathbf{k}\right| b^{\dagger }\left( \mathbf{k}\right) b\left( \mathbf{k}\right) d^{3}k\mid _{F_{2}}\right) \varphi \otimes \varphi _{j}\right) \geq  \)}\\
{\large }\\
{\large \( \geq \frac{4}{5}\left( E_{\mathbf{P},\varphi \otimes \varphi _{j}}-E_{\mathbf{P}}^{\epsilon }\right) +E_{\mathbf{P}}^{\epsilon }\geq \frac{3}{5}\epsilon \sqrt{\epsilon }+E_{\mathbf{P}}^{\epsilon } \)
}\\
{\large }\\
\emph{\large }\\
\emph{\large Discussion}{\large }\\
{\large }\\
{\large For the initial assumptions \( m>\left( \frac{75}{\sqrt{2}}\right) ^{2} \)
e \( 2\pi g^{2}\kappa <\frac{1}{4} \); the values of the variables of \( f \)
to be considered are: \( 0<c_{j}<5 \)~~~and ~~~\( E_{\mathbf{P},\varphi \otimes \varphi _{j}}\geq E_{\mathbf{P}}^{\epsilon }+\frac{3}{4}\epsilon \sqrt{\epsilon } \).
}\\
{\large }\\
{\large I will consider separately two cases:} \textbf{\large }\\
\textbf{\large }\\
{\large i) if} \textbf{\large \( E_{\mathbf{P},\varphi \otimes \varphi _{j}}\leq 1-2\pi g^{2}\kappa  \)}
{\large , }\\
{\large }\\
{\large then} \textbf{\large \( \frac{1}{5}c_{j}-3c_{j}\sqrt{\frac{2}{m}}\left( E_{\mathbf{P},\varphi \otimes \varphi _{j}}+2\pi g^{2}\kappa \right) ^{\frac{1}{2}}\geq 0 \)}
{\large  }\\
{\large }\\
{\large from which \( f\left( E_{\mathbf{P},\varphi \otimes \varphi _{j}},c_{j}\right) \geq \frac{1}{5}\left( E_{\mathbf{P},\varphi \otimes \varphi _{j}}-E_{\mathbf{P}}^{\epsilon }\right) >0 \);}\textbf{\large }\\
\textbf{\large }\\
\textbf{\large }\\
{\large ii) if} \textbf{\large \( E_{\mathbf{P},\varphi \otimes \varphi _{j}}>1-2\pi g^{2}\kappa  \)}
{\large ,}\\
{\large let us observe that \( \frac{\partial }{\partial E_{\mathbf{P},\varphi \otimes \varphi _{j}}}f\left( E_{\mathbf{P},\varphi \otimes \varphi _{j}},c_{j}\right) \equiv \frac{1}{5}-3c_{j}\frac{1}{\sqrt{2m}}\left( E_{\mathbf{P},\varphi \otimes \varphi _{j}}+2\pi g^{2}\kappa \right) ^{-\frac{1}{2}} \)
from which \( \frac{\partial }{\partial E_{\mathbf{P},\varphi \otimes \varphi _{j}}}f\left( E_{\mathbf{P},\varphi \otimes \varphi _{j}},c_{j}\right) \geq 0 \)
for \( E_{\mathbf{P},\varphi \otimes \varphi _{j}}\geq \left( 15c_{j}\frac{1}{\sqrt{2m}}\right) ^{2}-2\pi g^{2}\kappa  \),
that is a minimum value at fixed \( c_{j} \); since \( \left( 15c_{j}\frac{1}{\sqrt{2m}}\right) ^{2}<1 \),
by exploiting the result i), we arrive at \( f\left( E_{\mathbf{P},\varphi \otimes \varphi _{j}},c_{j}\right) \geq f\left( \left( 15c_{j}\frac{1}{\sqrt{2m}}\right) ^{2}-2\pi g^{2}\kappa ,c_{j}\right) >0 \)
.}\\
\textbf{\large }\\
{\large }\\
{\large }\\
{\large If \( E_{\mathbf{P}}^{\epsilon }+\frac{3}{5}\epsilon \sqrt{\epsilon }\leq 1-2\pi g^{2}\kappa  \),
the previous results tell us that }\\
{\large }\\
{\large \( \inf _{\varphi \otimes \varphi _{j}}\left( \varphi \otimes \varphi _{j},\: \left\{ \right. H_{\mathbf{P},\epsilon }\mid _{F_{2}}-\frac{1}{20}\int _{\epsilon \sqrt{\epsilon }}^{\epsilon }\left| \mathbf{k}\right| b^{\dagger }\left( \mathbf{k}\right) b\left( \mathbf{k}\right) d^{3}k\mid _{F_{2}}\left. \right\} \varphi \otimes \varphi _{j}\right) \geq E_{\mathbf{P}}^{\epsilon }+\frac{3}{5}\epsilon \sqrt{\epsilon } \)}
\textbf{\large }\\
{\large }\\
{\large }\\
\emph{\large Conclusion}{\large }\\
{\large }\\
{\large The results in} \textbf{\large a)} {\large and} \textbf{\large b)} {\large imply
that,} \textbf{\large if \( E_{\mathbf{P}}^{\epsilon }+\frac{3}{5}\epsilon \sqrt{\epsilon }\leq 1-2\pi g^{2}\kappa  \)
then the minimum value of the spectrum of \( H_{\mathbf{P},\epsilon }\mid _{F_{\epsilon \sqrt{\epsilon }}^{+}} \)is
\( E_{\mathbf{P}}^{\epsilon } \) and} \textbf{\large the gap is bigger} \textbf{\large or
equal to \( \frac{3}{5}\epsilon \sqrt{\epsilon } \)}{\large . For the lemma
1.1 the value of the coupling constant is such that \( E_{\mathbf{P}}^{\epsilon }<\frac{1}{4}\epsilon +\frac{\mathbf{P}^{2}}{2m} \).
Since \( \epsilon <\left( \frac{1}{4}\right) ^{16} \), \( 2\pi g^{2}\kappa <\frac{1}{4} \)
and} \textbf{\large \( \mathbf{P}\in \Sigma  \)}{\large , the condition \( E_{\mathbf{P}}^{\epsilon }+\frac{3}{5}\epsilon \sqrt{\epsilon }\leq 1-2\pi g^{2}\kappa  \)
is satisfied.}\\
{\large }\\
{\large }\\
{\large }\\
\textbf{\large Lemma 1.3}\\
\textbf{\large }\\
{\large The following relation between} \textbf{\large \( E_{\mathbf{P}}^{\epsilon },\: E_{\mathbf{P}}^{\epsilon \sqrt{\epsilon }} \)}
{\large (ground energy of \( H_{\mathbf{P},\epsilon \sqrt{\epsilon }}\mid _{F_{\epsilon \sqrt{\epsilon }}^{+}} \))
holds: }\\
{\large }\\
\textbf{\large 
\[
E_{\mathbf{P}}^{\epsilon }\geq E_{\mathbf{P}}^{\epsilon \sqrt{\epsilon }}\geq E_{\mathbf{P}}^{\epsilon }-40\pi g^{2}\epsilon \]
}{\large }\\
{\large }\\
{\large Proof.} {\large }\\
{\large }\\
{\large }\\
{\large \( H_{\mathbf{P},\epsilon \sqrt{\epsilon }}\mid _{F_{\epsilon \sqrt{\epsilon }}^{+}}=\frac{1}{2m}\left( \mathbf{P}^{mes}\mid ^{+\infty }_{\epsilon }\right) ^{2}\otimes 1+\int _{\epsilon }^{+\infty }\left| \mathbf{k}\right| b^{\dagger }\left( \mathbf{k}\right) b\left( \mathbf{k}\right) d^{3}k\otimes 1-\frac{\mathbf{P}\cdot \mathbf{P}^{mes}\mid ^{+\infty }_{\epsilon }}{m}\otimes 1+\frac{\mathbf{P}^{2}}{2m}\otimes 1+ \)}\\
\\
{\large \( +g\int _{\epsilon }^{\kappa }\left( b\left( \mathbf{k}\right) +b^{\dagger }\left( \mathbf{k}\right) \right) \frac{d^{3}\mathbf{k}}{\sqrt{2}\left| \mathbf{k}\right| ^{\frac{1}{2}}}\otimes 1+1\otimes \frac{1}{2m}\left( \mathbf{P}^{mes}\mid ^{\epsilon }_{\epsilon \sqrt{\epsilon }}\right) ^{2}+\mathbf{P}^{mes}\mid ^{+\infty }_{\epsilon }\otimes \frac{\mathbf{P}^{mes}\mid ^{\epsilon }_{\epsilon \sqrt{\epsilon }}}{m}+ \)}\\
{\large }\\
{\large \( -1\otimes \frac{\mathbf{P}\cdot \mathbf{P}^{mes}\mid ^{\epsilon }_{\epsilon \sqrt{\epsilon }}}{m}+1\otimes \int _{\epsilon \sqrt{\epsilon }}^{\epsilon }\left| \mathbf{k}\right| b^{\dagger }\left( \mathbf{k}\right) b\left( \mathbf{k}\right) d^{3}k+1\otimes g\int _{\epsilon \sqrt{\epsilon }}^{\epsilon }\left( b\left( \mathbf{k}\right) +b^{\dagger }\left( \mathbf{k}\right) \right) \frac{d^{3}\mathbf{k}}{\sqrt{2}\left| \mathbf{k}\right| ^{\frac{1}{2}}} \)}\\
{\small }\\
{\large The mean value of \( H_{\mathbf{P},\epsilon \sqrt{\epsilon }}\mid _{F_{\epsilon \sqrt{\epsilon }}^{+}} \)on
\( \psi _{\mathbf{P}}^{\epsilon }\otimes \psi _{0} \) (normalized) is \( E_{\mathbf{P}}^{\epsilon } \).
}\\
{\large By definition, \( E_{\mathbf{P}}^{\epsilon \sqrt{\epsilon }} \) is
the inf of the mean value of \( H_{\mathbf{P},\epsilon \sqrt{\epsilon }}\mid _{F_{\epsilon \sqrt{\epsilon }}^{+}} \)
on the normalized vectors in \( F_{\epsilon \sqrt{\epsilon }}^{+} \) belonging
to the operator domain.Therefore \( E_{\mathbf{P}}^{\epsilon \sqrt{\epsilon }}\leq E_{\mathbf{P}}^{\epsilon } \)
and in general for \( \sigma _{1}>\sigma _{2} \) ~~\( E_{\mathbf{P}}^{\sigma _{2}}\leq E_{\mathbf{P}}^{\sigma _{1}} \).}\\
{\large }\\
{\large Moreover, as proved in the previous lemma}\\
{\large 
\[
\inf spec\left( H_{\mathbf{P},\epsilon }\mid _{F_{\epsilon \sqrt{\epsilon }}^{+}}-\frac{1}{20}\int _{\epsilon \sqrt{\epsilon }}^{\epsilon }\left| \mathbf{k}\right| b^{\dagger }\left( \mathbf{k}\right) b\left( \mathbf{k}\right) d^{3}k\right) >E_{\mathbf{P}}^{\epsilon }\]
while completing the square}\\
{\large 
\[
\frac{1}{20}\int _{\epsilon \sqrt{\epsilon }}^{\epsilon }\left| \mathbf{k}\right| b^{\dagger }\left( \mathbf{k}\right) b\left( \mathbf{k}\right) d^{3}k+g\int ^{\epsilon }_{\epsilon \sqrt{\epsilon }}\left( b\left( \mathbf{k}\right) +b^{\dagger }\left( \mathbf{k}\right) \right) \frac{d^{3}k}{\sqrt{2}\left| \mathbf{k}\right| ^{\frac{1}{2}}}+40\pi g^{2}\epsilon \geq 0\]
}\\
{\large It follows that}\\
{\large }\\
{\large \( E_{\mathbf{P}}^{\epsilon \sqrt{\epsilon }}=\inf spec\left( H_{\mathbf{P},\epsilon \sqrt{\epsilon }}\mid _{F_{\epsilon \sqrt{\epsilon }}^{+}}\right) =\inf spec\left( H_{\mathbf{P},\epsilon }\mid _{F_{\epsilon \sqrt{\epsilon }}^{+}}+g\int ^{\epsilon }_{\epsilon \sqrt{\epsilon }}\left( b\left( \mathbf{k}\right) +b^{\dagger }\left( \mathbf{k}\right) \right) \frac{d^{3}k}{\sqrt{2}\left| \mathbf{k}\right| ^{\frac{1}{2}}}\right) \geq  \)}\\
{\large }\\
{\large \( \geq \inf spec\left( H_{\mathbf{P},\epsilon }\mid _{F_{\epsilon \sqrt{\epsilon }}^{+}}-\frac{1}{20}\int _{\epsilon \sqrt{\epsilon }}^{\epsilon }\left| \mathbf{k}\right| b^{\dagger }\left( \mathbf{k}\right) b\left( \mathbf{k}\right) d^{3}k-40\pi g^{2}\epsilon \right) \geq E_{\mathbf{P}}^{\epsilon }-40\pi g^{2}\epsilon  \)}\\
{\large }\\
{\large }\\
{\large }\\
\textbf{\emph{\large Coupling constant constraints.}}{\large }\\
{\large }\\
{\large In the following lemma 1.4 a value \( \overline{g} \) will be fixed.
It allows to estimate the difference between the ground states} {\large at different
and arbitrarily small cut-off} {\large \( \sigma _{j} \). The validity of lemma
1.2 at each step for fixed \( \overline{g} \) is necessary for the consistency
of the iteration. The relation in lemma 1.3 assures the validity of lemma 1.2}
\textbf{\large }{\large since:}\textbf{\large }\\
\textbf{\large 
\[
E_{\mathbf{P}}^{\epsilon ^{\frac{j+1}{2}}}+\frac{3}{5}\epsilon ^{\frac{j+2}{2}}\leq E_{\mathbf{P}}^{\epsilon }+\frac{3}{5}\epsilon \sqrt{\epsilon }<1-2\pi \overline{g}^{2}\kappa \quad \forall j\]
}{\large }\\
{\large }\\
{\large }\\
\textbf{\large Definitions}{\large }\\
{\large }\\
{\large \( H_{\mathbf{P},\epsilon \sqrt{\epsilon }}\mid _{F_{\epsilon \sqrt{\epsilon }}^{+}}=H_{\mathbf{P},\epsilon }\mid _{F_{\epsilon \sqrt{\epsilon }}^{+}}+\left( \Delta H_{\mathbf{P}}\right) ^{\epsilon }_{\epsilon \sqrt{\epsilon }} \)}\\
{\large }\\
{\large \( \left( \Delta H_{\mathbf{P}}\right) ^{\epsilon }_{\epsilon \sqrt{\epsilon }}\equiv g\int _{\epsilon \sqrt{\epsilon }}^{\epsilon }\left( b\left( \mathbf{k}\right) +b^{\dagger }\left( \mathbf{k}\right) \right) \frac{d^{3}\mathbf{k}}{\sqrt{2}\left| \mathbf{k}\right| } \)}\\
{\large }\\
{\large }\\
\textbf{\large Lemma 1.4} {\large }\\
{\large }\\
{\large For a properly small \( \overline{g} \) , \( \left( \Delta H_{\mathbf{P}}\right) ^{\epsilon }_{\epsilon \sqrt{\epsilon }} \)
is small of order \( 1 \) with respect to \( H_{\mathbf{P},\epsilon }\mid _{F_{\epsilon \sqrt{\epsilon }}^{+}} \),
in the following sense:}\\
{\large }\\
{\large given \( E \) such that\( \left| E-\inf H_{\mathbf{P},\epsilon }\mid _{F_{\epsilon \sqrt{\epsilon }}^{+}}\right| =\left| E-E_{\mathbf{P}}^{\epsilon }\right| =\frac{20\epsilon \sqrt{\epsilon }}{11} \)}\\
{\large }\\
{\large \( \frac{1}{H_{\mathbf{P},\epsilon }+\left( \Delta H_{\mathbf{P}}\right) ^{\epsilon }_{\epsilon \sqrt{\epsilon }}-E}\mid _{_{F_{\epsilon \sqrt{\epsilon }}^{+}}}=\frac{1}{H_{\mathbf{P},\epsilon }-E}\mid _{_{F_{\epsilon \sqrt{\epsilon }}^{+}}}+\frac{1}{H_{\mathbf{P},\epsilon }-E}\sum ^{+\infty }_{n=1}\left( -\left( \Delta H_{\mathbf{P}}\right) ^{\epsilon }_{\epsilon \sqrt{\epsilon }}\frac{1}{H_{\mathbf{P},\epsilon }\mid _{F_{\epsilon \sqrt{\epsilon }}^{+}}-E}\right) ^{n}\mid _{F_{\epsilon \sqrt{\epsilon }}^{+}} \)}\\
{\large }\\
{\large and }\\
{\large \( \left\Vert \frac{1}{H_{\mathbf{P},\epsilon }\mid _{F_{\epsilon \sqrt{\epsilon }}^{+}}-E}\left( -\left( \Delta H_{\mathbf{P}}\right) ^{\epsilon }_{\epsilon \sqrt{\epsilon }}\frac{1}{H_{\mathbf{P},\epsilon }\mid _{F_{\epsilon \sqrt{\epsilon }}^{+}}-E}\right) ^{n}\right\Vert _{F\epsilon \sqrt{\epsilon }^{+}}\leq \frac{20\left( C\left( \overline{g},\overline{m}\right) \right) ^{n}}{\epsilon \sqrt{\epsilon }} \)
}\\
{\large }\\
{\large where \( 0<C\left( \overline{g},m\right) <\frac{1}{12} \) .}\\
{\large }\\
 {\large }{\large \par}

\textbf{\large Theorem} {\large }\textbf{\large 1.5}{\large }\\
{\large }\\
{\large \( H_{\mathbf{P},\epsilon \sqrt{\epsilon }}\mid _{F_{\epsilon \sqrt{\epsilon }}^{+}} \)has
a unique ground eigenvector} \textbf{\large \( \psi _{\mathbf{P}}^{\epsilon \sqrt{\epsilon }} \)}
{\large of energy \( E_{\mathbf{P}}^{\epsilon \sqrt{\epsilon }} \) and gap
\( \frac{\epsilon \sqrt{\epsilon }}{2} \); the un-normalized vector} \textbf{\large \( \psi _{\mathbf{P}}^{\epsilon \sqrt{\epsilon }} \)}
{\large corresponds to}\\
{\large }\\
{\large 
\[
\psi _{\mathbf{P}}^{\epsilon \sqrt{\epsilon }}=-\frac{1}{2\pi i}\oint \frac{1}{H_{\mathbf{P},\epsilon \sqrt{\epsilon }}-E}dE\: \psi _{\mathbf{P}}^{\epsilon }\]
}\marginpar{
{\large (2)}{\large \par}
} {\large }\\
{\large }\\
{\large where \( E\in \mathcal{C} \) and \( \left| E-E^{\epsilon }_{\mathbf{P}}\right| =\frac{11}{20}\epsilon \sqrt{\epsilon } \).}\\
{\large }\\
{\large Proof. }\\
{\large }\\
{\large Continuity argument.}\\
{\large }\\
{\large I distinguish the coupling constant \( g \) in \( H_{\mathbf{P},\epsilon }\mid _{F_{\epsilon \sqrt{\epsilon }}^{+}} \)from
that one in \( \left( \Delta H_{\mathbf{P}}\right) ^{\epsilon }_{\epsilon \sqrt{\epsilon }} \),
and I call the last one \( g^{\sharp } \). Kato's theorem ensures that (2)
is verified for sufficiently small \( g^{\sharp } \), since the gap of \( H_{\mathbf{P},\epsilon }\mid _{F_{\epsilon \sqrt{\epsilon }}^{+}} \)is
bigger or equal to \( \frac{3}{5}\epsilon \sqrt{\epsilon } \) and \( \left( \Delta H_{\mathbf{P}}\left( g^{\sharp }\right) \right) ^{\epsilon }_{\epsilon \sqrt{\epsilon }} \)
is a small Kato perturbation with respect to \( H_{\mathbf{P},\epsilon }\mid _{F_{\epsilon \sqrt{\epsilon }}^{+}} \).
}\\
{\large Now look at the figure }\\
{\large \par}

\vspace{0.3cm}
{\centering \resizebox*{!}{0.3\textheight}{\includegraphics{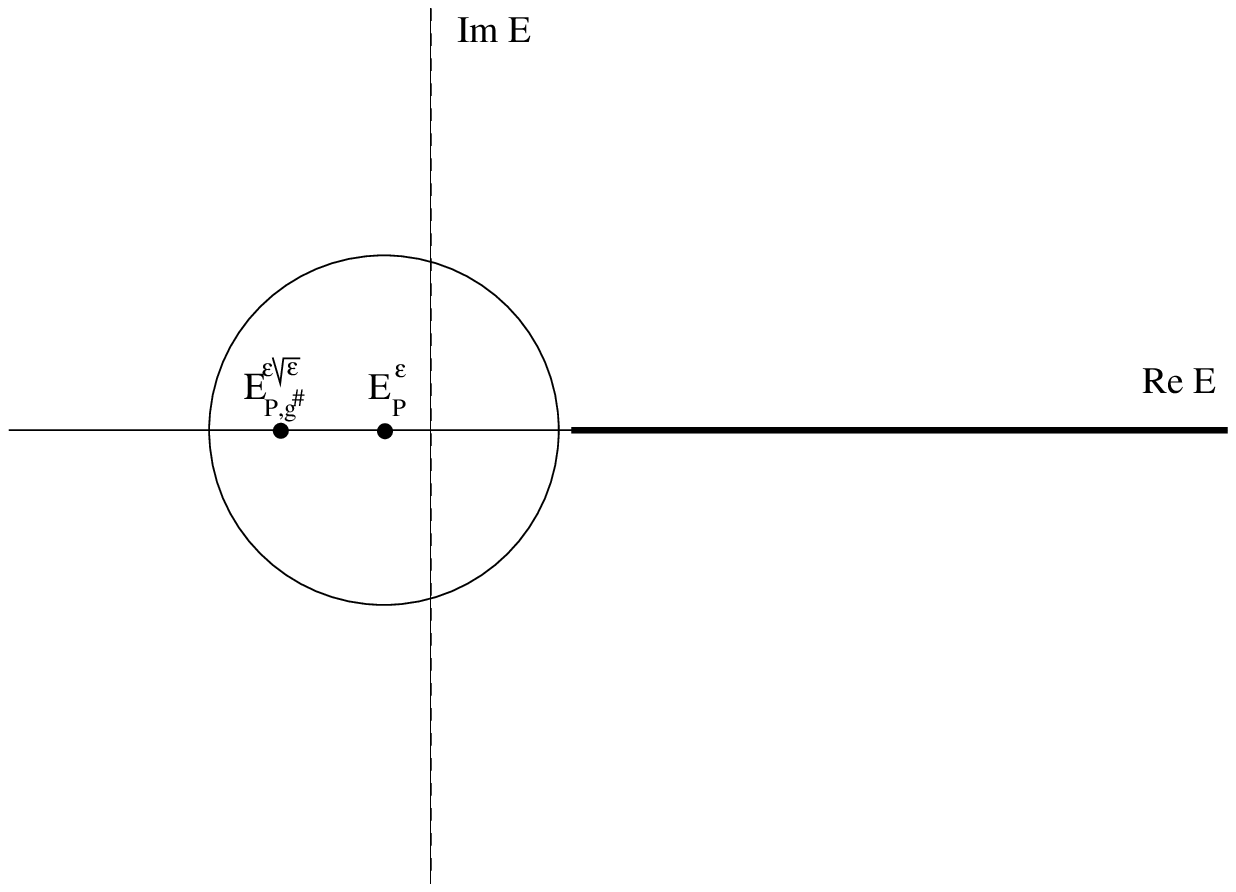}} \par}
\vspace{0.3cm}

{\large if \( g^{\sharp } \) increases, the equation (2) is valid till the
eigenvalue \( E_{\mathbf{P}g^{\sharp }}^{\epsilon \sqrt{\epsilon }} \) remains
inside the circle of integration and the remaining spectrum of \( H_{\mathbf{P},\epsilon \sqrt{\epsilon }}\mid _{F_{\epsilon \sqrt{\epsilon }}^{+}} \)
remains outside of the circle of integration; a limit value \( \overline{g^{\sharp }} \)
exists for which the expression \( -\frac{1}{2\pi i}\oint \left\Vert \frac{1}{H_{\mathbf{P},\epsilon \sqrt{\epsilon }}\mid _{F_{\epsilon \sqrt{\epsilon }}^{+}}-E}\right\Vert dE \)
diverges because the spectrum intersects the circle of integration. }\\
{\large }\\
{\large According to the estimates in lemma 1.4 we can conclude that: }\\
{\large }\\
\textbf{\large -} {\large the integral\( -\frac{1}{2\pi i}\oint \left\Vert \frac{1}{H_{\mathbf{P},\epsilon \sqrt{\epsilon }}\mid _{F_{\epsilon \sqrt{\epsilon }}^{+}}-E}\right\Vert dE \)
exists for \( 0\leq g^{\sharp }\leq g=\overline{g} \) and then \( \overline{g}<\overline{g^{\sharp }} \)
}\\
{\large }\\
\textbf{\large -} {\large the ground state of \( H_{\mathbf{P},\epsilon \sqrt{\epsilon }}\mid _{F_{\epsilon \sqrt{\epsilon }}^{+}} \)
is unique and it is not zero since}\\
{\large }\\
{\large \( \psi _{\mathbf{P}}^{\epsilon \sqrt{\epsilon }}=\psi _{\mathbf{P}}^{\epsilon }-\frac{1}{2\pi i}\sum ^{+\infty }_{n=1}\oint \frac{1}{H_{\mathbf{P},\epsilon }\mid _{F_{\epsilon \sqrt{\epsilon }}^{+}}-E}\left( -\left( \Delta H_{\mathbf{P}}\right) ^{\epsilon }_{\epsilon \sqrt{\epsilon }}\frac{1}{H_{\mathbf{P},\epsilon }\mid _{F_{\epsilon \sqrt{\epsilon }}^{+}}-E}\right) ^{n}dE\: \psi _{\mathbf{P}}^{\epsilon } \)}\marginpar{
{\large (3)}{\large \par}
}{\large }\\
{\large }\\
{\large where the norm of the remainder \( -\frac{1}{2\pi i}\sum ^{+\infty }_{n=1}\oint \frac{1}{H_{\mathbf{P},\epsilon }\mid _{F_{\epsilon \sqrt{\epsilon }}^{+}}-E}\left( -\left( \Delta H_{\mathbf{P}}\right) ^{\epsilon }_{\epsilon \sqrt{\epsilon }}\frac{1}{H_{\mathbf{P},\epsilon }\mid _{F_{\epsilon \sqrt{\epsilon }}^{+}}-E}\right) ^{n}dE\: \psi _{\mathbf{P}}^{\epsilon } \)
is less than \( \frac{11C\left( \overline{g},m\right) }{1-C\left( \overline{g},m\right) }\left\Vert \psi _{\mathbf{P}}^{\epsilon }\right\Vert  \).
Therefore \( \psi _{\mathbf{P}}^{\epsilon \sqrt{\epsilon }} \)is not zero since:
}\\
{\large }\\
{\large 
\[
\left\Vert \psi _{\mathbf{P}}^{\epsilon \sqrt{\epsilon }}\right\Vert \geq \left\Vert \psi _{\mathbf{P}}^{\epsilon }\right\Vert -\frac{11C\left( \overline{g},m\right) }{1-C\left( \overline{g},m\right) }\left\Vert \psi _{\mathbf{P}}^{\epsilon }\right\Vert \geq \frac{1-12C\left( \overline{g},m\right) }{1-C\left( \overline{g},m\right) }\left\Vert \psi _{\mathbf{P}}^{\epsilon }\right\Vert >0\]
}\\
{\large }\\
{\large }\\
\textbf{\large -} {\large since for lemma 1.3} \textbf{\large \( E^{\epsilon \sqrt{\epsilon }}_{\mathbf{P}}\leq E_{\mathbf{P}}^{\epsilon } \)}{\large ,}
\textbf{\large }{\large the} {\large gap is bigger than \( \frac{\epsilon \sqrt{\epsilon }}{2} \)
.}\\
{\large }\\
 {\large }{\large \par}

\textbf{\large Corollary 1.6}{\large }\\
{\large }\\
{\large Thanks to the results of theorem 1.5 about the existence of the ground
state of \( H_{\mathbf{P},\epsilon \sqrt{\epsilon }}\mid _{F_{\epsilon \sqrt{\epsilon }}^{+}} \)
and about the} \emph{\large gap} {\large ( bigger than \( \frac{\epsilon \sqrt{\epsilon }}{2} \)),
it is possible to iterate the procedure at fixed \( \overline{g} \), by applying
lemmas 1.2, 1.3 to \( H_{\mathbf{P},\epsilon \sqrt{\epsilon }}\mid _{F_{\epsilon ^{2}}^{+}} \)and
\( H_{\mathbf{P},\epsilon ^{2}}\mid _{F_{\epsilon ^{2}}^{+}} \) and by using
properly adapted versions of lemma 1.4 and of theorem 1.5 to compare \( H_{\mathbf{P},\epsilon \sqrt{\epsilon }}\mid _{F_{\epsilon ^{2}}^{+}} \)
and \( H_{\mathbf{P},\epsilon ^{2}}\mid _{F_{\epsilon ^{2}}^{+}} \). }\\
{\large Therefore the iteration is consistent and it does not end since the
vector obtained at the step \( j+1 \) has norm bigger than a fixed fraction
of the norm of the vector at the \( j \) step. At each step} {\large the cut-off}
{\large is reduced by a factor \( \epsilon ^{\frac{1}{2}} \), so that at the
\( j \) step we obtain: }\\
{\large \par}

{\large \( \psi _{\mathbf{P}}^{\epsilon ^{\frac{j+2}{2}}}=\psi _{\mathbf{P}}^{\epsilon ^{\frac{j+1}{2}}}-\frac{1}{2\pi i}\sum ^{+\infty }_{n=1}\oint \frac{1}{H_{\mathbf{P},\epsilon ^{\frac{j+1}{2}}}\mid _{F_{\epsilon ^{\frac{j+2}{2}}}^{+}}-E}\left( -\left( \Delta H_{\mathbf{P}}\right) ^{\epsilon ^{\frac{j+1}{2}}}_{\epsilon ^{\frac{j+2}{2}}}\frac{1}{H_{\mathbf{P},\epsilon ^{\frac{j+1}{2}}}\mid _{F_{\epsilon ^{\frac{j+2}{2}}}^{+}}-E}\right) ^{n}dE\: \psi _{\mathbf{P}}^{\epsilon ^{\frac{j+1}{2}}} \)}\\
 \textbf{\large }{\large }\\
{\large }\\
\textbf{\large Lemma 1.4} {\large }\\
{\large }\\
{\large For fixed and properly small \( \overline{g} \) , \( \left( \Delta H_{\mathbf{P}}\right) ^{\epsilon ^{\frac{j+1}{2}}}_{\epsilon ^{\frac{j+2}{2}}} \)
is small of order \( 1 \) with respect to \( H_{\mathbf{P},\epsilon ^{\frac{j+1}{2}}}\mid _{F_{\epsilon ^{\frac{j+2}{2}}}^{+}} \),
in the following sense:}\\
{\large }\\
{\large given \( E \) such that \( \left| E-\inf H_{\mathbf{P},\epsilon ^{\frac{j+1}{2}}}\mid _{F_{\epsilon ^{\frac{j+2}{2}}}^{+}}\right| =\left| E-E_{\mathbf{P}}^{\epsilon ^{\frac{j+1}{2}}}\right| =\frac{11\epsilon ^{\frac{j+2}{2}}}{20} \)}\\
{\large }\\
{\large \( \frac{1}{H_{\mathbf{P},\epsilon ^{\frac{j+1}{2}}}+\left( \Delta H_{\mathbf{P}}\right) ^{\epsilon ^{\frac{j+1}{2}}}_{\epsilon ^{\frac{j+2}{2}}}-E}\mid _{_{F_{\epsilon ^{\frac{j+2}{2}}}^{+}}}=\frac{1}{H_{\mathbf{P},\epsilon ^{\frac{j+1}{2}}}-E}\sum ^{+\infty }_{n=0}\left( -\left( \Delta H_{\mathbf{P}}\right) ^{\epsilon ^{\frac{j+1}{2}}}_{\epsilon ^{\frac{j+2}{2}}}\frac{1}{H_{\mathbf{P},\epsilon ^{\frac{j+1}{2}}}-E}\right) ^{n}\mid _{F_{\epsilon ^{\frac{j+2}{2}}}^{+}} \)}\\
{\large and \( \left\Vert \frac{1}{H_{\mathbf{P},\epsilon ^{\frac{j+1}{2}}}-E}\left( -\left( \Delta H_{\mathbf{P}}\right) ^{\epsilon ^{\frac{j+1}{2}}}_{\epsilon ^{\frac{j+2}{2}}}\frac{1}{H_{\mathbf{P},\epsilon ^{\frac{j+1}{2}}}-E}\right) ^{n}\mid _{F_{\epsilon ^{\frac{j+2}{2}}}^{+}}\right\Vert _{F_{\epsilon ^{\frac{j+2}{2}}}^{+}}\leq \frac{20\left( C\left( \overline{g},\overline{m}\right) \right) ^{n}}{\epsilon ^{\frac{j+2}{2}}} \)
}\\
{\large where \( 0<C\left( \overline{g},m\right) <\frac{1}{12} \) is a constant
independent of \( \epsilon ^{\frac{j+1}{2}} \).}\\
{\large }\\
{\large Proof}\\
{\large }\\
{\large Let us analyze the \( n^{th} \) term of the following sum }\\
{\large }\\
{\large 
\[
\frac{1}{H_{\mathbf{P},\epsilon ^{\frac{j+1}{2}}}-E}\sum ^{+\infty }_{n=1}\left( -\left( \Delta H_{\mathbf{P}}\right) ^{\epsilon ^{\frac{j+1}{2}}}_{\epsilon ^{\frac{j+2}{2}}}\frac{1}{H_{\mathbf{P},\epsilon ^{\frac{j+1}{2}}}-E}\right) \]
}\\
{\large }\\
{\large \( \left( -1\right) ^{n}\frac{1}{H_{\mathbf{P},\epsilon ^{\frac{j+1}{2}}}-E}\left( \Delta H_{\mathbf{P}}\right) ^{\epsilon ^{\frac{j+1}{2}}}_{\epsilon ^{\frac{j+2}{2}}}\frac{1}{H_{\mathbf{P},\epsilon ^{\frac{j+1}{2}}}-E}.......................\left( \Delta H_{\mathbf{P}}\right) ^{\epsilon ^{\frac{j+1}{2}}}_{\epsilon ^{\frac{j+2}{2}}}\frac{1}{H_{\mathbf{P},\epsilon ^{\frac{j+1}{2}}}-E}= \)}\\
{\large }\\
{\large \( =\left( -1\right) ^{n}\left( \frac{1}{H_{\mathbf{P},\epsilon ^{\frac{j+1}{2}}}-E}\right) ^{\frac{1}{2}}....\left( \frac{1}{H_{\mathbf{P},\epsilon ^{\frac{j+1}{2}}}-E}\right) ^{\frac{1}{2}}\left( \Delta H_{\mathbf{P}}\right) ^{\epsilon ^{\frac{j+1}{2}}}_{\epsilon ^{\frac{j+2}{2}}}\left( \frac{1}{H_{\mathbf{P},\epsilon ^{\frac{j+1}{2}}}-E}\right) ^{\frac{1}{2}}....\left( \frac{1}{H_{\mathbf{P},\epsilon ^{\frac{j+1}{2}}}-E}\right) ^{\frac{1}{2}} \)}\\
{\large }\\
{\large where \( \left( \frac{1}{H_{\mathbf{P},\epsilon ^{\frac{j+1}{2}}}-E}\right) ^{\frac{1}{2}} \)is
defined starting from the spectral representation of \( H_{\mathbf{P},\epsilon ^{\frac{j+1}{2}}} \)
by using the convention to take the branch of the square root with smaller argument
in \( (-\pi ,\pi ] \).}\\
{\large \par}

\textbf{\emph{\large Study of the norm of}} {\large \( \left( \frac{1}{H_{\mathbf{P},\epsilon ^{\frac{j+1}{2}}}-E}\right) ^{\frac{1}{2}}\left( \Delta H_{\mathbf{P}}\right) ^{\epsilon ^{\frac{j+1}{2}}}_{\epsilon ^{\frac{j+2}{2}}}\left( \frac{1}{H_{\mathbf{P},\epsilon ^{\frac{j+1}{2}}}-E}\right) ^{\frac{1}{2}}\mid _{F_{\epsilon ^{\frac{j+2}{2}}}^{+}} \).}\\
{\large }\\
{\large }\\
{\large \( \left\Vert \left( \frac{1}{H_{\mathbf{P},\epsilon ^{\frac{j+1}{2}}}-E}\right) ^{\frac{1}{2}}\left( \Delta H_{\mathbf{P}}\right) ^{\epsilon ^{\frac{j+1}{2}}}_{\epsilon ^{\frac{j+2}{2}}}\left( \frac{1}{H_{\mathbf{P},\epsilon ^{\frac{j+1}{2}}}-E}\right) ^{\frac{1}{2}}\right\Vert _{_{F_{\epsilon ^{\frac{j+2}{2}}}^{+}}}= \)}\\
{\large }\\
{\large \( =\left\Vert \left( \frac{1}{H_{\mathbf{P},\epsilon ^{\frac{j+1}{2}}}-E}\right) ^{\frac{1}{2}}\left( g\int ^{\epsilon ^{\frac{j+1}{2}}}_{\epsilon ^{\frac{j+2}{2}}}\frac{1}{\sqrt{2}\left| \mathbf{k}\right| ^{\frac{1}{2}}}\left( b\left( \mathbf{k}\right) +b^{\dagger }\left( \mathbf{k}\right) \right) d^{3}k\right) \left( \frac{1}{H_{\mathbf{P},\epsilon ^{\frac{j+1}{2}}}-E}\right) ^{\frac{1}{2}}\right\Vert _{_{F_{\epsilon ^{\frac{j+2}{2}}}^{+}}}= \)}\\
{\large }\\
{\large \( \leq 2g\left\Vert \left( \frac{1}{H_{\mathbf{P},\epsilon ^{\frac{j+1}{2}}}-E}\right) ^{\frac{1}{2}}\right\Vert _{_{F_{\epsilon ^{\frac{j+2}{2}}}^{+}}}\cdot \left\Vert \left( \int ^{\epsilon ^{\frac{j+1}{2}}}_{\epsilon ^{\frac{j+2}{2}}}\frac{1}{\sqrt{2}\left| \mathbf{k}\right| ^{\frac{1}{2}}}b\left( \mathbf{k}\right) d^{3}k\right) \left( \frac{1}{H_{\mathbf{P},\epsilon ^{\frac{j+1}{2}}}-E}\right) ^{\frac{1}{2}}\right\Vert _{_{F_{\epsilon ^{\frac{j+2}{2}}}^{+}}} \)}\\
{\large }\\
{\large if the above quantities exist. }\\
{\large }\\
{\large The following estimate holds:} {\large }\\
{\large }\\
{\large }\\
{\large \( \left\Vert \int ^{\epsilon ^{\frac{j+1}{2}}}_{\epsilon ^{\frac{j+2}{2}}}b\left( \mathbf{k}\right) \frac{d^{3}\mathbf{k}}{\sqrt{2\left| \mathbf{k}\right| }}\left( \frac{1}{H_{\mathbf{P},\epsilon ^{\frac{j+1}{2}}}-E}\right) ^{\frac{1}{2}}\right\Vert _{F^{+}_{\epsilon ^{\frac{j+2}{2}}}}\leq \sqrt{40\pi }\cdot \epsilon ^{\frac{j+1}{4}} \)}\marginpar{
{\large (4)}{\large \par}
}{\large }\\
{\large }\\
{\large Proof}\\
{\large }\\
{\large }\\
{\large Let us consider vectors \( \varphi \in D^{b}\bigcap F^{+}_{\epsilon ^{\frac{j+2}{2}}} \)
; then}\\
{\large }\\
{\large \( \left\Vert \int ^{\epsilon ^{\frac{j+1}{2}}}_{\epsilon ^{\frac{j+2}{2}}}b\left( \mathbf{k}\right) \frac{d^{3}\mathbf{k}}{\sqrt{2\left| \mathbf{k}\right| }}\left( \frac{1}{H_{\mathbf{P},\epsilon ^{\frac{j+1}{2}}}-E}\right) ^{\frac{1}{2}}\varphi \right\Vert ^{2}_{F^{+}_{\epsilon ^{\frac{j+2}{2}}}}= \)}\\
{\large }\\
{\large \( =\left( \left( \frac{1}{H_{\mathbf{P},\epsilon ^{\frac{j+1}{2}}}-E}\right) ^{\frac{1}{2}}\varphi \: ,\: \int ^{\epsilon ^{\frac{j+1}{2}}}_{\epsilon ^{\frac{j+2}{2}}}b^{\dagger }\left( \mathbf{k}\right) \frac{d^{3}\mathbf{k}}{\sqrt{2\left| \mathbf{k}\right| }}\int ^{\epsilon ^{\frac{j+1}{2}}}_{\epsilon ^{\frac{j+2}{2}}}b\left( \mathbf{k}\right) \frac{d^{3}\mathbf{k}}{\sqrt{2\left| \mathbf{k}\right| }}\left( \frac{1}{H_{\mathbf{P},\epsilon ^{\frac{j+1}{2}}}-E}\right) ^{\frac{1}{2}}\varphi \right)  \)}\\
{\large }\\
{\large In general}\\
{\large }\\
{\large \( \left( \varphi \: ,\: \int ^{\epsilon ^{\frac{j+1}{2}}}_{\epsilon ^{\frac{j+2}{2}}}\int ^{\epsilon ^{\frac{j+1}{2}}}_{\epsilon ^{\frac{j+2}{2}}}\frac{1}{\sqrt{2\left| \mathbf{k}\right| }}\frac{1}{\sqrt{2\left| \mathbf{q}\right| }}b^{\dagger }\left( \mathbf{k}\right) b\left( \mathbf{q}\right) d^{3}kd^{3}q\varphi \right) = \)}\\
{\large }\\
{\large \( =\int ^{\epsilon ^{\frac{j+1}{2}}}_{\epsilon ^{\frac{j+2}{2}}}\int ^{\epsilon ^{\frac{j+1}{2}}}_{\epsilon ^{\frac{j+2}{2}}}\frac{1}{\sqrt{2\left| \mathbf{k}\right| }}\frac{1}{\sqrt{2\left| \mathbf{q}\right| }}\left( \varphi \: ,\: b^{\dagger }\left( \mathbf{k}\right) b\left( \mathbf{q}\right) \varphi \right) d^{3}kd^{3}q\leq  \)}\\
{\large }\\
{\large \( \leq \int ^{\epsilon ^{\frac{j+1}{2}}}_{\epsilon ^{\frac{j+2}{2}}}\int ^{\epsilon ^{\frac{j+1}{2}}}_{\epsilon ^{\frac{j+2}{2}}}\frac{1}{\sqrt{2\left| \mathbf{k}\right| }}\frac{1}{\sqrt{2\left| \mathbf{q}\right| }}\left( \varphi \: ,\: b^{\dagger }\left( \mathbf{k}\right) b\left( \mathbf{k}\right) \varphi \right) ^{\frac{1}{2}}\left( \varphi \: ,\: b^{\dagger }\left( \mathbf{q}\right) b\left( \mathbf{q}\right) \varphi \right) ^{\frac{1}{2}}d^{3}kd^{3}q\leq  \)}\\
{\large \( \leq \left( \varphi ,H^{mes}\mid ^{\epsilon ^{\frac{j+1}{2}}}_{\epsilon ^{\frac{j+2}{2}}}\varphi \right) \cdot \left( \int ^{\epsilon ^{\frac{j+1}{2}}}_{\epsilon ^{\frac{j+2}{2}}}\frac{1}{2\left| \mathbf{k}\right| ^{2}}d^{3}k\right) ^{\frac{1}{2}}\cdot \left( \int ^{\epsilon ^{\frac{j+1}{2}}}_{\epsilon ^{\frac{j+2}{2}}}\frac{1}{2\left| \mathbf{q}\right| ^{2}}d^{3}q\right) ^{\frac{1}{2}}\leq  \)}\\
{\large }\\
{\large \( \leq 2\pi \epsilon ^{\frac{j+1}{2}}\left( \varphi ,H^{mes}\mid ^{\epsilon ^{\frac{j+1}{2}}}_{\epsilon ^{\frac{j+2}{2}}}\varphi \right)  \)}\marginpar{
{\large (5)}{\large \par}
}{\large }\\
{\large }\\
{\large and the inequality holds in the case of vectors in \( D^{b}\left( H^{mes}\right)  \)
too.}\\
{\large }\\
{\large Therefore we arrive at}\\
{\large }\\
{\large \( \left( \left( \frac{1}{H_{\mathbf{P},\epsilon ^{\frac{j+1}{2}}}-E}\right) ^{\frac{1}{2}}\varphi \: ,\: H^{mes}\mid ^{\epsilon ^{\frac{j+1}{2}}}_{\epsilon ^{\frac{j+2}{2}}}\left( \frac{1}{H_{\mathbf{P},\epsilon ^{\frac{j+1}{2}}}-E}\right) ^{\frac{1}{2}}\varphi \right) \leq  \)}\\
{\large }\\
{\large \( \leq \left\Vert \varphi \right\Vert \cdot \left\Vert H^{mes}\mid ^{\epsilon ^{\frac{j+1}{2}}}_{\epsilon ^{\frac{j+2}{2}}}\left[ \left( \frac{1}{H_{\mathbf{P},\epsilon ^{\frac{j+1}{2}}}-E}\right) ^{\frac{1}{2}}\right] ^{\dagger }\left( \frac{1}{H_{\mathbf{P},\epsilon ^{\frac{j+1}{2}}}-E}\right) ^{\frac{1}{2}}\varphi \right\Vert _{_{F_{\epsilon ^{\frac{j+2}{2}}}^{+}}} \)}\\
{\large }\\
{\large ( \( \left[ H_{\mathbf{P},\epsilon ^{\frac{j+1}{2}}},\int ^{\epsilon ^{\frac{j+1}{2}}}_{\epsilon ^{\frac{j+2}{2}}}\left| \mathbf{k}\right| b^{\dagger }\left( \mathbf{k}\right) b\left( \mathbf{k}\right) d^{3}k\right] =0 \))}\\
{\large }\\
{\large The operator norm of \( H^{mes}\mid ^{\epsilon ^{\frac{j+1}{2}}}_{\epsilon ^{\frac{j+2}{2}}}\cdot \left[ \left( \frac{1}{H^{w}_{\mathbf{P},\epsilon ^{\frac{j+1}{2}}}-E}\right) ^{\frac{1}{2}}\right] ^{\dagger }\cdot \left( \frac{1}{H^{w}_{\mathbf{P},\epsilon ^{\frac{j+1}{2}}}-E}\right) ^{\frac{1}{2}} \)
has to be studied separately on \( \psi _{\mathbf{P}}^{\epsilon ^{\frac{j+1}{2}}} \)and
on \( F_{\epsilon ^{\frac{j+2}{2}}}^{+}\ominus \psi _{\mathbf{P}}^{\epsilon ^{\frac{j+1}{2}}} \).
The operator vanishes on \( \psi _{\mathbf{P}}^{\epsilon ^{\frac{j+1}{2}}} \)(put
\( H^{mes}\mid ^{\epsilon ^{\frac{j+1}{2}}}_{\epsilon ^{\frac{j+2}{2}}} \)
on the right ). The discussion is restricted to the subspace \( F_{\epsilon ^{\frac{j+2}{2}}}^{+}\ominus \psi _{\mathbf{P}}^{\epsilon ^{\frac{j+1}{2}}} \)
 }\\
{\large }\\
{\large Moreover, as already seen in lemma 1.2.} {\large adapted to the hamiltonian\( H_{\mathbf{P},\epsilon ^{\frac{j+1}{2}}}\mid _{F_{\epsilon ^{\frac{j+2}{2}}}^{+}} \)
, we have}\\
{\large \par}

\begin{itemize}
\item {\large \( \inf spec\left( H_{\mathbf{P},\epsilon ^{\frac{j+1}{2}}}\mid _{F_{\epsilon ^{\frac{j+2}{2}}}^{+}}-\frac{1}{20}\int ^{\epsilon ^{\frac{j+1}{2}}}_{\epsilon ^{\frac{j+2}{2}}}\left| \mathbf{k}\right| b^{\dagger }\left( \mathbf{k}\right) b\left( \mathbf{k}\right) d^{3}k\mid _{F_{\epsilon ^{\frac{j+2}{2}}}^{+}}\right) \geq E_{\mathbf{P}}^{\epsilon ^{\frac{j+1}{2}}}+\frac{3}{5}\epsilon ^{\frac{j+2}{2}} \)}
{\small }\\
{\small \par}
\item {\large \( \inf spec\left( H_{\mathbf{P},\epsilon ^{\frac{j+1}{2}}}\mid _{F_{\epsilon ^{\frac{j+2}{2}}}^{+}}-\frac{1}{20}\int ^{\epsilon ^{\frac{j+1}{2}}}_{\epsilon ^{\frac{j+2}{2}}}\left| \mathbf{k}\right| b^{\dagger }\left( \mathbf{k}\right) b\left( \mathbf{k}\right) d^{3}k\mid _{F_{\epsilon ^{\frac{j+2}{2}}}^{+}}-ReE\right) \geq  \)}\\
{\large \( \geq \frac{3}{5}\epsilon ^{\frac{j+2}{2}}-\frac{11}{20}\epsilon ^{\frac{j+2}{2}}>0 \)}\\
{\large }\\
{\large \par}
\end{itemize}
{\large Going to the} \textbf{\large joined spectral representation} {\large of}
\textbf{\large \( H_{\mathbf{P},\epsilon ^{\frac{j+1}{2}}} \)}{\large and}
\textbf{\large \( H^{mes}\mid ^{\epsilon ^{\frac{j+1}{2}}}_{\epsilon ^{\frac{j+2}{2}}} \)}
{\large , we obtain}\\
{\large }\\
{\large 
\[
\left\Vert H^{mes}\mid _{\epsilon ^{\frac{j+2}{2}}}^{\epsilon ^{\frac{j+1}{2}}}\left[ \left( \frac{1}{H_{\mathbf{P},\epsilon ^{\frac{j+1}{2}}}-E}\right) ^{\frac{1}{2}}\right] ^{\dagger }\left( \frac{1}{H_{\mathbf{P},\epsilon ^{\frac{j+1}{2}}}-E}\right) ^{\frac{1}{2}}\right\Vert _{_{F_{\epsilon ^{\frac{j+2}{2}}}^{+}}}\leq 20\]
 }\marginpar{
{\large (6)}{\large \par}
}{\large }\\
{\large }\\
{\large Putting together the inequalities (5) and (6) we arrive at the inequality
(4).} \emph{\large }\\
\emph{\large }\\
\emph{\large Conclusion}{\large }\\
{\large }\\
{\large If \( g \) is less than a limit value \( \overline{g} \), the thesis
is proved since \( \left\Vert \left( \frac{1}{H_{\mathbf{P},\epsilon ^{\frac{j+1}{2}}}-E}\right) ^{\frac{1}{2}}\right\Vert _{_{F_{\epsilon ^{\frac{j+2}{2}}}^{+}}} \)is
of order \( \frac{1}{\epsilon ^{\frac{j+2}{4}}} \).}\\
{\large }\\
{\large \par}

\section{Convergence of the ground states of transformed \protect\( H_{\mathbf{P},\sigma _{j}}\protect \)
. }

{\large I conjecture that the hamiltonians \( H_{\mathbf{P}} \)} {\large have
a ground state for the representations of \( \left\{ b\left( \mathbf{k}\right) ,b^{\dagger }\left( \mathbf{k}\right) \right\}  \)
which are coherent in the infrared region (\( \mathbf{k}=0 \)). Then an argument
is developed which explicitly identifies the eventual coherent factor in the
case \( \mathbf{P}=0 \) and implicitly in the case \( \mathbf{P}\neq 0 \).
Such a heuristic information will be used in a rigorous proof which is based
on the iterative procedure of construction of the ground state. The two cases
\( \mathbf{P}=0 \) and \( \mathbf{P}\neq 0 \) are treated separately.}\\
\\
\textbf{\large Derivation of the coherent factor.}\\
{\large }\\
{\large Let us assume that \( \psi _{\mathbf{P}} \) is an eigenvector of \( H_{\mathbf{P}} \)
and that it is ``coherent in the infrared region'', which means \( b\left( \mathbf{k}\right) \psi _{\mathbf{P}}\approx f_{\mathbf{P}}\left( \mathbf{k}\right) \psi _{\mathbf{P}} \)
for \( \mathbf{k}\rightarrow 0 \), where the meaning of the limit is given
only ``a posteriori''. Then the coherent function \( f_{\mathbf{P}}\left( \mathbf{k}\right)  \)
has to satisfies the following relation, in the neighborhood of \( \mathbf{k}=0 \):}\\
{\large }\\
{\large \( \left( \psi _{\mathbf{P}},\left[ H_{\mathbf{P}},b\left( \mathbf{k}\right) \right] \psi _{\mathbf{P}}\right) =0\quad for\: \mathbf{k}\rightarrow 0 \)}
 {\large }\\
{\large }\\
{\large }\\
{\large \( \left( \psi _{\mathbf{P}},\left[ H_{\mathbf{P}},b\left( \mathbf{k}\right) \right] \psi _{\mathbf{P}}\right) = \)}\\
{\large }\\
{\large \( =\left( \psi _{\mathbf{P}},\left[ \frac{\left( \mathbf{P}^{mes}\right) ^{2}}{2m}-\frac{\mathbf{P}\cdot \mathbf{P}^{mes}}{m}+g\int _{0}^{\kappa }\left( b\left( \mathbf{k}\right) +b^{\dagger }\left( \mathbf{k}\right) \right) \frac{d^{3}\mathbf{k}}{\sqrt{2\left| \mathbf{k}\right| }}+H^{mes},b\left( \mathbf{k}\right) \right] \psi _{\mathbf{P}}\right) = \)}\\
{\large }\\
{\large \( =\left( \psi _{\mathbf{P}},\left[ \frac{\left( \mathbf{P}^{mes}\right) ^{2}}{2m}-\frac{\mathbf{P}\cdot \mathbf{P}^{mes}}{m},b\left( \mathbf{k}\right) \right] \psi _{\mathbf{P}}\right) +\left( \psi _{\mathbf{P}},\left[ g\int _{0}^{\kappa }\left( b\left( \mathbf{k}\right) +b^{\dagger }\left( \mathbf{k}\right) \right) \frac{d^{3}\mathbf{k}}{\sqrt{2\left| \mathbf{k}\right| }},b\left( \mathbf{k}\right) \right] \psi _{\mathbf{P}}\right) ++\left( \psi _{\mathbf{P}},\left[ H^{mes},b\left( \mathbf{k}\right) \right] \psi _{\mathbf{P}}\right) = \)}\\
{\large }\\
{\large \( =-\frac{\mathbf{k}\cdot \left( \psi _{\mathbf{P}},\mathbf{P}^{mes}b\left( \mathbf{k}\right) \psi _{\mathbf{P}}\right) }{2m}-\frac{\left( \psi _{\mathbf{P}},b\left( \mathbf{k}\right) \mathbf{P}^{mes}\psi _{\mathbf{P}}\right) \cdot \mathbf{k}}{2m}+\frac{\mathbf{P}\cdot \mathbf{k}}{m}\left( \psi _{\mathbf{P}},b\left( \mathbf{k}\right) \psi _{\mathbf{P}}\right) -\frac{g\cdot \chi ^{\kappa }_{0}\left( \mathbf{k}\right) }{\sqrt{2\left| \mathbf{k}\right| }}\left\Vert \psi _{\mathbf{P}}\right\Vert ^{2}-\left| \mathbf{k}\right| \left( \psi _{\mathbf{P}},b\left( \mathbf{k}\right) \psi _{\mathbf{P}}\right)  \)}
( \( \chi ^{\kappa }_{0}\left( \mathbf{k}\right)  \) characteristic function
of \( \left\{ \mathbf{k}:\left| \mathbf{k}\right| \leq \kappa \right\}  \)){\large }\\
{\large }\\
{\large then}\\
\\
{\large \( -\frac{\mathbf{k}\cdot \left( \psi _{\mathbf{P}},\mathbf{P}^{mes}b\left( \mathbf{k}\right) \psi _{\mathbf{P}}\right) }{2m}-\frac{\left( \psi _{\mathbf{P}},b\left( \mathbf{k}\right) \mathbf{P}^{mes}\psi _{\mathbf{P}}\right) \cdot \mathbf{k}}{2m}+\frac{\mathbf{P}\cdot \mathbf{k}}{m}\left( \psi _{\mathbf{P}},b\left( \mathbf{k}\right) \psi _{\mathbf{P}}\right) -\frac{g\cdot \chi ^{\kappa }_{0}\left( \mathbf{k}\right) }{\sqrt{2\left| \mathbf{k}\right| }}\left\Vert \psi _{\mathbf{P}}\right\Vert ^{2}-\left| \mathbf{k}\right| \left( \psi _{\mathbf{P}},b\left( \mathbf{k}\right) \psi _{\mathbf{P}}\right) = \)}\\
{\large }\\
{\large \( =0 \)}\\
{\large }\\
{\large \( -\frac{\mathbf{k}\cdot \left( \psi _{\mathbf{P}},\mathbf{P}^{mes}b\left( \mathbf{k}\right) \psi _{\mathbf{P}}\right) }{m}-\frac{\left( \psi _{\mathbf{P}},b\left( \mathbf{k}\right) \psi _{\mathbf{P}}\right) \left| \mathbf{k}\right| ^{2}}{2m}+\frac{\mathbf{P}\cdot \mathbf{k}}{m}\left( \psi _{\mathbf{P}},b\left( \mathbf{k}\right) \psi _{\mathbf{P}}\right) -\frac{g\cdot \chi ^{\kappa }_{0}\left( \mathbf{k}\right) }{\sqrt{2\left| \mathbf{k}\right| }}\left\Vert \psi _{\mathbf{P}}\right\Vert ^{2}-\left| \mathbf{k}\right| \left( \psi _{\mathbf{P}},b\left( \mathbf{k}\right) \psi _{\mathbf{P}}\right) = \)
}\\
{\large \( =0 \)}\\
{\large }\\
{\large }\\
{\large \( -\frac{\mathbf{k}\cdot \left( \psi _{\mathbf{P}},\mathbf{P}^{mes}\psi _{\mathbf{P}}\right) }{m}f_{\mathbf{P}}\left( \mathbf{k}\right) -\frac{\left| \mathbf{k}\right| ^{2}}{2m}\left\Vert \psi _{\mathbf{P}}\right\Vert ^{2}f_{\mathbf{P}}\left( \mathbf{k}\right) +\frac{\mathbf{P}\cdot \mathbf{k}}{m}\left\Vert \psi _{\mathbf{P}}\right\Vert ^{2}f_{\mathbf{P}}\left( \mathbf{k}\right) -\frac{g\cdot \chi ^{\kappa }_{0}\left( \mathbf{k}\right) }{\sqrt{2\left| \mathbf{k}\right| }}\left\Vert \psi _{\mathbf{P}}\right\Vert ^{2}-\left| \mathbf{k}\right| \left\Vert \psi _{\mathbf{P}}\right\Vert ^{2}f_{\mathbf{P}}\left( \mathbf{k}\right) = \)}\\
{\large }\\
{\large \( =0 \)}\\
{\large }\\
{\large Therefore the expected behavior is }\\
{\large }\\
{\large \( f_{\mathbf{P}}\left( \mathbf{k}\right) \approx _{\mathbf{k}\rightarrow 0}-\frac{g\cdot \chi ^{\kappa }_{0}\left( \mathbf{k}\right) }{\sqrt{2\left| \mathbf{k}\right| }}\cdot \frac{1}{\left( \left| \mathbf{k}\right| +\frac{\left| \mathbf{k}\right| ^{2}}{2m}-\frac{\mathbf{P}\cdot \mathbf{k}}{m}+\frac{\mathbf{k}\cdot \left( \psi _{\mathbf{P}},\mathbf{P}^{mes}\psi _{\mathbf{P}}\right) }{m\cdot \left\Vert \psi _{\mathbf{P}}\right\Vert ^{2}}\right) }=-\frac{g\cdot \chi ^{\kappa }_{0}\left( \mathbf{k}\right) }{\sqrt{2\left| \mathbf{k}\right| }}\cdot \frac{1}{\left( \left| \mathbf{k}\right| +\frac{\left| \mathbf{k}\right| ^{2}}{2m}-\frac{\mathbf{k}\cdot \left( \mathbf{P}-\overline{\mathbf{P}^{mes}}\right) }{m}\right) } \)}\\
{\large }\\
{\large and the coherent factor is labelled by \( \mathbf{P}_{1}=\mathbf{P}-\overline{\mathbf{P}^{mes}} \).
}\\
{\large }\\
{\large The argument proves that if the ground state is ``coherent in the infrared
region'', a non Fock state is necessarily. Starting from this result I operate
a proper coherent transformation on the variables \( \left\{ b\left( \mathbf{k}\right) ,b^{\dagger }\left( \mathbf{k}\right) \right\}  \)
of the hamiltonian \( H_{\mathbf{P}} \) and I look for a ground state of the
transformed hamiltonian in the Fock space with respect to \( \left\{ b\left( \mathbf{k}\right) ,b^{\dagger }\left( \mathbf{k}\right) \right\}  \).}\\
{\large }\\
\emph{\large Coherent transformation}{\large . }\\
{\large }\\
{\large \( b\left( \mathbf{k}\right) \longrightarrow b\left( \mathbf{k}\right) -\frac{g}{\sqrt{2}\left| \mathbf{k}\right| ^{\frac{3}{2}}\left( 1-\widehat{\mathbf{k}}\cdot \frac{\mathbf{P}_{1}}{m}\right) } \)
\( \quad  \)for \( \mathbf{k}: \)\( \quad 0\leq \left| \mathbf{k}\right| \leq \kappa  \)
}\\
{\large }\\
{\large using the} \emph{\large inter-twiner}{\large }\\
{\large }\\
\marginpar{

}{\large 
\[
\mathcal{W}\left( \mathbf{P}\right) =W\left( \frac{\mathbf{P}_{1}}{m}\right) =e^{-g\int _{0}^{\kappa }\frac{b\left( \mathbf{k}\right) -b^{\dagger }\left( \mathbf{k}\right) }{\left| \mathbf{k}\right| \left( 1-\widehat{\mathbf{k}}\cdot \frac{\mathbf{P}_{1}}{m}\right) }\frac{d^{3}k}{\sqrt{2\left| \mathbf{k}\right| }}}\qquad \]
}\marginpar{
{\large (7)}{\large \par}
} \\
\\
\emph{\large }\\
\emph{\large Transformed hamiltonian.}{\large }\\
{\large }\\
{\large I rewrite \( H_{\mathbf{P}} \), \( \mathbf{P}=\mathbf{P}_{1}+\mathbf{P}_{2} \)}
{\large , as}\\
{\large }\\
{\large \( H_{\mathbf{P}}=\frac{\left( \mathbf{P}_{1}+\mathbf{P}_{2}-\mathbf{P}^{mes}\right) ^{2}}{2m}+g\int _{0}^{\kappa }\left( b\left( \mathbf{k}\right) +b^{\dagger }\left( \mathbf{k}\right) \right) \frac{d^{3}\mathbf{k}}{\sqrt{2\left| \mathbf{k}\right| }}+H^{mes}= \)}\\
{\large }\\
{\large \( =\frac{\left( \mathbf{P}_{1}+\mathbf{P}_{2}\right) ^{2}}{2m}-\frac{\left( \mathbf{P}_{1}+\mathbf{P}_{2}\right) \cdot \mathbf{P}^{mes}}{m}+\frac{\mathbf{P}^{mes^{2}}}{2m}+g\int _{0}^{\kappa }\left( b\left( \mathbf{k}\right) +b^{\dagger }\left( \mathbf{k}\right) \right) \frac{d^{3}\mathbf{k}}{\sqrt{2\left| \mathbf{k}\right| }}+H^{mes}= \)}\\
{\large }\\
{\large \( =\frac{\left( \mathbf{P}_{1}+\mathbf{P}_{2}\right) ^{2}}{2m}+\frac{\mathbf{P}^{mes^{2}}}{2m}-\frac{\mathbf{P}_{2}\cdot \mathbf{P}^{mes}}{m}+\int ^{\infty }_{\kappa }\left( \left| \mathbf{k}\right| -\mathbf{k}\cdot \frac{\mathbf{P}_{1}}{m}\right) b^{\dagger }\left( \mathbf{k}\right) b\left( \mathbf{k}\right) d^{3}k+ \)}\\
{\large }\\
{\large \( +\int ^{\kappa }_{0}\left( \left| \mathbf{k}\right| -\mathbf{k}\cdot \frac{\mathbf{P}_{1}}{m}\right) \left( b^{\dagger }\left( \mathbf{k}\right) +\frac{g}{\sqrt{2}\left| \mathbf{k}\right| ^{\frac{3}{2}}\left( 1-\widehat{\mathbf{k}}\cdot \frac{\mathbf{P}_{1}}{m}\right) }\right) \left( b\left( \mathbf{k}\right) +\frac{g}{\sqrt{2}\left| \mathbf{k}\right| ^{\frac{3}{2}}\left( 1-\widehat{\mathbf{k}}\cdot \frac{\mathbf{P}_{1}}{m}\right) }\right) d^{3}k+ \)}\\
{\large }\\
{\large \( -g^{2}\int _{0}^{\kappa }\frac{1}{2\left| \mathbf{k}\right| ^{2}\left( 1-\widehat{\mathbf{k}}\cdot \frac{\mathbf{P}_{1}}{m}\right) }d^{3}k \)}\\
{\large }\\
{\large }\\
{\large and I act on it with the coherent transformation:}\\
{\large }\\
{\large }\\
{\large \( W\left( \frac{\mathbf{P}_{1}}{m}\right) H_{\mathbf{P}}W^{\dagger }\left( \frac{\mathbf{P}_{1}}{m}\right) = \)}\marginpar{
{\large (8)}{\large \par}
}{\large }\\
{\large }\\
{\large \( =\frac{\left( \mathbf{P}_{1}+\mathbf{P}_{2}\right) ^{2}}{2m}+\frac{1}{2m}\left( \mathbf{P}^{mes}-g\int _{0}^{\kappa }\frac{\mathbf{k}}{\sqrt{2}\left| \mathbf{k}\right| ^{\frac{3}{2}}\left( 1-\widehat{\mathbf{k}}\cdot \frac{\mathbf{P}_{1}}{m}\right) }\left( b\left( \mathbf{k}\right) +b^{\dagger }\left( \mathbf{k}\right) \right) d^{3}k+g^{2}\int _{0}^{\kappa }\frac{\mathbf{k}}{2\left| \mathbf{k}\right| ^{3}\left( 1-\widehat{\mathbf{k}}\cdot \frac{\mathbf{P}_{1}}{m}\right) ^{2}}d^{3}k\right) ^{2}+ \)}\\
{\large }\\
{\large \( -\frac{\mathbf{P}_{2}}{m}\left( \mathbf{P}^{mes}-g\int _{0}^{\kappa }\frac{\mathbf{k}}{\sqrt{2}\left| \mathbf{k}\right| ^{\frac{3}{2}}\left( 1-\widehat{\mathbf{k}}\cdot \frac{\mathbf{P}_{1}}{m}\right) }\left( b\left( \mathbf{k}\right) +b^{\dagger }\left( \mathbf{k}\right) \right) d^{3}k+g^{2}\int _{0}^{\kappa }\frac{\mathbf{k}}{2\left| \mathbf{k}\right| ^{3}\left( 1-\widehat{\mathbf{k}}\cdot \frac{\mathbf{P}_{1}}{m}\right) ^{2}}d^{3}k\right) + \)}\\
{\large }\\
{\large \( +\int ^{\infty }_{0}\left( \left| \mathbf{k}\right| -\mathbf{k}\cdot \frac{\mathbf{P}_{1}}{m}\right) b^{\dagger }\left( \mathbf{k}\right) b\left( \mathbf{k}\right) d^{3}k-g^{2}\int _{0}^{\kappa }\frac{1}{2\left| \mathbf{k}\right| ^{2}\left( 1-\widehat{\mathbf{k}}\cdot \frac{\mathbf{P}_{1}}{m}\right) }d^{3}k \)}\\
{\large }\\
{\large }\\
\emph{\large Remark}{\large }\\
{\large }\\
{\large Given \( \phi _{\mathbf{P}}=W\left( \frac{\mathbf{P}_{1}}{m}\right) \psi _{\mathbf{P}} \)
we have formally that}\\
{\large }\\
{\large \( \mathbf{P}_{2}=\mathbf{P}-\mathbf{P}_{1}=\overline{\mathbf{P}^{mes}}=\frac{\left( \psi _{\mathbf{P}},\mathbf{P}^{mes}\psi _{\mathbf{P}}\right) }{\left\Vert \psi _{\mathbf{P}}\right\Vert ^{2}}=\frac{\left( \phi _{\mathbf{P}},W\mathbf{P}^{mes}W^{\dagger }\phi _{\mathbf{P}}\right) }{\left\Vert \phi _{\mathbf{P}}\right\Vert ^{2}}= \)}\\
{\large }\\
{\large \( =\frac{1}{\left\Vert \phi _{\mathbf{P}}\right\Vert ^{2}}\left( \phi _{\mathbf{P}},\mathbf{P}^{mes}\phi _{\mathbf{P}}\right) -\frac{1}{\left\Vert \phi _{\mathbf{P}}\right\Vert ^{2}}\left( \phi _{\mathbf{P}},g\int _{0}^{\kappa }\frac{\mathbf{k}}{\sqrt{2}\left| \mathbf{k}\right| ^{\frac{3}{2}}\left( 1-\widehat{\mathbf{k}}\cdot \frac{\mathbf{P}_{1}}{m}\right) }\left( b\left( \mathbf{k}\right) +b^{\dagger }\left( \mathbf{k}\right) \right) d^{3}k\phi _{\mathbf{P}}\right) +g^{2}\int _{0}^{\kappa }\frac{\mathbf{k}}{2\left| \mathbf{k}\right| ^{3}\left( 1-\widehat{\mathbf{k}}\cdot \frac{\mathbf{P}_{1}}{m}\right) ^{2}}d^{3}k \)}\\
{\large }\\
{\large }\\
{\large Therefore \( \mathbf{P}_{1} \) has to satisfy the equation}\\
{\large }\\
{\large \( \mathbf{P}-\mathbf{P}_{1}= \)}\\
\\
\( =\frac{\left( \phi _{\mathbf{P}},\mathbf{P}^{mes}\phi _{\mathbf{P}}\right) }{\left\Vert \phi _{\mathbf{P}}\right\Vert ^{2}}-\frac{1}{\left\Vert \phi _{\mathbf{P}}\right\Vert ^{2}}\left( \phi _{\mathbf{P}},g\int _{0}^{\kappa }\frac{\mathbf{k}}{\sqrt{2}\left| \mathbf{k}\right| ^{\frac{3}{2}}\left( 1-\widehat{\mathbf{k}}\cdot \frac{\mathbf{P}_{1}}{m}\right) }\left( b\left( \mathbf{k}\right) +b^{\dagger }\left( \mathbf{k}\right) \right) d^{3}k\phi _{\mathbf{P}}\right) +g^{2}\int _{0}^{\kappa _{1}}\frac{\mathbf{k}}{2\left| \mathbf{k}\right| ^{3}\left( 1-\widehat{\mathbf{k}}\cdot \frac{\mathbf{P}_{1}}{m}\right) ^{2}}d^{3}k \)\marginpar{
{\large (9)}{\large \par}
}{\large }\\
{\large where \( \phi _{\mathbf{P}} \) is a ground state of the transformed
hamiltonian \( W\left( \frac{\mathbf{P}_{1}}{m}\right) H_{\mathbf{P}}W^{\dagger }\left( \frac{\mathbf{P}_{1}}{m}\right)  \).}\\
{\large }\\
{\large I define \( \Pi _{\mathbf{P}}=\mathbf{P}^{mes}-g\int _{0}^{\kappa }\frac{\mathbf{k}}{\sqrt{2}\left| \mathbf{k}\right| ^{\frac{3}{2}}\left( 1-\widehat{\mathbf{k}}\cdot \frac{\mathbf{P}_{1}}{m}\right) }\left( b\left( \mathbf{k}\right) +b^{\dagger }\left( \mathbf{k}\right) \right) d^{3}k \)
and by a substitution in the expressions (9) and (8) one arrives at}\\
{\large \par}

\begin{itemize}
\item {\large \( \mathbf{P}_{2}=\frac{1}{\left\Vert \phi _{\mathbf{P}}\right\Vert ^{2}}\left( \phi _{\mathbf{P}},\Pi _{\mathbf{P}}\phi _{\mathbf{P}}\right) +g^{2}\int _{0}^{\kappa }\frac{\mathbf{k}}{2\left| \mathbf{k}\right| ^{3}\left( 1-\widehat{\mathbf{k}}\cdot \frac{\mathbf{P}_{1}}{m}\right) ^{2}} \) }{\large \par}
\item {\large \( W\left( \frac{\mathbf{P}_{1}}{m}\right) H_{\mathbf{P}}W^{\dagger }\left( \frac{\mathbf{P}_{1}}{m}\right) = \)}\\
{\large \( =\frac{1}{2m}\Pi _{\mathbf{P}}^{2}+\frac{1}{2m}\left( g^{2}\int _{0}^{\kappa }\frac{\mathbf{k}}{2\left| \mathbf{k}\right| ^{3}\left( 1-\widehat{\mathbf{k}}\cdot \frac{\mathbf{P}_{1}}{m}\right) ^{2}}d^{3}k\right) ^{2}+\frac{1}{m}\Pi _{\mathbf{P}}\cdot \left( g^{2}\int _{0}^{\kappa }\frac{\mathbf{k}}{2\left| \mathbf{k}\right| ^{3}\left( 1-\widehat{\mathbf{k}}\cdot \frac{\mathbf{P}_{1}}{m}\right) ^{2}}d^{3}k\right) + \)}\\
{\large }\\
{\large \( -\frac{1}{m}\left( g^{2}\int _{0}^{\kappa }\frac{\mathbf{k}}{2\left| \mathbf{k}\right| ^{3}\left( 1-\widehat{\mathbf{k}}\cdot \frac{\mathbf{P}_{1}}{m}\right) ^{2}}d^{3}k\right) \cdot \Pi _{\mathbf{P}}-\frac{1}{m}\frac{\left( \phi _{\mathbf{P}},\Pi _{\mathbf{P}}\phi _{\mathbf{P}}\right) }{\left\Vert \phi _{\mathbf{P}}\right\Vert ^{2}}\cdot \Pi _{\mathbf{P}}+\int ^{\infty }_{0}\left( \left| \mathbf{k}\right| -\mathbf{k}\cdot \frac{\mathbf{P}_{1}}{m}\right) b^{\dagger }\left( \mathbf{k}\right) b\left( \mathbf{k}\right) d^{3}k+ \)}\\
{\large }\\
{\large \( +costants \)} \\

\end{itemize}
{\large The transformed hamiltonian corresponds to:}\\
{\large }\\
{\large \( H_{\mathbf{P}}^{w}=\frac{1}{2m}\left( \Pi _{\mathbf{P}}-\frac{\left( \phi _{\mathbf{P}},\Pi _{\mathbf{P}}\phi _{\mathbf{P}}\right) }{\left\Vert \phi _{\mathbf{P}}\right\Vert ^{2}}\right) ^{2}+\int ^{\infty }_{0}\left( \left| \mathbf{k}\right| -\mathbf{k}\cdot \frac{\mathbf{P}_{1}}{m}\right) b^{\dagger }\left( \mathbf{k}\right) b\left( \mathbf{k}\right) d^{3}k+costants \)}\marginpar{
{\large (10)}{\large \par}
}\\
{\large }\\
{\large Now I distinguish the case \( \mathbf{P}=0 \) from \( \mathbf{P}\neq 0 \)
in order to rigorously apply the iterative procedure to the transformed hamiltonians
with an infrared cut-off \( \sigma _{j} \), written in a canonic form analogous
to the expression (10). In the first case, the coherent representation is explicit,
since \( \overline{\mathbf{P}^{mes}}=0 \) by symmetry and then \( \mathbf{P}_{1}=0 \),
while in the second case the procedure is more lengthy.}\\
{\large \par}

\subsection{Case \protect\( \mathbf{P}=0\protect \). \\
 }

{\large I perform the coherent transformation (7) on the hamiltonian \( H_{\mathbf{P}=0,\sigma _{j}} \)
with infrared cut-off \( \sigma _{j} \) :}\\
{\large }\\
{\large \( H^{w}_{\epsilon ^{\frac{j+1}{2}}}\equiv e^{-g\int _{\epsilon ^{\frac{j+1}{2}}}^{\kappa }\frac{b\left( \mathbf{k}\right) -b^{\dagger }\left( \mathbf{k}\right) }{\left| \mathbf{k}\right| }\frac{d^{3}k}{\sqrt{2\left| \mathbf{k}\right| }}}H_{\mathbf{P}=0,\epsilon ^{\frac{j+1}{2}}}e^{g\int _{\epsilon ^{\frac{j+1}{2}}}^{\kappa }\frac{b\left( \mathbf{k}\right) -b^{\dagger }\left( \mathbf{k}\right) }{\left| \mathbf{k}\right| }\frac{d^{3}k}{\sqrt{2\left| \mathbf{k}\right| }}}= \)}\\
{\large }\\
{\large \( =\frac{1}{2m}\left( \mathbf{P}^{mes}-g\int _{\epsilon ^{\frac{j+1}{2}}}^{\kappa }\frac{\mathbf{k}}{\sqrt{2}\left| \mathbf{k}\right| ^{\frac{3}{2}}}\left( b\left( \mathbf{k}\right) +b^{\dagger }\left( \mathbf{k}\right) \right) d^{3}k\right) ^{2}+\int \left| \mathbf{k}\right| b^{\dagger }\left( \mathbf{k}\right) b\left( \mathbf{k}\right) d^{3}k+c\left( j+1\right)  \)}\\
{\large }\\
{\large where \( c\left( j+1\right) \equiv -g^{2}\int ^{\kappa }_{\epsilon ^{\frac{j+1}{2}}}\frac{1}{2\left| \mathbf{k}\right| ^{2}}d^{3}k \).}\\
{\large }\\
{\large The domain of selfadjointness (s.a.) of the transformed hamiltonian
\( H^{w}_{\epsilon ^{\frac{j+1}{2}}} \) coincides with \( D^{b}\left( H_{\mathbf{P}=0,\epsilon ^{\frac{j+1}{2}}}\right)  \)(see
an analogous proof in {[}1{]})}\textbf{\emph{\large .}}\\
\textbf{\emph{\large }}\\
\textbf{\emph{\large }}\\
\textbf{\emph{\large Preliminaries}}\emph{\large }\\
{\large }\\
{\large From the results of the previous chapter and by unitarity I can conclude
that the following properties hold \( \forall j \) (these properties are assumed
in lemma A1, Appendix A):}\\
{\large }\\
{\large i) \( H^{w}_{\epsilon ^{\frac{j+1}{2}}}\mid _{F_{\epsilon ^{\frac{j+1}{2}}}^{+}} \)
has ground eigenvalue \( E_{\mathbf{P}=0}^{\epsilon ^{\frac{j+1}{2}}} \) with
the corresponding gap bigger than \( \frac{\epsilon ^{\frac{j+1}{2}}}{2} \);}\\
{\large }\\
{\large ii) \( H^{w}_{\epsilon ^{\frac{j+1}{2}}}\mid _{F_{\epsilon ^{\frac{j+2}{2}}}^{+}} \)has
ground eigenvalue \( E_{\mathbf{P}=0}^{\epsilon ^{\frac{j+1}{2}}} \)with the
corresponding gap bigger than \( \frac{3\epsilon ^{\frac{j+2}{2}}}{5} \).}\\
{\large }\\
{\large }\\
{\large Now, for the values of \( g \) allowed by lemma A1 and on the basis
of the results of the lemma, starting from \( \phi ^{\epsilon }=e^{-g\int _{\epsilon }^{\kappa }\frac{b\left( \mathbf{k}\right) -b^{\dagger }\left( \mathbf{k}\right) }{\left| \mathbf{k}\right| }\frac{d^{3}k}{\sqrt{2\left| \mathbf{k}\right| }}}\psi _{\mathbf{P}=0}^{\epsilon } \)
and using an iteration analogous to the one of chapter 1,} {\large I construct}
{\large the vector \( \phi ^{\epsilon ^{\frac{j+2}{2}}} \)applying the spectral
projector to \( \phi ^{\epsilon ^{\frac{j+1}{2}}} \). We can point out that
in the difference}{\large }\\
{\large }\\
 {\large \( \phi ^{\epsilon ^{\frac{j+2}{2}}}-\phi ^{\epsilon ^{\frac{j+1}{2}}}=-\frac{1}{2\pi i}\oint \sum _{n}\frac{1}{H^{w}_{\epsilon ^{\frac{j+1}{2}}}-E\left( j+1\right) }\left( -\left( \Delta H^{w}\right) ^{\epsilon ^{\frac{j+1}{2}}}_{\epsilon ^{\frac{j+2}{2}}}\frac{1}{H^{w}_{\epsilon ^{\frac{j+1}{2}}}-E\left( j+1\right) }\right) ^{n}\phi ^{\epsilon ^{\frac{j+1}{2}}}dE\left( j+1\right)  \)} \marginpar{
{\large (11)}{\large \par}
}{\large }\\
{\large the terms that we cannot evaluate (in norm) with a power} {\large of
the cut-off} {\large \( \epsilon ^{\frac{j+1}{2}} \) with} {\large positive
exponent} {\large are vectors for which for all the n factors in the} {\large difference
\( \left( \Delta H^{w}\right) ^{\epsilon ^{\frac{j+1}{2}}}_{\epsilon ^{\frac{j+2}{2}}}= \)}\\
{\large \( =H^{w}_{\epsilon ^{\frac{j+2}{2}}}+c\left( j+1\right) -c\left( j+2\right) -H^{w}_{\epsilon ^{\frac{j+1}{2}}} \)}
{\large only the ``mixed'' terms} {\large are present:} {\large }\\
{\large }\\
{\large \( \frac{g}{2m}\int ^{\epsilon ^{\frac{j+1}{2}}}_{\epsilon ^{\frac{j+2}{2}}}\frac{\mathbf{k}}{\sqrt{2}\left| \mathbf{k}\right| ^{\frac{3}{2}}}\left( b\left( \mathbf{k}\right) +b^{\dagger }\left( \mathbf{k}\right) \right) d^{3}k\cdot \Pi _{\epsilon ^{\frac{j+1}{2}}}+\Pi _{\epsilon ^{\frac{j+1}{2}}}\cdot \frac{g}{2m}\int ^{\epsilon ^{\frac{j+1}{2}}}_{\epsilon ^{\frac{j+2}{2}}}\frac{\mathbf{k}}{\sqrt{2}\left| \mathbf{k}\right| ^{\frac{3}{2}}}\left( b\left( \mathbf{k}\right) +b^{\dagger }\left( \mathbf{k}\right) \right) d^{3}k \).}\\
{\large }\\
{\large If there are these terms, the estimate of the following norm, given
in lemma A1, }\\
{\large }\\
{\large \( \left\Vert \left( \frac{1}{H^{w}_{\epsilon ^{\frac{j+1}{2}}}-E\left( j+1\right) }\right) ^{\frac{1}{2}}\left( -\left( \Delta H^{w}\right) ^{\epsilon ^{\frac{j+1}{2}}}_{\epsilon ^{\frac{j+2}{2}}}\right) \left( \frac{1}{H^{w}_{\epsilon ^{\frac{j+1}{2}}}-E\left( j+1\right) }\right) ^{\frac{1}{2}}\right\Vert  \)}\\
{\large }\\
{\large is only of order 1.}\\
{\large }\\
{\large We can have a more precise estimate of the norm of \( \phi ^{\epsilon ^{\frac{j+2}{2}}}-\phi ^{\epsilon ^{\frac{j+1}{2}}} \)
examining the first factor on the right of the product (11)}\\
{\large }\\
{\large \( \left( \frac{1}{H^{w}_{\epsilon ^{\frac{j+1}{2}}}-E\left( j+1\right) }\right) ^{\frac{1}{2}}\left( -\left( \Delta H^{w}\right) ^{\epsilon ^{\frac{j+1}{2}}}_{\epsilon ^{\frac{j+2}{2}}}\right) \left( \frac{1}{H^{w}_{\epsilon ^{\frac{j+1}{2}}}-E\left( j+1\right) }\right) ^{\frac{1}{2}}\phi ^{\epsilon ^{\frac{j+1}{2}}} \)}\\
{\large }\\
{\large and noting that, if for the mixed terms }\\
{\large }\\
{\large \( \frac{g}{2m}\sum _{i}\left( \frac{1}{H^{w}_{\epsilon ^{\frac{j+1}{2}}}-E\left( j+1\right) }\right) ^{\frac{1}{2}}\left( \int ^{\epsilon ^{\frac{j+1}{2}}}_{\epsilon ^{\frac{j+2}{2}}}\frac{k^{i}}{\sqrt{2}\left| \mathbf{k}\right| ^{\frac{3}{2}}}\left( b\left( \mathbf{k}\right) +b^{\dagger }\left( \mathbf{k}\right) \right) d^{3}k\cdot \Pi ^{i}_{\epsilon ^{\frac{j+1}{2}}}\right) \left( \frac{1}{H^{w}_{\epsilon ^{\frac{j+1}{2}}}-E\left( j+1\right) }\right) ^{\frac{1}{2}}\phi ^{\epsilon ^{\frac{j+1}{2}}} \)}\\
{\large }\\
{\large the following inequality were true}\\
{\large }\\
{\large \( \frac{g}{2m}\left\Vert \sum _{i}\left( \frac{1}{H^{w}_{\epsilon ^{\frac{j+1}{2}}}-E\left( j+1\right) }\right) ^{\frac{1}{2}}\left( \int ^{\epsilon ^{\frac{j+1}{2}}}_{\epsilon ^{\frac{j+2}{2}}}\frac{k^{i}}{\sqrt{2}\left| \mathbf{k}\right| ^{\frac{3}{2}}}\left( b\left( \mathbf{k}\right) +b^{\dagger }\left( \mathbf{k}\right) \right) d^{3}k\cdot \Pi ^{i}_{\epsilon ^{\frac{j+1}{2}}}\right) \left( \frac{1}{H^{w}_{\epsilon ^{\frac{j+1}{2}}}-E\left( j+1\right) }\right) ^{\frac{1}{2}}\phi ^{\epsilon ^{\frac{j+1}{2}}}\right\Vert \leq  \)}\marginpar{

} {\large }\\
{\large \( \leq \frac{1}{4}\cdot \frac{\epsilon ^{\frac{j+1}{8}}}{2} \),}\\
{\large }\\
{\large then we would have an estimate of order \( \epsilon ^{\frac{j+1}{8}} \)
for \( \left\Vert \phi ^{\epsilon ^{\frac{j+2}{2}}}-\phi ^{\epsilon ^{\frac{j+1}{2}}}\right\Vert  \).
This is due to the fact that, using the lemma A1, the norm of the sum of the
quadratic terms (for a proper g) can be bounded by \( \frac{\epsilon ^{\frac{j+1}{8}}}{40} \)
while the norm of the other factors of the product is of order 1, in particular
less than \( \frac{1}{12} \). Therefore we would have}\\
{\large }\\
{\large \( \left\Vert \sum ^{\infty }_{n=1}\frac{1}{H^{w}_{\epsilon ^{\frac{j+1}{2}}}-E\left( j+1\right) }\left( -\left( \Delta H^{w}\right) ^{\epsilon ^{\frac{j+1}{2}}}_{\epsilon ^{\frac{j+2}{2}}}\frac{1}{H^{w}_{\epsilon ^{\frac{j+1}{2}}}-E\left( j+1\right) }\right) ^{n}\phi ^{\epsilon ^{\frac{j+1}{2}}}\right\Vert \leq  \)}\\
{\large \( \leq \left| \left( \frac{1}{E^{\epsilon ^{\frac{j+1}{2}}}-E}\right) \right| ^{\frac{1}{2}}\cdot \left\Vert \left( \frac{1}{H_{\epsilon ^{\frac{j+1}{2}}}^{w}-E\left( j+1\right) }\right) \right\Vert ^{\frac{1}{2}}_{F_{\epsilon ^{\frac{j+2}{2}}}^{+}}\left( \epsilon ^{\frac{j+1}{2}}\right) ^{\frac{1}{4}}\sum ^{\infty }_{n=1}\left( \frac{11}{40}\right) \left( \frac{1}{12}\right) ^{n-1}\leq  \)}\\
{\large }\\
{\large \( \leq \left| \left( \frac{1}{E^{\epsilon ^{\frac{j+1}{2}}}-E\left( j+1\right) }\right) \right| \cdot \left( \epsilon ^{\frac{j+1}{2}}\right) ^{\frac{1}{4}} \)}\\
{\large }\\
{\large }\\
{\large As it will be shown in corollary 2.4 (in the general case \( \mathbf{P}\in \Sigma  \)),
an estimate like \( \left\Vert \phi ^{\epsilon ^{\frac{j+2}{2}}}-\phi ^{\epsilon ^{\frac{j+1}{2}}}\right\Vert \leq \epsilon ^{\frac{j+1}{8}} \)
implies the convergence of the sequence \( \left\{ \phi ^{\epsilon ^{\frac{j+1}{2}}}\right\}  \).
Therefore it is crucial to prove the following inequality: }\\
{\large }\\
{\large \( \frac{g}{2m}\left\Vert \sum _{i}\left( \frac{1}{H^{w}_{\epsilon ^{\frac{j+1}{2}}}-E\left( j+1\right) }\right) ^{\frac{1}{2}}\left( \int ^{\epsilon ^{\frac{j+1}{2}}}_{\epsilon ^{\frac{j+2}{2}}}\frac{k^{i}}{\sqrt{2}\left| \mathbf{k}\right| ^{\frac{3}{2}}}\left( b\left( \mathbf{k}\right) +b^{\dagger }\left( \mathbf{k}\right) \right) d^{3}k\cdot \Pi ^{i}_{\epsilon ^{\frac{j+1}{2}}}\right) \left( \frac{1}{H^{w}_{\epsilon ^{\frac{j+1}{2}}}-E\left( j+1\right) }\right) ^{\frac{1}{2}}\phi ^{\epsilon ^{\frac{j+1}{2}}}\right\Vert = \)}\marginpar{

}{\large }\\
{\large \( =\frac{g}{2m}\left\Vert \sum _{i}\left( \frac{1}{H^{w}_{\epsilon ^{\frac{j+1}{2}}}-E\left( j+1\right) }\right) ^{\frac{1}{2}}\left( \int ^{\epsilon ^{\frac{j+1}{2}}}_{\epsilon ^{\frac{j+2}{2}}}\frac{k^{i}}{\sqrt{2}\left| \mathbf{k}\right| ^{\frac{3}{2}}}b^{\dagger }\left( \mathbf{k}\right) d^{3}k\cdot \Pi ^{i}_{\epsilon ^{\frac{j+1}{2}}}\right) \left( \frac{1}{H^{w}_{\epsilon ^{\frac{j+1}{2}}}-E\left( j+1\right) }\right) ^{\frac{1}{2}}\phi ^{\epsilon ^{\frac{j+1}{2}}}\right\Vert \leq  \)}\\
{\large }\\
{\large \( \leq \frac{1}{4}\cdot \frac{\epsilon ^{\frac{j+1}{8}}}{2} \).}\\
{\large }\\
{\large }\\
{\large }\\
{\large }\\
\textbf{\large Lemma 2.1}{\large }\\
{\large }\\
{\large The following inequalities hold:}\\
{\footnotesize }\\
\textbf{\large I)\( \left\Vert \left( \frac{1}{H^{w}_{\epsilon ^{\frac{j+1}{2}}}-E\left( j+1\right) }\right) ^{\frac{1}{2}}\int ^{\epsilon ^{\frac{j+1}{2}}}_{\epsilon ^{\frac{j+2}{2}}}\frac{k^{i}}{\sqrt{2}\left| \mathbf{k}\right| ^{\frac{3}{2}}}b^{\dagger }\left( \mathbf{k}\right) d^{3}k\cdot \Pi ^{i}_{\epsilon ^{\frac{j+1}{2}}}\phi ^{\epsilon ^{\frac{j+1}{2}}}\right\Vert ^{2}= \)}{\large }\\
{\large \( =\left( \int ^{\epsilon ^{\frac{j+1}{2}}}_{\epsilon ^{\frac{j+2}{2}}}\frac{k^{i}}{\sqrt{2}\left| \mathbf{k}\right| }b^{\dagger }\left( \mathbf{k}\right) d^{3}k\, \Pi _{\epsilon ^{\frac{j+1}{2}}}^{i}\phi ^{\epsilon ^{\frac{j+1}{2}}}\! ,\left| \frac{1}{H_{\epsilon ^{\frac{j+1}{2}}}^{w}-E\left( j+1\right) }\right| \! \int ^{\epsilon ^{\frac{j+1}{2}}}_{\epsilon ^{\frac{j+2}{2}}}\frac{k^{i}}{\sqrt{2}\left| \mathbf{k}\right| }b^{\dagger }\left( \mathbf{k}\right) d^{3}k\, \Pi _{\epsilon ^{\frac{j+1}{2}}}^{i}\phi ^{\epsilon ^{\frac{j+1}{2}}}\right) \leq  \)}\\
{\large }\\
{\large \( \leq \sqrt{112}\left| \left( \int ^{\epsilon ^{\frac{j+1}{2}}}_{\epsilon ^{\frac{j+2}{2}}}\frac{k^{i}}{\sqrt{2}\left| \mathbf{k}\right| }b^{\dagger }\left( \mathbf{k}\right) d^{3}k\, \Pi _{\epsilon ^{\frac{j+1}{2}}}^{i}\phi ^{\epsilon ^{\frac{j+1}{2}}}\! ,\! \frac{1}{H_{\epsilon ^{\frac{j+1}{2}}}^{w}-E\left( j+1\right) }\! \int ^{\epsilon ^{\frac{j+1}{2}}}_{\epsilon ^{\frac{j+2}{2}}}\frac{k^{i}}{\sqrt{2}\left| \mathbf{k}\right| }b^{\dagger }\left( \mathbf{k}\right) d^{3}k\, \Pi _{\epsilon ^{\frac{j+1}{2}}}^{i}\phi ^{\epsilon ^{\frac{j+1}{2}}}\right) \right|  \)}\\
{\large }\\
{\large }\\
\textbf{\large II)} {\large \( \left( \Pi _{\epsilon ^{\frac{j+1}{2}}}^{i}\phi ^{\epsilon ^{\frac{j+1}{2}}}\! ,\left| \frac{1}{H_{\epsilon ^{\frac{j+1}{2}}}^{w}+\left| \mathbf{k}\right| -E\left( j+1\right) }\right| \! \Pi _{\epsilon ^{\frac{j+1}{2}}}^{i}\phi ^{\epsilon ^{\frac{j+1}{2}}}\right) \leq  \)}\\
{\large \( \leq \sqrt{1+\left( \frac{11\sqrt{\epsilon }}{10-11\sqrt{\epsilon }}\right) ^{2}}\cdot \left| \left( \Pi _{\epsilon ^{\frac{j+1}{2}}}^{i}\phi ^{\epsilon ^{\frac{j+1}{2}}}\! ,\! \frac{1}{H_{\epsilon ^{\frac{j+1}{2}}}^{w}-E\left( j+1\right) }\! \Pi _{\epsilon ^{\frac{j+1}{2}}}^{i}\phi ^{\epsilon ^{\frac{j+1}{2}}}\right) \right|  \)}\\
\\
{\large }\\
{\large Proof}\\
{\large }\\
{\large I define the wave functions \( \zeta _{I}\left( z\right) ,\zeta _{II}\left( z\right)  \)
of \( \int ^{\epsilon ^{\frac{j+1}{2}}}_{\epsilon ^{\frac{j+2}{2}}}\frac{k^{i}}{\sqrt{2}\left| \mathbf{k}\right| }b^{\dagger }\left( \mathbf{k}\right) d^{3}k\, \Pi _{\epsilon ^{\frac{j+1}{2}}}^{i}\phi ^{\epsilon ^{\frac{j+1}{2}}} \)and
\( \Pi _{\epsilon ^{\frac{j+1}{2}}}^{i}\phi ^{\epsilon ^{\frac{j+1}{2}}} \),
in the spectral variable of \( H_{\epsilon ^{\frac{j+1}{2}}}^{w}-ReE\left( j+1\right)  \).
Note that: }\\
{\large }\\
{\large - the operator \( H_{\epsilon ^{\frac{j+1}{2}}}^{w}-ReE\left( j+1\right)  \),
applied to the vector \( \int ^{\epsilon ^{\frac{j+1}{2}}}_{\epsilon ^{\frac{j+2}{2}}}\frac{k^{i}}{\sqrt{2}\left| \mathbf{k}\right| }b^{\dagger }\left( \mathbf{k}\right) d^{3}k\, \Pi _{\epsilon ^{\frac{j+1}{2}}}^{i}\phi ^{\epsilon ^{\frac{j+1}{2}}} \),
takes values bigger or equal to \( \frac{1}{20}\epsilon ^{\frac{j+2}{2}\, } \)
\( \left( =\frac{3}{5}\epsilon ^{\frac{j+2}{2}}-\frac{11}{20}\epsilon ^{\frac{j+2}{2}}\right)  \)
because of lemma 1.2;}\\
{\large }\\
{\large - the operator \( H_{\epsilon ^{\frac{j+1}{2}}}^{w}-ReE\left( j+1\right)  \)
takes values bigger or equal to~~\( \frac{10-11\sqrt{\epsilon }}{20}\cdot \epsilon ^{\frac{j+1}{2}}\!  \)
\( \left( =\frac{1}{2}\epsilon ^{\frac{j+1}{2}}-\frac{11}{20}\epsilon ^{\frac{j+2}{2}}\right)  \)
if applied to the vector \( \Pi _{\epsilon ^{\frac{j+1}{2}}}^{i}\phi ^{\epsilon ^{\frac{j+1}{2}}} \)because
of theorem 1.5}\\
{\large }\\
{\large }\\
{\large I write the scalar products I) and II) using the spectral representation
of the operator \( H_{\epsilon ^{\frac{j+1}{2}}}^{w}-ReE\left( j+1\right)  \)
and getting rid of the remaining degrees of freedom. In the chosen spectral
representation, the following inequalities are evident:}\\
{\large }\\
{\large \( \left| \int \frac{\left| \zeta _{I,II}\left( z\right) \right| ^{2}}{z-iIm\left( E\left( j+1\right) \right) }dz\right| =\left\{ \left| \int \frac{z\left| \zeta _{I,II}\left( z\right) \right| ^{2}}{z^{2}+\left[ Im\left( E\left( j+1\right) \right) \right] ^{2}}dz\right| ^{2}+\left[ Im\left( E\left( j+1\right) \right) \right] ^{2}\left| \int \frac{\left| \zeta _{I,II}\left( z\right) \right| ^{2}}{z^{2}+\left[ Im\left( E\left( j+1\right) \right) \right] ^{2}}dz\right| ^{2}\right\} ^{\frac{1}{2}}\geq  \)}\\
{\large }\\
{\small \( \geq \left| \int \frac{\left| \zeta _{I,II}\left( z\right) \right| ^{2}}{\sqrt{z^{2}+\left[ Im\left( E\left( j+1\right) \right) \right] ^{2}}}\cdot \frac{z}{\sqrt{z^{2}+\left[ Im\left( E\left( j+1\right) \right) \right] ^{2}}}dz\right| \geq \left| \int \frac{\left| \zeta _{I,II}\left( z\right) \right| ^{2}}{\sqrt{z^{2}+\left[ Im\left( E\left( j+1\right) \right) \right] ^{2}}}\cdot \frac{z}{\sqrt{z^{2}+\left| E\left( j+1\right) \right| ^{2}}}dz\right| \geq  \)}{\large }\\
{\large }\\
{\large \( \geq \frac{z_{min}}{\sqrt{z^{2}_{min}+\left| E\left( j+1\right) \right| ^{2}}}\cdot \left| \int \frac{\left| \zeta _{I,II}\left( z\right) \right| ^{2}}{\sqrt{z^{2}+\left[ Im\left( E\left( j+1\right) \right) \right] ^{2}}}dz\right| \geq  \)}\\
{\large }\\
{\large \( \geq \frac{z_{min}}{\sqrt{z^{2}_{min}+\left| E\left( j+1\right) \right| ^{2}}}\cdot \left| \int \frac{\left| \zeta _{I,II}\left( z\right) \right| ^{2}}{\sqrt{z^{2}+\left| \mathbf{k}\right| ^{2}+2\left| \mathbf{k}\right| z+\left[ Im\left( E\left( j+1\right) \right) \right] ^{2}}}dz\right|  \)}\\
{\large }\\
{\large }\\
{\large It follows that:}\\
{\large \par}

\begin{itemize}
\item {\large in the case I), being \( z_{min}\geq \frac{1}{20}\epsilon ^{\frac{j+2}{2}} \),
\( \frac{z_{min}}{\sqrt{z^{2}_{min}+\left| E\left( j+1\right) \right| ^{2}}}\geq \frac{1}{\sqrt{112}} \)
}\\
{\large }\\
{\large \( \left| \int \frac{\left| \zeta _{I}\left( z\right) \right| ^{2}}{\sqrt{z^{2}+\left[ Im\left( E\left( j+1\right) \right) \right] ^{2}}}dz\right| \leq \sqrt{112}\cdot \left| \int \frac{\left| \zeta _{I}\left( z\right) \right| ^{2}}{z-iIm\left( E\left( j+1\right) \right) }dz\right|  \)}{\large \par}
\item {\large in the case II), being \( z_{min}\geq \frac{10-11\sqrt{\epsilon }}{20}\cdot \epsilon ^{\frac{j+1}{2}} \),
\( \frac{z_{min}}{\sqrt{z^{2}_{min}+\left| E\left( j+1\right) \right| ^{2}}}\geq \frac{1}{\sqrt{1+\left( \frac{11\sqrt{\epsilon }}{10-11\sqrt{\epsilon }}\right) ^{2}}} \)
}\\
{\large }\\
{\large \( \left| \int \frac{\left| \zeta _{II}\left( z\right) \right| ^{2}}{\sqrt{z^{2}+\left| \mathbf{k}\right| ^{2}+2\left| \mathbf{k}\right| z+\left[ Im\left( E\left( j+1\right) \right) \right] ^{2}}}dz\right| \leq \sqrt{1+\left( \frac{11\sqrt{\epsilon }}{10-11\sqrt{\epsilon }}\right) ^{2}}\cdot \left| \int \frac{\left| \zeta _{II}\left( z\right) \right| ^{2}}{z-iIm\left( E\left( j+1\right) \right) }dz\right| = \)}\\
{\large }\\
{\large \( =Q\left( \epsilon \right) \cdot \left| \int \frac{\left| \zeta _{II}\left( z\right) \right| ^{2}}{z-iIm\left( E\left( j+1\right) \right) }dz\right|  \)~~}
{\large where \( Q\left( \epsilon \right) \equiv \sqrt{1+\left( \frac{11\sqrt{\epsilon }}{10-11\sqrt{\epsilon }}\right) ^{2}} \)}.\\

\end{itemize}
\textbf{\large Lemma 2.2}{\large }\\
{\large }\\
{\large Taking into account lemma 2.1, we have that:}\\
{\footnotesize }\\
\textbf{\large \( \left\Vert \left( \frac{1}{H^{w}_{\epsilon ^{\frac{j+1}{2}}}-E\left( j+1\right) }\right) ^{\frac{1}{2}}\int ^{\epsilon ^{\frac{j+1}{2}}}_{\epsilon ^{\frac{j+2}{2}}}\frac{k^{i}}{\sqrt{2}\left| \mathbf{k}\right| ^{\frac{3}{2}}}b^{\dagger }\left( \mathbf{k}\right) d^{3}k\cdot \Pi ^{i}_{\epsilon ^{\frac{j+1}{2}}}\left( \frac{1}{H^{w}_{\epsilon ^{\frac{j+1}{2}}}-E\left( j+1\right) }\right) ^{\frac{1}{2}}\phi ^{\epsilon ^{\frac{j+1}{2}}}\right\Vert ^{2}\leq  \)}{\large }\\
{\large }\\
{\large \( \leq 2\cdot Q\left( \epsilon \right) \cdot \sqrt{112}\cdot \left| \frac{1}{E^{\epsilon ^{\frac{j+1}{2}}}-E\left( j+1\right) }\right| \cdot \int ^{\epsilon ^{\frac{j+1}{2}}}_{\epsilon ^{\frac{j+2}{2}}}\frac{k^{i^{2}}}{2\left| \mathbf{k}\right| ^{3}}\left| \left( \Pi _{\epsilon ^{\frac{j+1}{2}}}^{i}\phi ^{\epsilon ^{\frac{j+1}{2}}}\! ,\! \left( \frac{1}{H^{w}_{\epsilon ^{\frac{j+1}{2}}}-E\left( j+1\right) }\right) \! \Pi _{\epsilon ^{\frac{j+1}{2}}}^{i}\phi ^{\epsilon ^{\frac{j+1}{2}}}\right) \right| d^{3}k \)}\\
\\
{\large }\\
{\large Proof}\\
{\large }\\
\textbf{\large \( \left\Vert \left( \frac{1}{H^{w}_{\epsilon ^{\frac{j+1}{2}}}-E\left( j+1\right) }\right) ^{\frac{1}{2}}\int ^{\epsilon ^{\frac{j+1}{2}}}_{\epsilon ^{\frac{j+2}{2}}}\frac{k^{i}}{\sqrt{2}\left| \mathbf{k}\right| ^{\frac{3}{2}}}b^{\dagger }\left( \mathbf{k}\right) d^{3}k\cdot \Pi ^{i}_{\epsilon ^{\frac{j+1}{2}}}\left( \frac{1}{H^{w}_{\epsilon ^{\frac{j+1}{2}}}-E\left( j+1\right) }\right) ^{\frac{1}{2}}\phi ^{\epsilon ^{\frac{j+1}{2}}}\right\Vert ^{2}= \)}\\
\\
{\large \( =\left| \frac{1}{E^{\epsilon ^{\frac{j+1}{2}}}-E\left( j+1\right) }\right| \cdot \left( \left( \frac{1}{H_{\epsilon ^{\frac{j+1}{2}}}^{w}-E\left( j+1\right) }\right) ^{\frac{1}{2}}\int ^{\epsilon ^{\frac{j+1}{2}}}_{\epsilon ^{\frac{j+2}{2}}}\frac{k^{i}}{\sqrt{2}\left| \mathbf{k}\right| ^{\frac{3}{2}}}b^{\dagger }\left( \mathbf{k}\right) d^{3}k\Pi _{\epsilon ^{\frac{j+1}{2}}}^{i}\phi ^{\epsilon ^{\frac{j+1}{2}}},\right.  \)}\\
{\large }\\
{\large \( ,\left. \left( \frac{1}{H_{\epsilon ^{\frac{j+1}{2}}}^{w}-E\left( j+1\right) }\right) ^{\frac{1}{2}}\cdot \int ^{\epsilon ^{\frac{j+1}{2}}}_{\epsilon ^{\frac{j+2}{2}}}\frac{k^{i}}{\sqrt{2}\left| \mathbf{k}\right| ^{\frac{3}{2}}}b^{\dagger }\left( \mathbf{k}\right) d^{3}k\Pi _{\epsilon ^{\frac{j+1}{2}}}^{i}\phi ^{\epsilon ^{\frac{j+1}{2}}}\right) = \)}\\
{\large }\\
{\small \( =\left| \frac{1}{E^{\epsilon ^{\frac{j+1}{2}}}-E\left( j+1\right) }\right| \cdot \left( \int ^{\epsilon ^{\frac{j+1}{2}}}_{\epsilon ^{\frac{j+2}{2}}}\frac{k^{i}}{\sqrt{2}\left| \mathbf{k}\right| ^{\frac{3}{2}}}b^{\dagger }\left( \mathbf{k}\right) d^{3}k\Pi _{\epsilon ^{\frac{j+1}{2}}}^{i}\phi ^{\epsilon ^{\frac{j+1}{2}}},\left| \frac{1}{H_{\epsilon ^{\frac{j+1}{2}}}^{w}-E\left( j+1\right) }\right| \int ^{\epsilon ^{\frac{j+1}{2}}}_{\epsilon ^{\frac{j+2}{2}}}\frac{k^{i}}{\sqrt{2}\left| \mathbf{k}\right| ^{\frac{3}{2}}}b^{\dagger }\left( \mathbf{k}\right) d^{3}k\Pi _{\epsilon ^{\frac{j+1}{2}}}^{i}\phi ^{\epsilon ^{\frac{j+1}{2}}}\right) \leq  \)}{\large }\\
{\large }\\
{\small \( \leq \sqrt{112}\left| \frac{1}{E^{\epsilon ^{\frac{j+1}{2}}}-E\left( j+1\right) }\right| \cdot \left| \left( \int ^{\epsilon ^{\frac{j+1}{2}}}_{\epsilon ^{\frac{j+2}{2}}}\frac{k^{i}}{\sqrt{2}\left| \mathbf{k}\right| ^{\frac{3}{2}}}b^{\dagger }\left( \mathbf{k}\right) d^{3}k\Pi _{\epsilon ^{\frac{j+1}{2}}}^{i}\phi ^{\epsilon ^{\frac{j+1}{2}}}\, ,\, \left( \frac{1}{H_{\epsilon ^{\frac{j+1}{2}}}^{w}-E\left( j+1\right) }\right) \int ^{\epsilon ^{\frac{j+1}{2}}}_{\epsilon ^{\frac{j+2}{2}}}\frac{k^{i}}{\sqrt{2}\left| \mathbf{k}\right| ^{\frac{3}{2}}}b^{\dagger }\left( \mathbf{k}\right) d^{3}k\Pi _{\epsilon ^{\frac{j+1}{2}}}^{i}\phi ^{\epsilon ^{\frac{j+1}{2}}}\right) \right|  \)
\[
\]
}{\large }\\
{\large }\\
{\large Starting from the expression \( H^{w}_{\epsilon ^{\frac{j+1}{2}}}=\frac{1}{2m}\Pi ^{2}_{\epsilon ^{\frac{j+1}{2}}}+\int \left| \mathbf{k}\right| b^{\dagger }\left( \mathbf{k}\right) b\left( \mathbf{k}\right) d^{3}k+c\left( j+1\right)  \),
for \( \mathbf{k}:\, \left| \mathbf{k}\right| \leq \epsilon ^{\frac{j+1}{2}} \),
the following identity holds in distributional sense }\\
{\large }\\
{\large \( \left( \frac{1}{H^{w}_{\epsilon ^{\frac{j+1}{2}}}-E\left( j+1\right) }\right) b^{\dagger }\left( \mathbf{k}\right) =b^{\dagger }\left( \mathbf{k}\right) \left( \frac{1}{\frac{1}{2m}\left( \Pi _{\epsilon ^{\frac{j+1}{2}}}+\mathbf{k}\right) ^{2}+\int \left| \mathbf{k}\right| b^{\dagger }\left( \mathbf{k}\right) b\left( \mathbf{k}\right) +\left| \mathbf{k}\right| +c\left( j+1\right) -E\left( j+1\right) }\right) = \)}\\
{\large }\\
{\large \( =b^{\dagger }\left( \mathbf{k}\right) \left( \frac{1}{\frac{1}{2m}\left( \Pi ^{2}_{\epsilon ^{\frac{j+1}{2}}}+2\mathbf{k}\cdot \Pi _{\epsilon ^{\frac{j+1}{2}}}+\mathbf{k}^{2}\right) +\int \left| \mathbf{k}\right| b^{\dagger }\left( \mathbf{k}\right) b\left( \mathbf{k}\right) +\left| \mathbf{k}\right| +c\left( j+1\right) -E\left( j+1\right) }\right)  \)}\\
{\large }\\
{\large And also, for the assumptions made on \( m \), \( \epsilon  \) and
\( \left| E^{\epsilon ^{\frac{j+1}{2}}}-E\left( j+1\right) \right|  \), and
since \( \epsilon ^{\frac{j+2}{2}}\leq \left| \mathbf{k}\right| \leq \epsilon ^{\frac{j+1}{2}} \),}
{\large the following bound holds in the subspace \( F_{\epsilon ^{\frac{j+2}{2}}}^{+} \)}\\
{\large }\\
{\large \( \left\Vert \left( \frac{1}{H^{w}_{\epsilon ^{\frac{j+1}{2}}}+\left| \mathbf{k}\right| -E\left( j+1\right) }\right) ^{\frac{1}{2}}\left( \frac{1}{m}\mathbf{k}\cdot \Pi _{\epsilon ^{\frac{j+1}{2}}}+\frac{\mathbf{k}^{2}}{2m}\right) \left( \frac{1}{H^{w}_{\epsilon ^{\frac{j+1}{2}}}+\left| \mathbf{k}\right| -E\left( j+1\right) }\right) ^{\frac{1}{2}}\right\Vert _{F_{\epsilon ^{\frac{j+2}{2}}}^{+}}\leq \frac{1}{2} \)}\\
{\large }\\
{\large which well defines the series expansion}\\
\\
{\large \( b^{\dagger }\left( \mathbf{k}\right) \left( \frac{1}{\frac{1}{2m}\left( \Pi ^{2}_{\epsilon ^{\frac{j+1}{2}}}+2\mathbf{k}\cdot \Pi _{\epsilon ^{\frac{j+1}{2}}}+\mathbf{k}^{2}\right) +\int \left| \mathbf{k}\right| b^{\dagger }\left( \mathbf{k}\right) b\left( \mathbf{k}\right) +\left| \mathbf{k}\right| -E\left( j+1\right) }\right) = \)}\\
{\large }\\
{\large \( =b^{\dagger }\left( \mathbf{k}\right) \left( \left( \frac{1}{H^{w}_{\epsilon ^{\frac{j+1}{2}}}+\left| \mathbf{k}\right| -E\left( j+1\right) }\right) \sum ^{+\infty }_{n=0}\left( -\left( \frac{1}{m}\mathbf{k}\cdot \Pi _{\epsilon ^{\frac{j+1}{2}}}+\frac{\mathbf{k}^{2}}{2m}\right) \frac{1}{H^{w}_{\epsilon ^{\frac{j+1}{2}}}+\left| \mathbf{k}\right| -E\left( j+1\right) }\right) ^{n}\right)  \)}
{\large }{\large }\\
{\large from which}\\
{\large }\\
{\small \( \left( \int ^{\epsilon ^{\frac{j+1}{2}}}_{\epsilon ^{\frac{j+2}{2}}}\frac{k^{i}}{\sqrt{2}\left| \mathbf{k}\right| ^{\frac{3}{2}}}b^{\dagger }\left( \mathbf{k}\right) d^{3}k\Pi _{\epsilon ^{\frac{j+1}{2}}}^{i}\phi ^{\epsilon ^{\frac{j+1}{2}}}\, ,\, \left( \frac{1}{H_{\epsilon ^{\frac{j+1}{2}}}^{w}-E\left( j+1\right) }\right) \int ^{\epsilon ^{\frac{j+1}{2}}}_{\epsilon ^{\frac{j+2}{2}}}\frac{k^{i}}{\sqrt{2}\left| \mathbf{k}\right| ^{\frac{3}{2}}}b^{\dagger }\left( \mathbf{k}\right) d^{3}k\Pi _{\epsilon ^{\frac{j+1}{2}}}^{i}\phi ^{\epsilon ^{\frac{j+1}{2}}}\right) = \)}\marginpar{
{\large (12)}{\large \par}
}{\scriptsize }\\
{\small \( =\int ^{\epsilon ^{\frac{j+1}{2}}}_{\epsilon ^{\frac{j+2}{2}}}\frac{k^{i^{2}}}{2\left| \mathbf{k}\right| ^{3}}\left( \Pi _{\epsilon ^{\frac{j+1}{2}}}^{i}\phi ^{\epsilon ^{\frac{j+1}{2}}}\, ,\, \left( \frac{1}{H^{w}_{\epsilon ^{\frac{j+1}{2}}}+\left| \mathbf{k}\right| -E\left( j+1\right) }\right) \sum ^{\infty }_{n=0}\left( -\left( \frac{1}{m}\mathbf{k}\cdot \Pi _{\epsilon ^{\frac{j+1}{2}}}+\frac{\mathbf{k}^{2}}{2m}\right) \left( \frac{1}{H^{w}_{\epsilon ^{\frac{j+1}{2}}}+\left| \mathbf{k}\right| -E\left( j+1\right) }\right) \right) ^{n}\Pi _{\epsilon ^{\frac{j+1}{2}}}^{i}\phi ^{\epsilon ^{\frac{j+1}{2}}}\right) d^{3}k= \)}\\
{\small }\\
{\small \( =\sum ^{\infty }_{n=0}\int ^{\epsilon ^{\frac{j+1}{2}}}_{\epsilon ^{\frac{j+2}{2}}}\frac{k^{i^{2}}}{2\left| \mathbf{k}\right| ^{3}}\left( \Pi _{\epsilon ^{\frac{j+1}{2}}}^{i}\phi ^{\epsilon ^{\frac{j+1}{2}}}\, ,\, \left( \frac{1}{H^{w}_{\epsilon ^{\frac{j+1}{2}}}+\left| \mathbf{k}\right| -E\left( j+1\right) }\right) \left( -\left( \frac{1}{m}\mathbf{k}\cdot \Pi _{\epsilon ^{\frac{j+1}{2}}}+\frac{\mathbf{k}^{2}}{2m}\right) \left( \frac{1}{H^{w}_{\epsilon ^{\frac{j+1}{2}}}+\left| \mathbf{k}\right| -E\left( j+1\right) }\right) \right) ^{n}\Pi _{\epsilon ^{\frac{j+1}{2}}}^{i}\phi ^{\epsilon ^{\frac{j+1}{2}}}\right) d^{3}k \)}
{\small }{\small }\\
{\large }\\
{\large }\\
{\large I prove that the module of the \( n^{th} \) term of the series can
be reduced to the one of the first term, so that the whole sum is of the same
order of the term at \( n=0 \).}\\
{\large }\\
{\large Exploiting the Schwartz inequality and the identity} \\
\\
\( \left( \frac{1}{H^{w}_{\epsilon ^{\frac{j+1}{2}}}+\left| \mathbf{k}\right| -E\left( j+1\right) }\right) \left( \left( -\right) \left( \frac{1}{m}\mathbf{k}\cdot \Pi _{\epsilon ^{\frac{j+1}{2}}}+\frac{\mathbf{k}^{2}}{2m}\right) \left( \frac{1}{H^{w}_{\epsilon ^{\frac{j+1}{2}}}+\left| \mathbf{k}\right| -E\left( j+1\right) }\right) \right) ^{n}= \)\\
\\
\( =\left( \frac{1}{H^{w}_{\epsilon ^{\frac{j+1}{2}}}+\left| \mathbf{k}\right| -E\left( j+1\right) }\right) ^{\frac{1}{2}}\cdot  \)\\
\( \cdot \left( \left( \frac{1}{H^{w}_{\epsilon ^{\frac{j+1}{2}}}+\left| \mathbf{k}\right| -E\left( j+1\right) }\right) ^{\frac{1}{2}}\left( -\right) \left( \frac{1}{m}\mathbf{k}\cdot \Pi _{\epsilon ^{\frac{j+1}{2}}}+\frac{\mathbf{k}^{2}}{2m}\right) \left( \frac{1}{H^{w}_{\epsilon ^{\frac{j+1}{2}}}+\left| \mathbf{k}\right| -E\left( j+1\right) }\right) ^{\frac{1}{2}}\right) ^{n}\left( \frac{1}{H^{w}_{\epsilon ^{\frac{j+1}{2}}}+\left| \mathbf{k}\right| -E\left( j+1\right) }\right) ^{\frac{1}{2}} \)\\
{\large }\\
{\large we have that}\\
{\large }\\
{\small \( \left| \int ^{\epsilon ^{\frac{j+1}{2}}}_{\epsilon ^{\frac{j+2}{2}}}\frac{k^{i^{2}}}{2\left| \mathbf{k}\right| ^{3}}\left( \Pi ^{i}_{\epsilon ^{\frac{j+1}{2}}}\phi ^{\epsilon ^{\frac{j+1}{2}}}\! ,\! \left( \frac{1}{H^{w}_{\epsilon ^{\frac{j+1}{2}}}+\left| \mathbf{k}\right| -E\left( j+1\right) }\right) \left( \left( -\right) \left( \frac{1}{m}\mathbf{k}\cdot \Pi _{\epsilon ^{\frac{j+1}{2}}}+\frac{\mathbf{k}^{2}}{2m}\right) \left( \frac{1}{H^{w}_{\epsilon ^{\frac{j+1}{2}}}+\left| \mathbf{k}\right| -E\left( j+1\right) }\right) \right) ^{n}\Pi _{\epsilon ^{\frac{j+1}{2}}}\phi ^{\epsilon ^{\frac{j+1}{2}}}\right) d^{3}k\right| \leq  \)}\\
{\small \(  \)}\\
{\small \( \leq \int ^{\epsilon ^{\frac{j+1}{2}}}_{\epsilon ^{\frac{j+2}{2}}}\frac{k^{i^{2}}}{2\left| \mathbf{k}\right| ^{3}}\left\Vert \left( \frac{1}{H^{w}_{\epsilon ^{\frac{j+1}{2}}}+\left| \mathbf{k}\right| -E\left( j+1\right) }\right) ^{\frac{1}{2}}\! \Pi ^{i}_{\epsilon ^{\frac{j+1}{2}}}\phi ^{\epsilon ^{\frac{j+1}{2}}}\right\Vert ^{2}\left\Vert \left( \frac{1}{H^{w}_{\epsilon ^{\frac{j+1}{2}}}+\left| \mathbf{k}\right| -E\left( j+1\right) }\right) ^{\frac{1}{2}}\left( \frac{1}{m}\mathbf{k}\cdot \Pi _{\epsilon ^{\frac{j+1}{2}}}+\frac{\mathbf{k}^{2}}{2m}\right) \left( \frac{1}{H^{w}_{\epsilon ^{\frac{j+1}{2}}}+\left| \mathbf{k}\right| -E\left( j+1\right) }\right) ^{\frac{1}{2}}\right\Vert _{F_{\epsilon ^{\frac{j+1}{2}}}^{+}}^{n}d^{3}k \)}\\
{\small }\\
(remember that{\small \( \left\Vert \left[ \left( \frac{1}{H^{w}_{\epsilon ^{\frac{j+1}{2}}}+\left| \mathbf{k}\right| -E\left( j+1\right) }\right) ^{\frac{1}{2}}\right] ^{\dagger }\! \Pi ^{i}_{\epsilon ^{\frac{j+1}{2}}}\phi ^{\epsilon ^{\frac{j+1}{2}}}\right\Vert =\left\Vert \left( \frac{1}{H^{w}_{\epsilon ^{\frac{j+1}{2}}}+\left| \mathbf{k}\right| -E\left( j+1\right) }\right) ^{\frac{1}{2}}\! \Pi ^{i}_{\epsilon ^{\frac{j+1}{2}}}\phi ^{\epsilon ^{\frac{j+1}{2}}}\right\Vert  \)})\\
\\
{\large Therefore, thanks to the series expansion and to the lemma 2.1, the
module of the scalar product (12)} {\large is bounded by}\\
\\
{\large \( \sum ^{\infty }_{n=0}\left\{ \int ^{\epsilon ^{\frac{j+1}{2}}}_{\epsilon ^{\frac{j+2}{2}}}\frac{k^{i^{2}}}{2\left| \mathbf{k}\right| ^{3}}\left\Vert \left( \frac{1}{H^{w}_{\epsilon ^{\frac{j+1}{2}}}+\left| \mathbf{k}\right| -E\left( j+1\right) }\right) ^{\frac{1}{2}}\! \Pi ^{i}_{\epsilon ^{\frac{j+1}{2}}}\phi ^{\epsilon ^{\frac{j+1}{2}}}\right\Vert ^{2}d^{3}k\right\} \left( \frac{1}{2}\right) ^{n}\leq  \)
}\\
{\large \( \leq 2\int ^{\epsilon ^{\frac{j+1}{2}}}_{\epsilon ^{\frac{j+2}{2}}}\frac{k^{i^{2}}}{2\left| \mathbf{k}\right| ^{3}}\cdot \left\Vert \left( \frac{1}{H^{w}_{\epsilon ^{\frac{j+1}{2}}}+\left| \mathbf{k}\right| -E\left( j+1\right) }\right) ^{\frac{1}{2}}\! \Pi ^{i}_{\epsilon ^{\frac{j+1}{2}}}\phi ^{\epsilon ^{\frac{j+1}{2}}}\right\Vert ^{2}d^{3}k\leq  \)}\\
{\large }\\
{\large \( \leq 2\cdot Q\left( \epsilon \right) \cdot \int ^{\epsilon ^{\frac{j+1}{2}}}_{\epsilon ^{\frac{j+2}{2}}}\frac{k^{i^{2}}}{2\left| \mathbf{k}\right| ^{3}}\cdot \left| \left( \Pi _{\epsilon ^{\frac{j+1}{2}}}^{i}\phi ^{\epsilon ^{\frac{j+1}{2}}}\! ,\! \left( \frac{1}{H^{w}_{\epsilon ^{\frac{j+1}{2}}}-E\left( j+1\right) }\right) \! \Pi _{\epsilon ^{\frac{j+1}{2}}}^{i}\phi ^{\epsilon ^{\frac{j+1}{2}}}\right) \right| d^{3}k \)}\\
 \textbf{\emph{\large }}{\large }\\
\\
\textbf{\large Theorem 2.3}{\large }\\
{\large }\\
\textbf{\large \( \left\Vert \left( \frac{1}{H^{w}_{\epsilon ^{\frac{j+1}{2}}}-E\left( j+1\right) }\right) ^{\frac{1}{2}}\frac{g}{2m}\int ^{\epsilon ^{\frac{j+1}{2}}}_{\epsilon ^{\frac{j+2}{2}}}\frac{\mathbf{k}}{\sqrt{2}\left| \mathbf{k}\right| ^{\frac{3}{2}}}b^{\dagger }\left( \mathbf{k}\right) d^{3}k\cdot \Pi _{\epsilon ^{\frac{j+1}{2}}}\phi ^{\epsilon ^{\frac{j+1}{2}}}\left( \frac{1}{E^{\epsilon ^{\frac{j+1}{2}}}-E\left( j+1\right) }\right) ^{\frac{1}{2}}\right\Vert \leq \frac{\epsilon ^{\frac{j+1}{8}}}{8} \)} \marginpar{
{\large (13)}{\large \par}
}{\large }\\
{\large }\\
{\large }\\
{\large Proof}\\
{\large }\\
{\large Due to the results of the lemma 2.2 the inequality (13) is true if the
following estimate holds:} \textbf{\emph{\large }}{\large }\\
{\large }\\
\( 2\sqrt{112}\cdot Q\left( \epsilon \right) \left| \left( \frac{1}{E^{\epsilon ^{\frac{j+1}{2}}}-E\left( j+1\right) }\right) \right| \frac{g^{2}}{4m^{2}}\int ^{\epsilon ^{\frac{j+1}{2}}}_{\epsilon ^{\frac{j+2}{2}}}\frac{k^{i^{2}}}{2\left| \mathbf{k}\right| ^{3}}d^{3}k\left| \left( \Pi _{\epsilon ^{\frac{j+1}{2}}}^{i}\phi ^{\epsilon ^{\frac{j+1}{2}}}\! ,\! \left( \frac{1}{H^{w}_{\epsilon ^{\frac{j+1}{2}}}-E\left( j+1\right) }\right) \! \Pi _{\epsilon ^{\frac{j+1}{2}}}^{i}\phi ^{\epsilon ^{\frac{j+1}{2}}}\right) \right| \leq  \){\large }\\
{\large }\\
\( \leq \left( \frac{\epsilon ^{\frac{j+1}{8}}}{24}\right) ^{2} \){\small }\\
\\
\\
{\large For this purpose I prove by induction that}\\
{\large }\\
{\large \( \frac{g^{2}}{4m^{2}}\left| \left( \Pi _{\epsilon ^{\frac{j+1}{2}}}^{i}\phi ^{\epsilon ^{\frac{j+1}{2}}}\! ,\! \left( \frac{1}{H^{w}_{\epsilon ^{\frac{j+1}{2}}}-E\left( j+1\right) }\right) \! \Pi _{\epsilon ^{\frac{j+1}{2}}}^{i}\phi ^{\epsilon ^{\frac{j+1}{2}}}\right) \right| <\frac{11}{\sqrt{112}\cdot Q\left( \epsilon \right) 40\pi }\cdot \frac{1}{\left( 4^{4}\cdot 24\right) ^{2}\cdot \epsilon ^{\frac{1}{4}\left( j+1\right) }} \)}\marginpar{
{\large (14)}{\large \par}
}{\large }\\
{\large }\\
{\large I note that the inequality (14) implies the thesis of the theorem and
therefore, as discussed in pag.29, also the bound \( \left\Vert \phi ^{\epsilon ^{\frac{j+2}{2}}}-\phi ^{\epsilon ^{\frac{j+1}{2}}}\right\Vert \leq \epsilon ^{\frac{j+1}{8}} \).}\\
{\large }\\
{\large }\\
{\large In order to prove the inequality (14) I start analyzing}\\
{\large }\\
{\large \( \left( \Pi _{\epsilon ^{\frac{j+1}{2}}}^{i}\phi ^{\epsilon ^{\frac{j+1}{2}}}\! ,\! \left( \frac{1}{H^{w}_{\epsilon ^{\frac{j+1}{2}}}-E\left( j+1\right) }\right) \! \Pi _{\epsilon ^{\frac{j+1}{2}}}^{i}\phi ^{\epsilon ^{\frac{j+1}{2}}}\right) = \)}\marginpar{
{\large (15)}{\large \par}
} {\large }\\
{\large }\\
{\large \( =\left( \Pi _{\epsilon ^{\frac{j+1}{2}}}^{i}\phi ^{\epsilon ^{\frac{j+1}{2}}}\! ,\! \left( \frac{1}{H^{w}_{\epsilon ^{\frac{j+1}{2}}}-H^{w}_{\epsilon ^{\frac{j}{2}}}+H^{w}_{\epsilon ^{\frac{j}{2}}}-E\left( j\right) +E\left( j\right) -E\left( j+1\right) }\right) \! \Pi _{\epsilon ^{\frac{j+1}{2}}}^{i}\phi ^{\epsilon ^{\frac{j+1}{2}}}\right)  \)}\\
{\large }\\
{\large }\\
{\large In lemma A1 I defined} {\large \( \left( \Delta H_{\mathbf{P}}^{w}\right) ^{\epsilon ^{\frac{j}{2}}}_{\epsilon ^{\frac{j+1}{2}}}\equiv \widehat{H}^{w}_{\mathbf{P},\epsilon ^{\frac{j+1}{2}}}+c_{\mathbf{P}}\left( j\right) -\widehat{c}_{\mathbf{P}}\left( j+1\right) -H^{w}_{\mathbf{P},\epsilon ^{\frac{j}{2}}} \),}
{\large that for \( \mathbf{P}=0 \) becomes \( \left( \Delta H^{w}\right) ^{\epsilon ^{\frac{j}{2}}}_{\epsilon ^{\frac{j+1}{2}}}\equiv H^{w}_{\epsilon ^{\frac{j+1}{2}}}+c\left( j\right) -c\left( j+1\right) -H^{w}_{\epsilon ^{\frac{j}{2}}} \).}\\
{\large }\\
{\large Now I define \( \widetilde{\Delta }\left( H^{w}\right) ^{\epsilon ^{\frac{j}{2}}}_{\epsilon ^{\frac{j+1}{2}}}\equiv \Delta H^{w,\epsilon ^{\frac{j}{2}}}_{\epsilon ^{\frac{j+1}{2}}}-c\left( j\right) +c\left( j+1\right) +E\left( j\right) -E\left( j+1\right)  \)
and I observe that choosing a proper value \( \overline{g_{1}} \) for \( g \),
the following inequalities hold for all \( j \): \(  \)}\\
{\large }\\
{\large \par}

\begin{itemize}
\item {\large \( \left\Vert \left( \frac{1}{H_{\epsilon ^{\frac{j}{2}}}^{w}-E\left( j\right) }\right) ^{\frac{1}{2}}\left( -\right) \Delta \left( H^{w}\right) ^{\epsilon ^{\frac{j}{2}}}_{\epsilon ^{\frac{j+1}{2}}}\left( \frac{1}{H^{w}_{\epsilon ^{\frac{j}{2}}}-E\left( j\right) }\right) ^{\frac{1}{2}}\right\Vert _{F_{\epsilon ^{\frac{j+1}{2}}}^{+}}<\frac{1}{6} \)}\\
{\large \par}
\item {\large \( \left\Vert \left( \frac{1}{H_{\epsilon ^{\frac{j}{2}}}^{w}-E\left( j\right) }\right) ^{\frac{1}{2}}\left( -\right) \left( -c\left( j\right) +c\left( j+1\right) \right) \left( \frac{1}{H^{w}_{\epsilon ^{\frac{j}{2}}}-E\left( j\right) }\right) ^{\frac{1}{2}}\right\Vert _{F_{\epsilon ^{\frac{j+1}{2}}}^{+}}<\frac{1}{6} \)}\\
{\large \par}
\item {\large \( \left\Vert \left( \frac{1}{H_{\epsilon ^{\frac{j}{2}}}^{w}-E\left( j\right) }\right) ^{\frac{1}{2}}\left( -\right) \left( E\left( j\right) -E\left( j+1\right) \right) \left( \frac{1}{H^{w}_{\epsilon ^{\frac{j}{2}}}-E\left( j\right) }\right) ^{\frac{1}{2}}\right\Vert _{F_{\epsilon ^{\frac{j+1}{2}}}^{+}}<\frac{1}{6} \)}{\large \par}
\end{itemize}
\emph{\large Observation}{\large s}\\
{\large }\\
{\large The first inequality is the content of lemma A1. The second is true
as \( c\left( j+1\right) + \)}\\
{\large \( -c\left( j\right) =-2\pi g^{2}\epsilon ^{\frac{j}{2}}\left( 1-\sqrt{\epsilon }\right)  \).
Concerning the third inequality we arrive at the given estimate by re-conducting
\( \left| E\left( j\right) -E\left( j+1\right) \right|  \) to \( \left| E^{\epsilon ^{\frac{j+1}{2}}}-E^{\epsilon ^{\frac{j}{2}}}\right|  \)
and using the generalized lemma 1.3, for \( g \) sufficiently small and uniform
in \( j \) .}\\
{\large }\\
{\large In conclusion:}\\
{\large }\\
{\large \( \left\Vert \left( \frac{1}{H_{\epsilon ^{\frac{j}{2}}}^{w}-E\left( j\right) }\right) ^{\frac{1}{2}}\left( -\right) \widetilde{\Delta }\left( H^{w}\right) ^{\epsilon ^{\frac{j}{2}}}_{\epsilon ^{\frac{j+1}{2}}}\left( \frac{1}{H^{w}_{\epsilon ^{\frac{j}{2}}}-E\left( j\right) }\right) ^{\frac{1}{2}}\right\Vert _{F_{\epsilon ^{\frac{j+1}{2}}}^{+}}<\frac{1}{2} \)
}\\
{\large }\\
{\large }\\
{\large It is therefore possible to expand in series the expression (15):}\\
{\large }\\
{\large \( \left( \Pi _{\epsilon ^{\frac{j+1}{2}}}^{i}\phi ^{\epsilon ^{\frac{j+1}{2}}}\! ,\! \left( \frac{1}{H^{w}_{\epsilon ^{\frac{j}{2}}}-E\left( j\right) }\right) \sum _{n=0}\left( \left( -\right) \widetilde{\Delta }\left( H^{w}\right) ^{\epsilon ^{\frac{j}{2}}}_{\epsilon ^{\frac{j+1}{2}}}\left( \frac{1}{H_{\epsilon ^{\frac{j}{2}}}-E\left( j\right) }\right) \right) ^{n}\! \Pi _{\epsilon ^{\frac{j+1}{2}}}^{i}\phi ^{\epsilon ^{\frac{j+1}{2}}}\right)  \)}\\
{\large }\\
{\large \( =\sum _{n=0}\left( \left[ \left( \frac{1}{H^{w}_{\epsilon ^{\frac{j}{2}}}-E\left( j\right) }\right) ^{\frac{1}{2}}\right] ^{\dagger }\Pi _{\epsilon ^{\frac{j+1}{2}}}^{i}\phi ^{\epsilon ^{\frac{j+1}{2}}}\! ,\right.  \)}\\
{\large }\\
{\large \( ,\left. \! \left[ \left( \frac{1}{H^{w}_{\epsilon ^{\frac{j}{2}}}-E\left( j\right) }\right) ^{\frac{1}{2}}\left( -\right) \widetilde{\Delta }\left( H^{w}\right) ^{\epsilon ^{\frac{j}{2}}}_{\epsilon ^{\frac{j+1}{2}}}\left( \frac{1}{H^{w}_{\epsilon ^{\frac{j}{2}}}-E\left( j\right) }\right) ^{\frac{1}{2}}\right] ^{n}\left( \frac{1}{H^{w}_{\epsilon ^{\frac{j}{2}}}-E\left( j\right) }\right) ^{\frac{1}{2}}\Pi _{\epsilon ^{\frac{j+1}{2}}}^{i}\phi ^{\epsilon ^{\frac{j+1}{2}}}\right)  \)}\\
{\large }\\
{\large }\\
{\large }\\
{\large so that we obtain the bound }\\
{\large }\\
{\large \( \left| \left( \Pi _{\epsilon ^{\frac{j+1}{2}}}^{i}\phi ^{\epsilon ^{\frac{j+1}{2}}}\! ,\! \left( \frac{1}{H^{w}_{\epsilon ^{\frac{j+1}{2}}}-E\left( j+1\right) }\right) \! \! \Pi _{\epsilon ^{\frac{j+1}{2}}}^{i}\phi ^{\epsilon ^{\frac{j+1}{2}}}\right) \right| \leq 2\left| \left( \! \Pi _{\epsilon ^{\frac{j+1}{2}}}^{i}\phi ^{\epsilon ^{\frac{j+1}{2}}}\! ,\! \left| \frac{1}{H_{\epsilon ^{\frac{j}{2}}}^{w}-E\left( j\right) }\right| \: \Pi _{\epsilon ^{\frac{j+1}{2}}}^{i}\phi ^{\epsilon ^{\frac{j+1}{2}}}\right) \right|  \)}\\
\marginpar{
{\large (16)}{\large \par}
} {\large }\\
{\large }\\
{\large }\\
\textbf{\large Proof by induction of the inequality (14)}{\large }\\
{\large }\\
{\large For \( j=1,2,3,...,7 \) and \( g \) sufficiently small (that I can
always assume as bigger or equal to \( \overline{g}_{1} \), redefining \( \overline{g}_{1} \)),
the bound (14) is valid. If one assumes that it holds for \( j=1,....,j-1,j \)
e \( g=\overline{g}\equiv \min \left( \overline{g}_{1},\overline{g}_{2},1\right)  \)
where}\\
{\large }\\
{\large \( \overline{g}_{2}=\frac{2m^{2}}{W}\cdot \frac{1}{\sqrt{112}\cdot Q\left( \epsilon \right) }\frac{11}{40\pi S}\cdot \frac{1}{\left( 24\cdot 4^{4}\right) ^{2}\cdot \epsilon ^{\frac{1}{4}}} \)
~~~~and}{\large \par}

\begin{itemize}
\item {\large \( W=2\left\{ \left[ 2m\sqrt{2m}\cdot \left( \epsilon ^{-\frac{1}{4}}\sqrt{20}\right) +2m\left( \left( \epsilon ^{-\frac{1}{4}}\sqrt{20}\right) +2\sqrt{40\pi }\right) \right] \cdot \sqrt{\frac{11}{\sqrt{112}\cdot 40\pi }\cdot \frac{1}{\left( 24\cdot 4^{4}\right) ^{2}}}\right\} + \)}\\
{\large }\\
{\large \( +\left\{ 2m\left( \epsilon ^{-\frac{1}{4}}\sqrt{20}\right) ^{2}+\left( \left( \epsilon ^{-\frac{1}{4}}\sqrt{20\pi }\right) +2\sqrt{40\pi }\right) ^{2}+2\cdot \sqrt{2m}\cdot \left( \epsilon ^{-\frac{1}{4}}\sqrt{20}\right) \left( \left( \epsilon ^{-\frac{1}{4}}\sqrt{20\pi }\right) +2\sqrt{40\pi }\right) \right\}  \) }{\large \par}
\item {\large \( S=\sum ^{\infty }_{n=0}\left( 2Q\left( \epsilon \right) \cdot \epsilon ^{\frac{1}{4}}\right) ^{n}+\frac{20\epsilon ^{\frac{1}{4}}}{11W}\cdot \left( 2Q\left( \epsilon \right) \epsilon ^{\frac{1}{12}}\right) ^{6} \)} 
\end{itemize}
{\large then the bound (14) holds for \( j+1 \), where \( j\geq 7 \).}\\
{\large }\\
{\large }\\
{\large The starting point of the procedure is the inequality (16). Adding and
subtracting \( \Pi _{\epsilon ^{\frac{j}{2}}}^{i}\phi ^{\epsilon ^{\frac{j}{2}}} \)
on the left and on the right of the scalar product, I bound the new terms that
I get, using elementary properties of the scalar product:} {\small }\\
{\small }\\
{\small }\\
{\large \( 2\left| \left( \! \Pi _{\epsilon ^{\frac{j+1}{2}}}^{i}\phi ^{\epsilon ^{\frac{j+1}{2}}}\! ,\! \left| \frac{1}{H_{\epsilon ^{\frac{j}{2}}}^{w}-E\left( j\right) }\right| \, \Pi _{\epsilon ^{\frac{j+1}{2}}}^{i}\phi ^{\epsilon ^{\frac{j+1}{2}}}\right) \right| \leq  \)}\\
{\large }\\
{\large \( \leq 2\left| \left( \Pi _{\epsilon ^{\frac{j+1}{2}}}^{i}\phi ^{\epsilon ^{\frac{j+1}{2}}}-\Pi _{\epsilon ^{\frac{j}{2}}}^{i}\phi ^{\epsilon ^{\frac{j}{2}}}\! ,\! \left| \frac{1}{H_{\epsilon ^{\frac{j}{2}}}^{w}-E\left( j\right) }\right| \! \left( \Pi _{\epsilon ^{\frac{j+1}{2}}}^{i}\phi ^{\epsilon ^{\frac{j+1}{2}}}-\Pi _{\epsilon ^{\frac{j}{2}}}^{i}\phi ^{\epsilon ^{\frac{j}{2}}}\right) \right) \right| + \)}\marginpar{
{\large (17)}{\large \par}
}{\large }\\
{\large \( +4\left| \left( \Pi _{\epsilon ^{\frac{j+1}{2}}}^{i}\phi ^{\epsilon ^{\frac{j+1}{2}}}-\Pi _{\epsilon ^{\frac{j}{2}}}^{i}\phi ^{\epsilon ^{\frac{j}{2}}}\! ,\! \left| \frac{1}{H_{\epsilon ^{\frac{j}{2}}}^{w}-E\left( j\right) }\right| \! \Pi _{\epsilon ^{\frac{j}{2}}}^{i}\phi ^{\epsilon ^{\frac{j}{2}}}\right) \right| + \)}\marginpar{
{\large (18)}{\large \par}
}{\large }\\
{\large }\\
{\large \( +2\left| \left( \Pi _{\epsilon ^{\frac{j}{2}}}^{i}\phi ^{\epsilon ^{\frac{j}{2}}}\! ,\! \left| \frac{1}{H_{\epsilon ^{\frac{j}{2}}}^{w}-E\left( j\right) }\right| \! \Pi _{\epsilon ^{\frac{j}{2}}}^{i}\phi ^{\epsilon ^{\frac{j}{2}}}\right) \right|  \)}\marginpar{
{\large (19)}{\large \par}
}{\large }\\
{\large }\\
{\large }\\
\emph{\large Observation}{\large }\\
{\large }\\
{\large The above expression displays the inductive procedure to arrive at the
thesis of the theorem, as the quantity (19) has the same form of the (15), up
to a factor \( 2\cdot Q\left( \epsilon \right)  \), where the cut-off} {\large is}
{\large \( \epsilon ^{\frac{j}{2}} \) instead of \( \epsilon ^{\frac{j+1}{2}} \).
 }\\
{\large }\\
{\large }\\
{\large In the bound of the expressions (17) and (18) I will use: }\\
{\large }\\
\emph{\large 1)} {\large \( \Pi _{\epsilon ^{\frac{j+1}{2}}}^{i}\phi ^{\epsilon ^{\frac{j+1}{2}}}=\Pi _{\epsilon ^{\frac{j}{2}}}^{i}\phi ^{\epsilon ^{\frac{j+1}{2}}}-g\int ^{\epsilon ^{\frac{j}{2}}}_{\epsilon ^{\frac{j+1}{2}}}\frac{\mathbf{q}}{\sqrt{2}\left| \mathbf{q}\right| ^{\frac{3}{2}}}\left( b\left( \mathbf{q}\right) +b^{\dagger }\left( \mathbf{q}\right) \right) d^{3}q\phi ^{\epsilon ^{\frac{j+1}{2}}} \)}\\
{\large }\\
\emph{\large 2)} {\large \( \left\Vert \int ^{\epsilon ^{\frac{j}{2}}}_{\epsilon ^{\frac{j+1}{2}}}q^{i}b\left( \mathbf{k}\right) \frac{d^{3}q}{\left| \mathbf{q}\right| \sqrt{2\left| \mathbf{q}\right| }}\left| \frac{1}{H_{\epsilon ^{\frac{j}{2}}}^{w}-E\left( j\right) }\right| ^{\frac{1}{2}}\right\Vert _{F_{\epsilon ^{\frac{j+1}{2}}}^{+}}\leq \sqrt{40\pi }\cdot \epsilon ^{\frac{j}{4}} \)
}\\
{\large }\\
{\large (analogous to the (a1) of the lemma A1 ( Appendix A));}\\
{\large }\\
\emph{\large 3)}{\large }\\
{\large }\\
{\large \( \left\Vert \left| \frac{1}{H_{\epsilon ^{\frac{j}{2}}}^{w}-E\left( j\right) }\right| ^{\frac{1}{2}}\right\Vert _{F_{\epsilon ^{\frac{j+1}{2}}}^{+}}\leq \left( \epsilon ^{-\frac{1}{4}}\sqrt{20}\right) \cdot \epsilon ^{-\frac{j}{4}} \)}
{\large }\\
{\large }\\
\emph{\large 3bis)} {\large }\\
{\large }\\
{\large \( \left\Vert \int ^{\epsilon ^{\frac{j}{2}}}_{\epsilon ^{\frac{j+1}{2}}}k^{i}b^{\dagger }\left( \mathbf{k}\right) \frac{d^{3}q}{\left| \mathbf{k}\right| \sqrt{2\left| \mathbf{k}\right| }}\left| \frac{1}{H_{\epsilon ^{\frac{j}{2}}}^{w}-E\left( j\right) }\right| ^{\frac{1}{2}}\right\Vert _{F_{\epsilon ^{\frac{j+1}{2}}}^{+}}\leq \sqrt{\pi }\epsilon ^{\frac{j}{2}}\left\Vert \left| \frac{1}{H_{\epsilon ^{\frac{j}{2}}}^{w}-E\left( j\right) }\right| ^{\frac{1}{2}}\right\Vert _{F_{\epsilon ^{\frac{j+1}{2}}}^{+}}+ \)}\\
{\large }\\
{\large \( +\left\Vert \int ^{\epsilon ^{\frac{j}{2}}}_{\epsilon ^{\frac{j+1}{2}}}q^{i}b\left( \mathbf{q}\right) \frac{d^{3}q}{\left| \mathbf{q}\right| \sqrt{2\left| \mathbf{q}\right| }}\left| \frac{1}{H_{\epsilon ^{\frac{j}{2}}}^{w}-E\left( j\right) }\right| ^{\frac{1}{2}}\right\Vert _{F_{\epsilon ^{\frac{j+1}{2}}}^{+}}\leq  \)}
{\large \(  \)}\\
{\large }\\
{\large \( \leq \sqrt{\pi }\epsilon ^{\frac{j}{2}}\left( \epsilon ^{-\frac{1}{4}}\sqrt{20}\right) \cdot \epsilon ^{-\frac{j}{4}}+\sqrt{40\pi }\cdot \epsilon ^{\frac{j}{4}}\leq \left( \left( \epsilon ^{-\frac{1}{4}}\sqrt{20\pi }\right) +\sqrt{40\pi }\right) \cdot \epsilon ^{\frac{j}{4}} \)}\\
{\large }\\
{\large }\\
{\large from which}\\
{\large }\\
{\large \( \left\Vert \left[ \left| \frac{1}{H_{\epsilon ^{\frac{j}{2}}}^{w}-E\left( j\right) }\right| ^{\frac{1}{2}}\right] ^{\dagger }\int ^{\epsilon ^{\frac{j}{2}}}_{\epsilon ^{\frac{j+1}{2}}}k^{i}\left( b\left( \mathbf{k}\right) +b^{\dagger }\left( \mathbf{k}\right) \right) \frac{d^{3}q}{\left| \mathbf{k}\right| \sqrt{2\left| \mathbf{k}\right| }}\right\Vert _{F_{\epsilon ^{\frac{j+1}{2}}}^{+}}\leq \left( \left( \epsilon ^{-\frac{1}{4}}\sqrt{20\pi }\right) +2\sqrt{40\pi }\right) \cdot \epsilon ^{\frac{j}{4}} \)}\\
{\large }\\
{\large }\\
\emph{\large 4)} {\large for the inductive hypothesis we have that:}\\
{\small }\\
{\large i) \( \frac{g^{2}}{4m^{2}}\left\Vert \left| \frac{1}{H_{\epsilon ^{\frac{j}{2}}}^{w}-E\left( j\right) }\right| ^{\frac{1}{2}}d^{3}k\! \Pi ^{i}_{\epsilon ^{\frac{j}{2}}}\phi ^{\epsilon ^{\frac{j}{2}}}\right\Vert ^{2}\leq \frac{g^{2}}{4m^{2}}Q\left( \epsilon \right) \left| \left( \Pi _{\epsilon ^{\frac{j}{2}}}^{i}\phi ^{\epsilon ^{\frac{j}{2}}}\! ,\! \left( \frac{1}{H^{w}_{\epsilon ^{\frac{j}{2}}}-E\left( j\right) }\right) \! \Pi _{\epsilon ^{\frac{j}{2}}}^{i}\phi ^{\epsilon ^{\frac{j}{2}}}\right) \right| \leq  \)}\\
{\large }\\
{\large \( \leq \frac{11}{\sqrt{112}\cdot 40\pi }\cdot \frac{1}{\left( 24\cdot 4^{4}\right) ^{2}\cdot \epsilon ^{\frac{j}{4}}} \)}\\
{\large }\\
{\large ii) \( \left\Vert \phi ^{\epsilon ^{\frac{j+1}{2}}}-\phi ^{\epsilon ^{\frac{j}{2}}}\right\Vert <\epsilon ^{\frac{j}{8}}=\sigma _{j-1}^{\frac{1}{4}} \)}
{\large ~from which~\( \left\Vert \Pi _{\epsilon ^{\frac{j}{2}}}^{i}\phi ^{\epsilon ^{\frac{j+1}{2}}}-\Pi _{\epsilon ^{\frac{j}{2}}}^{i}\phi ^{\epsilon ^{\frac{j}{2}}}\right\Vert <\sqrt{2m}\cdot \left( E^{\epsilon ^{\frac{j}{2}}}-c\left( j\right) \right) ^{\frac{1}{2}}\epsilon ^{\frac{j}{8}}< \)}\\
{\large \( <\sqrt{2m}\cdot \epsilon ^{\frac{j}{8}} \)} \emph{\large }\\
\emph{\large }\\
{\large (for the initial assumptions ~\( \left( E^{\epsilon ^{\frac{j}{2}}}-c\left( j\right) \right) <1 \))}\emph{\large }\\
\emph{\large }\\
\emph{\large }\\
\emph{\large Bound of (17)} {\large }\\
{\large }\\
{\large }\\
{\large \( \left| \left( \Pi _{\epsilon ^{\frac{j+1}{2}}}^{i}\phi ^{\epsilon ^{\frac{j+1}{2}}}-\Pi _{\epsilon ^{\frac{j}{2}}}^{i}\phi ^{\epsilon ^{\frac{j}{2}}}\! ,\! \left| \frac{1}{H_{\epsilon ^{\frac{j}{2}}}^{w}-E\left( j\right) }\right| \! \left( \Pi _{\epsilon ^{\frac{j+1}{2}}}^{i}\phi ^{\epsilon ^{\frac{j+1}{2}}}-\Pi _{\epsilon ^{\frac{j}{2}}}^{i}\phi ^{\epsilon ^{\frac{j}{2}}}\right) \right) \right| \leq  \)}\\
{\large }\\
{\large \( =\left| \left( \Pi _{\epsilon ^{\frac{j}{2}}}^{i}\phi ^{\epsilon ^{\frac{j+1}{2}}}-\Pi _{\epsilon ^{\frac{j}{2}}}^{i}\phi ^{\epsilon ^{\frac{j}{2}}}\! ,\! \left| \frac{1}{H_{\epsilon ^{\frac{j}{2}}}^{w}-E\left( j\right) }\right| \! \left( \Pi _{\epsilon ^{\frac{j}{2}}}^{i}\phi ^{\epsilon ^{\frac{j+1}{2}}}-\Pi _{\epsilon ^{\frac{j}{2}}}^{i}\phi ^{\epsilon ^{\frac{j}{2}}}\right) \right) \right| + \)}\\
{\large }\\
{\large \( +\left| \left( g\int ^{\epsilon ^{\frac{j}{2}}}_{\epsilon ^{\frac{j+1}{2}}}\frac{\mathbf{k}}{\sqrt{2}\left| \mathbf{k}\right| ^{\frac{3}{2}}}\left( b\left( \mathbf{k}\right) +b^{\dagger }\left( \mathbf{k}\right) \right) d^{3}k\phi ^{\epsilon ^{\frac{j+1}{2}}}\! ,\! \left| \frac{1}{H_{\epsilon ^{\frac{j}{2}}}^{w}-E\left( j\right) }\right| \! g\int ^{\epsilon ^{\frac{j}{2}}}_{\epsilon ^{\frac{j+1}{2}}}\frac{\mathbf{k}}{\sqrt{2}\left| \mathbf{k}\right| ^{\frac{3}{2}}}\left( b\left( \mathbf{k}\right) +b^{\dagger }\left( \mathbf{k}\right) \right) d^{3}k\phi ^{\epsilon ^{\frac{j+1}{2}}}\right) \right| + \)}\\
{\large }\\
{\large \( +2\left| \left( g\int ^{\epsilon ^{\frac{j}{2}}}_{\epsilon ^{\frac{j+1}{2}}}\frac{\mathbf{k}}{\sqrt{2}\left| \mathbf{k}\right| ^{\frac{3}{2}}}\left( b\left( \mathbf{k}\right) +b^{\dagger }\left( \mathbf{k}\right) \right) d^{3}k\phi ^{\epsilon ^{\frac{j+1}{2}}}\! ,\! \left| \frac{1}{H_{\epsilon ^{\frac{j}{2}}}^{w}-E\left( j\right) }\right| \! \left( \Pi _{\epsilon ^{\frac{j}{2}}}^{i}\phi ^{\epsilon ^{\frac{j+1}{2}}}-\Pi _{\epsilon ^{\frac{j}{2}}}^{i}\phi ^{\epsilon ^{\frac{j}{2}}}\right) \right) \right|  \)}\\
{\large }\\
{\large }\\
{\large The (17) is therefore bounded by}\\
{\large }\\
\( 2\left\{ 2m\cdot \left( \epsilon ^{-\frac{1}{4}}\sqrt{20}\right) ^{2}+\left( \left( \epsilon ^{-\frac{1}{4}}\sqrt{20\pi }\right) +2\sqrt{40\pi }\right) ^{2}+2\cdot \left( \epsilon ^{-\frac{1}{4}}\sqrt{20}\right) \cdot \left( \left( \epsilon ^{-\frac{1}{4}}\sqrt{20\pi }\right) +2\sqrt{40\pi }\right) \right\} \cdot \epsilon ^{-\frac{j}{4}} \){\large }\\
{\large }\\
{\large }\\
\emph{\large Bound of (18).} {\large }\\
{\large }\\
{\large }\\
{\large \( \left| \left( \Pi _{\epsilon ^{\frac{j+1}{2}}}^{i}\phi ^{\epsilon ^{\frac{j+1}{2}}}-\Pi _{\epsilon ^{\frac{j}{2}}}^{i}\phi ^{\epsilon ^{\frac{j}{2}}}\! ,\! \left| \frac{1}{H_{\epsilon ^{\frac{j}{2}}}^{w}-E\left( j\right) }\right| \! \Pi _{\epsilon ^{\frac{j}{2}}}^{i}\phi ^{\epsilon ^{\frac{j}{2}}}\right) \right| \leq  \)}\\
{\large }\\
{\large \( \leq \left| \left( \Pi _{\epsilon ^{\frac{j}{2}}}^{i}\phi ^{\epsilon ^{\frac{j+1}{2}}}-\Pi _{\epsilon ^{\frac{j}{2}}}^{i}\phi ^{\epsilon ^{\frac{j}{2}}}\! ,\! \left| \frac{1}{H_{\epsilon ^{\frac{j}{2}}}^{w}-E\left( j\right) }\right| \! \Pi _{\epsilon ^{\frac{j}{2}}}^{i}\phi ^{\epsilon ^{\frac{j}{2}}}\right) \right| + \)}\marginpar{
{\large (18.1)}{\large \par}
}{\large }\\
{\large }\\
{\large \( +\left| \left( g\int ^{\epsilon ^{\frac{j}{2}}}_{\epsilon ^{\frac{j+1}{2}}}\frac{\mathbf{k}}{\sqrt{2}\left| \mathbf{k}\right| ^{\frac{3}{2}}}\left( b\left( \mathbf{k}\right) +b^{\dagger }\left( \mathbf{k}\right) \right) d^{3}k\phi ^{\epsilon ^{\frac{j+1}{2}}}\! ,\! \left| \frac{1}{H_{\epsilon ^{\frac{j}{2}}}^{w}-E\left( j\right) }\right| \! \Pi _{\epsilon ^{\frac{j}{2}}}^{i}\phi ^{\epsilon ^{\frac{j}{2}}}\right) \right|  \)}\marginpar{
{\large (18.2)}{\large \par}
}{\large }\\
{\large }\\
{\large }\\
{\large I examine the two terms on the right side: }\\
{\large }\\
{\large 18.1) }\\
{\large }\\
{\large \( \left| \left( \Pi _{\epsilon ^{\frac{j}{2}}}^{i}\phi ^{\epsilon ^{\frac{j+1}{2}}}-\Pi _{\epsilon ^{\frac{j}{2}}}^{i}\phi ^{\epsilon ^{\frac{j}{2}}}\! ,\! \left| \frac{1}{H_{\epsilon ^{\frac{j}{2}}}^{w}-E\left( j\right) }\right| \! \Pi _{\epsilon ^{\frac{j}{2}}}^{i}\phi ^{\epsilon ^{\frac{j}{2}}}\right) \right| = \)}\\
{\large }\\
{\large \( =\left| \left( \Pi _{\epsilon ^{\frac{j}{2}}}^{i}\phi ^{\epsilon ^{\frac{j+1}{2}}}-\Pi _{\epsilon ^{\frac{j}{2}}}^{i}\phi ^{\epsilon ^{\frac{j}{2}}}\! ,\! \left| \frac{1}{H_{\epsilon ^{\frac{j}{2}}}^{w}-E\left( j\right) }\right| ^{\frac{1}{2}}\left| \frac{1}{H_{\epsilon ^{\frac{j}{2}}}^{w}-E\left( j\right) }\right| ^{\frac{1}{2}}\! \Pi _{\epsilon ^{\frac{j}{2}}}^{i}\phi ^{\epsilon ^{\frac{j}{2}}}\right) \right| \leq  \)
}\\
{\large \( \leq \left\Vert \Pi _{\epsilon ^{\frac{j}{2}}}^{i}\phi ^{\epsilon ^{\frac{j+1}{2}}}-\Pi _{\epsilon ^{\frac{j}{2}}}^{i}\phi ^{\epsilon ^{\frac{j}{2}}}\right\Vert \cdot \left\Vert \left| \frac{1}{H_{\epsilon ^{\frac{j}{2}}}^{w}-E\left( j\right) }\right| ^{\frac{1}{2}}\right\Vert _{F_{\epsilon ^{\frac{j+1}{2}}}^{+}}\cdot \left\Vert \left| \frac{1}{H_{\epsilon ^{\frac{j}{2}}}^{w}-E\left( j\right) }\right| ^{\frac{1}{2}}d^{3}k\: \Pi ^{i}_{\epsilon ^{\frac{j}{2}}}\phi ^{\epsilon ^{\frac{j}{2}}}\right\Vert  \)}
{\large }{\large }\\
{\large \( \leq \epsilon ^{\frac{j}{8}}\cdot \sqrt{2m}\cdot \left( \epsilon ^{-\frac{1}{4}}\sqrt{20}\right) \cdot \epsilon ^{-\frac{j}{4}}\cdot \frac{2m}{g}\cdot \sqrt{\frac{11}{\sqrt{112}\cdot 40\pi }\cdot \frac{1}{\left( 24\cdot 4^{4}\right) ^{2}\cdot \epsilon ^{\frac{j}{4}}}} \)}\\
{\large }\\
{\large 18.2)}\\
{\large }\\
{\large \( \left| \left( g\int ^{\epsilon ^{\frac{j}{2}}}_{\epsilon ^{\frac{j+1}{2}}}\frac{k^{i}}{\sqrt{2}\left| \mathbf{k}\right| ^{\frac{3}{2}}}\left( b\left( \mathbf{k}\right) +b^{\dagger }\left( \mathbf{k}\right) \right) d^{3}k\phi ^{\epsilon ^{\frac{j+1}{2}}}\! ,\! \left| \frac{1}{H_{\epsilon ^{\frac{j}{2}}}^{w}-E\left( j\right) }\right| \! \Pi _{\epsilon ^{\frac{j}{2}}}^{i}\phi ^{\epsilon ^{\frac{j}{2}}}\right) \right| \leq  \)}\\
{\large }\\
{\large \( \leq \left\Vert g\int ^{\epsilon ^{\frac{j}{2}}}_{\epsilon ^{\frac{j+1}{2}}}k^{i}\left( b\left( \mathbf{k}\right) +b^{\dagger }\left( \mathbf{k}\right) \right) \frac{d^{3}\mathbf{k}}{\left| \mathbf{k}\right| \sqrt{2\left| \mathbf{k}\right| }}\left| \frac{1}{H_{\epsilon ^{\frac{j}{2}}}^{w}-E\left( j\right) }\right| ^{\frac{1}{2}}\right\Vert _{F_{\epsilon ^{\frac{j+1}{2}}}^{+}}\cdot \left\Vert \left| \frac{1}{H_{\epsilon ^{\frac{j}{2}}}^{w}-E\left( j\right) }\right| ^{\frac{1}{2}}d^{3}k\: \Pi ^{i}_{\epsilon ^{\frac{j}{2}}}\phi ^{\epsilon ^{\frac{j}{2}}}\right\Vert \leq  \)}\\
{\large }\\
{\large \( \leq g\cdot \left( \left( \epsilon ^{-\frac{1}{4}}\sqrt{20}\right) +2\sqrt{40\pi }\right) \cdot \epsilon ^{\frac{j}{4}}\cdot \frac{2m}{g}\cdot \sqrt{\frac{11}{\sqrt{112}\cdot 40\pi }\cdot \frac{1}{\left( 24\cdot 4^{4}\right) ^{2}\cdot \epsilon ^{\frac{j}{4}}}}\leq  \)}\\
{\large }\\
{\large \( \leq 2m\cdot \left( \left( \epsilon ^{-\frac{1}{4}}\sqrt{20}\right) +2\sqrt{40\pi }\right) \cdot \sqrt{\frac{11}{\sqrt{112}\cdot 40\pi }\cdot \frac{1}{\left( 24\cdot 4^{4}\right) ^{2}}}\cdot \epsilon ^{\frac{j}{8}} \)}
{\large }\\
{\large }\\
{\large We conclude that (18) is bounded by }\\
{\large }\\
{\large \( 4\left\{ \left[ \frac{2m}{g}\sqrt{2m}\cdot \left( \epsilon ^{-\frac{1}{4}}\sqrt{20}\right) +2m\left( \left( \epsilon ^{-\frac{1}{4}}\sqrt{20}\right) +2\sqrt{40\pi }\right) \right] \cdot \sqrt{\frac{11}{\sqrt{112}\cdot 40\pi }\cdot \frac{1}{\left( 24\cdot 4^{4}\right) ^{2}}}\right\} \cdot \epsilon ^{-\frac{j}{4}} \)}\\
{\large }\\
{\large }\\
{\large From the previous estimates we have that for \( g=\overline{g} \) (\( \overline{g}\leq 1 \)),
fixed from the beginning of the theorem, the sum of (17) and (18) times \( \frac{\overline{g}^{2}}{4m^{2}} \)
is less than}\\
{\large \( \frac{\overline{g}^{2}}{4m^{2}}\cdot \left[ 2\left\{ 2m\left( \epsilon ^{-\frac{1}{4}}\sqrt{20}\right) ^{2}+\left( \left( \epsilon ^{-\frac{1}{4}}\sqrt{20\pi }\right) +2\sqrt{40\pi }\right) ^{2}+2\cdot \sqrt{2m}\cdot \left( \epsilon ^{-\frac{1}{4}}\sqrt{20}\right) \cdot \left( \left( \epsilon ^{-\frac{1}{4}}\sqrt{20\pi }\right) +2\sqrt{40\pi }\right) \right\} +\right.  \)}\\
{\large }\\
{\large \( \left. +4\left\{ \left[ \frac{2m}{\overline{g}}\sqrt{2m}\cdot \left( \epsilon ^{-\frac{1}{4}}\sqrt{20}\right) +2m\left( \left( \epsilon ^{-\frac{1}{4}}\sqrt{20}\right) +2\sqrt{40\pi }\right) \right] \cdot \sqrt{\frac{11}{\sqrt{112}\cdot 40\pi }\cdot \frac{1}{\left( 24\cdot 4^{4}\right) ^{2}}}\right\} \right] \cdot \epsilon ^{-\frac{j}{4}}\leq  \)}\\
{\large }\\
\( \leq \frac{\overline{g}}{2m^{2}}\left[ \left\{ 2m\left( \epsilon ^{-\frac{1}{4}}\sqrt{20}\right) ^{2}+\left( \left( \epsilon ^{-\frac{1}{4}}\sqrt{20\pi }\right) +2\sqrt{40\pi }\right) ^{2}+2\cdot \sqrt{2m}\cdot \left( \epsilon ^{-\frac{1}{4}}\sqrt{20}\right) \cdot \left( \left( \epsilon ^{-\frac{1}{4}}\sqrt{20\pi }\right) +2\sqrt{40\pi }\right) \right\} +\right.  \){\large }\\
{\large }\\
{\large \( \left. +2\left\{ \left[ 2m\sqrt{2m}\cdot \left( \epsilon ^{-\frac{1}{4}}\sqrt{20}\right) +2m\left( \left( \epsilon ^{-\frac{1}{4}}\sqrt{20}\right) +2\sqrt{40\pi }\right) \right] \cdot \sqrt{\frac{11}{\sqrt{112}\cdot 40\pi }\cdot \frac{1}{\left( 24\cdot 4^{4}\right) ^{2}}}\right\} \right] \cdot \epsilon ^{-\frac{j}{4}}\leq  \)}\\
{\large }\\
{\large \( \leq \frac{11}{40\pi S}\cdot \frac{1}{\sqrt{112}\cdot Q\left( \epsilon \right) }\cdot \frac{1}{\left( 24\cdot 4^{4}\right) ^{2}\cdot \epsilon ^{\frac{1}{4}\left( j+1\right) }} \)
.  }\\
{\large }\\
{\large At this point the procedure has to be repeated nearly up to} {\large the
cut-off} \emph{\large ~}{\large \( \epsilon ^{\frac{j+1}{8}} \), precisely
the} {\large cut-off} {\large ~~\( \epsilon ^{\frac{\left[ \widehat{j}+1\right] +1}{2}} \)
~where \( \frac{\widehat{j}+1}{2}=\frac{j+1}{8} \) e~ \( \left[ \widehat{j}+1\right] = \)integer
part of \( \widehat{j}+1 \):}\\
{\large }\\
{\large }\\
{\large \( \frac{\overline{g}^{2}}{4m^{2}}\left| \left( \Pi _{\epsilon ^{\frac{j+1}{2}}}^{i}\phi ^{\epsilon ^{\frac{j+1}{2}}}\! ,\! \left( \frac{1}{H^{w}_{\epsilon ^{\frac{j+1}{2}}}-E\left( j+1\right) }\right) \! \Pi _{\epsilon ^{\frac{j+1}{2}}}^{i}\phi ^{\epsilon ^{\frac{j+1}{2}}}\right) \right| \leq  \)}\\
{\large }\\
{\large \( \leq \frac{11}{40\pi S}\cdot \frac{1}{\sqrt{112}\cdot Q\left( \epsilon \right) }\cdot \frac{1}{\left( 24\cdot 4^{4}\right) ^{2}\cdot \epsilon ^{\frac{1}{4}\left( j+1\right) }}+\frac{\overline{g}^{2}}{4m^{2}}\cdot 2\left| \left( \Pi _{\epsilon ^{\frac{j}{2}}}^{i}\phi ^{\epsilon ^{\frac{j+1}{2}}}\! ,\! \left| \left( \frac{1}{H_{\epsilon ^{\frac{j}{2}}}^{w}-E\left( j\right) }\right) \right| \! \Pi _{\epsilon ^{\frac{j}{2}}}^{i}\phi ^{\epsilon ^{\frac{j+1}{2}}}\right) \right| \leq  \)}\\
{\large }\\
{\large \( \leq \frac{11}{40\pi S}\cdot \frac{1}{\sqrt{112}\cdot Q\left( \epsilon \right) }\cdot \frac{1}{\left( 24\cdot 4^{4}\right) ^{2}\cdot \epsilon ^{\frac{1}{4}\left( j+1\right) }}+\frac{\overline{g}^{2}}{4m^{2}}\cdot 2Q\left( \epsilon \right) \left| \left( \Pi _{\epsilon ^{\frac{j}{2}}}^{i}\phi ^{\epsilon ^{\frac{j+1}{2}}}\! ,\! \left( \frac{1}{H_{\epsilon ^{\frac{j}{2}}}^{w}-E\left( j\right) }\right) \! \Pi _{\epsilon ^{\frac{j}{2}}}^{i}\phi ^{\epsilon ^{\frac{j+1}{2}}}\right) \right| \leq  \)}\\
{\large }\\
{\large \( \leq \frac{11}{40\pi S}\cdot \frac{1}{\sqrt{112}\cdot Q\left( \epsilon \right) }\cdot \frac{1}{\left( 24\cdot 4^{4}\right) ^{2}\cdot \epsilon ^{\frac{1}{4}\left( j+1\right) }}+\frac{\overline{g}^{2}}{4m^{2}}\cdot 2^{2}Q\left( \epsilon \right) \left| \left( \Pi _{\epsilon ^{\frac{j}{2}}}^{i}\phi ^{\epsilon ^{\frac{j+1}{2}}}\! ,\! \left| \left( \frac{1}{H_{\epsilon ^{\frac{j-1}{2}}}^{w}-E\left( j-1\right) }\right) \right| \! \Pi _{\epsilon ^{\frac{j}{2}}}^{i}\phi ^{\epsilon ^{\frac{j+1}{2}}}\right) \right| \leq  \)}\\
{\large }\\
{\large \( \leq \frac{11}{40\pi S}\cdot \frac{1}{\sqrt{112}\cdot Q\left( \epsilon \right) }\cdot \frac{1}{\left( 24\cdot 4^{4}\right) ^{2}\cdot \epsilon ^{\frac{1}{4}\left( j+1\right) }}\left( 1+\left( 2Q\left( \epsilon \right) \cdot \epsilon ^{\frac{1}{4}}\right) +.....+\left( 2Q\left( \epsilon \right) \cdot \epsilon ^{\frac{1}{4}}\right) ^{\frac{3j+3}{4}}\right) + \)}\\
{\large }\\
{\large \( +\frac{\overline{g}^{2}}{4m^{2}}\cdot \frac{20}{11}\cdot \frac{2\epsilon ^{\frac{1}{2}}}{\epsilon ^{\frac{1}{4}\left( j+1\right) }}\left( 2Q\left( \epsilon \right) \right) ^{\frac{3j+3}{4}}\cdot \epsilon ^{\frac{j-3}{8}} \)}\\
{\large }\\
{\large \(  \)}\\
{\large the formula above can be explained with:}{\large \par}

\begin{itemize}
\item {\large \( j-\left[ \widehat{j}+1\right] \leq j+1-\widehat{j}-1=\frac{3j+3}{4} \) }{\large \par}
\item {\large being \( Q\left( \epsilon \right) >1 \)~~~~~~~~~~~\( \left( 2Q\left( \epsilon \right) \right) ^{\frac{3j+3}{4}}\geq \left( 2Q\left( \epsilon \right) \right) ^{j-\left[ \widehat{j}+1\right] } \) }{\large \par}
\item {\large ~~\( \frac{1}{\epsilon ^{\frac{\left[ \widehat{j}+1\right] +1}{2}}}\leq \frac{1}{\epsilon ^{\frac{\frac{j+1}{4}+1}{2}}}=\frac{1}{\epsilon ^{\frac{j+5}{8}}}=\frac{1}{\epsilon ^{\frac{1}{4}\left( j+1\right) }}\cdot \epsilon ^{\frac{j-3}{8}} \)}{\large \par}
\end{itemize}
{\large then}\\
{\large \( \frac{\overline{g}^{2}}{4m^{2}}\left| \left( \Pi _{\epsilon ^{\frac{j+1}{2}}}^{i}\phi ^{\epsilon ^{\frac{j+1}{2}}}\! ,\! \left( \frac{1}{H^{w}_{\epsilon ^{\frac{j+1}{2}}}-E\left( j+1\right) }\right) \! \Pi _{\epsilon ^{\frac{j+1}{2}}}^{i}\phi ^{\epsilon ^{\frac{j+1}{2}}}\right) \right| \leq  \)}\\
{\large }\\
{\large \( \leq \frac{11}{40\pi S}\cdot \frac{1}{\sqrt{112}\cdot Q\left( \epsilon \right) }\cdot \frac{1}{\left( 24\cdot 4^{4}\right) ^{2}\cdot \epsilon ^{\frac{1}{4}\left( j+1\right) }}\left( 1+\left( 2Q\left( \epsilon \right) \cdot \epsilon ^{\frac{1}{4}}\right) ^{1}+.....+\left( 2Q\left( \epsilon \right) \cdot \epsilon ^{\frac{1}{4}}\right) ^{j-\left[ \widehat{j}+1\right] }\right) + \)}\\
{\large }\\
{\large \( +\frac{\overline{g}^{2}}{4m^{2}}\cdot \frac{40}{11}\cdot \frac{\epsilon ^{\frac{1}{2}}}{\epsilon ^{\frac{1}{4}\left( j+1\right) }}\left( 2Q\left( \epsilon \right) \epsilon ^{\frac{j-3}{6j+6}}\right) ^{\frac{3j+3}{4}}\leq  \)}\\
{\large }\\
\( \leq \frac{11}{40\pi S}\cdot \frac{1}{\sqrt{112}\cdot Q\left( \epsilon \right) }\cdot \frac{1}{\left( 24\cdot 4^{4}\right) ^{2}\cdot \epsilon ^{\frac{1}{4}\left( j+1\right) }}\left( 1+\left( 2Q\left( \epsilon \right) \cdot \epsilon ^{\frac{1}{4}}\right) ^{1}+..+\left( 2Q\left( \epsilon \right) \cdot \epsilon ^{\frac{1}{4}}\right) ^{j-\left[ \widehat{j}+1\right] }+\frac{20\cdot \epsilon ^{\frac{1}{4}}}{11\cdot W}\left( 2Q\left( \epsilon \right) \epsilon ^{\frac{j-3}{6j+6}}\right) ^{\frac{3j+3}{4}}\right)  \){\large }\\
{\large }\\
{\large \( \leq \frac{11}{40\pi }\cdot \frac{1}{\sqrt{112}\cdot Q\left( \epsilon \right) }\cdot \frac{1}{\left( 24\cdot 4^{4}\right) ^{2}\cdot \epsilon ^{\frac{1}{4}\left( j+1\right) }} \)}\\
{\large }\\
{\large }\\
\emph{\large Notes}{\large \par}

\begin{itemize}
\item {\large by hypothesis, \( 0<\overline{g}\leq \frac{2m^{2}}{W}\cdot \frac{1}{\sqrt{112}\cdot Q\left( \epsilon \right) }\cdot \frac{11}{40\pi S}\cdot \frac{1}{\left( 24\cdot 4^{4}\right) ^{2}\cdot \epsilon ^{\frac{1}{4}}} \)
~and \( \overline{g}\leq 1 \), therefore} {\large we have that \( \frac{\overline{g}^{2}}{4m^{2}}\cdot \frac{40\epsilon ^{\frac{1}{2}}}{11}\leq \frac{20\epsilon ^{\frac{1}{4}}}{11\cdot W}\cdot \frac{1}{\sqrt{112}\cdot Q\left( \epsilon \right) }\cdot \frac{11}{40\pi S}\cdot \frac{1}{\left( 24\cdot 4^{4}\right) ^{2}} \)}\\
 {\large }{\large \par}
\item {\large note that \( \left( 2Q\left( \epsilon \right) \epsilon ^{\frac{j-3}{6j+6}}\right) ^{\frac{3j+3}{4}}\leq \left( 2Q\left( \epsilon \right) \epsilon ^{\frac{1}{12}}\right) ^{6} \)
~~\( \forall j\geq 7 \) , as}\\
{\large \( 0<\epsilon <\left( \frac{1}{4}\right) ^{16} \), \( 1<Q\left( \epsilon \right) <2 \)~~~\( \Rightarrow  \)~~~~\( 2Q\left( \epsilon \right) \cdot \epsilon ^{\frac{1}{12}}<1 \)}\\
{\large \par}
\item {\large by definition \( S=\sum ^{\infty }_{n=0}\left( 2Q\left( \epsilon \right) \cdot \epsilon ^{\frac{1}{4}}\right) ^{n}+\frac{20\epsilon ^{\frac{1}{4}}}{11\cdot W}\cdot \left( 2Q\left( \epsilon \right) \epsilon ^{\frac{1}{12}}\right) ^{6} \)
}\\
{\large }\\
{\large }\\
{\large \par}
\end{itemize}

\subsection{Case \protect\( \mathbf{P}\neq 0\protect \).}

{\large I repeat on the hamiltonian \( H_{\mathbf{P},\sigma } \) the same operations
made for the canonic form (10) of the transformed hamiltonian \( H_{\mathbf{P}} \).
Unlike the case without infrared cut-off, the transformation of \( H_{\mathbf{P},\sigma } \)
is implemented by the unitary operator }\\
{\large }\\
{\large 
\[
\mathcal{W}_{\sigma }\left( \mathbf{P}\right) =W_{\sigma }\left( \frac{\mathbf{P}_{1}}{m}\right) =e^{-g\int _{\sigma }^{\kappa }\frac{b\left( \mathbf{k}\right) -b^{\dagger }\left( \mathbf{k}\right) }{\left| \mathbf{k}\right| \left( 1-\widehat{\mathbf{k}}\cdot \frac{\mathbf{P}_{1}}{m}\right) }\frac{d^{3}k}{\sqrt{2\left| \mathbf{k}\right| }}}\qquad \]
}\\
{\large where \( \mathbf{P}_{1}\equiv \mathbf{P}-\overline{\mathbf{P}^{mes}}=m\nabla E^{\sigma }\left( \mathbf{P}\right)  \)}.
{\large The last inequality follows from the perturbation of the isolated eigenvalue
\( E_{\mathbf{P}}^{0} \) of \( H_{\mathbf{P}}^{0}\mid _{F_{\sigma }^{+}} \)(see
{[}2{]})} {\small .}{\large }\\
{\large }\\
{\large \( \mathbf{P}_{1} \) verifies the equation:}\\
{\large }\\
{\small \( \mathbf{P}-\mathbf{P}_{1}=\frac{1}{\left\Vert \phi _{\mathbf{P}}^{\sigma }\right\Vert ^{2}}\left( \phi _{\mathbf{P}}^{\sigma },\mathbf{P}^{mes}\phi _{\mathbf{P}}^{\sigma }\right) -\frac{1}{\left\Vert \phi _{\mathbf{P}}^{\sigma }\right\Vert ^{2}}\left( \phi _{\mathbf{P}}^{\sigma },g\int _{\sigma }^{\kappa }\frac{\mathbf{k}}{\sqrt{2}\left| \mathbf{k}\right| ^{\frac{3}{2}}\left( 1-\widehat{\mathbf{k}}\cdot \frac{\mathbf{P}_{1}}{m}\right) }\left( b\left( \mathbf{k}\right) +b^{\dagger }\left( \mathbf{k}\right) \right) d^{3}k\phi _{\mathbf{P}}^{\sigma }\right) +g^{2}\int _{\sigma }^{\kappa }\frac{\mathbf{k}}{2\left| \mathbf{k}\right| ^{3}\left( 1-\widehat{\mathbf{k}}\cdot \frac{\mathbf{P}_{1}}{m}\right) ^{2}}d^{3}k \)}{\large }\\
{\large }\\
{\large as \( \phi _{\mathbf{P}}^{\sigma } \) is the ground state of the transformed
hamiltonian \( W_{\sigma }\left( \frac{\mathbf{P}_{1}}{m}\right) H_{\mathbf{P},\sigma }W_{\sigma }^{\dagger }\left( \frac{\mathbf{P}_{1}}{m}\right)  \)}\\
{\large }\\
{\large }\\
{\large }\\
 {\large }\emph{\large Transformed hamiltonian.}{\large }\\
{\large }\\
{\large I will write \( H_{\mathbf{P},\sigma } \), \( \mathbf{P}=\mathbf{P}_{1}+\mathbf{P}_{2} \)}
{\large , as}\\
{\large }\\
{\large \( H_{\mathbf{P},\sigma }=\frac{\left( \mathbf{P}_{1}+\mathbf{P}_{2}-\mathbf{P}^{mes}\right) ^{2}}{2m}+g\int _{\sigma }^{\kappa }\left( b\left( \mathbf{k}\right) +b^{\dagger }\left( \mathbf{k}\right) \right) \frac{d^{3}\mathbf{k}}{\sqrt{2\left| \mathbf{k}\right| }}+H^{mes}= \)}\\
{\large }\\
{\large \( =\frac{\left( \mathbf{P}_{1}+\mathbf{P}_{2}\right) ^{2}}{2m}-\frac{\left( \mathbf{P}_{1}+\mathbf{P}_{2}\right) \cdot \mathbf{P}^{mes}}{m}+\frac{\mathbf{P}^{mes^{2}}}{2m}+g\int _{\sigma }^{\kappa }\left( b\left( \mathbf{k}\right) +b^{\dagger }\left( \mathbf{k}\right) \right) \frac{d^{3}\mathbf{k}}{\sqrt{2\left| \mathbf{k}\right| }}+H^{mes}= \)}\\
{\large }\\
{\large \( =\frac{\left( \mathbf{P}_{1}+\mathbf{P}_{2}\right) ^{2}}{2m}+\frac{\mathbf{P}^{mes^{2}}}{2m}-\frac{\mathbf{P}_{2}\cdot \mathbf{P}^{mes}}{m}+\int ^{\infty }_{\kappa }\left( \left| \mathbf{k}\right| -\mathbf{k}\cdot \frac{\mathbf{P}_{1}}{m}\right) b^{\dagger }\left( \mathbf{k}\right) b\left( \mathbf{k}\right) d^{3}k+ \)}\\
{\large }\\
{\large \( +\int ^{\kappa }_{\sigma }\left( \left| \mathbf{k}\right| -\mathbf{k}\cdot \frac{\mathbf{P}_{1}}{m}\right) \left( b^{\dagger }\left( \mathbf{k}\right) +\frac{g}{\sqrt{2}\left| \mathbf{k}\right| ^{\frac{3}{2}}\left( 1-\widehat{\mathbf{k}}\cdot \frac{\mathbf{P}_{1}}{m}\right) }\right) \left( b\left( \mathbf{k}\right) +\frac{g}{\sqrt{2}\left| \mathbf{k}\right| ^{\frac{3}{2}}\left( 1-\widehat{\mathbf{k}}\cdot \frac{\mathbf{P}_{1}}{m}\right) }\right) d^{3}k \)}\\
{\large }\\
{\large \( +\int ^{\sigma }_{0}\left( \left| \mathbf{k}\right| -\mathbf{k}\cdot \frac{\mathbf{P}_{1}}{m}\right) b^{\dagger }\left( \mathbf{k}\right) b\left( \mathbf{k}\right) d^{3}k-g^{2}\int _{\sigma }^{\kappa }\frac{1}{2\left| \mathbf{k}\right| ^{2}\left( 1-\widehat{\mathbf{k}}\cdot \frac{\mathbf{P}_{1}}{m}\right) }d^{3}k \)}\\
{\large }\\
{\large }\\
{\large and I will perform the coherent transformation:}\\
{\large }\\
{\large }\\
{\large \( W_{\sigma }\left( \frac{\mathbf{P}_{1}}{m}\right) H_{\mathbf{P},\sigma }W_{\sigma }^{\dagger }\left( \frac{\mathbf{P}_{1}}{m}\right) = \)}{\large }\\
{\large }\\
{\large \( =\frac{\left( \mathbf{P}_{1}+\mathbf{P}_{2}\right) ^{2}}{2m}+\frac{1}{2m}\left( \mathbf{P}^{mes}-g\int _{\sigma }^{\kappa }\frac{\mathbf{k}}{\sqrt{2}\left| \mathbf{k}\right| ^{\frac{3}{2}}\left( 1-\widehat{\mathbf{k}}\cdot \frac{\mathbf{P}_{1}}{m}\right) }\left( b\left( \mathbf{k}\right) +b^{\dagger }\left( \mathbf{k}\right) \right) d^{3}k+g^{2}\int _{\sigma }^{\kappa }\frac{\mathbf{k}}{2\left| \mathbf{k}\right| ^{3}\left( 1-\widehat{\mathbf{k}}\cdot \frac{\mathbf{P}_{1}}{m}\right) ^{2}}d^{3}k\right) ^{2}+ \)}\\
{\large }\\
{\large \( -\frac{\mathbf{P}_{2}}{m}\left( \mathbf{P}^{mes}-g\int _{\sigma }^{\kappa }\frac{\mathbf{k}}{\sqrt{2}\left| \mathbf{k}\right| ^{\frac{3}{2}}\left( 1-\widehat{\mathbf{k}}\cdot \frac{\mathbf{P}_{1}}{m}\right) }\left( b\left( \mathbf{k}\right) +b^{\dagger }\left( \mathbf{k}\right) \right) d^{3}k+g^{2}\int _{\sigma }^{\kappa }\frac{\mathbf{k}}{2\left| \mathbf{k}\right| ^{3}\left( 1-\widehat{\mathbf{k}}\cdot \frac{\mathbf{P}_{1}}{m}\right) ^{2}}d^{3}k\right) + \)}\\
{\large }\\
{\large \( +\int ^{\infty }_{0}\left( \left| \mathbf{k}\right| -\mathbf{k}\cdot \frac{\mathbf{P}_{1}}{m}\right) b^{\dagger }\left( \mathbf{k}\right) b\left( \mathbf{k}\right) d^{3}k-g^{2}\int _{\sigma }^{\kappa }\frac{1}{2\left| \mathbf{k}\right| ^{2}\left( 1-\widehat{\mathbf{k}}\cdot \frac{\mathbf{P}_{1}}{m}\right) }d^{3}k \)}\\
{\large }\\
\emph{\large }\\
\emph{\large Considerations}  {\large }\\
{\large }\\
{\large Having defined \( \Pi _{\mathbf{P},\sigma }=\mathbf{P}^{mes}-g\int _{\sigma }^{\kappa }\frac{\mathbf{k}\left( b\left( \mathbf{k}\right) +b^{\dagger }\left( \mathbf{k}\right) \right) }{\sqrt{2}\left| \mathbf{k}\right| ^{\frac{3}{2}}\left( 1-\widehat{\mathbf{k}}\cdot \frac{\mathbf{P}_{1}}{m}\right) }d^{3}k \)
 it follows that}\\
{\large }\\
{\large \( \mathbf{P}_{2}=\frac{1}{\left\Vert \phi _{\mathbf{P}}^{\sigma }\right\Vert ^{2}}\left( \phi ^{\sigma }_{\mathbf{P}},\Pi _{\mathbf{P},\sigma }\phi ^{\sigma }_{\mathbf{P}}\right) +g^{2}\int _{\sigma }^{\kappa }\frac{\mathbf{k}}{2\left| \mathbf{k}\right| ^{3}\left( 1-\widehat{\mathbf{k}}\cdot \frac{\mathbf{P}_{1}}{m}\right) ^{2}}d^{3}k \)}\\
{\large }\\
{\large By substitution we obtain }\\
{\large }\\
{\large \( H_{\mathbf{P},\sigma }^{w}\equiv W_{\sigma }\left( \frac{\mathbf{P}_{1}}{m}\right) H_{\mathbf{P},\sigma }W_{\sigma }^{\dagger }\left( \frac{\mathbf{P}_{1}}{m}\right) = \)}\\
{\large }\\
{\large \( =\frac{\mathbf{P}^{2}}{2m}+\frac{1}{2m}\left( \Pi _{\mathbf{P},\sigma }+g^{2}\int _{\sigma }^{\kappa }\frac{\mathbf{k}}{2\left| \mathbf{k}\right| ^{3}\left( 1-\widehat{\mathbf{k}}\cdot \frac{\mathbf{P}_{1}}{m}\right) ^{2}}d^{3}k\right) ^{2}-\frac{\mathbf{P}_{2}}{m}\left( \Pi _{\mathbf{P},\sigma }+g^{2}\int _{\sigma }^{\kappa }\frac{\mathbf{k}}{2\left| \mathbf{k}\right| ^{3}\left( 1-\widehat{\mathbf{k}}\cdot \frac{\mathbf{P}_{1}}{m}\right) ^{2}}d^{3}k\right) + \)}\\
{\large }\\
{\large \( +\int ^{\infty }_{0}\left( \left| \mathbf{k}\right| -\mathbf{k}\cdot \frac{\mathbf{P}_{1}}{m}\right) b^{\dagger }\left( \mathbf{k}\right) b\left( \mathbf{k}\right) d^{3}k-g^{2}\int _{\sigma }^{\kappa }\frac{1}{2\left| \mathbf{k}\right| ^{2}\left( 1-\widehat{\mathbf{k}}\cdot \frac{\mathbf{P}_{1}}{m}\right) }d^{3}k= \)}\\
{\large }\\
{\large \( =\frac{1}{2m}\left( \Pi _{\mathbf{P},\sigma }\right) ^{2}+\frac{1}{2m}\left( g^{2}\int _{\sigma }^{\kappa }\frac{\mathbf{k}}{2\left| \mathbf{k}\right| ^{3}\left( 1-\widehat{\mathbf{k}}\cdot \frac{\mathbf{P}_{1}}{m}\right) ^{2}}d^{3}k\right) ^{2}+\frac{1}{m}\Pi _{\mathbf{P},\sigma }\cdot \left( g^{2}\int _{\sigma }^{\kappa }\frac{\mathbf{k}}{2\left| \mathbf{k}\right| ^{3}\left( 1-\widehat{\mathbf{k}}\cdot \frac{\mathbf{P}_{1}}{m}\right) ^{2}}d^{3}k\right) + \)}\\
{\large }\\
{\large \( -\frac{1}{m}\left( \frac{1}{\left\Vert \phi _{\mathbf{P}}^{\sigma }\right\Vert ^{2}}\cdot \left( \phi ^{\sigma }_{\mathbf{P}},\Pi _{\mathbf{P},\sigma }\phi ^{\sigma }_{\mathbf{P}}\right) +g^{2}\int _{\sigma }^{\kappa }\frac{\mathbf{k}}{2\left| \mathbf{k}\right| ^{3}\left( 1-\widehat{\mathbf{k}}\cdot \frac{\mathbf{P}_{1}}{m}\right) ^{2}}d^{3}k\right) \cdot \left( \Pi _{\mathbf{P},\sigma }+g^{2}\int _{\sigma }^{\kappa }\frac{\mathbf{k}}{2\left| \mathbf{k}\right| ^{3}\left( 1-\widehat{\mathbf{k}}\cdot \frac{\mathbf{P}_{1}}{m}\right) ^{2}}d^{3}k\right) + \)}\\
{\large }\\
{\large \( +\int ^{\infty }_{0}\left( \left| \mathbf{k}\right| -\mathbf{k}\cdot \frac{\mathbf{P}_{1}}{m}\right) b^{\dagger }\left( \mathbf{k}\right) b\left( \mathbf{k}\right) d^{3}k-g^{2}\int _{\sigma }^{\kappa }\frac{1}{2\left| \mathbf{k}\right| ^{2}\left( 1-\widehat{\mathbf{k}}\cdot \frac{\mathbf{P}_{1}}{m}\right) }d^{3}k+\frac{\mathbf{P}^{2}}{2m}= \)}
{\large }\\
{\large }\\
{\large \( =\frac{1}{2m}\left( \Pi _{\mathbf{P},\sigma }-\frac{1}{\left\Vert \phi _{\mathbf{P}}^{\sigma }\right\Vert ^{2}}\cdot \left( \phi _{\mathbf{P}}^{\sigma },\Pi _{\mathbf{P},\sigma }\phi _{\mathbf{P}}^{\sigma }\right) \right) ^{2}+\int ^{\infty }_{0}\left( \left| \mathbf{k}\right| -\mathbf{k}\cdot \nabla E^{\sigma }\left( \mathbf{P}\right) \right) b^{\dagger }\left( \mathbf{k}\right) b\left( \mathbf{k}\right) d^{3}k+ \)}\\
{\large }\\
{\large \( +\frac{1}{2m}\left( g^{2}\int _{\sigma }^{\kappa }\frac{\mathbf{k}}{2\left| \mathbf{k}\right| ^{3}\left( 1-\widehat{\mathbf{k}}\cdot \frac{\mathbf{P}_{1}}{m}\right) ^{2}}d^{3}k\right) ^{2}-\frac{1}{m}\left( g^{2}\int _{\sigma }^{\kappa }\frac{\mathbf{k}}{2\left| \mathbf{k}\right| ^{3}\left( 1-\widehat{\mathbf{k}}\cdot \frac{\mathbf{P}_{1}}{m}\right) ^{2}}d^{3}k\right) \cdot \left( g^{2}\int _{\sigma }^{\kappa }\frac{\mathbf{k}}{2\left| \mathbf{k}\right| ^{3}\left( 1-\widehat{\mathbf{k}}\cdot \frac{\mathbf{P}_{1}}{m}\right) ^{2}}d^{3}k\right) + \)}\\
{\large }\\
{\large \( -g^{2}\int _{\sigma }^{\kappa }\frac{1}{2\left| \mathbf{k}\right| ^{2}\left( 1-\widehat{\mathbf{k}}\cdot \frac{\mathbf{P}_{1}}{m}\right) }d^{3}k+\frac{\mathbf{P}^{2}}{2m}-\frac{1}{2m}\left( \frac{1}{\left\Vert \phi _{\mathbf{P}}^{\sigma }\right\Vert ^{2}}\cdot \left( \phi _{\mathbf{P}}^{\sigma },\Pi _{\mathbf{P},\sigma }\phi _{\mathbf{P}}^{\sigma }\right) \right) ^{2}+ \)}\\
{\large }\\
{\large \( -\frac{1}{m}\left( \frac{1}{\left\Vert \phi _{\mathbf{P}}^{\sigma }\right\Vert ^{2}}\cdot \left( \phi _{\mathbf{P}}^{\sigma },\Pi _{\mathbf{P},\sigma }\phi _{\mathbf{P}}^{\sigma }\right) \right) \cdot \left( g^{2}\int _{\sigma }^{\kappa }\frac{\mathbf{k}}{2\left| \mathbf{k}\right| ^{3}\left( 1-\widehat{\mathbf{k}}\cdot \frac{\mathbf{P}_{1}}{m}\right) ^{2}}d^{3}k\right) = \)}\\
{\large }\\
{\large \( =\frac{1}{2m}\left( \Pi _{\mathbf{P},\sigma }-\frac{1}{\left\Vert \phi _{\mathbf{P}}^{\sigma }\right\Vert ^{2}}\cdot \left( \phi _{\mathbf{P}}^{\sigma },\Pi _{\mathbf{P},\sigma }\phi _{\mathbf{P}}^{\sigma }\right) \right) ^{2}+\int ^{\infty }_{0}\left( \left| \mathbf{k}\right| -\mathbf{k}\cdot \nabla E^{\sigma }\left( \mathbf{P}\right) \right) b^{\dagger }\left( \mathbf{k}\right) b\left( \mathbf{k}\right) d^{3}k+c_{\mathbf{P}}\left( \sigma \right)  \)}\\
{\large }\\
{\large where}\\
{\large \( c_{\mathbf{P}}\left( \sigma \right) =-\frac{1}{2m}\left( g^{2}\int _{\sigma }^{\kappa }\frac{\mathbf{k}}{2\left| \mathbf{k}\right| ^{3}\left( 1-\widehat{\mathbf{k}}\cdot \frac{\mathbf{P}_{1}}{m}\right) ^{2}}d^{3}k\right) ^{2}-\frac{1}{2m}\left( \frac{1}{\left\Vert \phi _{\mathbf{P}}^{\sigma }\right\Vert ^{2}}\cdot \left( \phi _{\mathbf{P}}^{\sigma },\Pi _{\mathbf{P},\sigma }\phi _{\mathbf{P}}^{\sigma }\right) \right) ^{2}+ \)}\\
{\large }\\
{\large \( -\frac{1}{m}\left( \frac{1}{\left\Vert \phi _{\mathbf{P}}^{\sigma }\right\Vert ^{2}}\cdot \left( \phi _{\mathbf{P}}^{\sigma },\Pi _{\mathbf{P},\sigma }\phi _{\mathbf{P}}^{\sigma }\right) \right) \cdot \left( g^{2}\int _{\sigma }^{\kappa }\frac{\mathbf{k}}{2\left| \mathbf{k}\right| ^{3}\left( 1-\widehat{\mathbf{k}}\cdot \frac{\mathbf{P}_{1}}{m}\right) ^{2}}d^{3}k\right) -g^{2}\int _{\sigma }^{\kappa }\frac{1}{2\left| \mathbf{k}\right| ^{2}\left( 1-\widehat{\mathbf{k}}\cdot \frac{\mathbf{P}_{1}}{m}\right) }d^{3}k+\frac{\mathbf{P}^{2}}{2m}= \)}\\
{\large }\\
{\small \( =-\frac{1}{2m}\left[ \left( g^{2}\int _{\sigma }^{\kappa }\frac{\mathbf{k}}{2\left| \mathbf{k}\right| ^{3}\left( 1-\widehat{\mathbf{k}}\cdot \frac{\mathbf{P}_{1}}{m}\right) ^{2}}d^{3}k\right) +\left( \frac{1}{\left\Vert \phi _{\mathbf{P}}^{\sigma }\right\Vert ^{2}}\cdot \left( \phi _{\mathbf{P}}^{\sigma },\Pi _{\mathbf{P},\sigma }\phi _{\mathbf{P}}^{\sigma }\right) \right) \right] ^{2}-g^{2}\int _{\sigma }^{\kappa }\frac{1}{2\left| \mathbf{k}\right| ^{2}\left( 1-\widehat{\mathbf{k}}\cdot \frac{\mathbf{P}_{1}}{m}\right) }d^{3}k+ \)}{\large }\\
{\large }\\
\( +\frac{\mathbf{P}^{2}}{2m}=-\frac{1}{2m}\left[ \mathbf{P}-m\nabla E^{\sigma }\left( \mathbf{P}\right) \right] ^{2}-g^{2}\int _{\sigma }^{\kappa }\frac{1}{2\left| \mathbf{k}\right| ^{2}\left( 1-\widehat{\mathbf{k}}\cdot \nabla E^{\sigma }\right) }d^{3}k+\frac{\mathbf{P}^{2}}{2m} \){\large }\\
{\large }\\
{\large From lemma A2 (point 1), \( \left| c_{\mathbf{P}}\left( \sigma \right) \right|  \)
will be bounded uniformly in \( \sigma  \) and \( \left| \nabla E^{\sigma }\left( \mathbf{P}\right) \right| <1 \).}\\
{\large }\\
{\large }\\
{\large }\\
{\large Now, given the infrared cut-off \( \epsilon ^{\frac{j+1}{2}} \)and
\( \epsilon ^{\frac{j+2}{2}} \), I perform the unitary transformations \(  \)}\\
{\large }\\
{\large i) \( H_{\mathbf{P},\epsilon ^{\frac{j+1}{2}\quad }}\rightarrow \quad W_{\epsilon ^{\frac{j+1}{2}}}\left( \nabla E^{\epsilon ^{\frac{j+1}{2}}}\left( \mathbf{P}\right) \right) H_{\mathbf{P},\epsilon ^{\frac{j+1}{2}}}W^{\dagger }_{\epsilon ^{\frac{j+1}{2}}}\left( \nabla E^{\epsilon ^{\frac{j+1}{2}}}\left( \mathbf{P}\right) \right)  \)
}\\
{\large }\\
{\large ii) \( H_{\mathbf{P},\epsilon ^{\frac{j+2}{2}\quad }}\rightarrow \quad W_{\epsilon ^{\frac{j+2}{2}}}\left( \nabla E^{\epsilon ^{\frac{j+1}{2}}}\left( \mathbf{P}\right) \right) H_{\mathbf{P},\epsilon ^{\frac{j+2}{2}}}W^{\dagger }_{\epsilon ^{\frac{j+2}{2}}}\left( \nabla E^{\epsilon ^{\frac{j+1}{2}}}\left( \mathbf{P}\right) \right)  \)}\\
{\large }\\
\textbf{\large Note that the two transformations are different in the cut-off
but not in the coherent factor}{\large . }\\
{\large }\\
\emph{\large Note} {\large }\\
{\large }\\
In order to compress the formulas, from now on I will use the following\\
\\
\( \Gamma ^{i}_{\mathbf{P},\epsilon ^{\frac{j+1}{2}}}\equiv \Pi ^{i}_{\mathbf{P},\epsilon ^{\frac{j+1}{2}}}-\frac{\left( \phi _{\mathbf{P}}^{\epsilon ^{\frac{j+1}{2}}},\Pi ^{i}_{\mathbf{P},\epsilon ^{\frac{j+1}{2}}}\phi _{\mathbf{P}}^{\epsilon ^{\frac{j+1}{2}}}\right) }{\left\Vert \phi ^{\epsilon ^{\frac{j+1}{2}}}_{\mathbf{P}}\right\Vert ^{2}} \)\\
{\large }\\
{\large }\\
{\large We define:}\\
{\large }\\
{\large \( H^{w}_{\mathbf{P},\epsilon ^{\frac{j+1}{2}}}\equiv W_{\epsilon ^{\frac{j+1}{2}}}\left( \nabla E^{\epsilon ^{\frac{j+1}{2}}}\left( \mathbf{P}\right) \right) H_{\mathbf{P},\epsilon ^{\frac{j+1}{2}}}W^{\dagger }_{\epsilon ^{\frac{j+1}{2}}}\left( \nabla E^{\epsilon ^{\frac{j+1}{2}}}\left( \mathbf{P}\right) \right) = \)}\\
{\large }\\
{\large \( =\frac{1}{2m}\left( \Gamma _{\mathbf{P},\epsilon ^{\frac{j+1}{2}}}\right) ^{2}+\int ^{\infty }_{0}\left( \left| \mathbf{k}\right| -\mathbf{k}\cdot \nabla E^{\epsilon ^{\frac{j+1}{2}}}\left( \mathbf{P}\right) \right) b^{\dagger }\left( \mathbf{k}\right) b\left( \mathbf{k}\right) d^{3}k+c_{\mathbf{P}}\left( j+1\right)  \)
}\\
{\large }\\
{\large }\\
{\large \( \widehat{H}^{w}_{\mathbf{P},\epsilon ^{\frac{j+2}{2}}}\equiv W_{\epsilon ^{\frac{j+2}{2}}}\left( \nabla E^{\epsilon ^{\frac{j+1}{2}}}\left( \mathbf{P}\right) \right) H_{\mathbf{P},\epsilon ^{\frac{j+2}{2}}}W^{\dagger }_{\epsilon ^{\frac{j+2}{2}}}\left( \nabla E^{\epsilon ^{\frac{j+1}{2}}}\left( \mathbf{P}\right) \right) = \)}\\
{\small }\\
{\small \( =\frac{1}{2m}\left( \Gamma _{\mathbf{P},\epsilon ^{\frac{j+1}{2}}}-g\int _{\epsilon ^{\frac{j+2}{2}}}^{\epsilon ^{\frac{j+1}{2}}}\frac{\mathbf{k}}{\sqrt{2}\left| \mathbf{k}\right| ^{\frac{3}{2}}\left( 1-\widehat{\mathbf{k}}\cdot \nabla E^{\epsilon ^{\frac{j+1}{2}}}\left( \mathbf{P}\right) \right) }\left( b\left( \mathbf{k}\right) +b^{\dagger }\left( \mathbf{k}\right) \right) d^{3}k+g^{2}\int _{\epsilon ^{\frac{j+2}{2}}}^{\epsilon ^{\frac{j+1}{2}}}\frac{\mathbf{k}}{2\left| \mathbf{k}\right| ^{3}\left( 1-\widehat{\mathbf{k}}\cdot \nabla E^{\epsilon ^{\frac{j+1}{2}}}\left( \mathbf{P}\right) \right) ^{2}}d^{3}k\right) ^{2}+ \)}
{\large }\\
\( +\int ^{\infty }_{0}\left( \left| \mathbf{k}\right| -\mathbf{k}\cdot \nabla E^{\epsilon ^{\frac{j+1}{2}}}\left( \mathbf{P}\right) \right) b^{\dagger }\left( \mathbf{k}\right) b\left( \mathbf{k}\right) d^{3}k+\widehat{c}_{\mathbf{P}}\left( j+2\right)  \)
\\
{\large }\\
{\large where }\\
{\large }\\
\( \widehat{c}_{\mathbf{P}}\left( j+2\right) =-\frac{1}{2m}\left( \frac{1}{\left\Vert \phi _{\mathbf{P}}^{\epsilon ^{\frac{j+1}{2}}}\right\Vert ^{2}}\left( \phi ^{\epsilon ^{\frac{j+1}{2}}}_{\mathbf{P}},\Pi _{\mathbf{P},\epsilon ^{\frac{j+1}{2}}}\phi ^{\epsilon ^{\frac{j+1}{2}}}_{\mathbf{P}}\right) +g^{2}\int _{\epsilon ^{\frac{j+1}{2}}}^{\kappa }\frac{\mathbf{k}}{2\left| \mathbf{k}\right| ^{3}\left( 1-\widehat{\mathbf{k}}\cdot \nabla E^{\epsilon ^{\frac{j+1}{2}}}\left( \mathbf{P}\right) \right) ^{2}}d^{3}k\right) ^{2}+ \)\\
\\
\( -g^{2}\int _{\epsilon ^{\frac{j+2}{2}}}^{\kappa }\frac{1}{2\left| \mathbf{k}\right| ^{2}\left( 1-\widehat{\mathbf{k}}\cdot \nabla E^{\epsilon ^{\frac{j+1}{2}}}\left( \mathbf{P}\right) \right) }d^{3}k+\frac{\mathbf{P}^{2}}{2m} \).\\
\\
{\large }\\
{\large By construction, the following equality holds}\\
{\large }\\
{\large \( H^{w}_{\mathbf{P},\epsilon ^{\frac{j+2}{2}}}=W_{\epsilon ^{\frac{j+2}{2}}}\left( \nabla E^{\epsilon ^{\frac{j+2}{2}}}\left( \mathbf{P}\right) \right) H_{\mathbf{P},\epsilon ^{\frac{j+2}{2}}}W^{\dagger }_{\epsilon ^{\frac{j+2}{2}}}\left( \nabla E^{\epsilon ^{\frac{j+2}{2}}}\left( \mathbf{P}\right) \right) = \)}\\
{\large }\\
{\large \( =W_{\epsilon ^{\frac{j+2}{2}}}\left( \nabla E^{\epsilon ^{\frac{j+2}{2}}}\left( \mathbf{P}\right) \right) W^{\dagger }_{\epsilon ^{\frac{j+2}{2}}}\left( \nabla E^{\epsilon ^{\frac{j+1}{2}}}\left( \mathbf{P}\right) \right) \widehat{H}^{w}_{\mathbf{P},\epsilon ^{\frac{j+2}{2}}}W_{\epsilon ^{\frac{j+2}{2}}}\left( \nabla E^{\epsilon ^{\frac{j+1}{2}}}\left( \mathbf{P}\right) \right) W^{\dagger }_{\epsilon ^{\frac{j+2}{2}}}\left( \nabla E^{\epsilon ^{\frac{j+2}{2}}}\left( \mathbf{P}\right) \right)  \)}\\
{\large }\\
{\large (the hamiltonians \( H_{\mathbf{P},\epsilon ^{\frac{j+1}{2}}} \),\( H^{w}_{\mathbf{P},\epsilon ^{\frac{j+1}{2}}} \)
and \( \widehat{H}^{w}_{\mathbf{P},\epsilon ^{\frac{j+1}{2}}} \) are s.a. on
the same domain and the formal equalities are well defined from an operatorial
point of view).}\\
{\large }\\
\emph{\large Definitions} \\
\\
\( \widehat{\Pi }^{i}_{\mathbf{P},\epsilon ^{\frac{j+1}{2}}}\equiv W_{\epsilon ^{\frac{j+1}{2}}}\left( \nabla E^{\epsilon ^{\frac{j}{2}}}\left( \mathbf{P}\right) \right) W^{\dagger }_{\epsilon ^{\frac{j+1}{2}}}\left( \nabla E^{\epsilon ^{\frac{j+1}{2}}}\left( \mathbf{P}\right) \right) \Pi ^{i}_{\mathbf{P},\epsilon ^{\frac{j+1}{2}}}W_{\epsilon ^{\frac{j+1}{2}}}\left( \nabla E^{\epsilon ^{\frac{j+1}{2}}}\left( \mathbf{P}\right) \right) W^{\dagger }_{\epsilon ^{\frac{j+1}{2}}}\left( \nabla E^{\epsilon ^{\frac{j}{2}}}\left( \mathbf{P}\right) \right) = \)
{\large }\\
\( =\mathbf{P}^{mes^{i}}-g\int ^{\kappa }_{\epsilon ^{\frac{j+1}{2}}}\frac{k^{i}\left( b\left( \mathbf{k}\right) +b^{\dagger }\left( \mathbf{k}\right) \right) }{\sqrt{2}\left| \mathbf{k}\right| ^{\frac{3}{2}}\left( 1-\widehat{\mathbf{k}}\cdot \nabla E^{\epsilon ^{\frac{j}{2}}}_{\mathbf{P}}\right) }d^{3}k+\frac{g^{2}}{2}\int ^{\kappa }_{\epsilon ^{\frac{j+1}{2}}}\frac{k^{i}}{\left| \mathbf{k}\right| ^{3}\left( 1-\widehat{\mathbf{k}}\cdot \nabla E^{\epsilon ^{\frac{j}{2}}}_{\mathbf{P}}\right) ^{2}}d^{3}k-\frac{g^{2}}{2}\int ^{\kappa }_{\epsilon ^{\frac{j+1}{2}}}\frac{k^{i}}{\left| \mathbf{k}\right| ^{3}\left( 1-\widehat{\mathbf{k}}\cdot \nabla E^{\epsilon ^{\frac{j+1}{2}}}_{\mathbf{P}}\right) ^{2}}d^{3}k \);{\large }\\
{\large }\\
{\large \( \widehat{\Gamma }^{i}_{\mathbf{P},\epsilon ^{\frac{j+1}{2}}}\equiv \widehat{\Pi }^{i}_{\mathbf{P},\epsilon ^{\frac{j+1}{2}}}-\frac{\left( \widehat{\phi }_{\mathbf{P}}^{\epsilon ^{\frac{j+1}{2}}},\widehat{\Pi }^{i}_{\mathbf{P},\epsilon ^{\frac{j+1}{2}}}\widehat{\phi }_{\mathbf{P}}^{\epsilon ^{\frac{j+1}{2}}}\right) }{\left\Vert \widehat{\phi }^{\epsilon ^{\frac{j+1}{2}}}_{\mathbf{P}}\right\Vert ^{2}} \)}\\
{\large }\\
{\large }\\
{\large }\\
{\large }\\
\textbf{\large Theorem 2.3bis }\\
\textbf{\large }\\
{\large Convergence of the sequence of the ground states \( H^{w}_{\mathbf{P},\epsilon ^{\frac{j+1}{2}}} \).}\\
{\large }\\
{\large In order to arrive to a convergent sequence of ground eigenvectors,
one starts from the cut-off} {\large \( \epsilon  \) and from the (normalized)
vector \( W_{\epsilon }\left( \nabla E^{\epsilon }\left( \mathbf{P}\right) \right) \psi _{\mathbf{P}}^{\epsilon } \)
and proceeds with the same iteration, on the basis of the results of lemma A1.
Comparing the hamiltonians \( H^{w}_{\mathbf{P},\epsilon ^{\frac{j+1}{2}}} \)
and \( \widehat{H}^{w}_{\mathbf{P},\epsilon ^{\frac{j+2}{2}}} \) we can build
\( \widehat{\phi }_{\mathbf{P}}^{\epsilon ^{\frac{j+2}{2}}} \)in terms of \( \phi _{\mathbf{P}}^{\epsilon ^{\frac{j+1}{2}}} \)
and finally we define}\\
{\large \( \phi _{\mathbf{P}}^{\epsilon ^{\frac{j+2}{2}}}\equiv W_{\epsilon ^{\frac{j+2}{2}}}\left( \nabla E^{\epsilon ^{\frac{j+2}{2}}}\left( \mathbf{P}\right) \right) W^{\dagger }_{\epsilon ^{\frac{j+2}{2}}}\left( \nabla E^{\epsilon ^{\frac{j+1}{2}}}\left( \mathbf{P}\right) \right) \widehat{\phi }_{\mathbf{P}}^{\epsilon ^{\frac{j+2}{2}}} \).
}\\
{\large Like in the case \( \mathbf{P}=0 \), in order to prove the convergence
of the vectors \( \phi _{\mathbf{P}}^{\epsilon ^{\frac{j+1}{2}}} \), one needs
a more refined estimate of the contribution, to the difference between the generic
vectors \( \phi _{\mathbf{P}}^{\epsilon ^{\frac{j+1}{2}}} \)and \( \widehat{\phi }_{\mathbf{P}}^{\epsilon ^{\frac{j+2}{2}}} \),
given by the mixed terms containing the \( \Gamma _{\mathbf{P},\epsilon ^{\frac{j+1}{2}}}^{i} \).
For this purpose the procedure to follow is slightly more elaborated than in
the case where \( \mathbf{P}=0 \), because it is necessary to compare the following
vectors one after the other:}\\
{\large }\\
{\large }\\
{\large \( \phi _{\mathbf{P}}^{\epsilon ^{\frac{j}{2}}}\rightarrow \widehat{\phi }^{\epsilon ^{\frac{j+1}{2}}}_{\mathbf{P}}\rightarrow \phi _{\mathbf{P}}^{\epsilon ^{\frac{j+1}{2}}}\rightarrow \widehat{\phi }^{\epsilon ^{\frac{j+2}{2}}}_{\mathbf{P}}\rightarrow \phi _{\mathbf{P}}^{\epsilon ^{\frac{j+2}{2}}} \).}\\
{\large }\\
{\large I apply a proper adaptation of the lemmas 2.1 e 2.2. to estimate the
contribution to the norm of \( \phi _{\mathbf{P}}^{\epsilon ^{\frac{j+1}{2}}}-\widehat{\phi }^{\epsilon ^{\frac{j+2}{2}}}_{\mathbf{P}} \)given
by the examined mixed terms. Apart from numeric differences, even in this case
the improvement of the estimate consists in the bound of an expression like
(14) of theorem 2.3 with a quantity of the order \( \epsilon ^{-\frac{j+1}{4}} \)
(that means multiplicative constants that are uniform in \( j \)):} {\large }\\
{\large }\\
{\large 
\[
\frac{g^{2}}{4m^{2}}\left| \left( \Gamma ^{i}_{\mathbf{P},\epsilon ^{\frac{j+1}{2}}}\phi _{\mathbf{P}}^{\epsilon ^{\frac{j+1}{2}}}\! ,\! \left( \frac{1}{H^{w}_{\mathbf{P},\epsilon ^{\frac{j+1}{2}}}-E\left( j+1\right) }\right) \! \Gamma ^{i}_{\mathbf{P},\epsilon ^{\frac{j+1}{2}}}\phi _{\mathbf{P}}^{\epsilon ^{\frac{j+1}{2}}}\right) \right| \]
 where \( E\left( j+1\right)  \) is s.t. \( \left| E\left( j+1\right) -E_{\mathbf{P}}^{\epsilon ^{\frac{j+1}{2}}}\right| =\frac{11}{20}\epsilon ^{\frac{j+2}{2}} \).}\\
{\large }\\
\emph{\large Not}{\large e}\\
\\
In the adaptation to the examined case of the derivation of the (13) from the
(14), take care of lemma A1 and of the following difference:\\
\\
\( \widehat{c}_{\mathbf{P}}\left( j+2\right) -c_{\mathbf{P}}\left( j+1\right) =-g^{2}\int _{\epsilon ^{\frac{j+2}{2}}}^{\epsilon ^{\frac{j+1}{2}}}\frac{1}{2\left| \mathbf{k}\right| ^{2}\left( 1-\widehat{\mathbf{k}}\cdot \nabla E^{\epsilon ^{\frac{j+1}{2}}}\left( \mathbf{P}\right) \right) }d^{3}k \){\small }\\
{\large }\\
{\large }\\
{\large The bound for the module of the above scalar product is the result of
an inductive procedure like the one where \( \mathbf{P}=0 \); it is for this
reason that I will not repeat the entire proof but I will point out the key
ingredients without further details.}\\
{\large }\\
{\large I apply to both the factors of the scalar product the unitary operator}\\
{\large 
\[
W_{\epsilon ^{\frac{j+1}{2}}}\left( \nabla E^{\epsilon ^{\frac{j}{2}}}\left( \mathbf{P}\right) \right) W^{\dagger }_{\epsilon ^{\frac{j+1}{2}}}\left( \nabla E^{\epsilon ^{\frac{j+1}{2}}}\left( \mathbf{P}\right) \right) \]
} {\large }\\
{\large (such an operation is not required if \( \mathbf{P}=0 \)) and I obtain}\\
\\
{\large \( \left( \widehat{\Gamma }^{i}_{\mathbf{P},\epsilon ^{\frac{j+1}{2}}}\widehat{\phi }_{\mathbf{P}}^{\epsilon ^{\frac{j+1}{2}}}\! ,\! \left( \frac{1}{\widehat{H}^{w}_{\mathbf{P},\epsilon ^{\frac{j+1}{2}}}-E\left( j+1\right) }\right) \! \widehat{\Gamma }^{i}_{\mathbf{P},\epsilon ^{\frac{j+1}{2}}}\widehat{\phi }_{\mathbf{P}}^{\epsilon ^{\frac{j+1}{2}}}\right)  \)}\\
\\
{\large as in theorem 2.3 I go on with the following bound (from above)}\\
{\large }\\
{\large \( \left| \left( \widehat{\Gamma }^{i}_{\mathbf{P},\epsilon ^{\frac{j+1}{2}}}\widehat{\phi }_{\mathbf{P}}^{\epsilon ^{\frac{j+1}{2}}}\! ,\! \left( \frac{1}{\widehat{H}^{w}_{\mathbf{P},\epsilon ^{\frac{j+1}{2}}}-E\left( j+1\right) }\right) \! \widehat{\Gamma }^{i}_{\mathbf{P},\epsilon ^{\frac{j+1}{2}}}\widehat{\phi }_{\mathbf{P}}^{\epsilon ^{\frac{j+1}{2}}}\right) \right| \leq  \)}\\
{\large }\\
{\large \( \leq 2\left| \left( \widehat{\Gamma }^{i}_{\mathbf{P},\epsilon ^{\frac{j+1}{2}}}\widehat{\phi }_{\mathbf{P}}^{\epsilon ^{\frac{j+1}{2}}}\! ,\! \left| \left( \frac{1}{H_{\mathbf{P},\epsilon ^{\frac{j}{2}}}^{w}-E\left( j\right) }\right) \right| \! \widehat{\Gamma }^{i}_{\mathbf{P},\epsilon ^{\frac{j+1}{2}}}\widehat{\phi }_{\mathbf{P}}^{\epsilon ^{\frac{j+1}{2}}}\right) \right| \leq  \)
}\\
{\large }\\
{\large \( \leq 2\left| \left( \widehat{\Gamma }^{i}_{\mathbf{P},\epsilon ^{\frac{j+1}{2}}}\widehat{\phi }_{\mathbf{P}}^{\epsilon ^{\frac{j+1}{2}}}-\Gamma ^{i}_{\mathbf{P},\epsilon ^{\frac{j}{2}}}\phi _{\mathbf{P}}^{\epsilon ^{\frac{j}{2}}}\! ,\! \left| \left( \frac{1}{H_{\mathbf{P},\epsilon ^{\frac{j}{2}}}^{w}-E\left( j\right) }\right) \right| \! \left( \widehat{\Gamma }^{i}_{\mathbf{P},\epsilon ^{\frac{j+1}{2}}}\widehat{\phi }_{\mathbf{P}}^{\epsilon ^{\frac{j+1}{2}}}-\Gamma ^{i}_{\mathbf{P},\epsilon ^{\frac{j}{2}}}\phi _{\mathbf{P}}^{\epsilon ^{\frac{j}{2}}}\right) \right) \right| + \)}\\
\\
{\large \( +4\left| \left( \widehat{\Gamma }^{i}_{\mathbf{P},\epsilon ^{\frac{j+1}{2}}}\widehat{\phi }_{\mathbf{P}}^{\epsilon ^{\frac{j+1}{2}}}-\Gamma ^{i}_{\mathbf{P},\epsilon ^{\frac{j}{2}}}\phi _{\mathbf{P}}^{\epsilon ^{\frac{j}{2}}}\! ,\! \left| \left( \frac{1}{H_{\mathbf{P},\epsilon ^{\frac{j}{2}}}^{w}-E\left( j\right) }\right) \right| \! \left( \Gamma ^{i}_{\mathbf{P},\epsilon ^{\frac{j}{2}}}\phi _{\mathbf{P}}^{\epsilon ^{\frac{j}{2}}}\right) \right) \right| + \)}\\
\\
{\large \( +2\left| \left( \Gamma ^{i}_{\mathbf{P},\epsilon ^{\frac{j}{2}}}\phi _{\mathbf{P}}^{\epsilon ^{\frac{j}{2}}}\! ,\! \left| \left( \frac{1}{H_{\mathbf{P},\epsilon ^{\frac{j}{2}}}^{w}-E\left( j\right) }\right) \right| \! \Gamma ^{i}_{\mathbf{P},\epsilon ^{\frac{j}{2}}}\phi _{\mathbf{P}}^{\epsilon ^{\frac{j}{2}}}\right) \right|  \)}\\
{\large }\\
{\large }\\
{\large The proof goes on as in theorem 2.3 taking into account the following
information:}\\
{\large }\\
\emph{\large 1)} {\large }\\
{\large \( \widehat{\Pi }^{i}_{\mathbf{P},\epsilon ^{\frac{j+1}{2}}}-\Pi ^{i}_{\mathbf{P},\epsilon ^{\frac{j}{2}}}=-g\int ^{\epsilon ^{\frac{j}{2}}}_{\epsilon ^{\frac{j+1}{2}}}\frac{k^{i}\left( b\left( \mathbf{k}\right) +b^{\dagger }\left( \mathbf{k}\right) \right) }{\sqrt{2}\left| \mathbf{k}\right| ^{\frac{3}{2}}\left( 1-\widehat{\mathbf{k}}\nabla E^{\epsilon ^{\frac{j}{2}}}_{\mathbf{P}}\right) }d^{3}k+\frac{g^{2}}{2}\int ^{\kappa }_{\epsilon ^{\frac{j+1}{2}}}\frac{k^{i}}{\left| \mathbf{k}\right| ^{3}\left( 1-\widehat{\mathbf{k}}\cdot \nabla E^{\epsilon ^{\frac{j}{2}}}_{\mathbf{P}}\right) ^{2}}d^{3}k+ \)}\\
{\large }\\
{\large \( -\frac{g^{2}}{2}\int ^{\kappa }_{\epsilon ^{\frac{j+1}{2}}}\frac{k^{i}}{\left| \mathbf{k}\right| ^{3}\left( 1-\widehat{\mathbf{k}}\cdot \nabla E^{\epsilon ^{\frac{j+1}{2}}}_{\mathbf{P}}\right) ^{2}}d^{3}k \)}\\
{\scriptsize }\\
{\large }\\
{\large \( \left( \widehat{\phi }_{\mathbf{P}}^{\epsilon ^{\frac{j+1}{2}}},\widehat{\Pi }^{i}_{\mathbf{P},\epsilon ^{\frac{j+1}{2}}}\widehat{\phi }_{\mathbf{P}}^{\epsilon ^{\frac{j+1}{2}}}\right) -\left( \phi _{\mathbf{P}}^{\epsilon ^{\frac{j}{2}}},\Pi ^{i}_{\mathbf{P},\epsilon ^{\frac{j}{2}}}\phi _{\mathbf{P}}^{\epsilon ^{\frac{j}{2}}}\right) = \)}\\
{\large }\\
{\large \( =2Re\left( \Pi ^{i}_{\mathbf{P},\epsilon ^{\frac{j}{2}}}\widehat{\phi }_{\mathbf{P}}^{\epsilon ^{\frac{j}{2}}},\widehat{\phi }_{\mathbf{P}}^{\epsilon ^{\frac{j+1}{2}}}-\phi _{\mathbf{P}}^{\epsilon ^{\frac{j}{2}}}\right) +\left( \Pi ^{i}_{\mathbf{P},\epsilon ^{\frac{j}{2}}}\left( \phi _{\mathbf{P}}^{\epsilon ^{\frac{j}{2}}}-\widehat{\phi }_{\mathbf{P}}^{\epsilon ^{\frac{j+1}{2}}}\right) ,\widehat{\phi }_{\mathbf{P}}^{\epsilon ^{\frac{j+1}{2}}}-\phi _{\mathbf{P}}^{\epsilon ^{\frac{j}{2}}}\right) + \)}\\
{\large }\\
{\large \( +\left( \widehat{\phi }_{\mathbf{P}}^{\epsilon ^{\frac{j+1}{2}}},\left\{ \widehat{\Pi }^{i}_{\mathbf{P},\epsilon ^{\frac{j+1}{2}}}-\Pi ^{i}_{\mathbf{P},\epsilon ^{\frac{j}{2}}}\right\} \widehat{\phi }_{\mathbf{P}}^{\epsilon ^{\frac{j+1}{2}}}\right)  \)}\\
{\large }\\
{\large }\\
{\large }\\
\emph{\large 2)} {\large \( \left\Vert \int ^{\epsilon ^{\frac{j}{2}}}_{\epsilon ^{\frac{j+1}{2}}}q^{i}b\left( \mathbf{q}\right) \frac{d^{3}q}{\left| \mathbf{q}\right| \sqrt{2\left| \mathbf{q}\right| }\left( 1-\widehat{\mathbf{q}}\cdot \nabla E^{\epsilon ^{\frac{j}{2}}}_{\mathbf{P}}\right) }\left( \frac{1}{H^{w}_{\mathbf{P},\epsilon ^{\frac{j}{2}}}-E\left( j\right) }\right) ^{\frac{1}{2}}\right\Vert _{F_{\epsilon ^{\frac{j+1}{2}}}^{+}} \)is
of order \( \epsilon ^{\frac{j}{4}} \)}\\
{\large }\\
\emph{\large 2bis)} {\large \( \left\Vert \int ^{\epsilon ^{\frac{j}{2}}}_{\epsilon ^{\frac{j+1}{2}}}q^{i}b^{\dagger }\left( \mathbf{q}\right) \frac{d^{3}q}{\left| \mathbf{q}\right| \sqrt{2\left| \mathbf{q}\right| }\left( 1-\widehat{\mathbf{q}}\cdot \nabla E^{\epsilon ^{\frac{j}{2}}}_{\mathbf{P}}\right) }\left( \frac{1}{H^{w}_{\mathbf{P},\epsilon ^{\frac{j}{2}}}-E\left( j\right) }\right) ^{\frac{1}{2}}\right\Vert _{F_{\epsilon ^{\frac{j+1}{2}}}^{+}} \)is
of order \( \epsilon ^{\frac{j}{4}} \)}\\
{\large }\\
{\large }\\
{\large }\\
\emph{\large 3)} {\large inductive hypothesis for evaluating:}\\
{\large }\\
{\large i)\( \frac{g^{2}}{4m^{2}}\left\Vert \left| \frac{1}{H_{\mathbf{P},\epsilon ^{\frac{j}{2}}}^{w}-E\left( j\right) }\right| ^{\frac{1}{2}}d^{3}k\! \Gamma ^{i}_{\epsilon ^{\frac{j}{2}}}\phi ^{\epsilon ^{\frac{j}{2}}}\right\Vert ^{2} \)
of order \( \epsilon ^{-\frac{j}{4}} \)}\\
{\large }\\
{\large ii) \( \left\Vert \widehat{\phi }_{\mathbf{P}}^{\epsilon ^{\frac{j+1}{2}}}-\phi _{\mathbf{P}}^{\epsilon ^{\frac{j}{2}}}\right\Vert  \)}
{\large of order \( \epsilon ^{\frac{j}{8}}=\sigma _{j-1}^{\frac{1}{4}} \),
which implies}{\large \par}

\begin{itemize}
\item {\large \( \left\Vert \Pi _{\mathbf{P},\epsilon ^{\frac{j}{2}}}^{i}\widehat{\phi }_{\mathbf{P}}^{\epsilon ^{\frac{j+1}{2}}}-\Pi _{\mathbf{P},\epsilon ^{\frac{j}{2}}}^{i}\phi _{\mathbf{P}}^{\epsilon ^{\frac{j}{2}}}\right\Vert <\sqrt{2m}\left( E_{\mathbf{P}}^{\epsilon ^{\frac{j}{2}}}-c_{\mathbf{P}}\left( j\right) \right) ^{\frac{1}{2}}\left\Vert \widehat{\phi }_{\mathbf{P}}^{\epsilon ^{\frac{j+1}{2}}}-\phi _{\mathbf{P}}^{\epsilon ^{\frac{j}{2}}}\right\Vert  \)~then
of order \( \epsilon ^{\frac{j}{8}} \) (the uniformity in j of the bound of
\( \left( E_{\mathbf{P}}^{\epsilon ^{\frac{j}{2}}}-c_{\mathbf{P}}\left( j\right) \right)  \)
follows from lemma A2)}\\
 
\item {\large \( \left| \nabla E_{\mathbf{P}}^{\epsilon ^{\frac{j+1}{2}}}-\nabla E_{\mathbf{P}}^{\epsilon ^{\frac{j}{2}}}\right| \leq C\left( g\right) \epsilon ^{\frac{j}{8}} \)~
(see lemma A2)}\\
{\large }\\
{\large \par}
\end{itemize}
{\large As for the estimate of }\\
{\large }\\
{\large \( \left\Vert \widehat{\phi }_{\mathbf{P}}^{\epsilon ^{\frac{j+2}{2}}}-\phi _{\mathbf{P}}^{\epsilon ^{\frac{j+2}{2}}}\right\Vert =\left\Vert W_{\epsilon ^{\frac{j+2}{2}}}\left( \nabla E^{\epsilon ^{\frac{j+1}{2}}}\left( \mathbf{P}\right) \right) W^{\dagger }_{\epsilon ^{\frac{j+2}{2}}}\left( \nabla E^{\epsilon ^{\frac{j+2}{2}}}\left( \mathbf{P}\right) \right) \phi _{\mathbf{P}}^{\epsilon ^{\frac{j+2}{2}}}-\phi _{\mathbf{P}}^{\epsilon ^{\frac{j+2}{2}}}\right\Vert  \)}\\
{\large }\\
{\large note that the difference between \( \nabla E_{\mathbf{P}}^{\epsilon ^{\frac{j+1}{2}}} \)
and \( \nabla E_{\mathbf{P}}^{\epsilon ^{\frac{j+2}{2}}} \)is bounded by a
quantity of order \( \epsilon ^{\frac{j+1}{8}} \). This allows to neutralize
the logarithmic divergence in \( \epsilon ^{\frac{j+2}{2}} \) which arises
from the Weyl operators \( W_{\epsilon ^{\frac{j+2}{2}}}\left( \nabla E^{\epsilon ^{\frac{j+1}{2}}}\left( \mathbf{P}\right) \right) W^{\dagger }_{\epsilon ^{\frac{j+1}{2}}}\left( \nabla E^{\epsilon ^{\frac{j+2}{2}}}\left( \mathbf{P}\right) \right)  \).
}\\
{\large }\\
{\large In conclusion,} \textbf{\large tuning g uniformly in j} {\large as we
saw in theorem 2.3, it is possible to bound the norm difference between \( \phi _{\mathbf{P}}^{\epsilon ^{\frac{j+1}{2}}} \)and
\( \widehat{\phi }_{\mathbf{P}}^{\epsilon ^{\frac{j+2}{2}}} \)with \( \epsilon ^{\frac{j+1}{8}} \)
and the difference between \( \phi _{\mathbf{P}}^{\epsilon ^{\frac{j+1}{2}}} \)and
\( \phi _{\mathbf{P}}^{\epsilon ^{\frac{j+2}{2}}} \)with \( \epsilon ^{\frac{j+1}{16}} \)
.}\\
{\large }\\
{\large }\\
\emph{\large Observation}{\large }\\
{\large }\\
The logarithmic divergence in \( \epsilon ^{\frac{j+2}{2}} \) related to the
Weyl operators includes the one that follows from the first approximation of:\\
\\
\( \left( \int ^{\kappa }_{\epsilon ^{\frac{j+2}{2}}}\left\Vert b\left( \mathbf{k}\right) W^{b}_{\epsilon ^{\frac{j+2}{2}}}\left( \nabla E^{\epsilon ^{\frac{j+2}{2}}}\left( \mathbf{P}\right) \right) \psi _{\mathbf{P},\epsilon ^{\frac{j+2}{2}}}\right\Vert ^{2}d^{3}k\right) ^{\frac{1}{2}} \)\\
\\
being \( b\left( \mathbf{k}\right) \psi _{\mathbf{P},\epsilon ^{\frac{j+2}{2}}}=\frac{g}{\sqrt{2\left| \mathbf{k}\right| }}\left( \frac{1}{E^{\epsilon ^{\frac{j+2}{2}}}\left( \mathbf{P}\right) -\left| \mathbf{k}\right| -H_{\mathbf{P}-\mathbf{k},\epsilon ^{\frac{j+2}{2}}}}\right) \psi _{\mathbf{P},\epsilon ^{\frac{j+2}{2}}} \)
(see {[}2{]}) and \( \left\Vert \psi _{\mathbf{P},\epsilon ^{\frac{j+2}{2}}}\right\Vert <1 \).{\large }\\
{\large }\\
(see also lemma B2 in Appendix B){\large .}\\
{\large }\\
\textbf{\large Corollary 2.4}\\
\textbf{\large }\\
{\large The sequence \( \left\{ \phi _{\mathbf{P}}^{\epsilon ^{\frac{j+2}{2}}}\right\}  \)
(\( \phi _{\mathbf{P}}^{\epsilon } \) normalized vector) converges to a non-vanishing
vector when the value of the coupling constant is less or equal to the \( \overline{g} \)
determined by theorems 2.3 and 2.3bis, therefore such as that \( \left\Vert \phi _{\mathbf{P}}^{\epsilon ^{\frac{j+2}{2}}}-\phi _{\mathbf{P}}^{\epsilon ^{\frac{j+1}{2}}}\right\Vert \leq \epsilon ^{\frac{j+1}{16}} \).}\\
{\large }\\
{\large }\\
{\large Proof}\\
{\large }\\
{\large This is a Cauchy sequence because \( \forall l,j\: \: \: l\geq j \)}\textbf{\large }\\
\textbf{\large }\\
\textbf{\large \( \left\Vert \phi _{\mathbf{P}}^{\epsilon ^{\frac{l+1}{2}}}-\phi _{\mathbf{P}}^{\epsilon ^{\frac{j+1}{2}}}\right\Vert \leq \left\Vert \phi _{\mathbf{P}}^{\epsilon ^{\frac{l+1}{2}}}-\phi _{\mathbf{P}}^{\epsilon ^{\frac{l}{2}}}+\phi _{\mathbf{P}}^{\epsilon ^{\frac{l}{2}}}-.......-\phi _{\mathbf{P}}^{\epsilon ^{\frac{j+2}{2}}}+\phi _{\mathbf{P}}^{\epsilon ^{\frac{j+2}{2}}}-\phi _{\mathbf{P}}^{\epsilon ^{\frac{j+1}{2}}}\right\Vert \leq  \)}\\
\textbf{\large }\\
\textbf{\large \( \leq \epsilon ^{\frac{l}{16}}+\epsilon ^{\frac{l-1}{16}}+......+\epsilon ^{\frac{j+1}{16}}\leq \epsilon ^{\frac{j+1}{16}}\cdot \left( \frac{1}{1-\epsilon ^{\frac{1}{16}}}\right)  \)}\\
\textbf{\large }\\
{\large The limit does not vanish since the following inequality holds, uniformly
in \( j \):}\textbf{\large }\\
\textbf{\large }\\
\textbf{\large \( \left\Vert \phi _{\mathbf{P}}^{\epsilon ^{\frac{j+1}{2}}}\right\Vert \geq \left\Vert \phi _{\mathbf{P}}^{\epsilon }\right\Vert -\left( \epsilon ^{\frac{1}{16}}+\epsilon ^{\frac{2}{16}}+....+\epsilon ^{\frac{j}{16}}\right) \geq 1-\left( \frac{\epsilon ^{\frac{1}{16}}}{1-\epsilon ^{\frac{1}{16}}}\right) \geq \frac{1-2\epsilon ^{\frac{1}{16}}}{1-\epsilon ^{\frac{1}{16}}}>\frac{2}{3} \)}{\large .}\textbf{\large }\\
{\large \par}

\section{Spectral regularity.\\
}

{\large In this paragraph I will define a normalized vector \( \phi _{\mathbf{P}}^{\sigma } \),
that is the ground state of \( H^{w}_{\mathbf{P},\sigma }\mid _{F_{\sigma }^{+}} \)
( \( \sigma \leq \epsilon  \) ). It has a regularity property in \( \mathbf{P} \)
required for the construction of the scattering states in the next chapter (in
particular I will use the vectors \( W^{\dagger }_{\sigma }\left( \nabla E^{\sigma }\left( \mathbf{P}\right) \right) \phi _{\mathbf{P}}^{\sigma } \)).
}\\
{\large We arrive at the vector \( \phi _{\mathbf{P}}^{\sigma } \) through
an intermediate (not normalized) vector \( \widetilde{\phi }^{\sigma }_{\mathbf{P}} \)
from which it differs only for a phase term, apart from the normalization.}
{\large The choice of the right phase is aimed at two results:} {\large }\\
{\large }\\
{\large - the norm convergence of the vector \( \phi _{\mathbf{P}}^{\sigma } \)
, for \( \sigma \rightarrow 0 \), to a vector \( \phi _{\mathbf{P}} \);} {\large }\\
{\large }\\
{\large - the acquisition of the following Hoelder property, with respect to
\( \mathbf{P} \): }\\
{\large 
\[
\left\Vert \phi ^{\sigma }_{\mathbf{P}+\Delta \mathbf{P}}-\phi ^{\sigma }_{\mathbf{P}}\right\Vert \leq C\left| \Delta \mathbf{P}\right| ^{\frac{1}{32}}\]
~~~}\\
{\large where \( C \) is a uniform constant in \( \mathbf{P},\mathbf{P}+\Delta \mathbf{P}\in \Sigma  \)
(\( \Delta \mathbf{P}<\left( \frac{1}{6}\right) ^{4} \)) and in \( 0<\sigma \leq \epsilon  \)}\\
{\large }\\
\textbf{\large }\\
\textbf{\large Definition of} {\large \( \widetilde{\phi }^{\sigma }_{\mathbf{P}} \)}\\
{\large }\\
\emph{\large Initial conditions}{\large }\\
{\large }\\
{\large We start from the initial infrared cut-off \( \left\{ \epsilon ':\; \epsilon \geq \epsilon '\geq \epsilon \sqrt{\epsilon }\right\}  \)
and from a coupling constant \( \overline{g} \) such that, uniformly in \( \epsilon ' \)
and in \( \mathbf{P}\in \Sigma  \), it is possible to perform the iterative
procedure with the properties already shown when the starting cut-off is \( \epsilon  \),
in particular the validity of the theorem 2.3bis. It is also required that for
the chosen value \( \overline{g} \) we have \( \forall \epsilon ' \) and \( \forall \mathbf{P}\in \Sigma  \)\( \quad \left| \left( \phi ^{\epsilon '}_{\mathbf{P}},\psi _{0}\right) \right| >1-r\left( \overline{g}\right)  \)
where \( r\left( \overline{g}\right) <\frac{2}{3} \) .}\\
{\large }\\
{\large }\\
\emph{\large Procedure}{\large }\\
{\large }\\
{\large I perform the iteration shown in the previous chapter for each \( \epsilon ' \)
. Given a \( \sigma  \) ranging between \( \epsilon ^{\frac{j+2}{2}} \)and
\( \epsilon ^{\frac{j+3}{2}} \) we can always write it as \( \epsilon '^{\frac{j+2}{2}} \)
where \( \epsilon '=\epsilon '\left( \sigma \right) =\sigma ^{\frac{2}{j+2}} \).
I define }\\
{\large 
\[
\widetilde{\phi }^{\sigma }_{\mathbf{P}}\equiv \phi _{\mathbf{P}}^{\epsilon '\left( \sigma \right) ^{\frac{j+2}{2}}}\]
}\\
{\large }\\
{\large }\\
\textbf{\large Definition of} {\large \( \phi _{\mathbf{P}}^{\sigma } \).}\\
{\large }\\
{\large To go on with the definition I use the thesis of lemma 3.1 that will
be later proved:}\\
{\large }\\
{\large \( \left( \widetilde{\phi }^{\sigma }_{\mathbf{P}},\psi _{0}\right) \neq 0 \)
\( \quad \forall \! \! 0<\sigma \leq \epsilon ,\: \forall \mathbf{P}\in \Sigma  \).
}\\
{\large }\\
{\large Since \( \widetilde{\phi }^{\sigma }_{\mathbf{P}} \) is ground state
of \( H^{w}_{\mathbf{P},\sigma }\mid _{F_{\sigma }^{+}} \)with a} {\large gap}
{\large bigger than \( \frac{\sigma }{2} \) and because of previous result,
I realize that the normalized vector}\\
{\large }\\
{\large 
\[
\phi _{\mathbf{P}}^{\sigma }\equiv \frac{-\frac{1}{2\pi i}\oint \frac{1}{H^{w}_{\mathbf{P},\sigma }-E}dE\: \psi _{0}}{\left\Vert -\frac{1}{2\pi i}\oint \frac{1}{H^{w}_{\mathbf{P},\sigma }-E}dE\: \psi _{0}\right\Vert }\]
}\\
(where \( E\in \mathcal{C} \) and s.t. \( \left| E-E_{\mathbf{P}}^{\sigma }\right| =\frac{\sigma }{4} \))
{\large }\\
{\large is the ground state of \( H^{w}_{\mathbf{P},\sigma }\mid _{F_{\sigma }^{+}} \).}\\
{\large }\\
\textbf{\large Lemma 3.1}{\large }\\
{\large }\\
{\large \( \left( \widetilde{\phi }^{\sigma }_{\mathbf{P}},\psi _{0}\right) \neq 0 \)
\( \quad \forall \sigma \leq \epsilon ,\: \forall \mathbf{P}\in \Sigma  \)}\\
{\large }\\
{\large Proof}\\
{\large }\\
{\large Knowing that:}\\
{\large \par}

\begin{itemize}
\item {\large \( \left\Vert \phi _{\mathbf{P}}^{\epsilon '}-\phi ^{\epsilon '^{\frac{j+2}{2}}}_{\mathbf{P}}\right\Vert \leq \frac{\epsilon '^{\frac{1}{16}}}{1-\epsilon '^{\frac{1}{16}}}\leq \frac{1}{3} \)~~~~(corollary
2.4) }\\
{\large \par}
\item {\large \( \left( \widetilde{\phi }^{\sigma }_{\mathbf{P}},\psi _{0}\right) =\left( \widetilde{\phi }^{\sigma }_{\mathbf{P}}-\phi _{\mathbf{P}}^{\epsilon '\left( \sigma \right) }+\phi _{\mathbf{P}}^{\epsilon '\left( \sigma \right) },\psi _{0}\right) =\left( \widetilde{\phi }^{\sigma }_{\mathbf{P}}-\phi _{\mathbf{P}}^{\epsilon '\left( \sigma \right) },\psi _{0}\right) +\left( \phi _{\mathbf{P}}^{\epsilon '\left( \sigma \right) },\psi _{0}\right)  \)}\\
{\large \par}
\item {\large for the hypothesis in the definition of \( \widetilde{\phi }^{\sigma }_{\mathbf{P}} \)
, \( \left| \left( \phi _{\mathbf{P}}^{\epsilon '},\psi _{0}\right) \right| >1-r\left( \overline{g}\right)  \)
where \( r\left( \overline{g}\right) <\frac{2}{3} \)  }\\
{\large \par}
\end{itemize}
{\large we have that}\\
{\large }\\
{\large \( \left| \left( \widetilde{\phi }^{\sigma }_{\mathbf{P}},\psi _{0}\right) \right| \geq \left| \left| \left( \phi _{\mathbf{P}}^{\epsilon '\left( \sigma \right) },\psi _{0}\right) \right| -\left| \left( \widetilde{\phi }^{\sigma }_{\mathbf{P}}-\phi _{\mathbf{P}}^{\epsilon '\left( \sigma \right) },\psi _{0}\right) \right| \right| >1-r\left( \overline{g}\right) -\frac{1}{3}>0 \)}\\
{\large }\\
{\large }\\
\textbf{\large Theorem 3.2} {\large }\\
{\large }\\
{\large For \( \mathbf{P}\in \Sigma  \), the limits \( s-\lim _{\sigma \rightarrow 0}\phi _{\mathbf{P}}^{\sigma }\equiv \phi _{\mathbf{P}} \)
and \( \lim _{\sigma \rightarrow 0}E_{\mathbf{P}}^{\sigma }=E_{\mathbf{P}} \)
exist. }\\
{\large }\\
{\large Proof}\\
{\large }\\
{\large I will write again \( \widetilde{\phi }^{\sigma _{2}}_{\mathbf{P}}-\widetilde{\phi }^{\sigma _{1}}_{\mathbf{P}} \)
in the following way}\\
{\large }\\
{\large \( \widetilde{\phi }^{\sigma _{2}}_{\mathbf{P}}-\widetilde{\phi }^{\sigma _{1}}_{\mathbf{P}}=\widetilde{\phi }^{\sigma _{2}}_{\mathbf{P}}-\phi _{\mathbf{P}}^{\epsilon _{2}\left( \sigma _{2}\right) ^{\frac{l+2}{2}}}+\phi _{\mathbf{P}}^{\epsilon _{2}\left( \sigma _{2}\right) ^{\frac{l+2}{2}}}-\phi _{\mathbf{P}}^{\epsilon _{1}\left( \sigma _{1}\right) ^{\frac{m+2}{2}}}+\phi _{\mathbf{P}}^{\epsilon _{1}\left( \sigma _{1}\right) ^{\frac{m+2}{2}}}-\widetilde{\phi }^{\sigma _{1}}_{\mathbf{P}} \)
.}\\
{\large }\\
{\large Now, given an arbitrarily small \( \delta  \), there exist \( l\left( \delta \right) ,m\left( \delta \right)  \)
sufficiently large and a phase \( e^{i\eta \left( \delta \right) } \) in which}\\
{\large }\\
{\large \( \left\Vert \phi _{\mathbf{P}}^{\epsilon _{1}\left( \sigma _{1}\right) ^{\frac{m+2}{2}}}-e^{i\eta \left( \delta \right) }\phi _{\mathbf{P}}^{\epsilon _{2}\left( \sigma _{2}\right) ^{\frac{l+2}{2}}}\right\Vert \leq \delta  \).
}\\
{\large }\\
{\large This is possible essentially because of the convergence established
in theorems 2.3 and 2.3bis and because, by construction, the ground state is
unique until there is a cut-off.}\\
{\large }\\
{\large Therefore \( \left\Vert \widetilde{\phi }^{\sigma _{2}}_{\mathbf{P}}-e^{-i\eta \left( \delta \right) }\widetilde{\phi }^{\sigma _{1}}_{\mathbf{P}}\right\Vert  \)
can be bounded with a quantity of order \( \sigma ^{\frac{1}{8}}_{2}+\sigma ^{\frac{1}{8}}_{1}+\delta  \).}{\large }\\
{\large }\\
{\large moreover}\\
{\large }\\
{\large \( \left\Vert P_{\widetilde{\phi }_{\mathbf{P}}^{\sigma _{1}}}\psi _{0}-P_{\widetilde{\phi }^{\sigma _{2}}_{\mathbf{P}}}\psi _{0}\right\Vert =\left\Vert \widetilde{\phi }^{\sigma _{1}}_{\mathbf{P}}\left( \widetilde{\phi }^{\sigma _{1}}_{\mathbf{P}},\psi _{0}\right) -\widetilde{\phi }^{\sigma _{2}}_{\mathbf{P}}\left( \widetilde{\phi }^{\sigma _{2}}_{\mathbf{P}},\psi _{0}\right) \right\Vert \leq  \)}\\
{\large }\\
{\large \( \leq \left\Vert \widetilde{\phi }^{\sigma _{1}}_{\mathbf{P}}\left( \widetilde{\phi }^{\sigma _{1}}_{\mathbf{P}},\psi _{0}\right) -e^{i\eta }\widetilde{\phi }^{\sigma _{2}}_{\mathbf{P}}\left( \widetilde{\phi }^{\sigma _{1}}_{\mathbf{P}},\psi _{0}\right) \right\Vert +\left\Vert \widetilde{\phi }^{\sigma _{2}}_{\mathbf{P}}\left( e^{-i\eta }\widetilde{\phi }^{\sigma _{1}}_{\mathbf{P}},\psi _{0}\right) -\widetilde{\phi }^{\sigma _{2}}_{\mathbf{P}}\left( \widetilde{\phi }^{\sigma _{2}}_{\mathbf{P}},\psi _{0}\right) \right\Vert = \)}\\
{\large }\\
{\large \( \leq \left\Vert \widetilde{\phi }^{\sigma _{1}}_{\mathbf{P}}-e^{i\eta }\widetilde{\phi }^{\sigma _{2}}_{\mathbf{P}}\right\Vert \cdot \left| \left( \widetilde{\phi }^{\sigma _{1}}_{\mathbf{P}},\psi _{0}\right) \right| +\left\Vert \widetilde{\phi }^{\sigma _{2}}_{\mathbf{P}}\right\Vert \cdot \left\Vert \psi _{0}\right\Vert \cdot \left\Vert e^{-i\eta }\widetilde{\phi }_{\mathbf{P}}^{\sigma _{1}}-\widetilde{\phi }_{\mathbf{P}}^{\sigma _{2}}\right\Vert  \)}\\
{\large }\\
{\large It follows that \( \phi _{\mathbf{P}}^{\sigma }\equiv \frac{P_{\widetilde{\phi }_{\mathbf{P}}^{\sigma }}\psi _{0}}{\left\Vert P_{\widetilde{\phi }_{\mathbf{P}}^{\sigma }}\psi _{0}\right\Vert } \)
converges strongly to a vector \( \phi _{\mathbf{P}} \), with an error of order
\( \sigma ^{\frac{1}{8}} \)}{\large .}\\
{\large }\\
{\large }\\
{\large The convergence of \( E_{\mathbf{P}}^{\sigma } \) follows from the
estimates of the generalized version of lemma 1.3 and from the fact that the
hamiltonian \( H_{\mathbf{P}} \) is bounded from below.}\\
{\large }\\
\textbf{\large Lemma 3.3}{\large }\\
{\large }\\
{\large For \( \mathbf{P} \)} {\large and \( \mathbf{P}+\Delta \mathbf{P} \)
belonging to \( \Sigma \equiv \left\{ \mathbf{P}:\, \left| \mathbf{P}\right| \leq \sqrt{m}\right\}  \)}
{\large and \( m \) sufficiently large the following Hoelder estimate on the
energy gradient holds: }\\
{\large 
\[
\left| \nabla E^{\sigma }\left( \mathbf{P}\right) -\nabla E^{\sigma }\left( \mathbf{P}+\Delta \mathbf{P}\right) \right| \leq C\left| \Delta \mathbf{P}\right| ^{\frac{1}{16}}\]
where the constant \( C \) is uniform in \( 0\leq \sigma <\epsilon  \) , in
\( \mathbf{P},\mathbf{P}+\Delta \mathbf{P}\in \Sigma  \) where \( \left| \Delta \mathbf{P}\right| \leq \left( \frac{1}{2}\right) ^{\frac{4}{3}} \).
}\\
{\large }\\
{\large Proof}\\
{\large }\\
{\large The idea is to perturb, in \( \mathbf{P} \), the \( \nabla E^{\left| \Delta \mathbf{P}\right| ^{\frac{1}{4}}}\left( \mathbf{P}\right) \equiv \left( \psi ^{\left| \Delta \mathbf{P}\right| ^{\frac{1}{4}}}_{\mathbf{P}},\frac{\mathbf{P}-\mathbf{P}^{mes}}{m}\psi ^{\left| \Delta \mathbf{P}\right| ^{\frac{1}{4}}}_{\mathbf{P}}\right)  \)~where
~\( \psi ^{\left| \Delta \mathbf{P}\right| ^{\frac{1}{4}}}_{\mathbf{P}} \)is
the normalized ground state of \( H_{\mathbf{P},\left| \Delta \mathbf{P}\right| ^{\frac{1}{4}}} \).
}\\
{\large For this purpose I expand the resolvent}\\
{\large }\\
{\large 
\[
\frac{1}{H_{\mathbf{P}+\Delta \mathbf{P},\left| \Delta \mathbf{P}\right| ^{\frac{1}{4}}}\mid _{F_{\left| \Delta \mathbf{P}\right| ^{\frac{1}{4}}}^{+}}-E}\]
}\\
{\large }\\
{\large ( \( E\in \mathcal{C} \) and s.t. \( \left| E-E_{\mathbf{P}}^{\left| \Delta \mathbf{P}\right| ^{\frac{1}{4}}}\right| =\frac{\left| \Delta \mathbf{P}\right| ^{\frac{1}{4}}}{4} \)),
on the basis of the following information:}\\
{\large \par}

\begin{itemize}
\item {\large \( H_{\mathbf{P}+\Delta \mathbf{P},\left| \Delta \mathbf{P}\right| ^{\frac{1}{4}}}-H_{\mathbf{P},\left| \Delta \mathbf{P}\right| ^{\frac{1}{4}}}=-\frac{\Delta \mathbf{P}}{m}\cdot \mathbf{P}^{mes}+\frac{\Delta \mathbf{P}}{m}\cdot \mathbf{P}+\frac{\left| \Delta \mathbf{P}\right| ^{2}}{2m} \); }{\large \par}
\item {\large \( H_{\mathbf{P},\left| \Delta \mathbf{P}\right| ^{\frac{1}{4}}}\mid _{F_{\left| \Delta \mathbf{P}\right| ^{\frac{1}{4}}}^{+}} \)has
unique ground state} \textbf{\large \( \psi _{\mathbf{P}}^{\left| \Delta \mathbf{P}\right| ^{\frac{1}{4}}} \)}
{\large of energy \( E^{\left| \Delta \mathbf{P}\right| ^{\frac{1}{4}}}_{\mathbf{P}} \)
and corresponding gap bounded from below by \( \frac{\left| \Delta \mathbf{P}\right| ^{\frac{1}{4}}}{2} \)
( theorem 1.5 in the continuum case); \(  \)}{\large \par}
\item {\large the norm} \textbf{\large \( \left| \Delta P^{i}\right| ^{\frac{1}{8}}\left\Vert \frac{\left( P^{i}-P^{mes^{i}}\right) }{\sqrt{2m}}\left( \frac{1}{H_{\mathbf{P},\left| \Delta \mathbf{P}\right| ^{\frac{1}{4}}}\mid _{F_{\left| \Delta \mathbf{P}\right| ^{\frac{1}{4}}}^{+}}-E}\right) ^{\frac{1}{2}}\right\Vert  \)}
{\large is uniformly bounded in \( \left| \Delta \mathbf{P}\right|  \). Therefore,
for a sufficiently large \( m \)} \textbf{\large }{\large we have that}\textbf{\large }\\
\textbf{\large }\\
{\large \( \sum _{i}\left| \Delta P^{i}\right| ^{\frac{1}{4}}\left\Vert \left( \frac{1}{H_{\mathbf{P}+\Delta \mathbf{P},\left| \Delta \mathbf{P}\right| ^{\frac{1}{4}}}\mid _{F_{\left| \Delta \mathbf{P}\right| ^{\frac{1}{4}}}^{+}}-E}\right) ^{\frac{1}{2}}\sqrt{\frac{2}{m}}\cdot \frac{\left( P^{i}-P^{mes^{i}}\right) }{\sqrt{2m}}\left( \frac{1}{H_{\mathbf{P},\left| \Delta \mathbf{P}\right| ^{\frac{1}{4}}}\mid _{F_{\left| \Delta \mathbf{P}\right| ^{\frac{1}{4}}}^{+}}-E}\right) ^{\frac{1}{2}}\right\Vert <\frac{1}{4} \)}\\
{\large }\\
{\large \par}
\item {\large \( \left\Vert -\frac{1}{2\pi i}\oint \frac{1}{H_{\mathbf{P}+\Delta \mathbf{P},\left| \Delta \mathbf{P}\right| ^{\frac{1}{4}}}-E}dE\: \psi _{\mathbf{P}}^{\left| \Delta \mathbf{P}\right| ^{\frac{1}{4}}}\right\Vert \geq 1-\sum ^{\infty }_{n=1}\left( \frac{\left| \Delta \mathbf{P}\right| ^{\frac{3}{4}}}{2}\right) ^{n}\geq \frac{2}{3} \)}\\
{\large for the constraint \( \left| \Delta \mathbf{P}\right| \leq \left( \frac{1}{2}\right) ^{\frac{4}{3}} \)}{\large \par}
\end{itemize}
{\large it follows that: }{\large \par}

\begin{itemize}
\item {\large \( -\frac{1}{2\pi i}\oint \frac{1}{H_{\mathbf{P}+\Delta \mathbf{P},\left| \Delta \mathbf{P}\right| ^{\frac{1}{4}}}-E}dE\: \psi _{\mathbf{P}}^{\left| \Delta \mathbf{P}\right| ^{\frac{1}{4}}} \)~is
ground state of \( H_{\mathbf{P}+\Delta \mathbf{P},\left| \Delta \mathbf{P}\right| ^{\frac{1}{4}}} \)
;}{\large \par}
\item {\large \( \left\Vert \frac{-\frac{1}{2\pi i}\oint \frac{1}{H_{\mathbf{P}+\Delta \mathbf{P},\left| \Delta \mathbf{P}\right| ^{\frac{1}{4}}}-E}dE\: \psi _{\mathbf{P}}^{\left| \Delta \mathbf{P}\right| ^{\frac{1}{4}}}}{\left\Vert -\frac{1}{2\pi i}\oint \frac{1}{H_{\mathbf{P}+\Delta \mathbf{P},\left| \Delta \mathbf{P}\right| ^{\frac{1}{4}}}-E}dE\: \psi _{\mathbf{P}}^{\left| \Delta \mathbf{P}\right| ^{\frac{1}{4}}}\right\Vert }-\psi _{\mathbf{P}}^{\left| \Delta \mathbf{P}\right| ^{\frac{1}{4}}}\right\Vert \leq C'\left| \Delta \mathbf{P}\right| ^{\frac{3}{4}} \)}\\
{\large }\\
{\large where \( C' \) is a constant uniform in \( \mathbf{P}\in \Sigma  \)
, \( \mathbf{P}+\Delta \mathbf{P}\in \Sigma  \) where \( \Delta \mathbf{P}\in \left\{ \Delta \mathbf{P}:\, \left| \Delta \mathbf{P}\right| \leq \left( \frac{1}{2}\right) ^{\frac{4}{3}}\right\}  \)
.  }\\
{\large \par}
\end{itemize}
{\large Since: }\\
{\large }\\
{\large 1) \( \nabla E^{\left| \Delta \mathbf{P}\right| ^{\frac{1}{4}}}\left( \mathbf{P}+\Delta \mathbf{P}\right) -\nabla E^{\left| \Delta \mathbf{P}\right| ^{\frac{1}{4}}}\left( \mathbf{P}\right) =\left( \psi ^{\left| \Delta \mathbf{P}\right| ^{\frac{1}{4}}}_{\mathbf{P}+\mathbf{P}},\frac{\mathbf{P}+\Delta \mathbf{P}-\mathbf{P}^{mes}}{m}\psi ^{\left| \Delta \mathbf{P}\right| ^{\frac{1}{4}}}_{\mathbf{P}+\Delta \mathbf{P}}\right) -\left( \psi ^{\left| \Delta \mathbf{P}\right| ^{\frac{1}{4}}}_{\mathbf{P}},\frac{\mathbf{P}-\mathbf{P}^{mes}}{m}\psi ^{\left| \Delta \mathbf{P}\right| ^{\frac{1}{4}}}_{\mathbf{P}}\right)  \)}
{\large }\\
{\large 2) \( H_{\mathbf{P},\left| \Delta \mathbf{P}\right| ^{\frac{1}{4}}}+2\pi g^{2}\kappa -\frac{\left( \mathbf{P}^{mes}-\mathbf{P}\right) ^{2}}{2m}\geq 0 \)}\\
{\large }\\
{\large we can conclude that }\\
{\large }\\
{\large \( \left| \nabla E^{\left| \Delta \mathbf{P}\right| ^{\frac{1}{4}}}\left( \mathbf{P}\right) -\nabla E^{\left| \Delta \mathbf{P}\right| ^{\frac{1}{4}}}\left( \mathbf{P}+\Delta \mathbf{P}\right) \right| \leq C''\left| \Delta \mathbf{P}\right| ^{\frac{3}{4}} \)}\marginpar{
{\large (20)}{\large \par}
}{\large }\\
{\large }\\
{\large where \( C'' \) is constant uniform in \( \mathbf{P}\in \Sigma  \)
, \( \mathbf{P}+\Delta \mathbf{P}\in \Sigma  \) where \( \Delta \mathbf{P}\in \left\{ \Delta \mathbf{P}:\, \left| \Delta \mathbf{P}\right| \leq \left( \frac{1}{2}\right) ^{\frac{4}{3}}\right\}  \),
and of order \( \frac{1}{\sqrt{m}} \).  }\\
{\large }\\
{\large If \( \sigma <\left| \Delta \mathbf{P}\right| ^{\frac{1}{4}} \), in
order to obtain the thesis of the lemma, I take advantage of the result of theorem
2.3bis together with lemma A2 in Appendix A:}\\
{\large }\\
{\large \( \left| \nabla E^{\left| \Delta \mathbf{P}\right| ^{\frac{1}{4}}}\left( \mathbf{P}\right) -\nabla E^{\sigma }\left( \mathbf{P}\right) \right| \leq C\left| \Delta \mathbf{P}\right| ^{\frac{1}{16}} \)for
\( \mathbf{P}\in \Sigma  \).}\\
{\large }\\
{\large If \( \sigma \geq \left| \Delta \mathbf{P}\right| ^{\frac{1}{4}} \)
an estimate analogous to (20) holds.}\\
{\large }\\
{\large }\\
\textbf{\large Theorem 3.4} {\large }\\
{\large }\\
{\large Under the hypotheses of lemma 3.3, the norm difference between \( \phi ^{\sigma }_{\mathbf{P}} \)
and \( \phi ^{\sigma }_{\mathbf{P}+\Delta \mathbf{P}} \) is Hoelder with coefficient
\( \frac{1}{32} \) and multiplicative constant that is uniform in \( 0\leq \sigma <\epsilon  \),
in \( \mathbf{P},\mathbf{P}+\Delta \mathbf{P}\in \Sigma  \) where \( \left| \Delta \mathbf{P}\right| \leq \left( \frac{1}{6}\right) ^{4} \).}\textbf{\large }\\
\textbf{\large }\\
\textbf{\large }\\
{\large Proof}\\
{\large }\\
{\large \par}

{\large Preliminary definitions:}\\
{\large }\\
{\large \( H_{\mathbf{P},\left| \Delta \mathbf{P}\right| ^{\frac{1}{4}}}^{w}=\frac{1}{2m}\left( \mathbf{P}^{mes}-g\int ^{\kappa }_{\left| \Delta \mathbf{P}\right| ^{\frac{1}{4}}}\frac{\mathbf{k}}{\sqrt{2}\left| \mathbf{k}\right| ^{\frac{3}{2}}\left( 1-\widehat{\mathbf{k}}\cdot \nabla E^{\left| \Delta \mathbf{P}\right| ^{\frac{1}{4}}}\left( \mathbf{P}\right) \right) }\left( b\left( \mathbf{k}\right) +b^{\dagger }\left( \mathbf{k}\right) \right) d^{3}k\right) ^{2} \)}\\
{\large }\\
{\large \( +\int \left( \left| \mathbf{k}\right| -\mathbf{k}\cdot \nabla E^{\left| \Delta \mathbf{P}\right| ^{\frac{1}{4}}}\left( \mathbf{P}\right) \right) b^{\dagger }\left( \mathbf{k}\right) b\left( \mathbf{k}\right) d^{3}k+c_{\mathbf{P}}\left( \left| \Delta \mathbf{P}\right| ^{\frac{1}{4}}\right)  \)}
{\small }\\
{\small }\\
{\large \( H_{\mathbf{P}+\Delta \mathbf{P},\left| \Delta \mathbf{P}\right| ^{\frac{1}{4}}}^{w}=\frac{1}{2m}\left( \mathbf{P}^{mes}-g\int ^{\kappa }_{\left| \Delta \mathbf{P}\right| ^{\frac{1}{4}}}\frac{\mathbf{k}}{\sqrt{2}\left| \mathbf{k}\right| ^{\frac{3}{2}}\left( 1-\widehat{\mathbf{k}}\cdot \nabla E^{\left| \Delta \mathbf{P}\right| ^{\frac{1}{4}}}\left( \mathbf{P}+\Delta \mathbf{P}\right) \right) }\left( b\left( \mathbf{k}\right) +b^{\dagger }\left( \mathbf{k}\right) \right) d^{3}k\right) ^{2}+ \)}\\
\\
{\large \( +\int \left( \left| \mathbf{k}\right| -\mathbf{k}\cdot \nabla E^{\left| \Delta \mathbf{P}\right| ^{\frac{1}{4}}}\left( \mathbf{P}+\Delta \mathbf{P}\right) \right) b^{\dagger }\left( \mathbf{k}\right) b\left( \mathbf{k}\right) d^{3}k+c_{\mathbf{P}+\Delta \mathbf{P}}\left( \left| \Delta \mathbf{P}\right| ^{\frac{1}{4}}\right)  \)}\\
{\large }\\
{\large }\\
{\large \( H_{\mathbf{P}+\Delta \mathbf{P},\left| \Delta \mathbf{P}\right| ^{\frac{1}{4}}}^{w}-H_{\mathbf{P},\left| \Delta \mathbf{P}\right| ^{\frac{1}{4}}}^{w}=c_{\mathbf{P}+\Delta \mathbf{P}}\left( \left| \Delta \mathbf{P}\right| ^{\frac{1}{4}}\right) -c_{\mathbf{P}}\left( \left| \Delta \mathbf{P}\right| ^{\frac{1}{4}}\right) + \)}
{\large }\\
{\large }\\
{\large \( +\int \left( -\mathbf{k}\cdot \nabla E^{\left| \Delta \mathbf{P}\right| ^{\frac{1}{4}}}\left( \mathbf{P}+\Delta \mathbf{P}\right) +\mathbf{k}\cdot \nabla E^{\left| \Delta \mathbf{P}\right| ^{\frac{1}{4}}}\left( \mathbf{P}\right) \right) b^{\dagger }\left( \mathbf{k}\right) b\left( \mathbf{k}\right) d^{3}k+ \)
}\\
{\large }\\
\( +\frac{1}{2m}\left( \Pi _{\mathbf{P},\left| \Delta \mathbf{P}\right| ^{\frac{1}{4}}}+\int ^{\kappa }_{\left| \Delta \mathbf{P}\right| ^{\frac{1}{4}}}\left[ g\frac{\mathbf{k}}{\sqrt{2}\left| \mathbf{k}\right| ^{\frac{3}{2}}\left( 1-\widehat{\mathbf{k}}\cdot \nabla E^{\left| \Delta \mathbf{P}\right| ^{\frac{1}{4}}}\left( \mathbf{P}\right) \right) }-g\frac{\mathbf{k}}{\sqrt{2}\left| \mathbf{k}\right| ^{\frac{3}{2}}\left( 1-\widehat{\mathbf{k}}\cdot \nabla E^{\left| \Delta \mathbf{P}\right| ^{\frac{1}{4}}}\left( \mathbf{P}+\Delta \mathbf{P}\right) \right) }\right] \left( b+b^{\dagger }\right) d^{3}k\right) ^{2}+ \){\large }\\
{\large }\\
{\large \( -\frac{1}{2m}\left( \Pi _{\mathbf{P},\left| \Delta \mathbf{P}\right| ^{\frac{1}{4}}}\right) ^{2}= \)}\\
{\large }\\
{\large \( =c_{\mathbf{P}+\Delta \mathbf{P}}\left( \left| \Delta \mathbf{P}\right| ^{\frac{1}{4}}\right) -c_{\mathbf{P}}\left( \left| \Delta \mathbf{P}\right| ^{\frac{1}{4}}\right) + \)}\\
{\large }\\
{\large \( +\int \left( \mathbf{k}\cdot \left( \nabla E^{\left| \Delta \mathbf{P}\right| ^{\frac{1}{4}}}\left( \mathbf{P}\right) -\nabla E^{\left| \Delta \mathbf{P}\right| ^{\frac{1}{4}}}\left( \mathbf{P}+\Delta \mathbf{P}\right) \right) \right) b^{\dagger }\left( \mathbf{k}\right) b\left( \mathbf{k}\right) d^{3}k+ \)}\\
{\large }\\
{\large \( +\frac{1}{2m}\left( g\int ^{\kappa }_{\left| \Delta \mathbf{P}\right| ^{\frac{1}{4}}}\frac{\widehat{\mathbf{k}}\cdot \left( \nabla E^{\left| \Delta \mathbf{P}\right| ^{\frac{1}{4}}}\left( \mathbf{P}\right) -\nabla E^{\left| \Delta \mathbf{P}\right| ^{\frac{1}{4}}}\left( \mathbf{P}+\Delta \mathbf{P}\right) \right) }{\sqrt{2}\left| \mathbf{k}\right| ^{\frac{3}{2}}\left( 1-\widehat{\mathbf{k}}\cdot \nabla E^{\left| \Delta \mathbf{P}\right| ^{\frac{1}{4}}}\left( \mathbf{P}+\Delta \mathbf{P}\right) \right) \left( 1-\widehat{\mathbf{k}}\cdot \nabla E^{\left| \Delta \mathbf{P}\right| ^{\frac{1}{4}}}\left( \mathbf{P}\right) \right) }\left( b\left( \mathbf{k}\right) +b\left( \mathbf{k}\right) ^{\dagger }\right) d^{3}k\right) ^{2}+ \)}
{\small }\\
{\large \( \frac{1}{2m}\Pi _{\mathbf{P},\left| \Delta \mathbf{P}\right| ^{\frac{1}{4}}}\left( g\int ^{\kappa }_{\left| \Delta \mathbf{P}\right| ^{\frac{1}{4}}}\frac{\widehat{\mathbf{k}}\cdot \left( \nabla E^{\left| \Delta \mathbf{P}\right| ^{\frac{1}{4}}}\left( \mathbf{P}\right) -\nabla E^{\left| \Delta \mathbf{P}\right| ^{\frac{1}{4}}}\left( \mathbf{P}+\Delta \mathbf{P}\right) \right) }{\sqrt{2}\left| \mathbf{k}\right| ^{\frac{3}{2}}\left( 1-\widehat{\mathbf{k}}\cdot \nabla E^{\left| \Delta \mathbf{P}\right| ^{\frac{1}{4}}}\left( \mathbf{P}+\Delta \mathbf{P}\right) \right) \left( 1-\widehat{\mathbf{k}}\cdot \nabla E^{\left| \Delta \mathbf{P}\right| ^{\frac{1}{4}}}\left( \mathbf{P}\right) \right) }\left( b\left( \mathbf{k}\right) +b\left( \mathbf{k}\right) ^{\dagger }\right) d^{3}k\right) + \)}{\small }\\
{\small }\\
{\large \( +\left( g\int ^{\kappa }_{\left| \Delta \mathbf{P}\right| ^{\frac{1}{4}}}\frac{\widehat{\mathbf{k}}\cdot \left( \nabla E^{\left| \Delta \mathbf{P}\right| ^{\frac{1}{4}}}\left( \mathbf{P}\right) -\nabla E^{\left| \Delta \mathbf{P}\right| ^{\frac{1}{4}}}\left( \mathbf{P}+\Delta \mathbf{P}\right) \right) }{\sqrt{2}\left| \mathbf{k}\right| ^{\frac{3}{2}}\left( 1-\widehat{\mathbf{k}}\cdot \nabla E^{\left| \Delta \mathbf{P}\right| ^{\frac{1}{4}}}\left( \mathbf{P}+\Delta \mathbf{P}\right) \right) \left( 1-\widehat{\mathbf{k}}\cdot \nabla E^{\left| \Delta \mathbf{P}\right| ^{\frac{1}{4}}}\left( \mathbf{P}\right) \right) }\left( b\left( \mathbf{k}\right) +b\left( \mathbf{k}\right) ^{\dagger }\right) d^{3}k\right) \frac{1}{2m}\Pi _{\mathbf{P},\left| \Delta \mathbf{P}\right| ^{\frac{1}{4}}} \)}
{\small }{\large }\\
{\large }\\
{\large }\\
{\large Considering that:}\\
{\large \par}

\begin{itemize}
\item {\large the estimate (20) in lemma 3.3 holds:}\\
{\large \( \left| \nabla E^{\left| \Delta \mathbf{P}\right| ^{\frac{1}{4}}}\left( \mathbf{P}\right) -\nabla E^{\left| \Delta \mathbf{P}\right| ^{\frac{1}{4}}}\left( \mathbf{P}+\Delta \mathbf{P}\right) \right| \leq C''\left| \Delta \mathbf{P}\right| ^{\frac{3}{4}} \);}\\
{\large \par}
\item {\large the operator \( \frac{H_{\mathbf{P}+\Delta \mathbf{P},\left| \Delta \mathbf{P}\right| ^{\frac{1}{4}}}^{w}-H_{\mathbf{P},\left| \Delta \mathbf{P}\right| ^{\frac{1}{4}}}^{w}}{\left| \Delta \mathbf{P}\right| ^{\frac{3}{4}}} \)
is form-bounded with respect to \( H_{\mathbf{P},\left| \Delta \mathbf{P}\right| ^{\frac{1}{4}}}^{w} \)
;}{\large \par}
\item {\large the gap of \( E^{\left| \Delta \mathbf{P}\right| ^{\frac{1}{4}}}_{\mathbf{P}} \)(that
is the ground eigenvalue of \( H_{\mathbf{P},\left| \Delta \mathbf{P}\right| ^{\frac{1}{4}}}^{w}\mid _{F_{\left| \Delta \mathbf{P}\right| ^{\frac{1}{4}}}^{+}} \))
is bounded from below by \( \frac{\left| \Delta \mathbf{P}\right| ^{\frac{1}{4}}}{2} \)~~(generalized
version of theorem 1.5);}\\
{\large \par}
\end{itemize}
{\large for the Kato's theorem on the analytic perturbation, the vector} {\large }\\
{\large }\\
{\large \( \phi ^{\left| \Delta \mathbf{P}\right| ^{\frac{1}{4}}}_{\mathbf{P}+\Delta \mathbf{P}}\equiv \frac{-\frac{1}{2\pi i}\oint \frac{1}{H^{w}_{\mathbf{P}+\Delta \mathbf{P},\left| \Delta \mathbf{P}\right| ^{\frac{1}{4}}}-E}dE\: \psi _{0}}{\left\Vert -\frac{1}{2\pi i}\oint \frac{1}{H^{w}_{\mathbf{P}+\Delta \mathbf{P},\left| \Delta \mathbf{P}\right| ^{\frac{1}{4}}}-E}dE\: \psi _{0}\right\Vert } \)}\\
( \( E\in \mathcal{C} \) and s.t. \( \left| E-E_{\mathbf{P}+\Delta \mathbf{P}}^{\left| \Delta \mathbf{P}\right| ^{\frac{1}{4}}}\right| =\frac{\left| \Delta \mathbf{P}\right| ^{\frac{1}{4}}}{4} \))
{\large }\\
{\large can be obtained, for \( m \) sufficiently large, perturbing \( \phi ^{\left| \Delta \mathbf{P}\right| ^{\frac{1}{4}}}_{\mathbf{P}} \).}
{\large }\\
{\large }\\
{\large From the perturbation we have the estimate:}\\
{\large }\\
 {\large \( \left\Vert \phi ^{\left| \Delta \mathbf{P}\right| ^{\frac{1}{4}}}_{\mathbf{P}+\Delta \mathbf{P}}-\phi ^{\left| \Delta \mathbf{P}\right| ^{\frac{1}{4}}}_{\mathbf{P}}\right\Vert \leq C'''\left| \Delta \mathbf{P}\right| ^{\frac{1}{4}} \)}\marginpar{
{\large (20.1)}{\large \par}
} {\large }\\
{\large }\\
{\large where the constant \( C''' \) is uniform} {\large in \( \mathbf{P},\mathbf{P}+\Delta \mathbf{P}\in \Sigma  \)
where\( \left| \Delta \mathbf{P}\right| \leq \left( \frac{1}{6}\right) ^{4} \)}{\large .
}\\
{\large }\\
{\large In conclusion, for \( \sigma <\left| \Delta \mathbf{P}\right| ^{\frac{1}{4}} \),
using a generalized version of corollary 2.4 for \( \sigma  \) in the continuum,
we have that: }\\
{\large }\\
{\large \( \left\Vert \phi ^{\sigma }_{\mathbf{P}+\Delta \mathbf{P}}-\phi ^{\sigma }_{\mathbf{P}}\right\Vert = \)}\\
{\large }\\
{\large \( =\left\Vert \phi ^{\sigma }_{\mathbf{P}+\Delta \mathbf{P}}-\phi ^{\left| \Delta \mathbf{P}\right| ^{\frac{1}{4}}}_{\mathbf{P}+\Delta \mathbf{P}}+\phi ^{\left| \Delta \mathbf{P}\right| ^{\frac{1}{4}}}_{\mathbf{P}+\Delta \mathbf{P}}-\phi ^{\left| \Delta \mathbf{P}\right| ^{\frac{1}{4}}}_{\mathbf{P}}+\phi ^{\left| \Delta \mathbf{P}\right| ^{\frac{1}{4}}}_{\mathbf{P}}-\phi ^{\sigma }_{\mathbf{P}}\right\Vert \leq  \)}\\
{\large }\\
{\large \( \leq \left\Vert \phi ^{\sigma }_{\mathbf{P}+\Delta \mathbf{P}}-\phi ^{\left| \Delta \mathbf{P}\right| ^{\frac{1}{4}}}_{\mathbf{P}+\Delta \mathbf{P}}\right\Vert +\left\Vert \phi ^{\left| \Delta \mathbf{P}\right| ^{\frac{1}{4}}}_{\mathbf{P}+\Delta \mathbf{P}}-\phi ^{\left| \Delta \mathbf{P}\right| ^{\frac{1}{4}}}_{\mathbf{P}}\right\Vert +\left\Vert \phi ^{\left| \Delta \mathbf{P}\right| ^{\frac{1}{4}}}_{\mathbf{P}}-\phi ^{\sigma }_{\mathbf{P}}\right\Vert \leq  \)}\\
{\large }\\
{\large \( \leq \left| \Delta \mathbf{P}\right| ^{\frac{1}{32}}+\left| \Delta \mathbf{P}\right| ^{\frac{1}{32}}+C'''\left| \Delta \mathbf{P}\right| ^{\frac{1}{4}} \)}\\
{\large }\\
{\large If \( \sigma \geq \left| \Delta \mathbf{P}\right| ^{\frac{1}{4}} \)
an estimate analogous to (20.1) holds.}\\
{\large }\\
{\large }\\
{\large }\\
{\large }\\
{\large }\\
{\large }\\
{\large }\\
{\large }\\
{\large }\\
{\large }\\
{\large }\\
{\large }\\
{\large }\\
{\large }\\
{\large }\\
{\large }\\
{\large }\\
{\large }\\
{\large }\\
{\large }\\
{\large }\\
{\large }\\
{\large }\\
{\large }\\
{\large }\\
{\large \par}

\part{Scattering theory{\large .}\\
\large }

{\large The infrared features of the model produce some difficulties in understanding
the scattering: }\\
{\large - the arbitrarily large number of mesons involved in the scattering
makes the explanation of the asymptotic decoupling difficult. In particular
a problem of consistency arises between the total emission of an infinite number
of massless particles (mesons) and a ``free'' asymptotic dynamics for the
nucleon. In other words we can say that a definition of ``free'' dynamics
for the infra-particle (the nucleon) is required; }\\
{\large - the L.S.Z. Weyl meson operators (see below) do not converge on generic
vectors of the Hilbert space }\\
{\large }\\
{\large \( e^{iHt}e^{-iH^{mes.}t}e^{i\left( a\left( \varphi \right) +a^{\dagger }\left( \varphi \right) \right) }e^{iH^{mes}t}e^{-iHt} \)\( \qquad \varphi \left( \mathbf{y}\right) =\int e^{-i\mathbf{ky}}\widetilde{\varphi }\left( \mathbf{k}\right) d^{3}k\in S\left( R^{3}\right)  \)}
{\large }\\
{\large }\\
{\large at least if one uses the same technique of the proof in {[}2{]}.}\\
{\large }\\
{\large Due to the absence of one particle states, from a conceptual point of
view it is not possible to use the Haag and Ruelle theory for the construction
of scattering states. Nevertheless, the decoupling mechanism in the H. and R.
theory can be reproduced in terms of fixed time locality properties of the meson
field and of the ``current density field'' of the nucleon . }\\
{\large According to this interpretative scheme I review the construction of
the asymptotic nucleon, which was considered in {[}2{]}. My purposes are two:
}\\
{\large }\\
{\large - the first one is to give a minimal (with respect to the meson cloud)
description of the nucleon out of the scattering}\\
{\large }\\
{\large - the second one consists in finding a subspace of states which will
be used in order to prove the asymptotic convergence of the massless field.
This subspace is the analogous of the one particle subspace in the regularized
case (infrared cut-off in the interaction).}\\
{\large }\\
{\large }\\
{\large The starting point is the construction of the vector \( \psi _{G}\left( t\right)  \)},
{\large at time t, that will approximate the minimal asymptotic nucleon state
\( \psi _{G}^{out(in)} \)} {\large }{\large :} {\large }\\
{\large \( \psi _{G}\left( t\right)  \)} {\large is constructed starting from
a one particle state for the hamiltonian \( H_{\sigma _{t}} \)}, {\large of
wave function \( G \)} {\large in} \textbf{\large \( \mathbf{P}- \)}{\large variables,
to which a Weyl operator, in properly evolved meson variables} {\large (L.S.Z.
Weyl operators, see later),} {\large is applied. The smearing function in Weyl
operators has the right infrared behavior established in the spectral analysis.
Its spectral distribution, near} \textbf{\large \( \mathbf{k}=0 \)}{\large ,
is} \textbf{\large }{\large labelled by the asymptotic nucleon mean velocity
(constructed ``a posteriori''); its frequency support goes from \( \sigma _{t} \)}
{\large (\( \sigma _{t}\rightarrow 0 \) for \( \left| t\right| \rightarrow +\infty  \))
to an arbitrarily small \( \kappa _{1}\neq 0 \). }\\
{\large Intuitively, in the asymptotic limit it describes a ``free'' (one
particle state) nucleon plus a} \textbf{\large not totally removable} {\large cloud
of asymptotic mesons. The main difference with an analogous construction by
Froehlich {[}1{]} is that the infrared cut-off of the cloud of interacting mesons
is removed only asymptotically at a rate faster than} \( \frac{1}{t} \){\large (in
accordance with the indetermination principle). The advantage consists in the
fact that the construction of the vector \( \psi _{G}\left( t\right)  \)} {\large is
always inside the Hilbert space and it makes it simpler to use locality (by
the nucleon position \( \mathbf{x} \)) in proving the convergence. Moreover,
the removal of the infrared cut-off is an ``a posteriori'' result and a byproduct
of the decoupling.}{\large }\\
{\large I want to stress the fact that the coherent function in the definition
of the minimal asymptotic nucleon states is arbitrary except for the infrared
limit. Nevertheless the conceptual and mathematical role of the minimal asymptotic
nucleon states is very useful in developing the scattering theory.}\\
{\large }\\
{\large }\\
{\large The mathematical counterpart of what I said is the following.}\\
{\large We construct a set of states \( \psi _{G}\left( t\right)  \) (where
\( G \) is the \( \mathbf{P}- \)wave function of the one \( \sigma _{t}- \)particle
state) on which there exist the limits }\\
{\large 
\[
s-\lim _{t\rightarrow \pm \infty }\psi _{G}\left( t\right) \equiv \psi _{G}^{out\left( in\right) }\]
} {\large }\\
{\large Moreover on these states the functions \( f \), continuous and of compact
support, of the nucleon mean velocity have limit} {\large :}\\
{\large }\\
{\large \( s-\lim _{t\rightarrow \pm \infty }e^{iHt}f\left( \frac{\mathbf{x}}{t}\right) e^{-iHt}\psi _{G}^{out\left( in\right) }=s-\lim _{t\rightarrow \pm \infty }e^{iHt}f\left( \frac{\mathbf{x}}{t}\right) e^{-iHt}\psi _{G}\left( t\right) \cong \psi _{f\cdot G}^{out\left( in\right) } \)}\\
{\large }\\
{\large The norm closure of the finite linear combinations \( \left\{ \bigvee \psi _{G}^{out\left( in\right) }\right\}  \)
substitutes the subspace of states at one nucleon which we have in the regularized
case}\\
{\large }\\
{\large \( s-\lim _{t\rightarrow \pm \infty }e^{iHt}e^{-iH^{mes}t}e^{i\left( a\left( \varphi \right) +a^{\dagger }\left( \varphi \right) \right) }e^{iH^{mes}t}e^{-iHt}\psi _{G}^{out\left( in\right) }= \)}{\large }\\
{\large }\\
{\large \( =s-\lim _{t\rightarrow \pm \infty }e^{iHt}e^{-iH^{mes}t}e^{i\left( a\left( \varphi \right) +a^{\dagger }\left( \varphi \right) \right) }e^{iH^{mes}t}e^{-iHt}\psi _{G}\left( t\right) \equiv \psi _{G,\varphi }^{out\left( in\right) } \)}{\large }\\
(\( \varphi \left( \mathbf{y}\right) =\int e^{-i\mathbf{ky}}\widetilde{\varphi }\left( \mathbf{k}\right) d^{3}k,\, \widetilde{\varphi }\left( \mathbf{k}\right) \in C_{0}^{\infty }\left( R^{3}\setminus 0\right)  \)
){\large }\\
{\large }\\
{\large By analogy, I will define the invariant (under space-time translation)
subspaces}{\large ~H}{\large \( ^{1out\left( in\right) } \)\( \equiv \left\{ \overline{\bigvee \psi _{G}^{out\left( in\right) }},\: G\left( \mathbf{P}\right) \in C_{0}^{1}\left( R^{3}\setminus 0\right) ,\mathbf{P}\in \Sigma \right\}  \).
}\\
{\large Then I define}{\large  ~H}{\large \( ^{out\left( in\right) } \)\( \equiv \left\{ \overline{\bigvee \psi _{G,\varphi }^{out\left( in\right) }},\: G\in C_{0}^{1}\left( R^{3}\setminus 0\right) ,\widetilde{\varphi }\in C_{0}^{\infty }\left( R^{3}\setminus 0\right) \right\}  \).The
last one is obtained by the vectors \( \psi _{G,\varphi }^{out\left( in\right) } \),
starting from}{\large  ~H}{\large \( ^{1out\left( in\right) } \). }\\
{\large As I already said, these definitions are arbitrary in some sense. Nevertheless,
through the (artificial) separation between}{\large  ~H}{\large \( ^{1out\left( in\right) } \)
and}{\large ~H}{\large \( ^{out\left( in\right) } \) I want to point out that:
}\\
{\large }\\
{\large - from a technical point of view, my construction of the scattering
states is based on some (arbitrary)}{\large  ~H}{\large \( ^{1out\left( in\right) } \);}\\
{\large }\\
{\large - from a physical point of view, even if the meson cloud described
by smearing functions \( \varphi  \) is totally removable, the meson cloud
linked to the vectors in}{\large  ~H}{\large \( ^{1out\left( in\right) } \)
is not completely removable. Then all the scattering states} {\large always
contain}{\large n asymptotic mesons, those ones of the spaces}{\large  ~H}{\large \( ^{1out\left( in\right) } \)~involved
in the construction of the spaces}{\large  ~H}{\large \( ^{out\left( in\right) } \).}
{\large }\\
{\large }\\
{\large Now, we make an observation which regards the coherent ``static''
factor \( g\frac{1}{\left| \mathbf{k}\right| \sqrt{2\left| \mathbf{k}\right| }\left( 1-\widehat{\mathbf{k}}\cdot \nabla E\right) } \)
founded in the spectral analysis. If we consider a \( \sigma - \)cut-off dynamics,
we can easily compute the asymptotic nucleon mean velocity. This operator coincides
with \( \nabla E^{\sigma } \) if it is applied to the one nucleon particle
states. Through a non-rigorous removal of the cut-off, the coherent factor can
be thought in terms of the asymptotic nucleon mean velocity to be constructed.}\\
{\large }\\
{\large }\\
\emph{\large Content of the chapters}{\large .}\\
{\large }\\
{\large In chapter 4, I define the approximating vector \( \psi _{G}\left( t\right)  \)
for the generic state of minimal asymptotic nucleon. Then I study the norm in
time, and finally I prove its strong convergence (paragraphs 4.1 and 4.2). }\\
{\large In chapter 5, I construct the scattering subspaces}{\large ~ H}{\large \( ^{out\left( in\right) } \).
Then I prove the existence of the asymptotic limits of the functions, continuous
and of compact support, of the nucleon mean velocity. I also prove the existence
of the asymptotic limit of L.S.Z. meson field Weyl operators and I discuss their
commutation properties.}\\
\\
{\large The construction will be explicitly performed in the case ``out''.
The case ``in'' is completely analogous.}\\
{\large }\\
{\large \par}

\section{Approximating vector \protect\( \psi _{G}\left( t\right) \protect \).\\
}

{\large I will consider a cubic region of volume \( V=L^{3} \) inside} {\large \( \Sigma  \)}
{\large in the \( \mathbf{P}- \)space , for simplicity} {\large of construction.
For \( \mathbf{P} \) in this region, we have:}\\
{\large \par}

\begin{itemize}
\item {\large the existence of the ground state of \( H^{\sigma }_{\mathbf{P}} \)
with the properties which implicitly follow from the results in chapter 3;}{\large \par}
\item {\large small nucleon velocities \( \left| \nabla E^{\sigma }\left( \mathbf{P}\right) \right| <1\quad \forall \sigma  \)}
{\large (see lemma A2}{\large ); }{\large \par}
\item {\large \( \left| \nabla E^{\sigma }\left( \mathbf{P}+\mathbf{k}\right) \right| <1\quad \forall \sigma  \),
for \( \mathbf{k}: \) \( 0\leq \left| \mathbf{k}\right| \leq \kappa _{1}\ll 1 \)
(\( \kappa _{1} \) properly small).}{\large \par}
\end{itemize}
{\large I will assume that the following hypothesis holds in \( \Sigma  \)
:}\textbf{\large }\\
\textbf{\large }\\
\textbf{\large hypothesis B1~~~~\( \exists m_{r}>0 \)~ ~~~}{\large s.t.} \textbf{\large ~~\( \forall \sigma  \)~\( \forall \mathbf{P}\in \Sigma  \)~~\( \frac{\partial E^{\sigma }\left( \mathbf{P}\right) }{\partial \left| \mathbf{P}\right| }\geq \frac{\left| \mathbf{P}\right| }{m_{r}} \)~~~}{\large and}
\textbf{\large ~~~\( \frac{\partial ^{2}E^{\sigma }\left( \mathbf{P}\right) }{\partial ^{2}\left| \mathbf{P}\right| }\geq \frac{1}{m_{r}} \)}
{\large }\\
{\large }\\
{\large }\\
{\large I consider a time-dependent (\( t\gg 1 \)) cell partition of the volume
\( V=L^{3} \). The linear dimension of each cell is \( \frac{L}{2^{n}} \),
where \( n\in \aleph ,\quad n\geq 1 \) is such that \( \left( 2^{n}\right) ^{\frac{1}{\epsilon }}\leq t<\left( 2^{n+1}\right) ^{\frac{1}{\epsilon }} \),~~\( \epsilon >0 \).
The exponent \( \epsilon  \) will be fixed only ``a posteriori''. It follows
that the number of cells is \( N\left( t\right) =\left( 2^{n}\right) ^{3} \)
where \( n=\left[ \log _{2}t^{\epsilon }\right]  \). I will call \( \Gamma _{i} \)
the \( i^{th} \) cell, centered in \( \overline{\mathbf{P}_{i}} \)}\\
{\large }\\
\emph{\large Constructive prescription of \( \psi _{G}\left( t\right)  \)}{\large .}\\
{\large }\\
{\large 1) I will consider the vector \( \psi _{i,\sigma _{t}}^{\left( t\right) }=\int _{\Gamma _{i}}G\left( \mathbf{P}\right) \psi _{\mathbf{P},\sigma _{t}}d^{3}P \),
where:}\\
{\large }\\
{\large - \( G\left( \mathbf{P}\right) \in C_{0}^{1}\left( R^{3}\setminus 0\right)  \);}\\
{\large }\\
{\large - \( \psi ^{\sigma _{t}}_{\mathbf{P}}\equiv W^{b^{\dagger }}_{\sigma _{t_{2}}}\left( \nabla E^{\sigma _{t}}\right) \phi ^{\sigma _{t}}_{\mathbf{P}} \)
is the normalized ground state of \( H_{\mathbf{P},\sigma _{t}} \) (I will
use the index \( b \) in \( W^{b^{\dagger }}_{\sigma _{t_{2}}} \) in order
to distinguish it from Weyl operators in \( a,a^{\dagger } \) variables);}\\
{\large }\\
{\large - \( \left( t\right)  \) is referred to the partition at time \( t \);}\\
{\large }\\
{\large - \( \left\Vert \psi _{i,\sigma _{t}}^{\left( t\right) }\right\Vert =\left( \int _{\Gamma _{i}}\left| G\left( \mathbf{P}\right) \right| ^{2}d^{3}P\right) ^{\frac{1}{2}} \)is
of order \( \left( N\left( t\right) \right) ^{-\frac{1}{2}} \)}\\
{\large }\\
{\large 2) }\\
{\large I will dress each \( \psi _{i,\sigma _{t}}^{\left( t\right) } \) by
the proper \( e^{iHt}e^{-iH^{mes}t}W_{\sigma _{t}}\left( \mathbf{v}_{i}\right) e^{iH^{mes}t}e^{-iH_{\sigma _{t}}t} \).
In such a way, the vector will remain inside the Hilbert space~}{\large H} {\large under
removal of the} {\large cut-off}{\large . Therefore I will define: }\\
{\large \(  \)}\\
{\large \( \psi _{G}\left( t\right) \equiv e^{iHt}e^{-iH^{mes}t}\sum _{i=1}^{N\left( t\right) }W_{\sigma _{t}}\left( \mathbf{v}_{i}\right) e^{iH^{mes}t}e^{i\gamma _{\sigma _{t}}\left( \mathbf{v}_{i},\nabla E^{\sigma _{t}}\left( \mathbf{P}\right) ,t\right) }e^{-iE^{\sigma _{t}}\left( \mathbf{P}\right) t}\psi _{i,\sigma _{t}}^{\left( t\right) }= \)}\\
{\large }\\
{\large \( =e^{iHt}\sum _{i=1}^{N\left( t\right) }W_{\sigma _{t}}\left( \mathbf{v}_{i},t\right) e^{i\gamma _{\sigma _{t}}\left( \mathbf{v}_{i},\nabla E^{\sigma _{t}}\left( \mathbf{P}\right) ,t\right) }e^{-iE^{\sigma _{t}}t}\psi _{i,\sigma _{t}}^{\left( t\right) } \)}\\
{\large }\\
{\large being }{\large \par}

\begin{itemize}
\item {\large \( W_{\sigma _{t}}\left( \mathbf{v}_{i}\right) =e^{-g\int ^{\kappa _{1}}_{\sigma _{t}}\frac{a\left( \mathbf{k}\right) -a^{\dagger }\left( \mathbf{k}\right) }{\left| \mathbf{k}\right| \left( 1-\widehat{\mathbf{k}}\cdot \mathbf{v}_{i}\right) }\frac{d^{3}k}{\sqrt{2\left| \mathbf{k}\right| }}} \)
, where \( \mathbf{v}_{i} \) is the velocity corresponding to the cell center
\( \overline{\mathbf{P}}_{i} \) of \( \Gamma _{i} \), \( \mathbf{v}_{i}\equiv \nabla E^{\sigma _{t}}\left( \overline{\mathbf{P}}_{i}\right)  \),
\( \kappa _{1} \) (\( 0<\kappa _{1}\ll 1 \)) is the integration upper bound
for the frequency;}\\
{\large \par}
\item {\large \( W_{\sigma _{t}}\left( \mathbf{v}_{i},t\right) =e^{-iH^{mes}t}W_{\sigma _{t}}\left( \mathbf{v}_{i}\right) e^{iH^{mes}t}=e^{-g\int ^{\kappa _{1}}_{\sigma _{t}}\frac{a\left( \mathbf{k}\right) e^{i\left| \mathbf{k}\right| t}-a^{\dagger }\left( \mathbf{k}\right) e^{-i\left| \mathbf{k}\right| t}}{\left| \mathbf{k}\right| \left( 1-\widehat{\mathbf{k}}\cdot \mathbf{v}_{i}\right) }\frac{d^{3}k}{\sqrt{2\left| \mathbf{k}\right| }}} \)
;}\\
{\large }\\
{\large \par}
\item {\large \( \gamma _{\sigma _{t}}\left( \mathbf{v}_{i},\nabla E^{\sigma _{t}}\left( \mathbf{P}\right) ,t\right) =-\int ^{t}_{1}\left\{ g^{2}\int ^{\sigma ^{S}_{\tau }}_{\sigma _{t}}\int \frac{\cos \left( \mathbf{k}\cdot \nabla E^{\sigma _{t}}\left( \mathbf{P}\right) \tau -\left| \mathbf{k}\right| \tau \right) }{\left( 1-\widehat{\mathbf{k}}\cdot \mathbf{v}_{i}\right) }d\Omega d\left| \mathbf{k}\right| \right\} d\tau = \)}\\
{\large }\\
 {\large \( =-\int ^{t}_{1}\left\{ g^{2}\int ^{\tau \sigma ^{S}_{\tau }}_{\tau \sigma _{t}}\int \frac{\cos \left( \mathbf{q}\cdot \nabla E^{\sigma _{t}}\left( \mathbf{P}\right) -\left| \mathbf{q}\right| \right) }{\left( 1-\widehat{\mathbf{q}}\cdot \mathbf{v}_{i}\right) }d\Omega \frac{d\left| \mathbf{q}\right| }{\tau }\right\} d\tau  \)
}\\
{\large }\\
 {\large \( e^{i\gamma _{\sigma _{t}}\left( \mathbf{v}_{i},\nabla E^{\sigma _{t}}\left( \mathbf{P}\right) ,t\right) } \)
is a phase factor whose origin and definition will be clear later. In particular,
in order to implement the convergence of \( \psi _{G}\left( t\right)  \), we
will see that the integration bound \( \sigma ^{S}_{\tau } \) is a ``slow''
cut-off that will be substantially treated as a fixed cut-off, while the ``speed''
cut-off \( \sigma _{\tau } \) will require the phase term. For this purpose
it is necessary that \( \sigma ^{S}_{\tau } \) goes like \( \tau ^{-\alpha } \),
where \( \alpha  \) is a positive number sufficiently less than 1, and that
\( \sigma _{\tau } \) is of order \( \tau ^{-\beta } \), where \( \beta  \)
is sufficiently bigger than \( 1 \). In the proofs, I will take \( \alpha =\frac{39}{40} \)
and I will not fix the exponent \( \beta  \). On the basis of partial estimates,
eventually \( \beta  \) will be chosen equal to \( 128 \) in order to gain
the strong convergence of the vector \( \psi _{G}\left( t\right)  \).}\\
{\large }\\
{\large \par}
\end{itemize}

\subsection{Control of the norm of \protect\( \psi _{G}\left( t\right) \protect \).\\
}

{\large I will study the scalar product:}\\
{\large }\\
{\large \( \left( \psi _{G}\left( t\right) ,\psi _{G}\left( t\right) \right) = \)}\\
\\
{\large \( =\sum _{i,j=1}^{N\left( t\right) }\left( W_{\sigma _{t}}\left( \mathbf{v}_{i},t\right) e^{i\gamma _{\sigma _{t}}\left( \mathbf{v}_{i},\nabla E^{\sigma _{t}}\left( \mathbf{P}\right) ,t\right) }e^{-iH_{\sigma _{t}}t}\psi _{i,\sigma _{t}}^{\left( t\right) },W_{\sigma _{t}}\left( \mathbf{v}_{j},t\right) e^{i\gamma _{\sigma _{t}}\left( \mathbf{v}_{j},\nabla E^{\sigma _{t}}\left( \mathbf{P}\right) ,t\right) }e^{-iH_{\sigma _{t}}t}\psi _{j,\sigma _{t}}^{\left( t\right) }\right)  \)}\\
{\large }\\
{\large }\\
{\large The diagonal terms of the above sum are easily under control. In fact,
if \( i=j \) we obtain:}\\
\\
{\large \( \sum _{i=1}^{N\left( t\right) }\left( \psi _{i,\sigma _{t}}^{\left( t\right) },\psi _{i,\sigma _{t}}^{\left( t\right) }\right) =\sum _{i=1}^{N\left( t\right) }\int _{\Gamma _{i}}\left| G\left( \mathbf{P}\right) \right| ^{2}d^{3}P=\int _{V}\left| G\left( \mathbf{P}\right) \right| ^{2}d^{3}P \)
}\\
{\large }\\
{\large The following step consists in proving that each mixed term of the sum
\( \sum _{i,j=1}^{N\left( t\right) } \) vanishes asymptotically with an order
in \( t \)} {\large independent of the dimension of the cell. Therefore, by
properly choosing the exponent \( \epsilon  \) that determines the} {\large rate}
{\large of growth} \emph{\large }{\large of \( N\left( t\right) =\left[ t^{\epsilon }\right] ^{3} \)
, we obtain that the sum of mixed terms for \( t\rightarrow +\infty  \) vanishes.}\\
\textbf{\emph{\large }}\\
\textbf{\emph{\large Remarks}} {\large }\textbf{\emph{\large on the notations}}{\large }\\
{\large }\\
{\large In the estimates I will produce in these paragraphs, I will generically
call \( C \) the multiplicative constants which are uniform in the infrared
cut-off and in the cells partition of the volume \( V \).} {\large The bounds
are intended from above, up to a different explicit warning.} {\large The time
\( t \) is intended much greater than \( 1 \).}\\
{\large }\\
\emph{\large mixed terms}{\large }\\
{\large }\\
{\large If \( i\neq j \), I define: }\\
{\large }\\
{\large \( M_{i,j}\left( t\right) \equiv \left( e^{i\gamma _{\sigma _{t}}\left( \mathbf{v}_{i},\nabla E^{\sigma _{t}}\left( \mathbf{P}\right) ,t\right) }\psi _{i,\sigma _{t}}^{\left( t\right) },e^{iH_{\sigma _{t}}t}W_{\sigma _{t},i,j}\left( t\right) e^{i\gamma _{\sigma _{t}}\left( \mathbf{v}_{j},\nabla E^{\sigma _{t}}\left( \mathbf{P}\right) ,t\right) }e^{-iH_{\sigma _{t}}t}\psi _{j,\sigma _{t}}^{\left( t\right) }\right)  \)}
\\
{\large \( W_{\sigma _{t},i,j}\left( t\right) \equiv e^{-\int ^{\kappa _{1}}_{\sigma _{t}}\frac{a\left( \mathbf{k}\right) e^{i\left| \mathbf{k}\right| t}-a^{\dagger }\left( \mathbf{k}\right) e^{-i\left| \mathbf{k}\right| t}}{\left| \mathbf{k}\right| }\cdot h_{i,j}\left( \widehat{\mathbf{k}}\right) \frac{d^{3}k}{\sqrt{2\left| \mathbf{k}\right| }}} \)}
{\large ~where} {\large \( h_{i,j}\left( \widehat{\mathbf{k}}\right) =\frac{g\widehat{\mathbf{k}}\cdot \left( \mathbf{v}_{j}-\mathbf{v}_{i}\right) }{\left( 1-\widehat{\mathbf{k}}\cdot \mathbf{v}_{j}\right) \cdot \left( 1-\widehat{\mathbf{k}}\cdot \mathbf{v}_{i}\right) } \)}\\
{\large }\\
{\large Now, I consider \( M_{i,j}\left( t\right)  \) as a two-variable function,
by distinguishing the time variable \( t \) which parametrizes the partition
from the time variable \( s \) of the evolution. For this purpose, I define:}\\
{\large }\\
{\large \( \widehat{M}_{i,j}\left( t,s\right) \equiv \left( e^{i\gamma _{\sigma _{t}}\left( \mathbf{v}_{i},\nabla E^{\sigma _{t}}\left( \mathbf{P}\right) ,s\right) }\psi _{i,\sigma _{t}}^{\left( t\right) },e^{iH_{\sigma _{t}}s}W_{\sigma _{t},i,j}\left( s\right) e^{i\gamma _{\sigma _{t}}\left( \mathbf{v}_{j},\nabla E^{\sigma _{t}}\left( \mathbf{P}\right) ,s\right) }e^{-iH_{\sigma _{t}}s}\psi _{j,\sigma _{t}}^{\left( t\right) }\right)  \)}\\
{\large }\\
{\large with the constraint \( s>t \) and the obvious property \( \widehat{M}_{i,j}\left( t,t\right) =M_{i,j}\left( t\right)  \).}\\
{\large }\\
{\large I verify that:}\\
{\large }\\
\textbf{\large I)} {\large \( \widehat{M}_{i,j}\left( t,+\infty \right) =\lim _{s\rightarrow +\infty }\widehat{M}_{i,j}\left( t,s\right) =0 \)}\\
{\large }\\
\textbf{\large II)} {\large \( \left| M_{i,j}\left( t\right) \right| =\left| \widehat{M}_{i,j}\left( t,t\right) -\widehat{M}_{i,j}\left( t,+\infty \right) +\widehat{M}_{i,j}\left( t,+\infty \right) \right| =\left| \widehat{M}_{i,j}\left( t,t\right) -\widehat{M}_{i,j}\left( t,+\infty \right) \right| \leq  \)}\\
{\large }\\
{\large \( \leq \left| \int ^{+\infty }_{t}\frac{d}{ds}\left( e^{i\gamma _{\sigma _{t}}\left( \mathbf{v}_{i},\nabla E_{\sigma _{t}},s\right) }\psi _{i,\sigma _{t}}^{\left( t\right) },e^{iH_{\sigma _{t}}s}W_{\sigma ,i,j}\left( s\right) e^{i\gamma _{\sigma _{t}}\left( \mathbf{v}_{j},\nabla E_{\sigma _{t}},s\right) }e^{-iH_{\sigma _{t}}s}\psi _{j,\sigma _{t}}^{\left( t\right) }\right) ds\right| \leq  \)}
{\large }\\
{\large \( \leq C\cdot t^{2\delta -3\epsilon }\cdot t^{-\frac{1}{112}}\cdot \left( \ln \sigma _{t}\right) ^{2} \)
.}\\
\\
{\large }\\
\textbf{\large Proof of I)} {\large }\\
{\large }\\
{\large For \( s>t \)}\\
{\large }\\
{\large \( \widehat{M}_{i,j}\left( \lambda ,t,s\right) \equiv \left( e^{i\gamma _{\sigma _{t}}\left( \mathbf{v}_{i},\nabla E^{\sigma _{t}}\left( \mathbf{P}\right) ,s\right) }\psi _{i,\sigma _{t}}^{\left( t\right) }\: ,\, e^{iH_{\sigma _{t}}s}W^{\lambda }_{\sigma _{t},i,j}\left( s\right) e^{i\gamma _{\sigma _{t}}\left( \mathbf{v}_{j},\nabla E^{\sigma _{t}}\left( \mathbf{P}\right) ,s\right) }e^{-iH_{\sigma _{t}}s}\psi _{j,\sigma _{t}}\right)  \)}\\
{\large }\\
{\large where:}\\
{\large \par}

\begin{description}
\item [{\large -}]{\large \( W_{\sigma _{t},i,j}^{\lambda }\left( s\right) \equiv e^{-\lambda \int ^{\kappa _{1}}_{\sigma _{t}}\frac{a\left( \mathbf{k}\right) e^{i\left| \mathbf{k}\right| s}-a^{\dagger }\left( \mathbf{k}\right) e^{-i\left| \mathbf{k}\right| s}}{\left| \mathbf{k}\right| }\cdot h_{i,j}\left( \widehat{\mathbf{k}}\right) \frac{d^{3}k}{\sqrt{2\left| \mathbf{k}\right| }}} \),
\( \lambda  \) a real parameter;} \\

\item [-\( \begin{array}{cccc}
 &  & -\int ^{s}_{1}\left\{ g^{2}\int ^{\tau \cdot \tau ^{-\frac{39}{40}}}_{\tau \cdot \sigma _{t}}\int \frac{\cos \left( \mathbf{q}\cdot \nabla E^{\sigma _{t}}\left( \mathbf{P}\right) -\left| \mathbf{q}\right| \right) }{\left( 1-\widehat{\mathbf{q}}\cdot \mathbf{v}_{i}\right) }d\Omega \frac{d\left| \mathbf{q}\right| }{\tau }\right\} d\tau  & for\: s:\quad s^{-\frac{39}{40}}\geq \sigma _{t}\\
\gamma _{\sigma _{t}}\left( \mathbf{v}_{i},\nabla E^{\sigma _{t}}\left( \mathbf{P}\right) ,s\right) = & \left\{ \right.  &  & \\
 &  & \: \quad \gamma _{\sigma _{t}}\left( \mathbf{v}_{i},\nabla E^{\sigma _{t}}\left( \mathbf{P}\right) ,\sigma _{t}^{-\frac{40}{39}}\right) \qquad \qquad \qquad \qquad \qquad  & for\: s:\quad s^{-\frac{39}{40}}<\sigma _{t}
\end{array} \)]~
\end{description}
{\large From the derivation with respect to the real parameter \( \lambda  \),
we arrive at the differential equation:}\\
{\large }\\
{\large \par}

{\large \( \frac{d\widehat{M}_{i,j}\left( \lambda ,t,s\right) }{d\lambda }=-\lambda C_{i,j}\widehat{M}_{i,j}\left( \lambda ,t,s\right) +r_{\sigma }\left( \lambda ,t,s\right)  \)}\marginpar{
{\large (21)}{\large \par}
} {\large }\\
{\large }\\
{\large ( \( \psi _{i,\sigma _{t}}^{\left( t\right) }\in D\left( H_{\sigma _{t}}\right)  \),
then it belongs to \( D\left( a\left( f\right) \right)  \) and \( D\left( a^{\dagger }\left( f\right) \right)  \),
\( f\in L^{2}\left( R^{3}\right)  \), and the derivative with respect to \( \lambda  \)
is well defined)}\\
{\large }\\
{\large where }{\large \par}

\begin{description}
\item [\( C_{i,j}=\int ^{\kappa _{1}}_{\sigma _{t}}\left| h_{i,j}\left( \widehat{\mathbf{k}}\right) \right| ^{2}\frac{d^{3}k}{2\left| \mathbf{k}\right| ^{3}} \)]~{\large \par}
\item [{\large \( r_{\sigma _{t}}\left( \lambda ,t,s\right) = \)}]~
\end{description}
\( =-\left( W_{\sigma _{t},i,j}^{\lambda \dagger }\left( s\right) e^{i\gamma _{\sigma _{t}}\left( \mathbf{v}_{i},\nabla E^{\sigma _{t}}\left( \mathbf{P}\right) ,s\right) }e^{-iH_{\sigma _{t}}s}\psi _{i,\sigma _{t}}^{\left( t\right) }\: ,\, \int ^{\kappa _{1}}_{\sigma _{t}}\frac{a\left( \mathbf{k}\right) e^{i\left| \mathbf{k}\right| s}}{\left| \mathbf{k}\right| }\cdot h_{i,j}\left( \widehat{\mathbf{k}}\right) \frac{d^{3}k}{\sqrt{2\left| \mathbf{k}\right| }}e^{i\gamma _{\sigma _{t}}\left( \mathbf{v}_{j},\nabla E^{\sigma _{t}}\left( \mathbf{P}\right) ,s\right) }e^{-iH_{\sigma _{t}}s}\psi _{j,\sigma _{t}}^{\left( t\right) }\right) + \)\\
\\
\( +\left( \int ^{\kappa _{1}}_{\sigma _{t}}\frac{a\left( \mathbf{k}\right) e^{i\left| \mathbf{k}\right| s}}{\left| \mathbf{k}\right| }\cdot h_{i,j}\left( \widehat{\mathbf{k}}\right) \frac{d^{3}k}{\sqrt{2\left| \mathbf{k}\right| }}e^{i\gamma _{\sigma _{t}}\left( \mathbf{v}_{i},\nabla E^{\sigma _{t}}\left( \mathbf{P}\right) ,s\right) }e^{-iH_{\sigma _{t}}s}\psi _{i,\sigma _{t}}^{\left( t\right) }\: ,\, W_{\sigma _{t},i,j}^{\lambda }\left( s\right) e^{i\gamma _{\sigma _{t}}\left( \mathbf{v}_{j},\nabla E^{\sigma _{t}}\left( \mathbf{P}\right) ,s\right) }e^{-iH_{\sigma _{t}}s}\psi _{j,\sigma _{t}}^{\left( t\right) }\right)  \)\\
{\large }\\
{\large }\\
{\large from which \( \widehat{M}_{i,j}\left( 1,t,s\right) =e^{-\frac{C_{i,j}}{2}}\widehat{M}_{i,j}\left( 0,t,s\right) +\int ^{1}_{0}r_{\sigma _{t}}\left( \lambda ,t,s\right) d\lambda  \)}\\
{\large }\\
{\large Note that:\(  \)}\\
{\large \par}

\begin{itemize}
\item {\large \( \widehat{M}_{i,j}\left( 0,t,s\right) =0 \) \( \forall t,s \) since
the} \textbf{\large \( \mathbf{P}- \)}{\large supports of \( \psi _{j,\sigma _{t}}^{\left( t\right) } \)
,\( \psi _{i,\sigma _{t}}^{\left( t\right) } \) (\( i\neq j \)) are disjoint;}{\large \par}
\item {\large thanks to lemma B2 (Appendix B) and by the control of the derivative
with respect to \( s \) (theorem B4) one can verify the existence of}\\
\\
{\large \( s-\lim _{s\rightarrow +\infty }e^{iH_{\sigma _{t}}s}\int ^{\kappa _{1}}_{\sigma _{t}}\frac{a\left( \mathbf{k}\right) e^{i\left| \mathbf{k}\right| s}}{\left| \mathbf{k}\right| }\cdot h_{i,j}\left( \widehat{\mathbf{k}}\right) \frac{d^{3}k}{\sqrt{\left| \mathbf{k}\right| }}e^{-iH_{\sigma _{t}}s}e^{i\gamma _{\sigma _{t}}\left( \mathbf{v}_{j},\nabla E^{\sigma _{t}}\left( \mathbf{P}\right) ,s\right) }\psi _{j,\sigma _{t}}^{\left( t\right) }\equiv  \)}{\small }\\
{\small }\\
{\large \( \equiv a_{\sigma _{t}}^{out}\left( h\right) e^{i\gamma _{\sigma _{t}}\left( \mathbf{v}_{i},\nabla E^{\sigma _{t}}\left( \mathbf{P}\right) ,\sigma _{t}^{-\frac{40}{39}}\right) }\psi _{j,\sigma _{t}}^{\left( t\right) } \)}{\small }\\
{\small \par}
\item {\large since the vector \( \psi _{i,\sigma _{t}}^{\left( t\right) } \) is
a Fock state for \( \left\{ a_{\sigma _{t}}^{out}\left( \mathbf{k}\right) ,\, \mathbf{k}:\, 0\leq \left| \mathbf{k}\right| \leq \kappa _{1}\right\}  \)(see
theorem B5, Appendix B)  \( \lim _{s\rightarrow +\infty }r_{\sigma _{t}}\left( \lambda ,t,s\right) =0 \);
in fact}\\
{\large }\\
{\large \( \lim _{s\rightarrow +\infty }e^{iH_{\sigma _{t}}s}\int ^{\kappa _{1}}_{\sigma _{t}}\frac{a\left( \mathbf{k}\right) e^{i\left| \mathbf{k}\right| s}}{\left| \mathbf{k}\right| }\cdot h_{i,j}\left( \widehat{\mathbf{k}}\right) \frac{d^{3}k}{\sqrt{2\left| \mathbf{k}\right| }}e^{-iH_{\sigma _{t}}s}e^{i\gamma _{\sigma _{t}}\left( \mathbf{v}_{j},\nabla E^{\sigma _{t}}\left( \mathbf{P}\right) ,s\right) }\psi _{j,\sigma _{t}}^{\left( t\right) }= \)}\\
{\large }\\
{\large \( =\lim _{s\rightarrow +\infty }e^{iH_{\sigma _{t}}s}\int ^{\kappa _{1}}_{\sigma _{t}}\frac{a\left( \mathbf{k}\right) e^{i\left| \mathbf{k}\right| s}}{\left| \mathbf{k}\right| }\cdot h_{i,j}\left( \widehat{\mathbf{k}}\right) \frac{d^{3}k}{\sqrt{\left| \mathbf{k}\right| }}e^{-iH_{\sigma _{t}}s}e^{i\gamma _{\sigma _{t}}\left( \mathbf{v}_{i},\nabla E^{\sigma _{t}}\left( \mathbf{P}\right) ,\sigma _{t}^{-\frac{40}{39}}\right) }\psi _{j,\sigma _{t}}^{\left( t\right) }= \)}\\
{\large }\\
{\large \( =a_{\sigma _{t}}^{out}\left( h\right) e^{i\gamma _{\sigma _{t}}\left( \mathbf{v}_{i},\nabla E^{\sigma _{t}}\left( \mathbf{P}\right) ,\sigma _{t}^{-\frac{40}{39}}\right) }\psi _{j,\sigma _{t}}^{\left( t\right) }=0 \)}\\
{\large \par}
\end{itemize}
{\large Therefore, starting from the equation(21), we have}\\
{\large 
\[
\widehat{M}_{i,j}\left( t,+\infty \right) =\widehat{M}_{i,j}\left( \lambda =1,t,+\infty \right) =\int ^{1}_{0}r_{\sigma _{t}}\left( \lambda ,t,+\infty \right) d\lambda =0\]
} \textbf{\large }\\
\textbf{\large }\\
\textbf{\large }\\
\textbf{\large }\\
\textbf{\large Proof of II)}{\large }\\
{\large }\\
{\large Let us consider:}\\
{\large }\\
{\large \( \frac{d}{ds}\left( e^{iH_{\sigma _{t}}s}W_{\sigma _{t}}\left( \mathbf{v}_{i},s\right) e^{i\gamma _{\sigma _{t}}\left( \mathbf{v}_{i},\nabla E^{\sigma _{t}}\left( \mathbf{P}\right) ,s\right) }e^{-iH_{\sigma _{t}}s}\right) \psi _{i,\sigma _{t}}^{\left( t\right) }= \)}\\
{\large }\\
{\large \( =ie^{iH_{\sigma _{t}}s}W_{\sigma _{t}}\left( \mathbf{v}_{i},s\right) \left( \varphi _{\sigma _{t},\mathbf{v}_{i}}\left( \mathbf{x},s\right) +\frac{d\gamma _{\sigma _{t}}\left( \mathbf{v}_{i},\nabla E^{\sigma _{t}}\left( \mathbf{P}\right) ,s\right) }{ds}\right) e^{i\gamma _{\sigma _{t}}\left( \mathbf{v}_{i},\nabla E^{\sigma _{t}}\left( \mathbf{P}\right) ,s\right) }e^{-iH_{\sigma _{t}}s}\psi _{i,\sigma _{t}}^{\left( t\right) } \)}\marginpar{
{\large (22)}{\large \par}
}{\large }\\
{\large }\\
{\large where} {\large \( \varphi _{\sigma _{t},\mathbf{v}_{i}}\left( \mathbf{x},s\right) \equiv g^{2}\int ^{\kappa _{1}}_{\sigma _{t}}\int \frac{\cos \left( \mathbf{k}\cdot \mathbf{x}-\left| \mathbf{k}\right| s\right) }{\left( 1-\widehat{\mathbf{k}}\cdot \mathbf{v}_{i}\right) }d\Omega d\left| \mathbf{k}\right|  \)}\marginpar{
{\large (23)}{\large \par}
}{\large }\\
{\large }\\
\emph{\large Derivation of (22)}{\large .}\\
{\large }\\
{\large The term \( ie^{iH_{\sigma _{t}}s}W_{\sigma _{t}}\left( \mathbf{v}_{i},s\right) \varphi _{\sigma _{t},\mathbf{v}_{i}}\left( \mathbf{x},s\right) e^{i\gamma _{\sigma _{t}}\left( \mathbf{v}_{i},\nabla E^{\sigma _{t}},s\right) }e^{-iH_{\sigma _{t}}s} \)
is formally obtained from:}\\
{\large }\\
{\large \( ie^{iH_{\sigma _{t}}s}\left[ H_{\sigma _{t}}-H^{mes},e^{-g\int ^{\kappa _{1}}_{\sigma _{t}}\frac{a\left( \mathbf{k}\right) e^{i\left| \mathbf{k}\right| s}-a^{\dagger }\left( \mathbf{k}\right) e^{-i\left| \mathbf{k}\right| s}}{\left| \mathbf{k}\right| \left( 1-\widehat{\mathbf{k}}\cdot \mathbf{v}_{i}\right) }\frac{d^{3}k}{\sqrt{2\left| \mathbf{k}\right| }}}\right] e^{-iH_{\sigma _{t}}s}= \)}\marginpar{
{\large (24)}{\large \par}
}{\large }\\
{\large }\\
{\large \( =ie^{iH_{\sigma _{t}}s}\left[ H_{I,\sigma _{t}\, },e^{-g\int ^{\kappa _{1}}_{\sigma _{t}}\frac{a\left( \mathbf{k}\right) e^{i\left| \mathbf{k}\right| s}-a^{\dagger }\left( \mathbf{k}\right) e^{-i\left| \mathbf{k}\right| s}}{\left| \mathbf{k}\right| \left( 1-\widehat{\mathbf{k}}\cdot \mathbf{v}_{i}\right) }\frac{d^{3}k}{\sqrt{2\left| \mathbf{k}\right| }}}\right] e^{-iH_{\sigma _{t}}s}= \)}\\
{\large }\\
{\large \( =ie^{iH_{\sigma _{t}}s}\left[ g\int ^{\kappa _{1}}_{\sigma _{t}}\left( a\left( \mathbf{k}\right) e^{i\mathbf{k}\cdot \mathbf{x}}+a^{\dagger }\left( \mathbf{k}\right) e^{-i\mathbf{k}\cdot \mathbf{x}}\right) \frac{d^{3}\mathbf{k}}{\sqrt{2\left| \mathbf{k}\right| }},e^{-g\int ^{\kappa _{1}}_{\sigma _{t}}\frac{a\left( \mathbf{k}\right) e^{i\left| \mathbf{k}\right| s}-a^{\dagger }\left( \mathbf{k}\right) e^{-i\left| \mathbf{k}\right| s}}{\left| \mathbf{k}\right| \left( 1-\widehat{\mathbf{k}}\cdot \mathbf{v}_{i}\right) }\frac{d^{3}k}{\sqrt{2\left| \mathbf{k}\right| }}}\right] e^{-iH_{\sigma _{t}}s}= \)
}\\
{\large \( =ie^{iH_{\sigma _{t}}s}e^{-g\int ^{\kappa _{1}}_{\sigma _{t}}\frac{a\left( \mathbf{k}\right) e^{i\left| \mathbf{k}\right| s}-a^{\dagger }\left( \mathbf{k}\right) e^{-i\left| \mathbf{k}\right| s}}{\left| \mathbf{k}\right| \left( 1-\widehat{\mathbf{k}}\cdot \mathbf{v}_{i}\right) }\frac{d^{3}k}{\sqrt{2\left| \mathbf{k}\right| }}}g^{2}\int ^{\kappa _{1}}_{\sigma _{t}}\int \frac{\cos \left( \mathbf{k}\cdot \mathbf{x}-\left| \mathbf{k}\right| s\right) }{\left( 1-\widehat{\mathbf{k}}\cdot \mathbf{v}_{i}\right) }d\Omega d\left| \mathbf{k}\right| e^{-iH_{\sigma _{t}}s} \)}\\
{\large }\\
{\large the last step follows from:}\\
{\large }\\
{\large \( \left[ g\int ^{\kappa _{1}}_{\sigma _{t}}\left( a\left( \mathbf{k}\right) e^{i\mathbf{k}\cdot \mathbf{x}}+a^{\dagger }\left( \mathbf{k}\right) e^{-i\mathbf{k}\cdot \mathbf{x}}\right) \frac{d^{3}k}{\sqrt{2\left| \mathbf{k}\right| }},-g\int ^{\kappa _{1}}_{\sigma _{t}}\frac{a\left( \mathbf{k}\right) e^{i\left| \mathbf{k}\right| s}-a^{\dagger }\left( \mathbf{k}\right) e^{-i\left| \mathbf{k}\right| s}}{\left| \mathbf{k}\right| \left( 1-\widehat{\mathbf{k}}\cdot \mathbf{v}_{i}\right) }\frac{d^{3}k}{\sqrt{2\left| \mathbf{k}\right| }}\right] = \)}\\
{\large }\\
{\large \( =g^{2}\int ^{\kappa _{1}}_{\sigma _{t}}\int \frac{\cos \left( \mathbf{k}\cdot \mathbf{x}-\left| \mathbf{k}\right| s\right) }{\left( 1-\widehat{\mathbf{k}}\cdot \mathbf{v}_{i}\right) }d\Omega d\left| \mathbf{k}\right|  \)}\\
{\large }\\
{\large }\\
{\large The formal steps are well defined in \( D\left( H_{\sigma _{t}}\right)  \)
from the operatorial point of view, because:}{\large \par}

\begin{itemize}
\item {\large \( H_{\sigma _{t}} \), \( H^{mes} \), \( H_{0} \) and \( g\int ^{\kappa _{1}}_{\sigma _{t}}\left( a\left( \mathbf{k}\right) e^{i\mathbf{k}\cdot \mathbf{x}}+a^{\dagger }\left( \mathbf{k}\right) e^{-i\mathbf{k}\cdot \mathbf{x}}\right) \frac{d^{3}k}{\sqrt{2\left| \mathbf{k}\right| }} \)
have a common domain \( D \) of essential selfadjointness;}{\large \par}
\item {\large \( \frac{d\left( e^{iH^{mes}s}e^{-iH_{\sigma _{t}}s}\right) }{ds} \)
is closable;}{\large \par}
\item {\large the sequences, obtained from the formal calculus by approximating (in
the norm \( \left\Vert H_{0}\psi \right\Vert +\left\Vert \psi \right\Vert  \))
the vectors in \( D\left( H_{\sigma _{t}}\right)  \) with vectors in \( D \),
are convergent. }\\
{\large \par}
\end{itemize}
{\large On the other side \( \gamma _{\sigma _{t}}\left( \mathbf{v}_{i},\nabla E^{\sigma _{t}},s\right) =-\int ^{s}_{1}\left\{ g^{2}\int ^{\tau ^{\frac{1}{40}}}_{\sigma _{t}\cdot \tau }\int \frac{\cos \left( \mathbf{q}\cdot \nabla E^{\sigma _{t}}-\left| \mathbf{q}\right| \right) }{\left( 1-\widehat{\mathbf{q}}\cdot \mathbf{v}_{i}\right) }d\Omega \frac{d\left| \mathbf{q}\right| }{\tau }\right\} d\tau  \)
from which}\\
{\large }\\
{\large \( \frac{d\gamma _{\sigma _{t}}\left( \mathbf{v}_{i},\nabla E^{\sigma _{t}},s\right) }{ds}=-g^{2}\int ^{s^{\frac{1}{40}}}_{\sigma _{t}\cdot s}\int \frac{\cos \left( \mathbf{q}\cdot \nabla E^{\sigma _{t}}-\left| \mathbf{q}\right| \right) }{\left( 1-\widehat{\mathbf{q}}\cdot \mathbf{v}_{i}\right) }d\Omega \frac{d\left| \mathbf{q}\right| }{s} \)}\\
{\large }\\
{\large \( \frac{d\gamma _{\sigma _{t}}}{ds}\left( \mathbf{v}_{i},\frac{\mathbf{x}}{s},s\right) \equiv -g^{2}\int ^{s^{\frac{1}{40}}}_{\sigma _{t}\cdot s}\int \frac{\cos \left( \mathbf{q}\cdot \frac{\mathbf{x}}{s}-\left| \mathbf{q}\right| \right) }{\left( 1-\widehat{\mathbf{q}}\cdot \mathbf{v}_{i}\right) }d\Omega \frac{d\left| \mathbf{q}\right| }{s} \)}\\
{\large }\\
{\large }\\
{\large I decompose \( \varphi _{\sigma _{t},\mathbf{v}_{i}}\left( \mathbf{x},s\right)  \)
(formula (23)) in \( \varphi ^{1}_{\sigma _{t},\mathbf{v}_{i}}\left( \mathbf{x},s\right) +\varphi ^{2}_{\sigma _{t},\mathbf{v}_{i}}\left( \mathbf{x},s\right)  \),
which are so defined}\\
{\large }\\
{\large \( \varphi ^{1}_{\sigma _{t},\mathbf{v}_{i}}\left( \mathbf{x},s\right) \equiv  \)}\\
{\large }\\
{\large \( \begin{array}{cccc}
 &  & g^{2}\int ^{s^{\frac{1}{40}}}_{s\cdot \sigma _{t}}\int \frac{\cos \left( \mathbf{k}\cdot \frac{\mathbf{x}}{s}-\left| \mathbf{k}\right| \right) }{\left( 1-\widehat{\mathbf{k}}\cdot \mathbf{v}_{i}\right) }d\Omega \frac{d\left| \mathbf{k}\right| }{s} & for\quad s\leq \sigma _{t}^{-\frac{40}{39}}\\
= & \left\{ \right.  &  & \\
 &  & \quad \quad 0\qquad \qquad \qquad \qquad  & for\quad s>\sigma ^{-\frac{40}{39}}_{t}
\end{array} \)}\\
{\large }\\
{\large \( \begin{array}{ccc}
 &  & \\
 & g^{2}\int ^{\kappa _{1}}_{s^{-\frac{39}{40}}}\int \frac{\cos \left( \mathbf{k}\cdot \mathbf{x}-\left| \mathbf{k}\right| s\right) }{\left( 1-\widehat{\mathbf{k}}\cdot \mathbf{v}_{i}\right) }d\Omega d\left| \mathbf{k}\right|  & for\: s\leq \sigma ^{-\frac{40}{39}}_{t}\\
\varphi ^{2}_{\sigma _{t},\mathbf{v}_{i}}\left( \mathbf{x},s\right) \equiv \varphi _{\sigma _{t},\mathbf{v}_{i}}\left( \mathbf{x},s\right) -\varphi ^{1}_{\sigma _{t},\mathbf{v}_{i}}\left( \mathbf{x},s\right) =\left\{ \right.  &  & \\
 & \varphi _{\sigma _{t},\mathbf{v}_{i}}\left( \mathbf{x},s\right) \qquad \qquad \qquad \qquad  & for\: s>\sigma ^{-\frac{40}{39}}_{t}
\end{array} \) }\\
{\large }\\
{\large }\\
{\large The estimate of the norm of the expression (22) requires the introduction
of some functions \( \chi ^{\left( t\right) }_{\mathbf{v}_{i}}\left( \nabla E^{\sigma _{t}},s\right)  \)
defined in Appendix B. These functions tend, for \( s\rightarrow +\infty  \),
to the characteristic functions of the transformed \( \Gamma _{i} \) cells,
under the application \( \mathbf{P}\rightarrow \nabla E^{\sigma _{t}}\left( \mathbf{P}\right)  \).
Such a regularization is necessary to the estimates discussed in Appendix B.
As we can see in the definition (b1) in Appendix B, an exponent \( \delta  \)
is introduced to which a scale length \( t^{-\frac{\delta }{6}} \) corresponds.
This scale length has to be less than the (\( \epsilon - \)dependent) linear
dimension of the cell \( \Gamma _{i} \) in order to have consistency.  }\\
{\large }\\
{\large }\\
\emph{\large Analysis of the expression (22) :}{\large }\\
{\large }\\
{\large \( \varphi _{\sigma _{t},\mathbf{v}_{i}}\left( \mathbf{x},s\right) e^{-iE^{\sigma _{t}}\left( \mathbf{P}\right) s}e^{i\gamma _{\sigma _{t}}\left( \mathbf{v}_{i},\nabla E^{\sigma _{t}},s\right) }\psi _{i,\sigma _{t}}^{\left( t\right) }+\frac{d\gamma _{\sigma _{t}}\left( \mathbf{v}_{i},\nabla E^{\sigma _{t}}\left( \mathbf{P}\right) ,s\right) }{ds}e^{-iE^{\sigma _{t}}\left( \mathbf{P}\right) s}e^{i\gamma _{\sigma _{t}}\left( \mathbf{v}_{i},\nabla E^{\sigma _{t}},s\right) }\psi _{i,\sigma _{t}}^{\left( t\right) }= \)}
{\large }\\
{\large }\\
{\large \( =\varphi _{\sigma _{t},\mathbf{v}_{i}}\left( \mathbf{x},s\right) e^{-iE^{\sigma _{t}}\left( \mathbf{P}\right) s}e^{i\gamma _{\sigma _{t}}\left( \mathbf{v}_{i},\nabla E^{\sigma _{t}},s\right) }\left( 1_{\Gamma _{i}}\left( \mathbf{P}\right) -\chi ^{\left( t\right) }_{\mathbf{v}_{i}}\left( \nabla E^{\sigma _{t}}\left( \mathbf{P}\right) ,s\right) \right) \psi _{i,\sigma _{t}}^{\left( t\right) }+ \)}\\
\marginpar{
{\large (25.0)}{\large \par}
}{\large }\\
{\large \( +\frac{d\gamma _{\sigma _{t}}\left( \mathbf{v}_{i},\nabla E^{\sigma _{t}}\left( \mathbf{P}\right) ,s\right) }{ds}e^{i\gamma _{\sigma _{t}}\left( \mathbf{v}_{i},\nabla E^{\sigma _{t}},s\right) }e^{-iE^{\sigma _{t}}\left( \mathbf{P}\right) s}\left( 1_{\Gamma _{i}}\left( \mathbf{P}\right) -\chi ^{\left( t\right) }_{\mathbf{v}_{i}}\left( \nabla E^{\sigma _{t}}\left( \mathbf{P}\right) ,s\right) \right) \psi _{i,\sigma _{t}}^{\left( t\right) }+ \)
}\\
{\large }\\
{\large \( +\varphi _{\sigma _{t},\mathbf{v}_{i}}\left( \mathbf{x},s\right) \left( \chi ^{\left( t\right) }_{\mathbf{v}_{i}}\left( \nabla E^{\sigma _{t}}\left( \mathbf{P}\right) ,s\right) -\chi ^{\left( t\right) }_{\mathbf{v}_{i}}\left( \frac{\mathbf{x}}{s},s\right) \right) e^{-iE^{\sigma _{t}}\left( \mathbf{P}\right) s}e^{i\gamma _{\sigma _{t}}\left( \mathbf{v}_{i},\nabla E^{\sigma _{t}},s\right) }\psi _{i,\sigma _{t}}^{\left( t\right) }+ \)}\marginpar{
{\large (25.1)}{\large \par}
}{\large }\\
{\large }\\
{\large \( +\varphi ^{2}_{\sigma _{t},\mathbf{v}_{i}}\left( \mathbf{x},s\right) \chi ^{\left( t\right) }_{\mathbf{v}_{i}}\left( \frac{\mathbf{x}}{s},s\right) e^{-iE^{\sigma _{t}}\left( \mathbf{P}\right) s}e^{i\gamma _{\sigma _{t}}\left( \mathbf{v}_{i},\nabla E^{\sigma _{t}},s\right) }\psi _{i,\sigma _{t}}^{\left( t\right) }+ \)}\marginpar{
{\large (25.2)}{\large \par}
}{\large }\\
{\large }\\
{\large \( +\left( \varphi ^{1}_{\sigma _{t},\mathbf{v}_{i}}\left( \mathbf{x},s\right) \chi ^{\left( t\right) }_{\mathbf{v}_{i}}\left( \frac{\mathbf{x}}{s},s\right) +\frac{d\gamma _{\sigma _{t}}}{ds}\left( \mathbf{v}_{i},\frac{\mathbf{x}}{s},s\right) \chi ^{\left( t\right) }_{\mathbf{v}_{i}}\left( \frac{\mathbf{x}}{s},s\right) \right) e^{-iE^{\sigma _{t}}\left( \mathbf{P}\right) s}e^{i\gamma _{\sigma _{t}}\left( \mathbf{v}_{i},\nabla E^{\sigma _{t}},s\right) }\psi _{i,\sigma _{t}}^{\left( t\right) }+ \)}\marginpar{
{\large (25.3)}{\large \par}
}{\large }\\
{\large }\\
{\large \( -\frac{d\gamma _{\sigma _{t}}}{ds}\left( \mathbf{v}_{i},\frac{\mathbf{x}}{s},s\right) \left( \chi ^{\left( t\right) }_{\mathbf{v}_{i}}\left( \frac{\mathbf{x}}{s},s\right) -\chi ^{\left( t\right) }_{\mathbf{v}_{i}}\left( \nabla E^{\sigma _{t}}\left( \mathbf{P}\right) ,s\right) \right) e^{-iE^{\sigma _{t}}\left( \mathbf{P}\right) s}e^{i\gamma _{\sigma _{t}}\left( \mathbf{v}_{i},\nabla E^{\sigma _{t}},s\right) }\psi _{i,\sigma _{t}}^{\left( t\right) }+ \)}\marginpar{
{\large (25.4)}{\large \par}
}{\large }\\
{\large }\\
{\large \( +\left( -\frac{d\gamma _{\sigma _{t}}}{ds}\left( \mathbf{v}_{i},\frac{\mathbf{x}}{s},s\right) +\frac{d\gamma _{\sigma _{t}}\left( \mathbf{v}_{i},\nabla E^{\sigma _{t}}\left( \mathbf{P}\right) ,s\right) }{ds}\right) e^{i\gamma _{\sigma _{t}}\left( \mathbf{v}_{i},\nabla E^{\sigma _{t}},s\right) }e^{-iE^{\sigma _{t}}\left( \mathbf{P}\right) s}\chi ^{\left( t\right) }_{\mathbf{v}_{i}}\left( \nabla E^{\sigma _{t}}\left( \mathbf{P}\right) ,s\right) \psi _{i,\sigma _{t}}^{\left( t\right) } \)}\marginpar{
{\large (25.5)}{\large \par}
}{\large }\\
{\large }\\
\emph{\large Not}{\large e}\\
{\large As regards the function \( 1_{\Gamma _{i}}\left( \mathbf{P}\right)  \),
it is assumed that \( \Gamma _{i}\bigcap suppG\neq \emptyset  \).}\\
{\large }\\
{\large }\\
{\large The norm of each term on the right hand side is less than}\\
{\large 
\[
C\cdot s^{-1}\cdot s^{2\delta -\frac{3\epsilon }{2}}\cdot s^{-\frac{1}{112}}\cdot \left( \ln \sigma _{t}\right) ^{2}\]
}\marginpar{
{\large (26)}{\large \par}
} {\large }\\
{\large }\\
{\large (25.0) }{\large \par}

\begin{itemize}
\item {\large because of lemma B1 and of lemma B3 (}\emph{\large Observation} {\large 1)
this term is bounded by \( C\cdot \left| \frac{\ln \left( \sigma _{t}\right) }{s}\right| \cdot \frac{1}{s^{\frac{\delta }{12}}}\cdot t^{-\epsilon } \)~~\( \left( \delta >24\epsilon \right)  \); }{\large \par}
\end{itemize}
{\large (25.1) }{\large \par}

\begin{itemize}
\item {\large \( \left| \varphi _{\sigma _{t},\mathbf{v}_{i}}\left( \mathbf{x},s\right) \right| \leq C\cdot \left| \frac{\ln \left( \sigma _{t}\right) }{s}\right|  \)}
\emph{\large ~}{\large (}\emph{\large Observation} {\large 1}\emph{\large ,}
{\large lemma} {\large }{\large B3); }\\
{\large \( \left\Vert \left( \chi ^{\left( t\right) }_{\mathbf{v}_{i}}\left( \nabla E^{\sigma _{t}}\left( \mathbf{P}\right) ,s\right) \psi _{i,\sigma _{t}}^{\left( t\right) }-\chi ^{\left( t\right) }_{\mathbf{v}_{i}}\left( \frac{\mathbf{x}}{s},s\right) \right) e^{-iE^{\sigma _{t}}\left( \mathbf{P}\right) s}e^{i\gamma _{\sigma _{t}}\left( \mathbf{v}_{i},\nabla E^{\sigma _{t}},s\right) }\psi _{i,\sigma _{t}}^{\left( t\right) }\right\Vert \leq  \)}\\
{\large \( \leq C\cdot s^{-\frac{1}{112}}\cdot s^{2\delta }\cdot \left| \ln \sigma _{t}\right| \cdot t^{-\frac{3\epsilon }{2}} \)
(see lemma B2);}{\large \par}
\end{itemize}
{\large (25.2) }{\large \par}

\begin{itemize}
\item {\large \( \left| \varphi ^{2}_{\sigma _{t},\mathbf{v}_{i}}\left( \mathbf{x},s\right) \chi ^{\left( t\right) }_{\mathbf{v}_{i}}\left( \frac{\mathbf{x}}{s},s\right) \right| \leq C\cdot s^{-2}\cdot s^{\frac{39}{40}} \)
as the support of \( \chi ^{\left( t\right) }_{\mathbf{v}_{i}}\left( \frac{\mathbf{x}}{s},s\right)  \)
leads to the estimate contained in} \emph{\large Observation 2} {\large , lemma
B3;}{\large \par}
\end{itemize}
{\large (25.3)}{\large \par}

\begin{itemize}
\item {\large since by definition of \( \gamma _{\sigma _{t}}\left( \mathbf{v}_{i},\, .\, ,s\right)  \)
we have \( \frac{d\gamma _{\sigma _{t}}}{ds}\left( \mathbf{v}_{i},\frac{\mathbf{x}}{s},s\right) \equiv -\varphi ^{1}_{\sigma _{t},\mathbf{v}_{i}}\left( \mathbf{x},s\right)  \),
this term is zero ;}{\large \par}
\end{itemize}
{\large (25.4)}{\large \par}

\begin{itemize}
\item {\large by lemma B3 \( \left| \frac{d\gamma _{\sigma _{t}}}{ds}\left( \mathbf{v}_{i},\frac{\mathbf{x}}{s},s\right) \right| =\left| \varphi ^{1}_{\sigma _{t},\mathbf{v}_{i}}\left( \mathbf{x},s\right) \right| \leq C\cdot \left| \frac{\ln \left( \sigma _{t}\right) }{s}\right|  \).
Besides, as already seen,} {\large \( \left\Vert \left( \chi ^{\left( t\right) }_{\mathbf{v}_{i}}\left( \nabla E^{\sigma _{t}}\left( \mathbf{P}\right) ,s\right) \psi _{i,\sigma _{t}}^{\left( t\right) }-\chi ^{\left( t\right) }_{\mathbf{v}_{i}}\left( \frac{\mathbf{x}}{s},s\right) \right) e^{-iE^{\sigma _{t}}\left( \mathbf{P}\right) s}e^{i\gamma _{\sigma _{t}}\left( \mathbf{v}_{i},\nabla E^{\sigma _{t}},s\right) }\psi _{i,\sigma _{t}}^{\left( t\right) }\right\Vert \leq  \)}\\
{\large \( \leq C\cdot s^{-\frac{1}{112}}\cdot s^{2\delta }\cdot \left| \ln \sigma _{t}\right| \cdot t^{-\frac{3\epsilon }{2}} \)}{\large \par}
\end{itemize}
{\large (25.5)}{\large \par}

\begin{itemize}
\item {\large from Corollary B2 a bound of order \( s^{-1}\cdot s^{-\frac{1}{112}}\cdot s^{2\delta }\cdot \left| \ln \sigma _{t}\right| \cdot t^{-\frac{3\epsilon }{2}} \)follows}\\
{\large }\\
{\large \par}
\end{itemize}
{\large Taking into account the estimate (26) and assuming the constraints}\\
{\large \par}

\begin{itemize}
\item {\large \( \delta >24\epsilon  \) }{\large \par}
\item {\large \( 2\delta +3\epsilon <\frac{1}{112} \) }{\large \par}
\end{itemize}
{\large I consider that:}\\
{\large }\\
{\large \( \left| M_{i,j}\left( t\right) \right| =\left| \widehat{M}_{i,j}\left( t,t\right) -\widehat{M}_{i,j}\left( t,+\infty \right) +\widehat{M}_{i,j}\left( t,+\infty \right) \right| =\left| \widehat{M}_{i,j}\left( t,t\right) -\widehat{M}_{i,j}\left( t,+\infty \right) \right| \leq  \)}\\
{\large }\\
{\large \( \leq \left| \int ^{+\infty }_{t}\frac{d}{ds}\left( e^{i\gamma _{\sigma _{t}}\left( \mathbf{v}_{i},\nabla E^{\sigma _{t}}\left( \mathbf{P}\right) ,s\right) }\psi _{i,\sigma _{t}}^{\left( t\right) }\, ,\, e^{iH_{\sigma _{t}}s}W_{\sigma ,i,j}\left( s\right) e^{-iH_{\sigma _{t}}s}e^{i\gamma _{\sigma _{t}}\left( \mathbf{v}_{j},\nabla E^{\sigma _{t}}\left( \mathbf{P}\right) ,s\right) }\psi _{j,\sigma _{t}}^{\left( t\right) }\right) ds\right| \leq  \)}\\
{\large }\\
{\large \( \leq \int ^{+\infty }_{t}2\left\Vert \frac{d}{ds}\left\{ e^{iH_{\sigma _{t}}s}W_{\sigma _{t}}\left( \mathbf{v}_{i},s\right) e^{i\gamma _{\sigma _{t}}\left( \mathbf{v}_{i},\nabla E^{\sigma _{t}}\left( \mathbf{P}\right) ,s\right) }e^{-iH_{\sigma _{t}}s}\psi _{i,\sigma _{t}}^{\left( t\right) }\right\} \right\Vert \left\Vert \psi ^{\left( t\right) }_{i,\sigma _{t}}\right\Vert ds\leq  \)}\\
{\large }\\
{\large \( \leq C\cdot t^{-\frac{1}{112}}\cdot t^{2\delta -3\epsilon }\cdot \left( \ln \sigma _{t}\right) ^{2} \)}
\\
\\
{\large Therefore, the sum of mixed terms is bounded by \( C\cdot t^{-\frac{1}{112}}\cdot t^{2\delta +3\epsilon }\cdot \left( \ln \sigma _{t}\right) ^{2} \)
(where \( \sigma _{t}=t^{\beta } \) , \( \beta >1 \), by hypothesis) which
vanishes for \( t\rightarrow +\infty  \) if \( \delta  \), \( \epsilon  \)
are such that}{\large \par}

\begin{itemize}
\item {\large \( \delta >24\epsilon  \) }{\large \par}
\item {\large \( 2\delta +3\epsilon <\frac{1}{112} \) }{\large \par}
\end{itemize}
{\large I will assume these constraints in the following paragraphs.}\\
{\large }\\
{\large }\\
{\large \par}

\subsection{Strong convergence of \protect\( \psi _{G}\left( t\right) \protect \) for
\protect\( t\rightarrow \infty \protect \).\\
}

{\large I will display the Cauchy property of} \( \psi _{G}\left( t\right)  \),
{\large by studying the norm of :}\\
{\large }\\
{\large \( \psi _{G}\left( t_{2}\right) -\psi _{G}\left( t_{1}\right) = \)}
{\large }\\
{\large }\\
{\large \( =e^{iHt_{2}}\sum _{i=1}^{N\left( t_{1}\right) }\sum _{l\left( i\right) }W_{\sigma _{t_{2}}}\left( \mathbf{v}_{l\left( i\right) },t_{2}\right) e^{i\gamma _{\sigma _{t_{2}}}\left( \mathbf{v}_{l\left( i\right) },\nabla E^{\sigma _{t_{2}}}\left( \mathbf{P}\right) ,t_{2}\right) }e^{-iH_{\sigma _{t_{2}}}t_{2}}\psi _{l\left( i\right) ,\sigma _{t_{2}}}^{\left( t_{2}\right) }+ \)}\\
\marginpar{
{\large (27)}{\large \par}
}{\large }\\
{\large \( -e^{iHt_{1}}\sum _{i=1}^{N\left( t_{1}\right) }W_{\sigma _{t_{1}}}\left( \mathbf{v}_{i},t_{1}\right) e^{i\gamma _{\sigma _{t_{1}}}\left( \mathbf{v}_{i},\nabla E^{\sigma _{t_{1}}}\left( \mathbf{P}\right) ,t_{1}\right) }e^{-iH_{\sigma _{t_{1}}}t_{1}}\psi _{i,\sigma _{t_{1}}}^{\left( t_{1}\right) } \)}\textbf{\large }\\
\textbf{\large }\\
\textbf{\large }\\
\textbf{\large Notations}{\large }\\
{\large }\\
{\large - If \( t_{2}>t_{1}\gg 1 \) and \( N\left( t_{2}\right) \neq N\left( t_{1}\right)  \),
then \( \sum _{j=1}^{N\left( t_{2}\right) }\equiv \sum _{i=1}^{N\left( t_{1}\right) }\sum _{l\left( i\right) } \)
where \( l\left( i\right)  \) is the index which counts the sub-cells relative
to \( t_{2}- \)partition, contained in the \( i^{th} \) cell of the \( t_{1}- \)partition.
Therefore we have \( 1\leq l\left( i\right) \leq \frac{N\left( t_{2}\right) }{N\left( t_{1}\right) } \);
}\\
{\large }\\
{\large - adding and subtracting the same quantities, I will rewrite the expression
(27) as the sum of three contributions} \textbf{\large A1}{\large ,}\textbf{\large A2}{\large ,}\textbf{\large B}{\large .}
\textbf{\large A1} {\large is a variation of the ``dressing'' \( W_{\sigma _{t_{2}}}\left( \mathbf{v}_{i},t_{2}\right)  \)
at fixed partition (that one relative to \( t_{2} \)),} \textbf{\large A2}
{\large is the variation of the partition, the other variables remain fixed}
{\large , while} \textbf{\large B} {\large is the variation from \( t_{2} \)
to \( t_{1} \), at fixed partition (the one relative to \( t_{1} \)).}\\
{\large }\\
{\large \( \psi _{G}\left( t_{2}\right) -\psi _{G}\left( t_{1}\right) = \)}\\
{\large }\\
{\large \( =e^{iHt_{2}}\sum _{i=1}^{N\left( t_{1}\right) }\sum _{l\left( i\right) }W_{\sigma _{t_{2}}}\left( \mathbf{v}_{l\left( i\right) },t_{2}\right) e^{i\gamma _{\sigma _{t_{2}}}\left( \mathbf{v}_{l\left( i\right) },\nabla E^{\sigma _{t_{2}}}\left( \mathbf{P}\right) ,t_{2}\right) }e^{-iH_{\sigma _{t_{2}}}t_{2}}\psi _{l\left( i\right) ,\sigma _{t_{2}}}^{\left( t_{2}\right) }+ \)}\\
 \marginpar{
\textbf{\large A1}{\large \par}
} {\large }\\
{\large \( -e^{iHt_{2}}\sum _{i=1}^{N\left( t_{1}\right) }\sum _{l\left( i\right) }W_{\sigma _{t_{2}}}\left( \mathbf{v}_{i},t_{2}\right) e^{i\gamma _{\sigma _{t_{2}}}\left( \mathbf{v}_{i},\nabla E^{\sigma _{t_{2}}}\left( \mathbf{P}\right) ,t_{2}\right) }e^{-iH_{\sigma _{t_{2}}}t_{2}}\psi _{l\left( i\right) ,\sigma _{t_{2}}}^{\left( t_{2}\right) }+ \)}\\
{\large }\\
{\large \( +e^{iHt_{2}}\sum _{i=1}^{N\left( t_{1}\right) }\sum _{l\left( i\right) }W_{\sigma _{t_{2}}}\left( \mathbf{v}_{i},t_{2}\right) e^{i\gamma _{\sigma _{t_{2}}}\left( \mathbf{v}_{i},\nabla E^{\sigma _{t_{2}}}\left( \mathbf{P}\right) ,t_{2}\right) }e^{-iH_{\sigma _{t_{2}}}t_{2}}\psi _{l\left( i\right) ,\sigma _{t_{2}}}^{\left( t_{2}\right) }+ \)}
{\large }\\
\marginpar{
\textbf{\large A2}{\large \par}
}{\large }\\
{\large \( -e^{iHt_{2}}\sum _{i=1}^{N\left( t_{1}\right) }W_{\sigma _{t_{2}}}\left( \mathbf{v}_{i},t_{2}\right) e^{i\gamma _{\sigma _{t_{2}}}\left( \mathbf{v}_{i},\nabla E^{\sigma _{t_{2}}}\left( \mathbf{P}\right) ,t_{2}\right) }e^{-iH_{\sigma _{t_{2}}}t_{2}}\psi _{i,\sigma _{t_{2}}}^{\left( t_{1}\right) }+ \)}\\
{\large }\\
{\large \( +e^{iHt_{2}}\sum _{i=1}^{N\left( t_{1}\right) }W_{\sigma _{t_{2}}}\left( \mathbf{v}_{i},t_{2}\right) e^{i\gamma _{\sigma _{t_{2}}}\left( \mathbf{v}_{i},\nabla E^{\sigma _{t_{2}}}\left( \mathbf{P}\right) ,t_{2}\right) }e^{-iH_{\sigma _{t_{2}}}t_{2}}\psi _{i,\sigma _{t_{2}}}^{\left( t_{1}\right) }+ \)}\\
\marginpar{
\textbf{\large B}{\large \par}
}{\large }\\
{\large \( -e^{iHt_{1}}\sum _{i=1}^{N\left( t_{1}\right) }W_{\sigma _{t_{1}}}\left( \mathbf{v}_{i},t_{1}\right) e^{i\gamma _{\sigma _{t_{1}}}\left( \mathbf{v}_{i},\nabla E^{\sigma _{t_{1}}}\left( \mathbf{P}\right) ,t_{1}\right) }e^{-iH_{\sigma _{t_{1}}}t_{1}}\psi _{i,\sigma _{t_{1}}}^{\left( t_{1}\right) } \)}\\
{\large }\\
{\large In the following paragraphs I will study the norm} \textbf{\large A1}{\large ,}\textbf{\large A2}{\large ,}\textbf{\large B}
{\large in terms of \( t_{1} \) and \( t_{2} \).}\\
{\large }\\
\textbf{\emph{\large analysis of A}}\textbf{\large 1}{\large }\\
{\large }\\
{\large Let us examine the square norm:}\\
\\
{\small 
\[
\left\Vert e^{iH_{\sigma _{t_{2}}}t_{2}}\sum _{i=1}^{N\left( t_{1}\right) }\sum _{l\left( i\right) }\left( W_{\sigma _{t_{2}}}\left( \mathbf{v}_{l\left( i\right) },t_{2}\right) e^{i\gamma _{\sigma _{t_{2}}}\left( \mathbf{v}_{l\left( i\right) },\nabla E^{\sigma _{t_{2}}}\left( \mathbf{P}\right) ,t_{2}\right) }-W_{\sigma _{t_{2}}}\left( \mathbf{v}_{i},t_{2}\right) e^{i\gamma _{\sigma _{t_{2}}}\left( \mathbf{v}_{i},\nabla E^{\sigma _{t_{2}}}\left( \mathbf{P}\right) ,t_{2}\right) }\right) e^{-iH_{\sigma _{t_{2}}}t_{2}}_{}\psi _{l\left( i\right) ,\sigma _{t_{2}}}^{\left( t_{2}\right) }\right\Vert ^{2}=\]
} {\small }\\
{\small 
\[
=\sum _{i,i\prime =1}^{N\left( t_{1}\right) }\sum _{l\left( i\right) ,l\prime \left( i\prime \right) }\left( e^{-iH_{\sigma _{t_{2}}}t_{2}}\psi _{l\left( i\right) ,\sigma _{t_{2}}}^{\left( t_{2}\right) },\left( W^{\dagger }_{\sigma _{t_{2}}}\left( \mathbf{v}_{l\left( i\right) },t_{2}\right) -W^{\dagger }_{\sigma _{t_{2}}}\left( \mathbf{v}_{i},t_{2}\right) \right) \left( W_{\sigma _{t_{2}}}\left( \mathbf{v}_{l'\left( i'\right) },t_{2}\right) -W_{\sigma _{t_{2}}}\left( \mathbf{v}_{i'},t_{2}\right) \right) e^{-iH_{\sigma _{t_{2}}}t_{2}}\psi _{l\prime \left( \prime i\right) ,\sigma _{t_{2}}}^{\left( t_{2}\right) }\right) \]
}{\large }\\
(the phases \( e^{i\gamma } \) are omitted for reasons of space){\large }\\
{\large }\\
{\large }\\
{\large the sum of the terms where \( i'\neq i \) and \( l'\left( i\right) \neq l\left( i\right)  \)
vanish, for \( t_{2}\rightarrow +\infty  \). Its rate is surely bounded (from
above) by of a quantity of order \( t_{2}^{-\frac{1}{112}}\cdot t_{2}^{2\delta +3\epsilon }\cdot \left( \ln \sigma _{t_{2}}\right) ^{2} \)
as we can estimate by the same procedure used in the norm control. The remaining
terms are of this type}\\
{\large }\\
\( \left( e^{-iH_{\sigma _{t_{2}}}t_{2}}\psi _{l\left( i\right) ,\sigma _{t_{2}}}^{\left( t_{2}\right) },\left( W^{\dagger }_{\sigma _{t_{2}}}\left( \mathbf{v}_{l\left( i\right) },t_{2}\right) -W^{\dagger }_{\sigma _{t_{2}}}\left( \mathbf{v}_{i},t_{2}\right) \right) \left( W_{\sigma _{t_{2}}}\left( \mathbf{v}_{l\left( i\right) },t_{2}\right) -W_{\sigma _{t_{2}}}\left( \mathbf{v}_{i},t_{2}\right) \right) e^{-iH_{\sigma _{t_{2}}}t_{2}}\psi _{l\left( i\right) ,\sigma _{t_{2}}}^{\left( t_{2}\right) }\right) = \){\large }\\
{\large }\\
\( =\left( e^{-iH_{\sigma _{t_{2}}}t_{2}}\psi _{l\left( i\right) ,\sigma _{t_{2}}}^{\left( t_{2}\right) },\left( 2-W^{\dagger }_{\sigma _{t_{2}}}\left( \mathbf{v}_{i},t_{2}\right) W_{\sigma _{t_{2}}}\left( \mathbf{v}_{l\left( i\right) },t_{2}\right) -W^{\dagger }_{\sigma _{t_{2}}}\left( \mathbf{v}_{l\left( i\right) },t_{2}\right) W_{\sigma _{t_{2}}}\left( \mathbf{v}_{i},t_{2}\right) \right) e^{-iH_{\sigma _{t_{2}}}t_{2}}\psi _{l\left( i\right) ,\sigma _{t_{2}}}^{\left( t_{2}\right) }\right)  \)
{\large }\\
{\large }\\
{\large For example, I examine:} \\
\\
\( \left( e^{-iH_{\sigma _{t_{2}}}t_{2}}\psi _{l\left( i\right) ,\sigma _{t_{2}}}^{\left( t_{2}\right) },e^{i\gamma _{\sigma _{t_{2}}}\left( \mathbf{v}_{i},\nabla E^{\sigma _{t_{2}}},t_{2}\right) }W^{\dagger }_{\sigma _{t_{2}}}\left( \mathbf{v}_{i},t_{2}\right) W_{\sigma _{t_{2}}}\left( \mathbf{v}_{l\left( i\right) },t_{2}\right) e^{i\gamma _{\sigma _{t_{2}}}\left( \mathbf{v}_{l\left( i\right) },\nabla E^{\sigma _{t_{2}}},t_{2}\right) }e^{-iH_{\sigma _{t_{2}}}t_{2}}\psi _{l\left( i\right) ,\sigma _{t_{2}}}^{\left( t_{2}\right) }\right)  \){\large }\\
{\large }\\
{\large }\\
{\large As in the control of the norm, one considers}\\
{\large }\\
{\large \( \left( e^{-iH_{\sigma _{t_{2}}}s}\psi _{l\left( i\right) ,\sigma _{t_{2}}}^{\left( t_{2}\right) },e^{-i\gamma _{\sigma _{t_{2}}}\left( \mathbf{v}_{i},\nabla E^{\sigma _{t_{2}}},s\right) }W^{\dagger }_{\sigma _{t_{2}}}\left( \mathbf{v}_{i},s\right) W_{\sigma _{t_{2}}}\left( \mathbf{v}_{l\left( i\right) },s\right) e^{i\gamma _{\sigma _{t_{2}}}\left( \mathbf{v}_{l\left( i\right) },\nabla E^{\sigma _{t_{2}}},s\right) }e^{-iH_{\sigma _{t_{2}}}s}\psi _{l\left( i\right) ,\sigma _{t_{2}}}^{\left( t_{2}\right) }\right)  \)}\\
{\large }\\
{\large The limit for \( s\rightarrow +\infty  \) of the above expression is:}\\
{\large }\\
{\large 
\[
e^{-\frac{C_{l\left( i\right) ,i}}{2}}\left( e^{i\gamma _{\sigma _{t_{2}}}\left( \mathbf{v}_{l\left( i\right) },\nabla E^{\sigma _{t_{2}}}\left( \mathbf{P}\right) ,\sigma ^{-\frac{40}{39}}_{t_{2}}\right) }\psi _{l\left( i\right) ,\sigma _{t_{2}}}^{\left( t_{2}\right) },e^{i\gamma _{\sigma _{t_{2}}}\left( \mathbf{v}_{i},\nabla E^{\sigma _{t_{2}}}\left( \mathbf{P}\right) ,\sigma _{t_{2}}^{-\frac{40}{39}}\right) }\psi _{l\left( i\right) ,\sigma _{t_{2}}}^{\left( t_{2}\right) }\right) \]
}\\
{\large }\\
{\large where \( C_{l\left( i\right) ,i}=\int ^{\kappa _{1}}_{\sigma _{t_{2}}}\left| h_{l\left( i\right) ,i}\left( \widehat{\mathbf{k}}\right) \right| ^{2}\frac{d^{3}k}{2\left| \mathbf{k}\right| ^{3}} \)
is bounded by \( C\cdot t^{-\epsilon }_{1}\cdot \left| \ln \left( \sigma _{t_{2}}\right) \right|  \)
. }\\
{\large Summing on the cells, the total error is bounded by a quantity of order }{\large \par}

{\large 
\[
t_{2}^{-\frac{1}{112}}\cdot t_{2}^{2\delta }\cdot \left( \ln \sigma _{t_{2}}\right) ^{2}\]
}\\
{\large }\\
{\large Then the discussion is restricted to }\\
{\large }\\
{\large 
\[
\sum _{i=1}^{N\left( t_{1}\right) }\sum _{l\left( i\right) }\left( 2-e^{-\frac{C_{l\left( i\right) ,i}}{2}}-e^{-\frac{C_{i,l\left( i\right) }}{2}}\right) \left( \psi _{l\left( i\right) ,\sigma _{t_{2}}}^{\left( t_{2}\right) },\psi _{l\left( i\right) ,\sigma _{t_{2}}}^{\left( t_{2}\right) }\right) \]
}\\
{\large }\\
{\large that we can control by ~ \( \sum _{i=1}^{N\left( t_{1}\right) }\sum _{l\left( i\right) }\left( \left\Vert \psi _{l\left( i\right) ,\sigma _{t_{2}}}^{\left( t_{2}\right) }\right\Vert ^{2}\cdot C\cdot t^{-\epsilon }_{1}\cdot \left| \ln \left( \sigma _{t_{2}}\right) \right| \right) \leq C\cdot t^{-\epsilon }_{1}\cdot \left| \ln \left( \sigma _{t_{2}}\right) \right|  \)}\\
{\large }\\
{\large }\\
\textbf{\emph{\large analysis of}} \textbf{\large A2} {\large }\\
{\large }\\
{\large \( \left\Vert e^{iHt_{2}}\sum _{i=1}^{N\left( t_{1}\right) }\sum _{l\left( i\right) }W_{\sigma _{t_{2}}}\left( \mathbf{v}_{i},t_{2}\right) e^{i\gamma _{\sigma _{t_{2}}}\left( \mathbf{v}_{i},\nabla E^{\sigma _{t_{2}}}\left( \mathbf{P}\right) ,t_{2}\right) }e^{-iH_{\sigma _{t_{2}}}t_{2}}\left[ \psi _{l\left( i\right) ,\sigma _{t_{2}}}^{\left( t_{2}\right) }-\psi _{i,\sigma _{t_{2}}}^{\left( t_{1}\right) }\right] \right\Vert =0 \)}
{\large }\\
{\large }\\
{\large because} {\large \( \psi _{i,\sigma _{t_{2}}}^{\left( t_{1}\right) }=\int _{\Gamma _{i}}G\left( \mathbf{P}\right) \psi _{\mathbf{P},\sigma _{t_{2}}}d^{3}P=\sum _{l\left( i\right) }\int _{\Gamma _{l\left( i\right) }}G\left( \mathbf{P}\right) \psi _{\mathbf{P},\sigma _{t_{2}}}d^{3}P \)
by definition}{\large .}\\
{\large }\\
\textbf{\emph{\large }}\\
\textbf{\emph{\large }}\\
\textbf{\emph{\large analysis of B}}{\large }\\
{\large }\\
{\large In order to study the} \textbf{\large B} {\large term, I decompose it
into five contributions and I estimate their norms.}\\
{\large }\\
\( \sum _{i}e^{iHt_{2}}W_{\sigma _{t_{2}}}\left( \mathbf{v}_{i},t_{2}\right) W^{b^{\dagger }}_{\sigma _{t_{2}}}\left( \mathbf{v}_{i}\right) W^{b}_{\sigma _{t_{2}}}\left( \mathbf{v}_{i}\right) e^{i\gamma _{\sigma _{t_{2}}}\left( \mathbf{v}_{i},\nabla E^{\sigma _{t_{2}}},t_{2}\right) }W^{b^{\dagger }}_{\sigma _{t_{2}}}\left( \nabla E^{\sigma _{t_{2}}}\right) e^{-iE^{\sigma _{t_{2}}}t_{2}}W^{b}_{\sigma _{t_{2}}}\left( \nabla E^{\sigma _{t_{2}}}\right) \psi _{i,\sigma _{t_{2}}}^{\left( t_{1}\right) }+ \)\\
\\
\( -\sum _{i}e^{iHt_{1}}W_{\sigma _{t_{1}}}\left( \mathbf{v}_{i},t_{1}\right) W^{b^{\dagger }}_{\sigma _{t_{1}}}\left( \mathbf{v}_{i}\right) W^{b}_{\sigma _{t_{1}}}\left( \mathbf{v}_{i}\right) e^{i\gamma _{\sigma _{t_{1}}}\left( \mathbf{v}_{i},\nabla E^{\sigma _{t_{1}}},t_{1}\right) }W^{b^{\dagger }}_{\sigma _{t_{1}}}\left( \nabla E^{\sigma _{t_{1}}}\right) e^{-iE^{\sigma _{t_{1}}}t_{1}}W^{b}_{\sigma _{t_{1}}}\left( \nabla E^{\sigma _{t_{1}}}\right) \psi _{i,\sigma _{t_{1}}}^{\left( t_{1}\right) }= \)\\
\\
\( =\sum _{i}e^{iHt_{2}}W_{\sigma _{t_{2}}}\left( \mathbf{v}_{i},t_{2}\right) W^{b^{\dagger }}_{\sigma _{t_{2}}}\left( \mathbf{v}_{i}\right) W^{b}_{\sigma _{t_{2}}}\left( \mathbf{v}_{i}\right) W^{b^{\dagger }}_{\sigma _{t_{2}}}\left( \nabla E^{\sigma _{t_{2}}}\right) e^{i\gamma _{\sigma _{t_{2}}}\left( \mathbf{v}_{i},\nabla E^{\sigma _{t_{2}}},t_{2}\right) }e^{-iE^{\sigma _{t_{2}}}t_{2}}W^{b}_{\sigma _{t_{2}}}\left( \nabla E^{\sigma _{t_{2}}}\right) \psi _{i,\sigma _{t_{2}}}^{\left( t_{1}\right) }+ \)\\
\\
\marginpar{
\textbf{\large BI)}{\large \par}
}\\
\\
\( -\sum _{i}e^{iHt_{1}}W_{\sigma _{t_{2}}}\left( \mathbf{v}_{i},t_{1}\right) W^{b^{\dagger }}_{\sigma _{t_{2}}}\left( \mathbf{v}_{i}\right) W^{b}_{\sigma _{t_{2}}}\left( \mathbf{v}_{i}\right) W^{b^{\dagger }}_{\sigma _{t_{2}}}\left( \nabla E^{\sigma _{t_{2}}}\right) e^{i\gamma _{\sigma _{t_{2}}}\left( \mathbf{v}_{i},\nabla E^{\sigma _{t_{2}}},t_{1}\right) }e^{-iE^{\sigma _{t_{2}}}t_{1}}W^{b}_{\sigma _{t_{2}}}\left( \nabla E^{\sigma _{t_{2}}}\right) \psi _{i,\sigma _{t_{2}}}^{\left( t_{1}\right) }+ \)
\\
\\
\( +\sum _{i}e^{iHt_{1}}W_{\sigma _{t_{2}}}\left( \mathbf{v}_{i},t_{1}\right) W^{b^{\dagger }}_{\sigma _{t_{2}}}\left( \mathbf{v}_{i}\right) W^{b}_{\sigma _{t_{2}}}\left( \mathbf{v}_{i}\right) W^{b^{\dagger }}_{\sigma _{t_{2}}}\left( \nabla E^{\sigma _{t_{2}}}\right) e^{i\gamma _{\sigma _{t_{2}}}\left( \mathbf{v}_{i},\nabla E^{\sigma _{t_{2}}},t_{1}\right) }e^{-iE^{\sigma _{t_{2}}}t_{1}}W^{b}_{\sigma _{t_{2}}}\left( \nabla E^{\sigma _{t_{2}}}\right) \psi _{i,\sigma _{t_{2}}}^{\left( t_{1}\right) }+ \)\\
\\
\marginpar{
\textbf{\large BII)}{\large \par}
}\\
\\
\( -\sum _{i}e^{iHt_{1}}W_{\sigma _{t_{2}}}\left( \mathbf{v}_{i},t_{1}\right) W^{b^{\dagger }}_{\sigma _{t_{2}}}\left( \mathbf{v}_{i}\right) W^{b}_{\sigma _{t_{2}}}\left( \mathbf{v}_{i}\right) W^{b^{\dagger }}_{\sigma _{t_{2}}}\left( \nabla E^{\sigma _{t_{2}}}\right) e^{i\gamma _{\sigma _{t_{1}}}\left( \mathbf{v}_{i},\nabla E^{\sigma _{t_{1}}},t_{1}\right) }e^{-iE^{\sigma _{t_{1}}}t_{1}}W^{b}_{\sigma _{t_{1}}}\left( \nabla E^{\sigma _{t_{1}}}\right) \psi _{i,\sigma _{t_{1}}}^{\left( t_{1}\right) }+ \)\\
\\
\( +\sum _{i}e^{iHt_{1}}W_{\sigma _{t_{2}}}\left( \mathbf{v}_{i},t_{1}\right) W^{b^{\dagger }}_{\sigma _{t_{2}}}\left( \mathbf{v}_{i}\right) W^{b}_{\sigma _{t_{2}}}\left( \mathbf{v}_{i}\right) W^{b^{\dagger }}_{\sigma _{t_{2}}}\left( \nabla E^{\sigma _{t_{2}}}\right) e^{i\gamma _{\sigma _{t_{1}}}\left( \mathbf{v}_{i},\nabla E^{\sigma _{t_{1}}},t_{1}\right) }e^{-iE^{\sigma _{t_{1}}}t_{1}}W^{b}_{\sigma _{t_{1}}}\left( \nabla E^{\sigma _{t_{1}}}\right) \psi _{i,\sigma _{t_{1}}}^{\left( t_{1}\right) }+ \)\\
\\
\marginpar{
\textbf{\large BIII)}{\large \par}
}\\
\\
\( -\sum _{i}e^{iHt_{1}}W_{\sigma _{t_{1}}}\left( \mathbf{v}_{i},t_{1}\right) W^{b^{\dagger }}_{\sigma _{t_{1}}}\left( \mathbf{v}_{i}\right) W^{b}_{\sigma _{t_{2}}}\left( \mathbf{v}_{i}\right) W^{b^{\dagger }}_{\sigma _{t_{2}}}\left( \nabla E^{\sigma _{t_{2}}}\right) e^{i\gamma _{\sigma _{t_{1}}}\left( \mathbf{v}_{i},\nabla E^{\sigma _{t_{1}}},t_{1}\right) }e^{-iE^{\sigma _{t_{1}}}t_{1}}W^{b}_{\sigma _{t_{1}}}\left( \nabla E^{\sigma _{t_{1}}}\right) \psi _{i,\sigma _{t_{1}}}^{\left( t_{1}\right) }+ \)\\
\\
\( +\sum _{i}e^{iHt_{1}}W_{\sigma _{t_{1}}}\left( \mathbf{v}_{i},t_{1}\right) W^{b^{\dagger }}_{\sigma _{t_{1}}}\left( \mathbf{v}_{i}\right) W^{b}_{\sigma _{t_{2}}}\left( \mathbf{v}_{i}\right) W^{b^{\dagger }}_{\sigma _{t_{2}}}\left( \nabla E^{\sigma _{t_{2}}}\right) e^{i\gamma _{\sigma _{t_{1}}}\left( \mathbf{v}_{i},\nabla E^{\sigma _{t_{1}}},t_{1}\right) }e^{-iE^{\sigma _{t_{1}}}t_{1}}W^{b}_{\sigma _{t_{1}}}\left( \nabla E^{\sigma _{t_{1}}}\right) \psi _{i,\sigma _{t_{1}}}^{\left( t_{1}\right) }+ \)\\
\\
\marginpar{
\textbf{\large BIV)}{\large \par}
} \\
\\
\( -\sum _{i}e^{iHt_{1}}W_{\sigma _{t_{1}}}\left( \mathbf{v}_{i},t_{1}\right) W^{b^{\dagger }}_{\sigma _{t_{1}}}\left( \mathbf{v}_{i}\right) W^{b}_{\sigma _{t_{2}}}\left( \mathbf{v}_{i}\right) W^{b^{\dagger }}_{\sigma _{t_{2}}}\left( \nabla E^{\sigma _{t_{1}}}\right) e^{i\gamma _{\sigma _{t_{1}}}\left( \mathbf{v}_{i},\nabla E^{\sigma _{t_{1}}},t_{1}\right) }e^{-iE^{\sigma _{t_{1}}}t_{1}}W^{b}_{\sigma _{t_{1}}}\left( \nabla E^{\sigma _{t_{1}}}\right) \psi _{i,\sigma _{t_{1}}}^{\left( t_{1}\right) }+ \)\\
\\
\( +\sum _{i}e^{iHt_{1}}W_{\sigma _{t_{1}}}\left( \mathbf{v}_{i},t_{1}\right) W^{b^{\dagger }}_{\sigma _{t_{1}}}\left( \mathbf{v}_{i}\right) W^{b}_{\sigma _{t_{2}}}\left( \mathbf{v}_{i}\right) W^{b^{\dagger }}_{\sigma _{t_{2}}}\left( \nabla E^{\sigma _{t_{1}}}\right) e^{i\gamma _{\sigma _{t_{1}}}\left( \mathbf{v}_{i},\nabla E^{\sigma _{t_{1}}},t_{1}\right) }e^{-iE^{\sigma _{t_{1}}}t_{1}}W^{b}_{\sigma _{t_{1}}}\left( \nabla E^{\sigma _{t_{1}}}\right) \psi _{i,\sigma _{t_{1}}}^{\left( t_{1}\right) }+ \)\\
\\
\marginpar{
\textbf{\large BV)}{\large \par}
}\\
\\
\( -\sum _{i}e^{iHt_{1}}W_{\sigma _{t_{1}}}\left( \mathbf{v}_{i},t_{1}\right) W^{b^{\dagger }}_{\sigma _{t_{1}}}\left( \mathbf{v}_{i}\right) W^{b}_{\sigma _{t_{1}}}\left( \mathbf{v}_{i}\right) W^{b^{\dagger }}_{\sigma _{t_{1}}}\left( \nabla E^{\sigma _{t_{1}}}\right) e^{i\gamma _{\sigma _{t_{1}}}\left( \mathbf{v}_{i},\nabla E^{\sigma _{t_{1}}},t_{1}\right) }e^{-iE^{\sigma _{t_{1}}}t_{1}}W^{b}_{\sigma _{t_{1}}}\left( \nabla E^{\sigma _{t_{1}}}\right) \psi _{i,\sigma _{t_{1}}}^{\left( t_{1}\right) } \)\\
{\large }\\
\textbf{\emph{\large control of BI}}\textbf{\large )}{\large }\\
{\large }\\
{\large The term} \textbf{\large BI)} {\large corresponds to} \\
\\
\marginpar{

}\( \sum _{i}\left\{ e^{iHt_{2}}W_{\sigma _{t_{2}}}\left( \mathbf{v}_{i},t_{2}\right) e^{i\gamma _{\sigma _{t_{2}}}\left( \mathbf{v}_{i},\nabla E^{\sigma _{t_{2}}},t_{2}\right) }e^{-iE^{\sigma _{t_{2}}}t_{2}}\psi _{i,\sigma _{t_{2}}}^{\left( t_{1}\right) }-e^{iHt_{1}}W_{\sigma _{t_{2}}}\left( \mathbf{v}_{i},t_{1}\right) e^{i\gamma _{\sigma _{t_{2}}}\left( \mathbf{v}_{i},\nabla E^{\sigma _{t_{2}}},t_{1}\right) }e^{-iE^{\sigma _{t_{2}}}t_{1}}\psi _{i,\sigma _{t_{2}}}^{\left( t_{1}\right) }\right\}  \){\large }\\
{\large }\\
{\large I estimate the contribution of the single cell by expressing the difference
between times} {\large \( t_{1} \) and \( t_{2} \)} {\large as the integral
of its derivative. Then I estimate the norm of the derivative}{\large :}\\
{\large }\\
{\large \( \frac{d}{ds}\left\{ e^{iHs}W_{\sigma _{t_{2}}}\left( \mathbf{v}_{i},s\right) e^{i\gamma _{\sigma _{t_{2}}}\left( \mathbf{v}_{i},\nabla E^{\sigma _{t_{2}}},s\right) }e^{-iE^{\sigma _{t_{2}}}s}\psi _{i,\sigma _{t_{2}}}^{\left( t_{1}\right) }\right\} ds= \)}\\
{\large }\\
{\large \( =\frac{d}{ds}\left\{ e^{iHs}W_{\sigma _{t_{2}}}\left( \mathbf{v}_{i},s\right) e^{i\gamma _{\sigma _{t_{2}}}\left( \mathbf{v}_{i},\nabla E^{\sigma _{t_{2}}},s\right) }e^{-iE^{\sigma _{t_{2}}}s}\psi _{i,\sigma _{t_{2}}}^{\left( t_{1}\right) }\right\} = \)
}\\
{\large }\\
{\large \( =ie^{iHs}W_{\sigma _{t_{2}}}\left( \mathbf{v}_{i},s\right) \left( \varphi _{\sigma _{t_{2}},\mathbf{v}_{i}}\left( \mathbf{x},s\right) +\frac{d\gamma _{\sigma _{t_{2}}}\left( \mathbf{v}_{i},\nabla E^{\sigma _{t_{2}}}\left( \mathbf{P}\right) ,s\right) }{ds}\right) e^{i\gamma _{\sigma _{t_{2}}}\left( \mathbf{v}_{i},\nabla E^{\sigma _{t_{2}}},t_{2}\right) }e^{-iE^{\sigma _{t_{2}}}s}\psi _{i,\sigma _{t_{2}}}^{\left( t_{1}\right) }+ \)}\marginpar{
{\large (28)}{\large \par}
} {\large }\\
{\large }\\
{\large \( +ie^{iHs}W_{\sigma _{t_{2}}}\left( \mathbf{v}_{i},s\right) \left( H-H_{\sigma _{t_{2}}}\right) e^{i\gamma _{\sigma _{t_{2}}}\left( \mathbf{v}_{i},\nabla E^{\sigma _{t_{2}}},s\right) }e^{-iE^{\sigma _{t_{2}}}s}\psi _{i,\sigma _{t_{2}}}^{\left( t_{1}\right) } \)}\marginpar{
{\large (29)}{\large \par}
} {\large }{\large }\\
{\large }\\
{\large The formal steps are well defined as \( \psi _{i,\sigma _{t_{2}}}^{\left( t_{1}\right) }\in D\left( H_{\sigma _{t_{2}}}\right) \equiv D\left( H\right)  \).}\\
{\large }\\
\emph{\large estimate of (28)}{\large }\\
{\large }\\
{\large Analogously to the proof of} \textbf{\large II)} {\large in paragraph
4.1, I decompose \( \varphi _{\sigma _{t_{2}},\mathbf{v}_{i}}\left( \mathbf{x},s\right)  \)
in \( \varphi ^{1}_{\sigma _{t_{2}},\mathbf{v}_{i}}\left( \mathbf{x},s\right) + \)}\\
{\large \( +\varphi ^{2}_{\sigma _{t_{2}},\mathbf{v}_{i}}\left( \mathbf{x},s\right)  \),
having defined: }\\
{\large }\\
{\large \( \varphi ^{1}_{\sigma _{t_{2}},\mathbf{v}_{i}}\left( \mathbf{x},s\right) \equiv g^{2}\int ^{s^{-\frac{39}{40}}}_{\sigma _{t_{2}}}\int \frac{\cos \left( \mathbf{k}\cdot \mathbf{x}-\left| \mathbf{k}\right| s\right) }{\left( 1-\widehat{\mathbf{k}}\cdot \mathbf{v}_{i}\right) }d\Omega d\left| \mathbf{k}\right|  \)}\\
{\large \( \varphi ^{2}_{\sigma _{t_{2}},\mathbf{v}_{i}}\left( \mathbf{x},s\right) \equiv g^{2}\int ^{\kappa _{1}}_{s^{-\frac{39}{40}}}\int \frac{\cos \left( \mathbf{k}\cdot \mathbf{x}-\left| \mathbf{k}\right| s\right) }{\left( 1-\widehat{\mathbf{k}}\cdot \mathbf{v}_{i}\right) }d\Omega d\left| \mathbf{k}\right|  \)}
{\large }\\
{\large }\\
{\large by steps analogous to the ones used in} \textbf{\large II)} {\large of
paragraph 4.1, we obtain the following estimate of the norm of the expression
(28):}\\
{\large }\\
{\large 
\[
C\cdot s^{-1}\cdot s^{-\frac{1}{112}}\cdot s^{2\delta }\cdot t_{1}^{-\frac{3\epsilon }{2}}\cdot \left( \ln \sigma _{t_{2}}\right) ^{2}\]
}\\
{\large }\\
\emph{\large estimate of (29)}{\large }\\
{\large }\\
{\large The norm of the vector }\\
{\large }\\
{\large \( \left( H-H_{\sigma _{t_{2}}}\right) e^{i\gamma _{\sigma _{t_{2}}}\left( \mathbf{v}_{i},\nabla E^{\sigma _{t_{2}}},t_{2}\right) }\psi _{i,\sigma _{t_{2}}}^{\left( t_{1}\right) }=e^{i\gamma _{\sigma _{t_{2}}}\left( \mathbf{v}_{i},\nabla E^{\sigma _{t_{2}}},t_{2}\right) }g\int ^{\sigma _{t_{2}}}_{0}\left( b\left( \mathbf{k}\right) +b^{\dagger }\left( \mathbf{k}\right) \right) \frac{d^{3}k}{\sqrt{2\left| \mathbf{k}\right| }}\psi _{i,\sigma _{t_{2}}}^{\left( t_{1}\right) } \)
}\\
{\large can be estimated in the following way:}\\
{\large }\\
{\large \( \left\Vert g\int ^{\sigma _{t_{2}}}_{0}\left( b\left( \mathbf{k}\right) +b^{\dagger }\left( \mathbf{k}\right) \right) \frac{d^{3}k}{\sqrt{2\left| \mathbf{k}\right| }}\psi _{i,\sigma _{t_{2}}}^{\left( t_{1}\right) }\right\Vert \leq \left\Vert g\int ^{\sigma _{t_{2}}}_{0}b\left( \mathbf{k}\right) \frac{d^{3}k}{\sqrt{2\left| \mathbf{k}\right| }}\psi _{i,\sigma _{t_{2}}}^{\left( t_{1}\right) }\right\Vert +\left\Vert g\int ^{\sigma _{t_{2}}}_{0}b^{\dagger }\left( \mathbf{k}\right) \frac{d^{3}k}{\sqrt{2\left| \mathbf{k}\right| }}\psi _{i,\sigma _{t_{2}}}^{\left( t_{1}\right) }\right\Vert \leq  \)}\\
{\large }\\
{\large \( \leq \left( g^{2}\int _{0}^{\sigma _{t_{2}}}\frac{d^{3}k}{2\left| \mathbf{k}\right| }\right) ^{\frac{1}{2}}\cdot \left\Vert \psi _{i,\sigma _{t_{2}}}^{\left( t_{1}\right) }\right\Vert \leq C\cdot \left( \sigma _{t_{2}}\right) \cdot t^{-\frac{3\epsilon }{2}}_{1} \)}\\
{\large }\\
{\large ( I used the fact that \( b\left( \mathbf{k}\right) \psi _{i,\sigma _{t_{2}}}^{\left( t_{1}\right) }=0 \)
for \( \mathbf{k}:\quad  \)\( \left| \mathbf{k}\right| \leq \sigma _{t_{2}} \))}
{\large }\\
{\large }\\
{\large Therefore the norm of the term} \textbf{\large BI)} {\large is bounded
by a quantity of order }\\
{\large 
\[
t^{-\frac{1}{112}}_{1}\cdot t^{2\delta +\frac{3\epsilon }{2}}_{1}\cdot \left( \ln \sigma _{t_{2}}\right) ^{2}+t_{2}\cdot \left( \sigma _{t_{2}}\right) \cdot t^{\frac{3\epsilon }{2}}_{1}\]
}\\
{\large (remember that the constraint \( 2\delta +3\epsilon <\frac{1}{112} \)
is assumed)}\\
{\large }\\
\textbf{\emph{\large control of BII)}}{\large }\\
{\large }\\
{\large The contribution of a single cell vanishes with a rate related to the
decrease of the cut-off. In fact}\\
{\large }\\
{\large \( \left\Vert e^{i\gamma _{\sigma _{t_{2}}}\left( \mathbf{v}_{i},\nabla E^{\sigma _{t_{2}}},t_{1}\right) }e^{-iE^{\sigma _{t_{2}}}t_{1}}W^{b}_{\sigma _{t_{2}}}\left( \nabla E^{\sigma _{t_{2}}}\right) \psi _{i,\sigma _{t_{2}}}^{\left( t_{1}\right) }-e^{i\gamma _{\sigma _{t_{1}}}\left( \mathbf{v}_{i},\nabla E^{\sigma _{t_{1}}},t_{1}\right) }e^{-iE^{\sigma _{t_{1}}}t_{1}}W^{b}_{\sigma _{t_{1}}}\left( \nabla E^{\sigma _{t_{1}}}\right) \psi _{i,\sigma _{t_{1}}}^{\left( t_{1}\right) }\right\Vert \leq  \)}\\
\\
{\large \( \leq \left\Vert e^{i\gamma _{\sigma _{t_{2}}}\left( \mathbf{v}_{i},\nabla E^{\sigma _{t_{2}}},t_{1}\right) }e^{-iE^{\sigma _{t_{2}}}t_{1}}W^{b}_{\sigma _{t_{2}}}\left( \nabla E^{\sigma _{t_{2}}}\right) \psi _{i,\sigma _{t_{2}}}^{\left( t_{1}\right) }-e^{i\gamma _{\sigma _{t_{1}}}\left( \mathbf{v}_{i},\nabla E^{\sigma _{t_{1}}},t_{1}\right) }e^{-iE^{\sigma _{t_{2}}}t_{1}}W^{b}_{\sigma _{t_{2}}}\left( \nabla E^{\sigma _{t_{2}}}\right) \psi _{i,\sigma _{t_{2}}}^{\left( t_{1}\right) }\right\Vert + \)}\\
\\
{\large \( +\left\Vert e^{i\gamma _{\sigma _{t_{1}}}\left( \mathbf{v}_{i},\nabla E^{\sigma _{t_{1}}},t_{1}\right) }e^{-iE^{\sigma _{t_{2}}}t_{1}}W^{b}_{\sigma _{t_{2}}}\left( \nabla E^{\sigma _{t_{2}}}\right) \psi _{i,\sigma _{t_{2}}}^{\left( t_{1}\right) }-e^{i\gamma _{\sigma _{t_{1}}}\left( \mathbf{v}_{i},\nabla E^{\sigma _{t_{1}}},t_{1}\right) }e^{-iE^{\sigma _{t_{1}}}t_{1}}W^{b}_{\sigma _{t_{2}}}\left( \nabla E^{\sigma _{t_{2}}}\right) \psi _{i,\sigma _{t_{2}}}^{\left( t_{1}\right) }\right\Vert + \)}\\
\\
{\large \( +\left\Vert e^{i\gamma _{\sigma _{t_{1}}}\left( \mathbf{v}_{i},\nabla E^{\sigma _{t_{1}}},t_{1}\right) }e^{-iE^{\sigma _{t_{1}}}t_{1}}W^{b}_{\sigma _{t_{2}}}\left( \nabla E^{\sigma _{t_{2}}}\right) \psi _{i,\sigma _{t_{2}}}^{\left( t_{1}\right) }-e^{i\gamma _{\sigma _{t_{1}}}\left( \mathbf{v}_{i},\nabla E^{\sigma _{t_{1}}},t_{1}\right) }e^{-iE^{\sigma _{t_{1}}}t_{1}}W^{b}_{\sigma _{t_{1}}}\left( \nabla E^{\sigma _{t_{1}}}\right) \psi _{i,\sigma _{t_{1}}}^{\left( t_{1}\right) }\right\Vert  \)}\\
{\large }\\
{\large }\\
\emph{\large first term of the sum}{\large }\\
{\large }\\
{\large The module of \( e^{i\gamma _{\sigma _{t_{2}}}\left( \mathbf{v}_{i},\nabla E^{\sigma _{t_{2}}},t_{1}\right) }-e^{i\gamma _{\sigma _{t_{1}}}\left( \mathbf{v}_{i},\nabla E^{\sigma _{t_{1}}},t_{1}\right) } \)
can be estimated in terms of the module of the difference between the arguments
of the exponentials:}\\
{\large }\\
{\large \( \gamma _{\sigma _{t_{2}}}\left( \mathbf{v}_{i},\nabla E^{\sigma _{t_{2}}}\left( \mathbf{P}\right) ,t_{1}\right) -\gamma _{\sigma _{t_{1}}}\left( \mathbf{v}_{i},\nabla E^{\sigma _{t_{1}}}\left( \mathbf{P}\right) ,t_{1}\right) = \)}
{\large }\\
{\large }\\
{\large \( =-\int ^{t_{1}}_{1}\left\{ g^{2}\int ^{\sigma ^{L}_{\tau }}_{\sigma _{t_{1}}}\int \frac{\cos \left( \mathbf{k}\cdot \nabla E^{\sigma _{t_{2}}}\tau -\left| \mathbf{k}\right| \tau \right) -\cos \left( \mathbf{k}\cdot \nabla E^{\sigma _{t_{1}}}\tau -\left| \mathbf{k}\right| \tau \right) }{\left( 1-\widehat{\mathbf{k}}\cdot \mathbf{v}_{i}\right) }d\Omega d\left| \mathbf{k}\right| \right\} d\tau + \)}\\
{\large }\\
{\large \( -\int ^{t_{1}}_{1}\left\{ g^{2}\int ^{\sigma _{t_{1}}}_{\sigma _{t_{2}}}\int \frac{\cos \left( \mathbf{k}\cdot \nabla E^{\sigma _{t_{2}}}\tau -\left| \mathbf{k}\right| \tau \right) }{\left( 1-\widehat{\mathbf{k}}\cdot \mathbf{v}_{i}\right) }d\Omega d\left| \mathbf{k}\right| \right\} d\tau = \)}\\
{\large }\\
{\large \( =2\int ^{t_{1}}_{1}\left\{ g^{2}\int ^{\sigma ^{L}_{\tau }}_{\sigma _{t_{1}}}\int \frac{\sin \left( \frac{\mathbf{k}\cdot \left( \nabla E^{\sigma _{t_{2}}}-\nabla E^{\sigma _{t_{1}}}\right) \tau }{2}\right) \sin \left( \frac{\mathbf{k}\cdot \left( \nabla E^{\sigma _{t_{2}}}+\nabla E^{\sigma _{t_{1}}}\right) \tau -2\left| \mathbf{k}\right| \tau }{2}\right) }{\left( 1-\widehat{\mathbf{k}}\cdot \mathbf{v}_{i}\right) }d\Omega d\left| \mathbf{k}\right| \right\} d\tau + \)}\\
{\large }\\
{\large \( -\int ^{t_{1}}_{1}\left\{ g^{2}\int ^{\sigma _{t_{1}}}_{\sigma _{t_{2}}}\int \frac{\cos \left( \mathbf{k}\cdot \nabla E^{\sigma _{t_{2}}}\tau -\left| \mathbf{k}\right| \tau \right) }{\left( 1-\widehat{\mathbf{k}}\cdot \mathbf{v}_{i}\right) }d\Omega d\left| \mathbf{k}\right| \right\} d\tau  \)}\\
{\large }\\
{\large then}\\
{\large }\\
{\large \( \left| \gamma _{\sigma _{t_{2}}}\left( \mathbf{v}_{i},\nabla E^{\sigma _{t_{2}}}\left( \mathbf{P}\right) ,t_{1}\right) -\gamma _{\sigma _{t_{1}}}\left( \mathbf{v}_{i},\nabla E^{\sigma _{t_{1}}}\left( \mathbf{P}\right) ,t_{1}\right) \right| \leq  \)}\\
{\large }\\
{\large \( \leq C\cdot \left| \nabla E^{\sigma _{t_{2}}}\left( \mathbf{P}\right) -\nabla E^{\sigma _{t_{1}}}\left( \mathbf{P}\right) \right| \cdot \int ^{t_{1}}_{1}\left( \sigma ^{L}_{\tau }\right) ^{2}\cdot \tau d\tau +C\cdot t_{1}\cdot \sigma _{t_{1}}\leq  \)}\\
{\large }\\
{\large \( \leq C\cdot \left( \sigma _{t_{1}}\right) ^{\frac{1}{4}}\int ^{t_{1}}_{1}\tau ^{-\frac{19}{20}}d\tau +C\cdot t_{1}\cdot \sigma _{t_{1}} \)}\\
{\large }\\
{\large (the bound of the module \( \nabla E_{\mathbf{P}}^{\sigma _{t_{2}}}-\nabla E_{\mathbf{P}}^{\sigma _{t_{1}}} \)
comes substantially from lemma A2, \( \sigma _{t_{1}}=t^{-\beta }_{1} \) where
\( \beta >1 \), by hypothesis). }\\
{\large }\\
{\large Then we have a bound by a quantity of order \( t_{1}\cdot \left( \sigma _{t_{1}}\right) ^{\frac{1}{4}} \).}\\
{\large }\\
\emph{\large second term of the sum}{\large }\\
{\large }\\
{\large Taking in account the iterative procedure, the module \( \left| e^{-iE^{\sigma _{t_{2}}}t_{1}}-e^{-iE^{\sigma _{t_{1}}}t_{1}}\right|  \)
is bounded by a quantity of order \( \left( \sigma _{t_{1}}\right) \cdot t_{1} \)}
{\large }\\
{\large }\\
\emph{\large third term of the sum}{\large }\\
{\large }\\
{\large \( \left\Vert e^{i\gamma _{\sigma _{t_{1}}}\left( \mathbf{v}_{i},\nabla E^{\sigma _{t_{1}}},t_{1}\right) }e^{-iE^{\sigma _{t_{1}}}t_{1}}W^{b}_{\sigma _{t_{2}}}\left( \nabla E^{\sigma _{t_{2}}}\right) \psi _{i,\sigma _{t_{2}}}^{\left( t_{1}\right) }-e^{i\gamma _{\sigma _{t_{1}}}\left( \mathbf{v}_{i},\nabla E^{\sigma _{t_{1}}},t_{1}\right) }e^{-iE^{\sigma _{t_{1}}}t_{1}}W^{b}_{\sigma _{t_{1}}}\left( \nabla E^{\sigma _{t_{1}}}\right) \psi _{i,\sigma _{t_{1}}}^{\left( t_{1}\right) }\right\Vert = \)}\\
{\large }\\
{\large \( =\left\Vert \int _{\Gamma _{i}}G\left( \mathbf{P}\right) W^{b}_{\sigma _{t_{2}}}\left( \nabla E^{\sigma _{t_{2}}}\left( \mathbf{P}\right) \right) \psi _{\mathbf{P},\sigma _{t_{2}}}d^{3}P-\int _{\Gamma _{i}}G\left( \mathbf{P}\right) W^{b}_{\sigma _{t_{1}}}\left( \nabla E^{\sigma _{t_{1}}}\left( \mathbf{P}\right) \right) \psi _{\mathbf{P},\sigma _{t_{1}}}d^{3}P\right\Vert \leq  \)}\\
{\large }\\
{\large \( \leq C\cdot \left( \sigma _{t_{1}}\right) ^{\frac{1}{8}}\cdot t^{-\frac{3\epsilon }{2}}_{1} \)
(see theorem 3.2).}\\
{\large }\\
{\large }\\
{\large Therefore the norm of} \textbf{\large BII)} {\large is surely bounded
by: }\\
{\large 
\[
C\cdot t^{\frac{3\epsilon }{2}+1}_{1}\cdot \left( \sigma _{t_{1}}\right) ^{\frac{1}{8}}\]
}\\
{\large }\\
\textbf{\emph{\large control of BIII)}}\emph{\large }\\
{\large }\\
\( \sum _{i}e^{iHt_{1}}W_{\sigma _{t_{2}}}\left( \mathbf{v}_{i},t_{1}\right) W^{b^{\dagger }}_{\sigma _{t_{2}}}\left( \mathbf{v}_{i}\right) W^{b}_{\sigma _{t_{2}}}\left( \mathbf{v}_{i}\right) W^{b^{\dagger }}_{\sigma _{t_{2}}}\left( \nabla E^{\sigma _{t_{2}}}\right) e^{i\gamma _{\sigma _{t_{1}}}\left( \mathbf{v}_{i},\nabla E^{\sigma _{t_{1}}},t_{1}\right) }e^{-iE^{\sigma _{t_{1}}}t_{1}}W^{b}_{\sigma _{t_{1}}}\left( \nabla E^{\sigma _{t_{1}}}\right) \psi _{i,\sigma _{t_{1}}}^{\left( t_{1}\right) }+ \)\\
\\
\( -\sum _{i}e^{iHt_{1}}W_{\sigma _{t_{1}}}\left( \mathbf{v}_{i},t_{1}\right) W^{b^{\dagger }}_{\sigma _{t_{1}}}\left( \mathbf{v}_{i}\right) W^{b}_{\sigma _{t_{2}}}\left( \mathbf{v}_{i}\right) W^{b^{\dagger }}_{\sigma _{t_{2}}}\left( \nabla E^{\sigma _{t_{2}}}\right) e^{i\gamma _{\sigma _{t_{1}}}\left( \mathbf{v}_{i},\nabla E^{\sigma _{t_{1}}},t_{1}\right) }e^{-iE^{\sigma _{t_{1}}}t_{1}}W^{b}_{\sigma _{t_{1}}}\left( \nabla E^{\sigma _{t_{1}}}\right) \psi _{i,\sigma _{t_{1}}}^{\left( t_{1}\right) } \)\\
{\large }\\
{\large Having defined \( \varphi _{i}=W^{b}_{\sigma _{t_{2}}}\left( \mathbf{v}_{i}\right) W^{b^{\dagger }}_{\sigma _{t_{2}}}\left( \nabla E^{\sigma _{t_{2}}}\right) e^{i\gamma _{\sigma _{t_{1}}}\left( \mathbf{v}_{i},\nabla E^{\sigma _{t_{1}}},t_{1}\right) }e^{-iE^{\sigma _{t_{1}}}t_{1}}W^{b}_{\sigma _{t_{1}}}\left( \nabla E^{\sigma _{t_{1}}}\right) \psi _{i,\sigma _{t_{1}}}^{\left( t_{1}\right) } \)}\\
{\large the} \textbf{\large III)} {\large is written as:}\\
{\large }\\
{\large 
\[
\sum _{i}e^{iHt_{1}}\left( W_{\sigma _{t_{2}}}\left( \mathbf{v}_{i},t_{1}\right) W^{b^{\dagger }}_{\sigma _{t_{2}}}\left( \mathbf{v}_{i}\right) -W_{\sigma _{t_{1}}}\left( \mathbf{v}_{i},t_{1}\right) W^{b^{\dagger }}_{\sigma _{t_{1}}}\left( \mathbf{v}_{i}\right) \right) \varphi _{i}\]
}\\
{\large }\\
{\large I restrict the analysis to the single cell. For this purpose I examine}\\
{\large }\\
{\large \( W_{\sigma _{t_{2}}}\left( \mathbf{v}_{i},t_{1}\right) W^{b^{\dagger }}_{\sigma _{t_{2}}}\left( \mathbf{v}_{i}\right) -W_{\sigma _{t_{1}}}\left( \mathbf{v}_{i},t_{1}\right) W^{b^{\dagger }}_{\sigma _{t_{1}}}\left( \mathbf{v}_{i}\right) = \)}
{\large }\\
{\large }\\
{\large \( =W^{b^{\dagger }}_{\kappa _{1}}\left( \mathbf{v}_{i}\right) e^{-g\int ^{\kappa _{1}}_{\sigma _{t_{2}}}\frac{a\left( \mathbf{k}\right) \left( e^{i\left| \mathbf{k}\right| t_{1}}-e^{i\mathbf{k}\cdot \mathbf{x}}\right) -a^{\dagger }\left( \mathbf{k}\right) \left( e^{-i\left| \mathbf{k}\right| t_{1}}-e^{-i\mathbf{k}\cdot \mathbf{x}}\right) }{\left| \mathbf{k}\right| \left( 1-\widehat{\mathbf{k}}\cdot \mathbf{v}_{i}\right) }\frac{d^{3}k}{\sqrt{2\left| \mathbf{k}\right| }}}-W^{b^{\dagger }}_{\kappa _{1}}\left( \mathbf{v}_{i}\right) e^{-g\int ^{\kappa _{1}}_{\sigma _{t_{1}}}\frac{a\left( \mathbf{k}\right) \left( e^{i\left| \mathbf{k}\right| t_{1}}-e^{i\mathbf{k}\cdot \mathbf{x}}\right) -c.c.}{\left| \mathbf{k}\right| \left( 1-\widehat{\mathbf{k}}\cdot \mathbf{v}_{i}\right) }\frac{d^{3}k}{\sqrt{2\left| \mathbf{k}\right| }}} \)}\\
{\large }\\
{\large \( =W_{\sigma _{t_{1}}}\left( \mathbf{v}_{i},t_{1}\right) W^{b^{\dagger }}_{\sigma _{t_{1}}}\left( \mathbf{v}_{i}\right) \left( e^{-g\int ^{\sigma _{t_{1}}}_{\sigma _{t_{2}}}\frac{a\left( \mathbf{k}\right) \left( e^{i\left| \mathbf{k}\right| t_{1}}-e^{i\mathbf{k}\cdot \mathbf{x}}\right) -a^{\dagger }\left( \mathbf{k}\right) \left( e^{-i\left| \mathbf{k}\right| t_{1}}-e^{-i\mathbf{k}\cdot \mathbf{x}}\right) }{\left| \mathbf{k}\right| \left( 1-\widehat{\mathbf{k}}\cdot \mathbf{v}_{i}\right) }\frac{d^{3}k}{\sqrt{2\left| \mathbf{k}\right| }}}-1\right)  \)}\\
{\large }\\
{\large }\\
{\large On the vectors belonging to the domain of the generator of the exponential
we have}\\
{\large }\\
{\large \( e^{-g\int ^{\sigma _{t_{1}}}_{\sigma _{t_{2}}}\frac{a\left( \mathbf{k}\right) \left( e^{i\left| \mathbf{k}\right| t_{1}}-e^{i\mathbf{k}\cdot \mathbf{x}}\right) -a^{\dagger }\left( \mathbf{k}\right) \left( e^{-i\left| \mathbf{k}\right| t_{1}}-e^{-i\mathbf{k}\cdot \mathbf{x}}\right) }{\left| \mathbf{k}\right| \left( 1-\widehat{\mathbf{k}}\cdot \mathbf{v}_{i}\right) }\frac{d^{3}k}{\sqrt{2\left| \mathbf{k}\right| }}}-1= \)}
{\large }\\
{\large }\\
{\large \( =-\int ^{1}_{0}e^{-g\lambda \int ^{\sigma _{t_{1}}}_{\sigma _{t_{2}}}\frac{a\left( \mathbf{k}\right) \left( e^{i\left| \mathbf{k}\right| t_{1}}-e^{i\mathbf{k}\cdot \mathbf{x}}\right) -a^{\dagger }\left( \mathbf{k}\right) \left( e^{-i\left| \mathbf{k}\right| t_{1}}-e^{-i\mathbf{k}\cdot \mathbf{x}}\right) }{\left| \mathbf{k}\right| \left( 1-\widehat{\mathbf{k}}\cdot \mathbf{v}_{i}\right) }\frac{d^{3}k}{\sqrt{2\left| \mathbf{k}\right| }}}d\lambda \cdot g\int ^{\sigma _{t_{1}}}_{\sigma _{t_{2}}}\frac{a\left( \mathbf{k}\right) \left( e^{i\left| \mathbf{k}\right| t_{1}}-e^{i\mathbf{k}\cdot \mathbf{x}}\right) -c.c.}{\left| \mathbf{k}\right| \left( 1-\widehat{\mathbf{k}}\cdot \mathbf{v}_{i}\right) }\frac{d^{3}k}{\sqrt{2\left| \mathbf{k}\right| }} \)}\\
{\large }\\
{\large Since the vector \( \varphi _{i} \) belongs to the domain of the generator,
the following bound holds}\\
{\large }\\
{\large \( \left\Vert g\int ^{\sigma _{t_{1}}}_{\sigma _{t_{2}}}\frac{a\left( \mathbf{k}\right) \left( e^{i\left| \mathbf{k}\right| t_{1}}-e^{i\mathbf{k}\cdot \mathbf{x}}\right) -c.c.}{\left| \mathbf{k}\right| \left( 1-\widehat{\mathbf{k}}\cdot \mathbf{v}_{i}\right) }\frac{d^{3}k}{\sqrt{2\left| \mathbf{k}\right| }}\varphi _{i}\right\Vert \leq \left\Vert g\int ^{\sigma _{t_{1}}}_{\sigma _{t_{2}}}\frac{a\left( \mathbf{k}\right) \left( e^{i\left| \mathbf{k}\right| t_{1}}-e^{i\mathbf{k}\cdot \mathbf{x}}\right) }{\left| \mathbf{k}\right| \left( 1-\widehat{\mathbf{k}}\cdot \mathbf{v}_{i}\right) }\frac{d^{3}k}{\sqrt{2\left| \mathbf{k}\right| }}\varphi _{i}\right\Vert + \)}\\
{\large }\\
{\large \( +\left\Vert g\int ^{\sigma _{t_{1}}}_{\sigma _{t_{2}}}\frac{a^{\dagger }\left( \mathbf{k}\right) \left( e^{-i\left| \mathbf{k}\right| t_{1}}-e^{-i\mathbf{k}\cdot \mathbf{x}}\right) }{\left| \mathbf{k}\right| \left( 1-\widehat{\mathbf{k}}\cdot \mathbf{v}_{i}\right) }\frac{d^{3}k}{\sqrt{2\left| \mathbf{k}\right| }}\varphi _{i}\right\Vert \leq  \)}
{\large }\\
{\large }\\
{\large \( \leq 2\left\Vert g\int ^{\sigma _{t_{1}}}_{\sigma _{t_{2}}}\frac{b\left( \mathbf{k}\right) \left( e^{i\left| \mathbf{k}\right| t_{1}-i\mathbf{k}\cdot \mathbf{x}}-1\right) }{\left| \mathbf{k}\right| \left( 1-\widehat{\mathbf{k}}\cdot \mathbf{v}_{i}\right) }\frac{d^{3}k}{\sqrt{2\left| \mathbf{k}\right| }}\varphi _{i}\right\Vert +\left( \varphi _{i},g^{2}\int ^{\sigma _{t_{1}}}_{\sigma _{t_{2}}}\frac{\left| e^{i\left| \mathbf{k}\right| t_{1}}-e^{i\mathbf{k}\cdot \mathbf{x}}\right| ^{2}}{2\left| \mathbf{k}\right| ^{3}\left( 1-\widehat{\mathbf{k}}\cdot \mathbf{v}_{i}\right) ^{2}}d^{3}k\varphi _{i}\right) ^{\frac{1}{2}} \)}\\
{\large }\\
{\large }\\
\textbf{\emph{\large Analysis of}}{\large \( \left\Vert g\int ^{\sigma _{t_{1}}}_{\sigma _{t_{2}}}\frac{b\left( \mathbf{k}\right) \left( e^{i\left| \mathbf{k}\right| t_{1}-i\mathbf{k}\cdot \mathbf{x}}-1\right) }{\left| \mathbf{k}\right| \left( 1-\widehat{\mathbf{k}}\cdot \mathbf{v}_{i}\right) }\frac{d^{3}k}{\sqrt{2\left| \mathbf{k}\right| }}\varphi _{i}\right\Vert  \).}\\
{\large }\\
{\large }\\
{\large }\\
\textbf{\large 1)} {\large the expression \( b\left( \mathbf{k}\right) W^{b}_{\sigma _{t_{2}}}\left( \mathbf{v}_{i}\right) W^{b^{\dagger }}_{\sigma _{t_{2}}}\left( \nabla E^{\sigma _{t_{2}}}\right) e^{i\gamma _{\sigma _{t_{1}}}\left( \mathbf{v}_{i},\nabla E^{\sigma _{t_{1}}},t_{1}\right) }e^{-iE^{\sigma _{t_{1}}}t_{1}}W^{b}_{\sigma _{t_{1}}}\left( \nabla E^{\sigma _{t_{1}}}\right) \psi _{i,\sigma _{t_{1}}}^{\left( t_{1}\right) } \)
}\\
{\large is a well defined vector in~}{\large  H} {\large and it is strongly
continuos in} \textbf{\large \( \mathbf{k} \)}{\large . Therefore:}\\
\\
\( \left\Vert \int ^{\sigma _{t_{1}}}_{\sigma _{t_{2}}}\frac{b\left( \mathbf{k}\right) \left( e^{i\left| \mathbf{k}\right| t_{1}-i\mathbf{k}\cdot \mathbf{x}}-1\right) }{\left| \mathbf{k}\right| \left( 1-\widehat{\mathbf{k}}\cdot \mathbf{v}_{i}\right) }\frac{d^{3}k}{\sqrt{2\left| \mathbf{k}\right| }}W^{b}_{\sigma _{t_{2}}}\left( \mathbf{v}_{i}\right) W^{b^{\dagger }}_{\sigma _{t_{2}}}\left( \nabla E^{\sigma _{t_{2}}}\right) e^{i\gamma _{\sigma _{t_{1}}}\left( \mathbf{v}_{i},\nabla E^{\sigma _{t_{1}}},t_{1}\right) }e^{-iE^{\sigma _{t_{1}}}t_{1}}W^{b}_{\sigma _{t_{1}}}\left( \nabla E^{\sigma _{t_{1}}}\right) \psi _{i,\sigma _{t_{1}}}^{\left( t_{1}\right) }\right\Vert \leq  \){\large }\\
{\large }\\
\( \leq \int ^{\sigma _{t_{1}}}_{\sigma _{t_{2}}}\frac{\left| 1-\widehat{k}\cdot \mathbf{v}_{i}\right| ^{-1}}{\left| \mathbf{k}\right| ^{\frac{3}{2}}\sqrt{2}}\left\Vert b\left( \mathbf{k}\right) \left( e^{i\left| \mathbf{k}\right| t_{1}}e^{-i\mathbf{k}\cdot \mathbf{x}}-1\right) W^{b}_{\sigma _{t_{2}}}\left( \mathbf{v}_{i}\right) W^{b^{\dagger }}_{\sigma _{t_{2}}}\left( \nabla E^{\sigma _{t_{2}}}\right) e^{i\gamma _{\sigma _{t_{1}}}\left( \mathbf{v}_{i},\nabla E^{\sigma _{t_{1}}},t_{1}\right) }e^{-iE^{\sigma _{t_{1}}}t_{1}}W^{b}_{\sigma _{t_{1}}}\left( \nabla E_{\sigma _{t_{1}}}\right) \psi _{i,\sigma _{t_{1}}}^{\left( t_{1}\right) }\right\Vert d^{3}k \)
{\large }\\
{\large }\\
\textbf{\large 2)} {\large estimate of}\\
{\large }\\
{\large \( \left\Vert b\left( \mathbf{k}\right) \left( e^{i\left| \mathbf{k}\right| t_{1}}e^{-i\mathbf{k}\cdot \mathbf{x}}-1\right) W^{b}_{\sigma _{t_{2}}}\left( \mathbf{v}_{i}\right) W^{b^{\dagger }}_{\sigma _{t_{2}}}\left( \nabla E^{\sigma _{t_{2}}}\right) e^{i\gamma _{\sigma _{t_{1}}}\left( \mathbf{v}_{i},\nabla E^{\sigma _{t_{1}}},t_{1}\right) }e^{-iE^{\sigma _{t_{1}}}t_{1}}W^{b}_{\sigma _{t_{1}}}\left( \nabla E^{\sigma _{t_{1}}}\right) \psi _{i,\sigma _{t_{1}}}^{\left( t_{1}\right) }\right\Vert  \)}
{\large }\\
{\large 
\[
\Leftrightarrow \]
}\\
{\large }\\
\textbf{\large 1)} {\large In distributional sense, the following equality is
valid:}\\
{\large }\\
{\large \( b\left( \mathbf{k}\right) W^{b}_{\sigma _{t_{2}}}\left( \mathbf{v}_{i}\right) W^{b^{\dagger }}_{\sigma _{t_{2}}}\left( \nabla E^{\sigma _{t_{2}}}\left( \mathbf{P}\right) \right) =W^{b}_{\sigma _{t_{2}}}\left( \mathbf{v}_{i}\right) W^{b^{\dagger }}_{\sigma _{t_{2}}}\left( \nabla E^{\sigma _{t_{2}}}\left( \mathbf{P}\right) \right) \left( b\left( \mathbf{k}\right) +f\left( \mathbf{k},\mathbf{v}_{i},\mathbf{P}\right) \right)  \)
}\\
{\large }\\
{\large where \( f\left( \mathbf{k},\mathbf{v}_{i},\mathbf{P}\right) =-\frac{g\chi ^{\kappa }_{\sigma _{t_{2}}}\left( \mathbf{k}\right) }{\left| \mathbf{k}\right| ^{\frac{3}{2}}\left( 1-\widehat{\mathbf{k}}\cdot \mathbf{v}_{i}\right) }+\frac{g\chi ^{\kappa }_{\sigma _{t_{2}}}\left( \mathbf{k}\right) }{\left| \mathbf{k}\right| ^{\frac{3}{2}}\left( 1-\widehat{\mathbf{k}}\cdot \nabla E^{\sigma _{t_{2}}}\left( \mathbf{P}\right) \right) } \)}
{\large ~~~~~~}\\
{\large ( \( \chi ^{\kappa }_{\sigma _{t_{2}}} \)characteristic function of
the set \( \left\{ \mathbf{k}:\! \quad \sigma _{t_{2}}\leq \left| \mathbf{k}\right| \leq \kappa \right\}  \));}\\
{\large }\\
{\large }\\
{\large moreover, being} {\large \( \left| \mathbf{k}\right| \leq \sigma _{t_{1}} \),~~~~
\( b\left( \mathbf{k}\right) W^{b}_{\sigma _{t_{1}}}\left( \nabla E^{\sigma _{t_{1}}}\right) \psi _{i,\sigma _{t_{1}}}^{\left( t_{1}\right) }=0 \)}\\
{\large }\\
{\large Therefore for \( \mathbf{k} \) such that \( \left| \mathbf{k}\right| \leq \sigma _{t_{1}} \)}\\
{\large }\\
{\large \( b\left( \mathbf{k}\right) W^{b}_{\sigma _{t_{2}}}\left( \mathbf{v}_{i}\right) W^{b^{\dagger }}_{\sigma _{t_{2}}}\left( \nabla E^{\sigma _{t_{2}}}\right) e^{i\gamma _{\sigma _{t_{1}}}\left( \mathbf{v}_{i},\nabla E^{\sigma _{t_{1}}},t_{1}\right) }e^{-iE^{\sigma _{t_{1}}}t_{1}}W^{b}_{\sigma _{t_{1}}}\left( \nabla E^{\sigma _{t_{1}}}\right) \psi _{i,\sigma _{t_{1}}}^{\left( t_{1}\right) }= \)
}\\
{\large }\\
{\large \( =f\left( \mathbf{k},\mathbf{v}_{i},\mathbf{P}\right) W^{b}_{\sigma _{t_{2}}}\left( \mathbf{v}_{i}\right) W^{b^{\dagger }}_{\sigma _{t_{2}}}\left( \nabla E^{\sigma _{t_{2}}}\right) e^{i\gamma _{\sigma _{t_{1}}}\left( \mathbf{v}_{i},\nabla E^{\sigma _{t_{1}}},t_{1}\right) }e^{-iE^{\sigma _{t_{1}}}t_{1}}W^{b}_{\sigma _{t_{1}}}\left( \nabla E^{\sigma _{t_{1}}}\right) \psi _{i,\sigma _{t_{1}}}^{\left( t_{1}\right) } \).}\\
{\large }\\
{\large The strong continuity in \( \mathbf{k} \) , \( \sigma _{t_{2}}\leq \left| \mathbf{k}\right| \leq \sigma _{t_{1}} \),
comes from the continuity of the function \( f\left( \mathbf{k},\mathbf{v}_{i},\mathbf{P}\right)  \).}\\
{\large }\\
{\large }\\
{\large }\\
\textbf{\large 2)} {\large Estimate of}\\
{\large }\\
{\large \( \left\Vert b\left( \mathbf{k}\right) \left( e^{i\left| \mathbf{k}\right| t_{1}}e^{-i\mathbf{k}\cdot \mathbf{x}}-1\right) W^{b}_{\sigma _{t_{2}}}\left( \mathbf{v}_{i}\right) W^{b^{\dagger }}_{\sigma _{t_{2}}}\left( \nabla E^{\sigma _{t_{2}}}\right) e^{i\gamma _{\sigma _{t_{1}}}\left( \mathbf{v}_{i},\nabla E^{\sigma _{t_{1}}},t_{1}\right) }e^{-iE^{\sigma _{t_{1}}}t_{1}}W^{b}_{\sigma _{t_{1}}}\left( \nabla E^{\sigma _{t_{1}}}\right) \psi _{i,\sigma _{t_{1}}}^{\left( t_{1}\right) }\right\Vert  \)}\\
{\large }\\
{\large Starting from the identity:}\\
\\
\( e^{i\left| \mathbf{k}\right| t_{1}}W^{b}_{\sigma _{t_{2}}}\left( \mathbf{v}_{i}\right) W^{b^{\dagger }}_{\sigma _{t_{2}}}\left( \nabla E_{\mathbf{P}+\mathbf{k}}^{\sigma _{t_{2}}}\right) e^{i\gamma _{\sigma _{t_{1}}}\left( \mathbf{v}_{i},\nabla E_{\mathbf{P}+\mathbf{k}}^{\sigma _{t_{1}}},t_{1}\right) }e^{-iE_{\mathbf{P}+\mathbf{k}}^{\sigma _{t_{1}}}t_{1}}f\left( \mathbf{k},\mathbf{v}_{i},\mathbf{P}+\mathbf{k}\right) e^{-i\mathbf{k}\cdot \mathbf{x}}W^{b}_{\sigma _{t_{1}}}\left( \nabla E_{\mathbf{P}}^{\sigma _{t_{1}}}\right) \psi _{i,\sigma _{t_{1}}}^{\left( t_{1}\right) }+ \)\\
\\
\( -W^{b}_{\sigma _{t_{2}}}\left( \mathbf{v}_{i}\right) W^{b^{\dagger }}_{\sigma _{t_{2}}}\left( \nabla E^{\sigma _{t_{2}}}\left( \mathbf{P}\right) \right) e^{i\gamma _{\sigma _{t_{1}}}\left( \mathbf{v}_{i},\nabla E^{\sigma _{t_{1}}}\left( \mathbf{P}\right) ,t_{1}\right) }e^{-iE_{\mathbf{P}}^{\sigma _{t_{1}}}t_{1}}f\left( \mathbf{k},\mathbf{v}_{i},\mathbf{P}\right) W^{b}_{\sigma _{t_{1}}}\left( \nabla E^{\sigma _{t_{1}}}\left( \mathbf{P}\right) \right) \psi _{i,\sigma _{t_{1}}}^{\left( t_{1}\right) }= \)\\
\\
\( =e^{i\left| \mathbf{k}\right| t_{1}}W^{b}_{\sigma _{t_{2}}}\left( \mathbf{v}_{i}\right) W^{b^{\dagger }}_{\sigma _{t_{2}}}\left( \nabla E_{\mathbf{P}+\mathbf{k}}^{\sigma _{t_{2}}}\right) e^{i\gamma _{\sigma _{t_{1}}}\left( \mathbf{v}_{i},\nabla E_{\mathbf{P}+\mathbf{k}}^{\sigma _{t_{1}}},t_{1}\right) }e^{-iE_{\mathbf{P}+\mathbf{k}}^{\sigma _{t_{1}}}t_{1}}f\left( \mathbf{k},\mathbf{v}_{i},\mathbf{P}+\mathbf{k}\right) e^{-i\mathbf{k}\cdot \mathbf{x}}W^{b}_{\sigma _{t_{1}}}\left( \nabla E^{\sigma _{t_{1}}}\left( \mathbf{P}\right) \right) \psi _{i,\sigma _{t_{1}}}^{\left( t_{1}\right) }+ \)\\
\marginpar{
{\large (30)}{\large \par}
}\\
\\
\( -W^{b}_{\sigma _{t_{2}}}\left( \mathbf{v}_{i}\right) W^{b^{\dagger }}_{\sigma _{t_{2}}}\left( \nabla E^{\sigma _{t_{2}}}\left( \mathbf{P}\right) \right) e^{i\gamma _{\sigma _{t_{1}}}\left( \mathbf{v}_{i},\nabla E^{\sigma _{t_{1}}}\left( \mathbf{P}\right) ,t_{1}\right) }e^{-iE^{\sigma _{t_{1}}}\left( \mathbf{P}\right) t_{1}}f\left( \mathbf{k},\mathbf{v}_{i},\mathbf{P}\right) e^{-i\mathbf{k}\cdot \mathbf{x}}W^{b}_{\sigma _{t_{1}}}\left( \nabla E^{\sigma _{t_{1}}}\right) \psi _{i,\sigma _{t_{1}}}^{\left( t_{1}\right) }+ \)\\
\\
\( +W^{b}_{\sigma _{t_{2}}}\left( \mathbf{v}_{i}\right) W^{b^{\dagger }}_{\sigma _{t_{2}}}\left( \nabla E^{\sigma _{t_{2}}}\left( \mathbf{P}\right) \right) e^{i\gamma _{\sigma _{t_{1}}}\left( \mathbf{v}_{i},\nabla E^{\sigma _{t_{1}}}\left( \mathbf{P}\right) ,t_{1}\right) }e^{-iE^{\sigma _{t_{1}}}\left( \mathbf{P}\right) t_{1}}f\left( \mathbf{k},\mathbf{v}_{i},\mathbf{P}\right) e^{-i\mathbf{k}\cdot \mathbf{x}}W^{b}_{\sigma _{t_{1}}}\left( \nabla E^{\sigma _{t_{1}}}\right) \psi _{i,\sigma _{t_{1}}}^{\left( t_{1}\right) }+ \)\\
\marginpar{
{\large (31)}{\large \par}
}\\
\( -W^{b}_{\sigma _{t_{2}}}\left( \mathbf{v}_{i}\right) W^{b^{\dagger }}_{\sigma _{t_{2}}}\left( \nabla E^{\sigma _{t_{2}}}\left( \mathbf{P}\right) \right) e^{i\gamma _{\sigma _{t_{1}}}\left( \mathbf{v}_{i},\nabla E^{\sigma _{t_{1}}},t_{1}\right) }e^{-iE^{\sigma _{t_{1}}}\left( \mathbf{P}\right) t_{1}}f\left( \mathbf{k},\mathbf{v}_{i},\mathbf{P}\right) W^{b}_{\sigma _{t_{1}}}\left( \nabla E^{\sigma _{t_{1}}}\right) \psi _{i,\sigma _{t_{1}}}^{\left( t_{1}\right) } \){\large }\\
{\large }\\
{\large I estimate the expression (30) and the expression (31).}\\
{\large }\\
\emph{\large Estimate of (30)}{\large }\\
{\large }\\
{\large Considering that:} 

\begin{enumerate}
\item {\large \( \left\Vert \left( W^{b^{\dagger }}_{\sigma _{t_{2}}}\left( \nabla E_{\mathbf{P}+\mathbf{k}}^{\sigma _{t_{2}}}\right) -W^{b^{\dagger }}_{\sigma _{t_{2}}}\left( \nabla E_{\mathbf{P}}^{\sigma _{t_{2}}}\right) \right) f\left( \mathbf{k},\mathbf{v}_{i},\mathbf{P}+\mathbf{k}\right) e^{-i\mathbf{k}\cdot \mathbf{x}}W^{b}_{\sigma _{t_{1}}}\left( \nabla E_{\mathbf{P}}^{\sigma _{t_{1}}}\right) \psi _{i,\sigma _{t_{1}}}^{\left( t_{1}\right) }\right\Vert \leq  \)}
{\large \( \leq C\cdot \left( \ln \left( \sigma _{t_{2}}\right) \cdot \ln \left( \sigma _{t_{1}}\right) \right) ^{\frac{1}{2}}\cdot t^{-\frac{3\epsilon }{2}}_{1}\cdot \left| \mathbf{k}\right| ^{\frac{1}{16}}\cdot \left| \mathbf{k}\right| ^{-\frac{3}{2}} \)}
{\large (it is proved starting from an estimate analogous to the (b4) in lemma
B2 and from the fact that \( \left| \mathbf{k}\right| \leq \sigma _{t_{1}}\ll 1 \));}\\
 {\large }{\large \par}
\item {\large \( \left| e^{i\gamma _{\sigma _{t_{1}}}\left( \mathbf{v}_{i},\nabla E^{\sigma _{t_{1}}}\left( \mathbf{P}+\mathbf{k}\right) ,t_{1}\right) }-e^{i\gamma _{\sigma _{t_{1}}}\left( \mathbf{v}_{i},\nabla E^{\sigma _{t_{1}}}\left( \mathbf{P}\right) ,t_{1}\right) }\right| \leq C\cdot \left| \mathbf{k}\right| ^{\frac{1}{16}}\cdot t_{1}^{\frac{1}{20}} \)}\\
{\large }\\
{\large by comparing the arguments of the exponentials as in the discussion
of} \textbf{\emph{\large BII)}}{\large ;}\\
{\large }\\
{\large \par}
\item {\large \( \left| e^{i\left| \mathbf{k}\right| t_{1}}-1\right| <2t_{1}\cdot \left| \mathbf{k}\right|  \)}\\
{\large }\\
{\large \par}
\item {\large \( \left| f\left( \mathbf{k},\mathbf{v}_{i},\mathbf{P}\right) -f\left( \mathbf{k},\mathbf{v}_{i},\mathbf{P}+\mathbf{k}\right) \right| \leq C\cdot \frac{\left| \mathbf{k}\right| ^{\frac{1}{16}}}{\left| \mathbf{k}\right| ^{\frac{3}{2}}} \)}
{\large ~~~~~~}\\
{\large }\\
{\large \par}
\item {\large \( \left| e^{-iE^{\sigma _{t_{1}}}\left( \mathbf{P}+\mathbf{k}\right) t_{1}}-e^{-iE^{\sigma _{t_{1}}}\left( \mathbf{P}\right) t_{1}}\right| \leq C\cdot \left| \mathbf{k}\right| ^{\frac{1}{16}}\cdot t_{1} \)}{\large \par}
\end{enumerate}
{\large being \( \left| \mathbf{k}\right| \leq \sigma _{t_{1}}\ll 1 \), one
can conclude that (30) is bounded by}\\
{\large 
\[
C\cdot \left| \mathbf{k}\right| ^{\frac{1}{16}}\cdot \left| \mathbf{k}\right| ^{-\frac{3}{2}}\cdot t_{1}\cdot t^{-\frac{3\epsilon }{2}}_{1}\cdot \left| \ln \sigma _{t_{2}}\right| \]
}\\
\\
\emph{\large Estimate of (31)}{\large }\\
{\large }\\
{\large The norm of the expression (31) is equal to}\\
{\large }\\
{\large \( \left\Vert e^{-i\mathbf{k}\cdot \mathbf{x}}W^{b}_{\sigma _{t_{1}}}\left( \nabla E^{\sigma _{t_{1}}}\right) \psi _{i,\sigma _{t_{1}}}^{\left( t_{1}\right) }-W^{b}_{\sigma _{t_{1}}}\left( \nabla E^{\sigma _{t_{1}}}\right) \psi _{i,\sigma _{t_{1}}}^{\left( t_{1}\right) }\right\Vert = \)}
{\large }\\
{\large }\\
{\large \( =\left\Vert e^{-i\mathbf{k}\cdot \mathbf{x}}\int _{\Gamma _{i}}G\left( \mathbf{P}\right) W^{b}_{\sigma _{t_{1}}}\left( \nabla E^{\sigma _{t_{1}}}\left( \mathbf{P}\right) \right) \psi _{\mathbf{P},\sigma _{t_{1}}}d^{3}P-\int _{\Gamma _{i}}G\left( \mathbf{P}\right) W^{b}_{\sigma _{t_{1}}}\left( \nabla E^{\sigma _{t_{1}}}\left( \mathbf{P}\right) \right) \psi _{\mathbf{P},\sigma _{t_{1}}}d^{3}P\right\Vert  \)}\\
{\large }\\
{\large }\\
{\large I observe that \( e^{-i\mathbf{k}\cdot \mathbf{x}}W^{b}_{\sigma _{t_{1}}}\left( \nabla E^{\sigma _{t_{1}}}\left( \mathbf{P}\right) \right) \psi _{\mathbf{P},\sigma _{t_{1}}}\in H_{\mathbf{P}-\mathbf{k}} \)
and that the following equality holds in \( F^{b} \):}\\
{\large }\\
{\large 
\[
I_{\mathbf{P}-\mathbf{k}}\left( e^{-i\mathbf{k}\cdot \mathbf{x}}W^{b}_{\sigma _{t_{1}}}\left( \nabla E^{\sigma _{t_{1}}}\left( \mathbf{P}\right) \right) \psi _{\mathbf{P},\sigma _{t_{1}}}\right) =I_{\mathbf{P}}\left( W^{b}_{\sigma _{t_{1}}}\left( \nabla E^{\sigma _{t_{1}}}\left( \mathbf{P}\right) \right) \psi _{\mathbf{P},\sigma _{t_{1}}}\right) \]
} {\large }\\
{\large }\\
{\large Then }\\
{\large }\\
{\large \( e^{-i\mathbf{k}\cdot \mathbf{x}}\int _{\Gamma _{i}}G\left( \mathbf{P}\right) W^{b}_{\sigma _{t_{1}}}\left( \nabla E^{\sigma _{t_{1}}}\left( \mathbf{P}\right) \right) \psi _{\mathbf{P},\sigma _{t_{1}}}d^{3}P-\int _{\Gamma _{i}}G\left( \mathbf{P}\right) W^{b}_{\sigma _{t_{1}}}\left( \nabla E^{\sigma _{t_{1}}}\left( \mathbf{P}\right) \right) \psi _{\mathbf{P},\sigma _{t_{1}}}d^{3}P= \)}\\
{\large }\\
{\large \( =\int _{\Gamma _{i}}G\left( \mathbf{P}\right) e^{-i\mathbf{k}\cdot \mathbf{x}}\left\{ W^{b}_{\sigma _{t_{1}}}\left( \nabla E^{\sigma _{t_{1}}}\left( \mathbf{P}\right) \right) \psi _{\mathbf{P},\sigma _{t_{1}}}\right\} d^{3}P-\int _{\Gamma _{i}}G\left( \mathbf{P}\right) W^{b}_{\sigma _{t_{1}}}\left( \nabla E^{\sigma _{t_{1}}}\left( \mathbf{P}-\mathbf{k}\right) \right) \psi _{\mathbf{P}-\mathbf{k},\sigma _{t_{1}}}d^{3}P+ \)}\\
\marginpar{
{\large (32.1)}{\large \par}
} {\large }\\
{\large }\\
{\large \( +\int _{\Gamma _{i}}G\left( \mathbf{P}\right) W^{b}_{\sigma _{t_{1}}}\left( \nabla E^{\sigma _{t_{1}}}\left( \mathbf{P}-\mathbf{k}\right) \right) \psi _{\mathbf{P}-\mathbf{k},\sigma _{t_{1}}}d^{3}P-\int _{\Gamma _{i}}G\left( \mathbf{P}-\mathbf{k}\right) W^{b}_{\sigma _{t_{1}}}\left( \nabla E^{\sigma _{t_{1}}}\left( \mathbf{P}-\mathbf{k}\right) \right) \psi _{\mathbf{P}-\mathbf{k},\sigma _{t_{1}}}d^{3}P+ \)}{\small }\\
\marginpar{
{\large (32.2)}{\large \par}
}{\large }\\
{\large }\\
{\large \( +\int _{\Gamma _{i}}G\left( \mathbf{P}-\mathbf{k}\right) W^{b}_{\sigma _{t_{1}}}\left( \nabla E^{\sigma _{t_{1}}}\left( \mathbf{P}-\mathbf{k}\right) \right) \psi _{\mathbf{P}-\mathbf{k},\sigma _{t_{1}}}d^{3}P-\int _{\Gamma _{i}}G\left( \mathbf{P}\right) W^{b}_{\sigma _{t_{1}}}\left( \nabla E^{\sigma _{t_{1}}}\left( \mathbf{P}\right) \right) \psi _{\mathbf{P},\sigma _{t_{1}}}d^{3}P \)}\\
\marginpar{
{\large (32.3)}{\large \par}
}\\
{\large }\\
{\large The term (32.1) can be estimated starting from }\\
{\large }\\
{\large \( \left\Vert I_{\mathbf{P}-\mathbf{k}}\left( e^{-i\mathbf{k}\cdot \mathbf{x}}\left\{ W^{b}_{\sigma _{t_{1}}}\left( \nabla E^{\sigma _{t_{1}}}\left( \mathbf{P}\right) \right) \psi _{\mathbf{P},\sigma _{t_{1}}}\right\} \right) -I_{\mathbf{P}-\mathbf{k}}\left( W^{b}_{\sigma _{t_{1}}}\left( \nabla E^{\sigma _{t_{1}}}\left( \mathbf{P}-\mathbf{k}\right) \right) \psi _{\mathbf{P}-\mathbf{k},\sigma _{t_{1}}}\right) \right\Vert _{F} \)}\\
{\large }\\
{\large and then it is norm bounded by \( C\cdot \left| \mathbf{k}\right| ^{\frac{1}{32}}\cdot t^{-\frac{3\epsilon }{2}}_{1} \),
for theorem 3.4.}\\
{\large }\\
{\large Being \( G\in C_{0}^{1}\left( R^{3}\setminus 0\right)  \), the norm
of the term (32.2) is bounded by \( C\cdot \left| \mathbf{k}\right| \cdot t^{-\frac{3\epsilon }{2}}_{1} \).}\\
{\large }\\
{\large After having estimated a volume difference, the norm of the term (32.3)
is bounded by a quantity of order~ \( \left| \mathbf{k}\right| ^{\frac{1}{2}}\cdot t^{-\epsilon }_{1} \).}
{\large }\\
{\large }\\
{\large In conclusion the norm of the (31) is bounded by \( C\cdot \left| \mathbf{k}\right| ^{\frac{1}{32}}\cdot t^{-\epsilon }_{1} \)
( \( \left| \mathbf{k}\right| \ll 1 \)).}\\
{\large }\\
{\large }\\
{\large Then \( \left\Vert g\int ^{\sigma _{t_{1}}}_{\sigma _{t_{2}}}\frac{b\left( \mathbf{k}\right) \left( e^{i\left| \mathbf{k}\right| t_{1}-i\mathbf{k}\cdot \mathbf{x}}-1\right) }{\left| \mathbf{k}\right| \left( 1-\widehat{\mathbf{k}}\cdot \mathbf{v}_{i}\right) }\frac{d^{3}k}{\sqrt{2\left| \mathbf{k}\right| }}\varphi _{i}\right\Vert \leq C\cdot t_{1}\cdot t^{-\epsilon }_{1}\cdot \left| \ln \sigma _{t_{2}}\right| \cdot \int ^{\sigma _{t_{1}}}_{0}\frac{\left| \mathbf{k}\right| ^{\frac{1}{32}}}{\left| \mathbf{k}\right| }d\left| \mathbf{k}\right| = \)}\\
{\large }\\
{\large \( =C\cdot t_{1}\cdot t^{-\epsilon }_{1}\cdot \left| \ln \sigma _{t_{2}}\right| \cdot \left( \sigma _{t_{1}}\right) ^{\frac{1}{32}} \)}\\
{\large }\\
\emph{\large Note}\\
\\
{\large By the same steps, one obtains that}\\
{\large }\\
{\large 
\[
\left\Vert \left( e^{i\left| \mathbf{k}\right| t_{1}}e^{-i\mathbf{k}\cdot \mathbf{x}}-1\right) W^{b}_{\sigma _{t_{2}}}\left( \mathbf{v}_{i}\right) W^{b^{\dagger }}_{\sigma _{t_{2}}}\left( \nabla E^{\sigma _{t_{2}}}\right) e^{i\gamma _{\sigma _{t_{1}}}\left( \mathbf{v}_{i},\nabla E^{\sigma _{t_{1}}},t_{1}\right) }e^{-iE^{\sigma _{t_{1}}}t_{1}}W^{b}_{\sigma _{t_{1}}}\left( \nabla E^{\sigma _{t_{1}}}\right) \psi _{i,\sigma _{t_{1}}}^{\left( t_{1}\right) }\right\Vert \]
}\\
{\large is bounded by a quantity of order~~~ \( \left| \mathbf{k}\right| ^{\frac{1}{32}}\cdot t_{1}\cdot t^{-\epsilon }_{1}\cdot \left| \ln \sigma _{t_{2}}\right|  \)}.{\large }\\
{\large }\\
\textbf{\emph{\large Analysis of \( \left( \varphi _{i},g^{2}\int ^{\sigma _{t_{1}}}_{\sigma _{t_{2}}}\frac{\left| e^{i\left| \mathbf{k}\right| t_{1}}-e^{i\mathbf{k}\cdot \mathbf{x}}\right| ^{2}}{2\left| \mathbf{k}\right| ^{3}\left( 1-\widehat{\mathbf{k}}\cdot \mathbf{v}_{i}\right) ^{2}}d^{3}k\varphi _{i}\right) ^{\frac{1}{2}} \)
}} {\large }\\
{\large }\\
{\large Note that \( \left| e^{i\left| \mathbf{k}\right| t_{1}}-e^{i\mathbf{k}\cdot \mathbf{x}}\right| ^{2}=2-e^{-i\left| \mathbf{k}\right| t_{1}}e^{i\mathbf{k}\cdot \mathbf{x}}-e^{i\left| \mathbf{k}\right| t_{1}}e^{-i\mathbf{k}\cdot \mathbf{x}} \).
Taking into account the result of the previous note, we have that:}\\
{\large }\\
{\large \( \left( \varphi _{i},g^{2}\int ^{\sigma _{t_{1}}}_{\sigma _{t_{2}}}\frac{\left| e^{i\left| \mathbf{k}\right| t_{1}}-e^{i\mathbf{k}\cdot \mathbf{x}}\right| ^{2}}{2\left| \mathbf{k}\right| ^{3}\left( 1-\widehat{\mathbf{k}}\cdot \mathbf{v}_{i}\right) ^{2}}d^{3}k\varphi _{i}\right) ^{\frac{1}{2}}\leq C\cdot t^{\frac{1}{2}}_{1}\cdot t^{-\frac{5\epsilon }{4}}_{1}\cdot \left( \left| \ln \sigma _{t_{2}}\right| \right) ^{\frac{1}{2}}\cdot \left( \sigma _{t_{1}}\right) ^{\frac{1}{64}} \)}\\
{\large }\\
{\large }\\
{\large Therefore the norm of the term} \textbf{\large BIII)} {\large is surely
bounded by a quantity of order:}\\
{\large }\\
{\large 
\[
t_{1}\cdot t^{2\epsilon }_{1}\cdot \left| \ln \sigma _{t_{2}}\right| \cdot \left( \sigma _{t_{1}}\right) ^{\frac{1}{64}}\]
}\\
{\large }\\
{\large }\\
\textbf{\emph{\large control of BIV)}}\\
\\
{\large \( \sum _{i}e^{iHt_{1}}W_{\sigma _{t_{1}}}\left( \mathbf{v}_{i},t_{1}\right) W^{b^{\dagger }}_{\sigma _{t_{1}}}\left( \mathbf{v}_{i}\right) W^{b}_{\sigma _{t_{2}}}\left( \mathbf{v}_{i}\right) W^{b^{\dagger }}_{\sigma _{t_{2}}}\left( \nabla E^{\sigma _{t_{2}}}\right) e^{i\gamma _{\sigma _{t_{1}}}\left( \mathbf{v}_{i},\nabla E^{\sigma _{t_{1}}},t_{1}\right) }e^{-iE^{\sigma _{t_{1}}}t_{1}}W^{b}_{\sigma _{t_{1}}}\left( \nabla E^{\sigma _{t_{1}}}\right) \psi _{i,\sigma _{t_{1}}}^{\left( t_{1}\right) }+ \)}\\
{\large }\\
{\large \( -\sum _{i}e^{iHt_{1}}W_{\sigma _{t_{1}}}\left( \mathbf{v}_{i},t_{1}\right) W^{b^{\dagger }}_{\sigma _{t_{1}}}\left( \mathbf{v}_{i}\right) W^{b}_{\sigma _{t_{2}}}\left( \mathbf{v}_{i}\right) W^{b^{\dagger }}_{\sigma _{t_{2}}}\left( \nabla E^{\sigma _{t_{1}}}\right) e^{i\gamma _{\sigma _{t_{1}}}\left( \mathbf{v}_{i},\nabla E^{\sigma _{t_{1}}},t_{1}\right) }e^{-iE^{\sigma _{t_{1}}}t_{1}}W^{b}_{\sigma _{t_{1}}}\left( \nabla E^{\sigma _{t_{1}}}\right) \psi _{i,\sigma _{t_{1}}}^{\left( t_{1}\right) } \)}\\
\\
{\large For a single cell }\\
\\
{\large \( \left\Vert \left( W^{b^{\dagger }}_{\sigma _{t_{2}}}\left( \nabla E^{\sigma _{t_{2}}}\right) -W^{b^{\dagger }}_{\sigma _{t_{2}}}\left( \nabla E^{\sigma _{t_{1}}}\right) \right) W^{b}_{\sigma _{t_{1}}}\left( \nabla E^{\sigma _{t_{1}}}\right) \psi _{i,\sigma _{t_{1}}}^{\left( t_{1}\right) }\right\Vert \leq C\cdot \left( \sigma _{t_{1}}\right) ^{\frac{1}{4}}\cdot \left| \ln \sigma _{t_{2}}\right| \cdot t^{-\frac{3\epsilon }{2}}_{1} \)};\\
{\large therefore the norm of} \textbf{\large BIV}{\large ) is bounded by \( C\cdot t_{1}^{\frac{3\epsilon }{2}}\cdot \left( \sigma _{t_{1}}\right) ^{\frac{1}{4}}\cdot \left| \ln \sigma _{t_{2}}\right|  \)}\\
\\
\\
\textbf{\emph{\large control of BV)}}\\
\\
{\large \( \sum _{i}e^{iHt_{1}}W_{\sigma _{t_{1}}}\left( \mathbf{v}_{i},t_{1}\right) W^{b^{\dagger }}_{\sigma _{t_{1}}}\left( \mathbf{v}_{i}\right) W^{b}_{\sigma _{t_{2}}}\left( \mathbf{v}_{i}\right) W^{b^{\dagger }}_{\sigma _{t_{2}}}\left( \nabla E^{\sigma _{t_{1}}}\right) e^{i\gamma _{\sigma _{t_{1}}}\left( \mathbf{v}_{i},\nabla E_{\sigma _{t_{1}}},t_{1}\right) }e^{-iE^{\sigma _{t_{1}}}t_{1}}W^{b}_{\sigma _{t_{1}}}\left( \nabla E^{\sigma _{t_{1}}}\right) \psi _{i,\sigma _{t_{1}}}^{\left( t_{1}\right) }+ \)}\\
{\large }\\
{\large \( -\sum _{i}e^{iHt_{1}}W_{\sigma _{t_{1}}}\left( \mathbf{v}_{i},t_{1}\right) W^{b^{\dagger }}_{\sigma _{t_{1}}}\left( \mathbf{v}_{i}\right) W^{b}_{\sigma _{t_{1}}}\left( \mathbf{v}_{i}\right) W^{b^{\dagger }}_{\sigma _{t_{1}}}\left( \nabla E^{\sigma _{t_{1}}}\right) e^{i\gamma _{\sigma _{t_{1}}}\left( \mathbf{v}_{i},\nabla E^{\sigma _{t_{1}}},t_{1}\right) }e^{-iE^{\sigma _{t_{1}}}t_{1}}W^{b}_{\sigma _{t_{1}}}\left( \nabla E^{\sigma _{t_{1}}}\right) \psi _{i,\sigma _{t_{1}}}^{\left( t_{1}\right) }= \)}\\
\\
{\large \( =\sum _{i}e^{iHt_{1}}W_{\sigma _{t_{1}}}\left( \mathbf{v}_{i},t_{1}\right) W^{b}\mid ^{\sigma _{t_{1}}}_{\sigma _{t_{2}}}\left( \mathbf{v}_{i}\right) W^{b^{\dagger }}\mid ^{\sigma _{t_{1}}}_{\sigma _{t_{2}}}\left( \nabla E^{\sigma _{t_{1}}}\right) e^{i\gamma _{\sigma _{t_{1}}}\left( \mathbf{v}_{i},\nabla E^{\sigma _{t_{1}}},t_{1}\right) }e^{-iE^{\sigma _{t_{1}}}t_{1}}\psi _{i,\sigma _{t_{1}}}^{\left( t_{1}\right) }+ \)}\\
{\large }\\
{\large \( -\sum _{i}e^{iHt_{1}}W_{\sigma _{t_{1}}}\left( \mathbf{v}_{i},t_{1}\right) e^{i\gamma _{\sigma _{t_{1}}}\left( \mathbf{v}_{i},\nabla E^{\sigma _{t_{1}}},t_{1}\right) }e^{-iE^{\sigma _{t_{1}}}t_{1}}\psi _{i,\sigma _{t_{1}}}^{\left( t_{1}\right) }= \)}\\
{\large }\\
{\large \( =\sum _{i}e^{iHt_{1}}W_{\sigma _{t_{1}}}\left( \mathbf{v}_{i},t_{1}\right) \left( W^{b}\mid ^{\sigma _{t_{1}}}_{\sigma _{t_{2}}}\left( \mathbf{v}_{i}\right) W^{b^{\dagger }}\mid ^{\sigma _{t_{1}}}_{\sigma _{t_{2}}}\left( \nabla E^{\sigma _{t_{1}}}\right) -1\right) e^{i\gamma _{\sigma _{t_{1}}}\left( \mathbf{v}_{i},\nabla E_{\sigma _{t_{1}}},t_{1}\right) }e^{-iE_{\sigma _{t_{1}}}t_{1}}\psi _{i,\sigma _{t_{1}}}^{\left( t_{1}\right) } \)}\\
{\large }\\
{\large with the definitions}\\
{\large \par}

\begin{itemize}
\item {\large \( W^{b}\mid ^{\sigma _{t_{1}}}_{\sigma _{t_{2}}}\left( \mathbf{v}_{i}\right) \equiv W^{b^{\dagger }}_{\sigma _{t_{1}}}\left( \mathbf{v}_{i}\right) W^{b}_{\sigma _{t_{2}}}\left( \mathbf{v}_{i}\right) =e^{g\int _{\sigma _{t_{1}}}^{\kappa }\frac{b\left( \mathbf{k}\right) -b^{\dagger }\left( \mathbf{k}\right) }{\left| \mathbf{k}\right| \left( 1-\widehat{\mathbf{k}}\cdot \mathbf{v}_{i}\right) }\frac{d^{3}k}{\sqrt{2\left| \mathbf{k}\right| }}}e^{-g\int _{\sigma _{t_{2}}}^{\kappa }\frac{b\left( \mathbf{k}\right) -b^{\dagger }\left( \mathbf{k}\right) }{\left| \mathbf{k}\right| \left( 1-\widehat{\mathbf{k}}\cdot \mathbf{v}_{i}\right) }\frac{d^{3}k}{\sqrt{2\left| \mathbf{k}\right| }}}= \)}\\
{\large }\\
{\large \( =e^{-g\int _{\sigma _{t_{2}}}^{\sigma _{t_{1}}}\frac{b\left( \mathbf{k}\right) -b^{\dagger }\left( \mathbf{k}\right) }{\left| \mathbf{k}\right| \left( 1-\widehat{\mathbf{k}}\cdot \mathbf{v}_{i}\right) }\frac{d^{3}k}{\sqrt{2\left| \mathbf{k}\right| }}} \)}\\
{\large \par}
\item {\large \( W^{b^{\dagger }}\mid ^{\sigma _{t_{2}}}_{\sigma _{t_{1}}}\left( \nabla E^{\sigma _{t_{1}}}\right) \equiv W^{b^{\dagger }}_{\sigma _{t_{2}}}\left( \nabla E^{\sigma _{t_{1}}}\right) W^{b}_{\sigma _{t_{1}}}\left( \nabla E^{\sigma _{t_{1}}}\right) =e^{g\int _{\sigma _{t_{2}}}^{\sigma _{t_{1}}}\frac{b\left( \mathbf{k}\right) -b^{\dagger }\left( \mathbf{k}\right) }{\left| \mathbf{k}\right| \left( 1-\widehat{\mathbf{k}}\cdot \nabla E^{\sigma _{t_{1}}}\right) }\frac{d^{3}k}{\sqrt{2\left| \mathbf{k}\right| }}} \)}\\
{\large }\\
{\large }\\
{\large \par}
\end{itemize}
{\large The discussion of this contribution requires the study of the squared
norm and the control of mixed terms in the scalar product. In order to do it
I will verify that the sum of the mixed terms vanishes and I will determine
the rate. Then I will examine the diagonals terms.}\\
\\
\emph{\large mixed term}s\\
\\
{\large As in par.4.1} \textbf{\large ``Control of the norm..''}{\large ,
I will consider the generic i-j term. I observe that it is possible to reply
the same procedure since the operators, obtained from derivation with respect
to the parameter \( \lambda  \), commute with \( \left( W^{b}\mid ^{\sigma _{t_{1}}}_{\sigma _{t_{2}}}\left( \mathbf{v}_{i}\right) W^{b^{\dagger }}\mid ^{\sigma _{t_{1}}}_{\sigma _{t_{2}}}\left( \nabla E^{\sigma _{t_{1}}}\right) -1\right)  \).
Then the result is analogous}: {\large the sum of the modules of the mixed terms
is bounded by \( C\cdot t_{1}^{-\frac{1}{112}}\cdot t_{1}^{2\delta +3\epsilon }\cdot \left( \ln \sigma _{t_{1}}\right) ^{2} \)}
.\\
{\large }\\
\\
\emph{\large diagonal terms}\\
{\large }\\
{\large Considering that:}\\
{\large \par}

\begin{itemize}
\item {\large the norm \( \left\Vert \left( W^{b}\mid ^{\sigma _{t_{1}}}_{\sigma _{t_{2}}}\left( \mathbf{v}_{i}\right) W^{b}\mid ^{\sigma _{t_{1}}}_{\sigma _{t_{2}}}\left( \nabla E^{\sigma _{t_{1}}}\left( \mathbf{P}\right) \right) -1\right) \psi _{i,\sigma _{t_{1}}}^{\left( t_{1}\right) }\right\Vert  \)
can be estimated by a quantity of order \( \sup _{\mathbf{P}\in \Gamma _{i}}\left| \mathbf{v}_{i}-\nabla E^{\sigma _{t_{1}}}\left( \mathbf{P}\right) \right| \cdot \left( \left| \ln \sigma _{t_{2}}\right| \right) ^{\frac{1}{2}}\cdot t_{1}^{-\frac{3\epsilon }{2}} \)
}\\
{\large }\\
{\large \par}
\item {\large \( \sup _{\mathbf{P}\in \Gamma _{i}}\left| \mathbf{v}_{i}-\nabla E^{\sigma _{t_{1}}}\left( \mathbf{P}\right) \right| \leq \sup _{\mathbf{P}\in \Gamma _{i}}\left| \nabla E^{\sigma _{t_{1}}}\left( \mathbf{P}\right) -\nabla E^{\sigma _{t_{1}}}\left( \overline{\mathbf{P}_{i}}\right) \right| \leq  \)}\\
{\large }\\
{\large \( \leq C\cdot \sup _{\mathbf{P}\in \Gamma _{i}}\left| \mathbf{P}-\overline{\mathbf{P}_{i}}\right| ^{\frac{1}{16}}\leq C\cdot t^{-\frac{\epsilon }{16}}_{1} \)
(for the last step, see lemma 3.3)}\\
 {\large }{\large \par}
\end{itemize}
{\large the sum of the diagonal terms gives a contribution bounded by }\\
{\large 
\[
C\cdot t^{-\frac{\epsilon }{32}}_{1}\cdot \left( \ln \left( \sigma _{t_{2}}\right) \right) ^{\frac{1}{4}}\]
}\\
{\large }\\
{\large }\\
{\large Therefore it follows that the norm of the term} \textbf{\large BV)}
{\large is bounded by}\\
{\large 
\[
C\cdot \max \left( t^{-\frac{\epsilon }{32}}_{1},\left( t_{1}^{-\frac{1}{112}}\cdot t_{1}^{2\delta +3\epsilon }\right) ^{\frac{1}{2}}\right) \cdot \left| \ln \sigma _{t_{2}}\right| \]
}\\
{\large }\\
\emph{\large }\\
\textbf{\large Theorem 4.1} \emph{\large }\\
\emph{\large }\\
{\large The vector \( \psi _{G}\left( t\right)  \) converges strongly, for
\( t\rightarrow +\infty  \), with an error of order \( \frac{1}{t^{\rho }} \)
~where \( \rho >0 \) is a properly small coefficient.}\\
{\large }\\
{\large Proof}\\
{\large }\\
{\large We look at the bounds obtained for the norms of A1, A2 and B }\\
{\large }\\
{\large A1) \( C\cdot \left\{ \left( t^{-\epsilon }_{1}\cdot \left| \ln \sigma _{t_{2}}\right| \right) +t_{2}^{-\frac{1}{112}}\cdot t_{2}^{2\delta +3\epsilon }\cdot \left( \ln \sigma _{t_{2}}\right) ^{2}\right\} ^{\frac{1}{2}} \)}\\
{\large }\\
{\large A2) \( 0 \)}\\
{\large }\\
{\large BI) \( C\cdot t^{-\frac{1}{112}}_{1}\cdot t^{2\delta +\frac{3\epsilon }{2}}_{1}\cdot \left( \ln \sigma _{t_{2}}\right) ^{2}+C\cdot t_{2}\cdot \left( \sigma _{t_{2}}\right) \cdot t^{\frac{3\epsilon }{2}}_{1} \)}\\
{\large }\\
{\large BII) \( C\cdot t^{\frac{3\epsilon }{2}}_{1}\cdot t_{1}\cdot \left( \sigma _{t_{1}}\right) ^{\frac{1}{8}} \)}\\
{\large }\\
{\large BIII) \( C\cdot t_{1}\cdot t^{2\epsilon }_{1}\cdot \left| \ln \sigma _{t_{2}}\right| \cdot \left( \sigma _{t_{1}}\right) ^{\frac{1}{64}} \)}\\
{\large }\\
{\large BIV) \( C\cdot t_{1}^{\frac{3\epsilon }{2}}\cdot \left( \sigma _{t_{1}}\right) ^{\frac{1}{4}}\cdot \left| \ln \sigma _{t_{2}}\right|  \)}\\
{\large }\\
{\large BV) \( C\cdot \max \left( t^{-\frac{\epsilon }{32}}_{1},\left( t_{1}^{-\frac{1}{112}}\cdot t_{1}^{2\delta +3\epsilon }\right) ^{\frac{1}{2}}\right) \cdot \left| \ln \sigma _{t_{2}}\right|  \)}\\
{\large }\\
{\large where we have tuned the time scales related to \( \epsilon  \) e \( \delta  \)
in accordance to the constraints~} 

\begin{itemize}
\item {\large ~\( 24\epsilon <\delta  \) .~}{\large \par}
\item {\large ~\( 2\delta +3\epsilon <\frac{1}{112} \)}{\large \par}
\end{itemize}
{\large and we have chosen, for example, \( \sigma _{t}=t^{-128} \). Thus we
can observe that the time \( t_{2} \) appears in the numerators only in \( \ln \left( \sigma _{t_{2}}\right)  \)
and that we can estimate }\\
 {\large 
\[
\left\Vert \psi _{G}\left( t_{2}\right) -\psi _{G}\left( t_{1}\right) \right\Vert <\frac{C}{M}\cdot \left( \frac{\ln \left( t_{2}\right) }{t^{3\rho }_{1}}\right) ^{2}\]
}\\
{\large where \( \rho >0 \) e \( \frac{C}{M}>0 \) are independent of \( t_{1} \)e
\( t_{2} \)(for \( t_{2}\geq t_{1}>\overline{t}\gg 1 \)).}\\
{\large }\\
{\large Now, let us consider the sequence \( t_{1},t^{2}_{1},t^{3}_{1},.....t^{n}_{1},.... \)~and
put ~\( t^{n}_{1}\leq t_{2}<t^{n+1}_{1} \). Due to the norm properties, it
follows that:}\\
{\large }\\
{\large \( \left\Vert \psi _{G}\left( t_{2}\right) -\psi _{G}\left( t_{1}\right) \right\Vert \leq \left\Vert \psi _{G}\left( t^{2}_{1}\right) -\psi _{G}\left( t_{1}\right) \right\Vert +\left\Vert \psi _{G}\left( t_{1}^{3}\right) -\psi _{G}\left( t^{2}_{1}\right) \right\Vert +.....+\left\Vert \psi _{G}\left( t_{2}\right) -\psi _{G}\left( t^{n}_{1}\right) \right\Vert \leq  \)}\\
{\large }\\
{\large \( \leq \frac{C}{M}\cdot \left\{ \left( \frac{2\ln \left( t_{1}\right) }{t^{3\rho }_{1}}\right) ^{2}+\left( \frac{3\ln \left( t_{1}\right) }{t^{2\cdot 3\rho }_{1}}\right) ^{2}+\left( \frac{4\ln \left( t_{1}\right) }{t^{3\cdot 3\rho }_{1}}\right) ^{2}+.......+\left( \frac{\left( n+1\right) \ln \left( t_{1}\right) }{t^{n\cdot 3\rho }_{1}}\right) ^{2}\right\} = \)}
{\large }\\
{\large \( \leq \frac{C}{M\cdot t^{\rho }_{1}}\cdot \left\{ \left( \frac{2}{t^{\rho }_{1}}\cdot \frac{\ln \left( t_{1}\right) }{t^{\rho }_{1}}\right) ^{2}+\left( \frac{3}{t^{2\rho }_{1}}\cdot \frac{\ln \left( t_{1}\right) }{t^{\rho }_{1}}\right) ^{2}+\left( \frac{4}{t^{3\rho }_{1}}\cdot \frac{\ln \left( t_{1}\right) }{t^{\rho }_{1}}\right) ^{2}+.......+\left( \frac{n+1}{t^{n\rho }_{1}}\cdot \frac{\ln \left( t_{1}\right) }{t^{\rho }_{1}}\right) ^{2}\right\}  \)}\\
{\large }\\
{\large For \( t_{1} \) sufficiently large \( t_{1}\geq \widehat{t_{1}}>\overline{t}\gg 1 \)
the sequence inside the brackets is convergent, and it is limited by a constant
\( M \).}\\
{\large I can conclude that \( \forall t_{1},t_{2} \) where \( t_{2}\geq t_{1}\geq \widehat{t_{1}} \)
}\\
{\large }\\
{\large 
\[
\left\Vert \psi _{G}\left( t_{2}\right) -\psi _{G}\left( t_{1}\right) \right\Vert \leq \frac{C}{t^{\rho }_{1}}\]
} {\large }{\large }\\
{\large \par}

\section{Scattering subspaces and asymptotic observables.\\
}

{\large In this chapter I begin by constructing the subspace}{\large ~H}{\large \( ^{1out\left( in\right) } \)
as the norm closure of the finite linear combinations of the vectors \( \psi _{G}^{out\left( in\right) } \).
The invariance requirement under space -time translations for the space}{\large  ~H}{\large \( ^{1out\left( in\right) } \)
implies a more general definition of the vector \( \psi _{G}^{out\left( in\right) } \).
Therefore it will be labelled as a \( \psi _{G,\kappa _{1},\tau ,\mathbf{a}}^{out\left( in\right) } \).
It corresponds to the evolution of \( \psi _{G}^{out\left( in\right) } \) in
the time \( \tau  \) and to the translation of a quantity \( \mathbf{a} \).
}\\
{\large I will verify that on}{\large  ~H}{\large \( ^{1out\left( in\right) } \)\( \equiv \overline{\bigvee \psi _{G,\kappa _{1},\tau ,\mathbf{a}}^{out\left( in\right) }} \)
the strong limits of the functions, continuous and of compact support, of nucleon
mean velocity and the strong limits of the L.S.Z. Weyl operator associated to
the meson field exist.} {\large Due to these results, we can define the vectors
\( \psi _{G,\varphi }^{out\left( in\right) } \) (omitting \( \kappa _{1},\tau ,\mathbf{a} \))
obtained by applying the L.S.Z. Weyl operators, with smearing function \( \varphi  \),
to the total set of ~}{\large  H}{\large \( ^{1out\left( in\right) } \)}{\large .
The norm closure of finite linear combinations} {\large of the} {\large \( \psi _{G,\varphi }^{out\left( in\right) } \)}
{\large is a reasonable candidate for the scattering subspaces}{\large  ~H}{\large \( ^{out\left( in\right) } \).
The meaning of this definition is in the characterization of the}{\large  ~H}{\large \( ^{out\left( in\right) } \)
states, in terms of quantum numbers of the asymptotic variables which are defined
on them: the asymptotic meson Weyl operators and the asymptotic nucleon mean
velocity.} {\large }\\
{\large The spectral restriction on nucleon velocity (strictly less than 1)
and the consequent restriction of}{\large  ~H}{\large \( ^{out\left( in\right) } \)}
{\large as subspaces of}{\large ~H} {\large are in accordance with the partial
non-relativistic character of the model}{\large .} \textbf{\large }\\
{\large In theorem 5.1 the vectors} {\large \( \psi _{G,\varphi }^{out\left( in\right) } \)}
{\large are constructed. In theorem 5.2 we prove the convergence of continuous
and of compact support functions of nucleon mean velocity on the vectors of}{\large  ~H}{\large \( ^{out\left( in\right) } \).
The corollary 5.3 is a check of the fact that the strong limits on}{\large  ~H}{\large \( ^{out\left( in\right) } \)
of the L.S.Z. Weyl operators generate a canonic Weyl algebra} {\large \( \mathcal{A}^{out\left( in\right) } \)}
{\large , to which a free massless scalar field is associated .}\\
{\large }\\
{\large }\\
\textbf{\large Definition of the vector} {\large \( \psi _{G_{\tau },\kappa _{1},\tau ,\mathbf{a}}^{out\left( in\right) } \).}\\
{\small }\\
{\large I apply \( e^{-i\mathbf{a}\cdot \mathbf{P}}e^{-iH\tau } \)to the generic
vector} \( \psi _{G}^{out} \) {\large , constructed in the paragraphs 4.1,4.2}{\small ,}{\large :}{\small }\\
{\large }\\
{\large \( e^{-i\mathbf{a}\cdot \mathbf{P}}e^{-iH\tau }\psi _{G}^{out}= \)}{\small }\\
{\small }\\
{\large \( =s-\lim _{t\rightarrow +\infty }e^{-i\mathbf{a}\cdot \mathbf{P}}e^{-iH\tau }e^{iHt}\sum _{i=1}^{N\left( t\right) }W_{\sigma _{t}}\left( \mathbf{v}_{i},t\right) e^{i\gamma _{\sigma _{t}}\left( \mathbf{v}_{i},\nabla E^{\sigma _{t}}\left( \mathbf{P}\right) ,t\right) }e^{-iE^{\sigma _{t}}\left( t-\tau \right) }e^{-iE^{\sigma _{t}}\tau }\psi _{i,\sigma _{t}}^{\left( t\right) }= \)}\\
\\
{\small \( =s-\lim _{t\rightarrow +\infty }e^{iHt}\sum _{i=1}^{N\left( t\right) }e^{-g\int ^{\kappa _{1}}_{\sigma _{t+\tau }}\frac{a\left( \mathbf{k}\right) e^{i\mathbf{k}\cdot \mathbf{a}}e^{i\left| \mathbf{k}\right| \left( t+\tau \right) }-c.c.}{\left| \mathbf{k}\right| \left( 1-\widehat{\mathbf{k}}\cdot \mathbf{v}_{i}\right) }\frac{d^{3}k}{\sqrt{2\left| \mathbf{k}\right| }}}e^{i\gamma _{\sigma _{t+\tau }}\left( \mathbf{v}_{i},\nabla E^{\sigma _{t+\tau }}\left( \mathbf{P}\right) ,t+\tau \right) }e^{-i\mathbf{a}\cdot \mathbf{P}}e^{-iE^{\sigma _{t+\tau }}t}e^{-iE^{\sigma _{t+\tau }}\tau }\psi _{i,\sigma _{t+\tau }}^{\left( t+\tau \right) } \)}\\
{\large }\\
{\large The limit exists and the proof is similar to the one given for \( \psi _{G}^{out} \),}
{\large apart from some little marginal differences. Then I define}\\
{\large }\\
{\large 
\[
\psi _{G,\kappa _{1},\tau ,\mathbf{a}}^{out\left( in\right) }\equiv e^{-i\mathbf{a}\cdot \mathbf{P}}e^{-iH\tau }\psi _{G}^{out\left( in\right) }\]
}\\
{\large The subspace of minimal asymptotic nucleon states is}{\large  ~H}{\large \( ^{1out\left( in\right) } \)\( \equiv \overline{\bigvee \psi _{G,\kappa _{1},\tau ,\mathbf{a}}^{out\left( in\right) }} \).
Later on I will simply call \( \left\{ \psi _{G}^{out\left( in\right) }\right\}  \)
the total set that generates}{\large  ~H}{\large \( ^{1out\left( in\right) } \).}\\
{\large }\\
\textbf{\large Theorem 5.1}{\large }\\
{\large }\\
{\large The strong limits}\\
{\large 
\[
s-lim_{t\rightarrow +\infty }e^{iHt}e^{i\left( a\left( \varphi _{t}\right) +a^{\dagger }\left( \varphi _{t}\right) \right) }e^{-iHt}\psi _{G}^{out}\equiv \psi _{G,\varphi }^{out}\]
} {\large }\\
{\large exist,where \( \widetilde{\varphi }\left( \mathbf{k}\right) \in C_{0}^{\infty }\left( R^{3}\setminus 0\right)  \),
\( \varphi _{t}\left( \mathbf{y}\right) =\int e^{-i\mathbf{ky}+i\left| \mathbf{k}\right| t}\widetilde{\varphi }\left( \mathbf{k}\right) d^{3}k \)
and}\\
{\large \( a^{\dagger }\left( \varphi \right) \equiv \left( a\left( \varphi \right) \right) ^{\dagger }=\left( \int a\left( \mathbf{k}\right) \widetilde{\varphi }\left( \mathbf{k}\right) d^{3}k\right) ^{\dagger } \)}\\
{\large }\\
{\large Proof}\\
{\large }\\
{\large The convergence follows from the integrability of the norm}\\
{\large }\\
{\large \( \left\Vert \frac{d\left( e^{iHt}e^{i\left( a\left( \varphi _{t}\right) +a^{\dagger }\left( \varphi _{t}\right) \right) }e^{-iHt}\right) }{dt}\psi _{G}^{out}\right\Vert = \)}\\
{\large }\\
{\large \( =\left\Vert e^{i\left( a\left( \varphi _{t}\right) +a^{\dagger }\left( \varphi _{t}\right) \right) }\varphi \left( \mathbf{x},t\right) e^{-iHt}\psi _{G}^{out}\right\Vert \leq  \)
}\\
{\large }\\
{\large \( \leq \left\Vert e^{i\left( a\left( \varphi _{t}\right) +a^{\dagger }\left( \varphi _{t}\right) \right) }\varphi \left( \mathbf{x},t\right) e^{-iHt}\psi _{G}^{out}-e^{i\left( a\left( \varphi _{t}\right) +a^{\dagger }\left( \varphi _{t}\right) \right) }\varphi \left( \mathbf{x},t\right) e^{-iHt}\psi _{G}\left( t\right) \right\Vert + \)}\\
{\large }\\
{\large \( +\left\Vert e^{i\left( a\left( \varphi _{t}\right) +a^{\dagger }\left( \varphi _{t}\right) \right) }\varphi \left( \mathbf{x},t\right) e^{-iHt}\psi _{G}\left( t\right) \right\Vert \leq  \)}\\
{\large }\\
{\large \( \leq \left\Vert \varphi \left( \mathbf{x},t\right) e^{-iHt}\left( \psi _{G}^{out}-\psi _{G}\left( t\right) \right) \right\Vert +\left\Vert \varphi \left( \mathbf{x},t\right) e^{-iHt}\psi _{G}\left( t\right) \right\Vert  \)}\\
{\large }\\
{\large where \( \varphi \left( \mathbf{x},t\right) \equiv -ig\int \left( \widetilde{\varphi }\left( \mathbf{k}\right) e^{i\left| \mathbf{k}\right| t-i\mathbf{k}\cdot \mathbf{x}}+\overline{\widetilde{\varphi }}\left( \mathbf{k}\right) e^{-i\left| \mathbf{k}\right| t+i\mathbf{k}\cdot \mathbf{x}}\right) \frac{d^{3}k}{\sqrt{2\left| \mathbf{k}\right| }} \)}\\
{\large }\\
{\large }\\
{\large Both norms on the right hand side are bounded by quantities of order
\( \frac{1}{t^{1+\rho '}} \)~, \( \rho '>0 \):}\\
{\large \par}

\begin{itemize}
\item {\large the first one because \( \sup _{\mathbf{x}}\left| \varphi \left( \mathbf{x},t\right) \right| \leq \frac{M}{t} \)
and \( \left\Vert \left( \psi _{G}^{out}-\psi _{G}\left( t\right) \right) \right\Vert \leq \frac{C}{t^{\rho }} \)~~~~(\( M \)
and \( C \) are constants)}\\
{\large \par}
\item {\large as regards the second one, starting from the identity:}\\
{\large }\\
{\large \( \varphi \left( \mathbf{x},t\right) e^{-iHt}\psi _{G}\left( t\right) =\varphi \left( \mathbf{x},t\right) \sum _{i=1}^{N\left( t\right) }W_{\sigma _{t}}\left( \mathbf{v}_{i},t\right) e^{i\gamma _{\sigma _{t}}\left( \mathbf{v}_{i},\nabla E^{\sigma _{t}}\left( \mathbf{P}\right) ,t\right) }e^{-iE^{\sigma _{t}}t}\psi _{i,\sigma _{t}}^{\left( t\right) }= \)}\\
{\large }\\
{\large \( =\sum _{i=1}^{N\left( t\right) }W_{\sigma _{t}}\left( \mathbf{v}_{i},t\right) \varphi \left( \mathbf{x},t\right) e^{i\gamma _{\sigma _{t}}\left( \mathbf{v}_{i},\nabla E^{\sigma _{t}}\left( \mathbf{P}\right) ,t\right) }e^{-iE^{\sigma _{t}}t}\int _{\Gamma _{i}}G\left( \mathbf{P}\right) \psi _{\mathbf{P},\sigma _{t}}d^{3}P= \)}\\
{\large }\\
{\large \( =\sum _{i=1}^{N\left( t\right) }W_{\sigma _{t}}\left( \mathbf{v}_{i},t\right) \varphi \left( \mathbf{x},t\right) \left( 1_{\Gamma _{i}}\left( \mathbf{P}\right) -\chi _{\mathbf{v}_{i}}^{\left( t\right) }\left( \nabla E^{\sigma _{t}},t\right) \right) e^{i\gamma _{\sigma _{t}}\left( \mathbf{v}_{i},\nabla E^{\sigma _{t}}\left( \mathbf{P}\right) ,t\right) }e^{-iE^{\sigma _{t}}t}\int _{\Gamma _{i}}G\left( \mathbf{P}\right) \psi _{\mathbf{P},\sigma _{t}}d^{3}P+ \)}\\
\marginpar{
{\large (33.1)}{\large \par}
}{\large }\\
{\large }\\
{\large \( +\sum _{i=1}^{N\left( t\right) }W_{\sigma _{t}}\left( \mathbf{v}_{i},t\right) \varphi \left( \mathbf{x},t\right) \left( \chi _{\mathbf{v}_{i}}^{\left( t\right) }\left( \nabla E^{\sigma _{t}},t\right) -\chi ^{\left( t\right) }_{\mathbf{v}_{i}}\left( \frac{\mathbf{x}}{t},t\right) \right) e^{i\gamma _{\sigma _{t}}\left( \mathbf{v}_{i},\nabla E^{\sigma _{t}}\left( \mathbf{P}\right) ,t\right) }e^{-iE^{\sigma _{t}}t}\int _{\Gamma _{i}}G\left( \mathbf{P}\right) \psi _{\mathbf{P},\sigma _{t}}d^{3}P+ \)}{\small }\\
\marginpar{
{\large (33.2)}{\large \par}
}{\large }\\
{\large }\\
{\large \( +\sum _{i=1}^{N\left( t\right) }W_{\sigma _{t}}\left( \mathbf{v}_{i},t\right) \varphi \left( \mathbf{x},t\right) \chi ^{\left( t\right) }_{\mathbf{v}_{i}}\left( \frac{\mathbf{x}}{t},t\right) e^{i\gamma _{\sigma _{t}}\left( \mathbf{v}_{i},\nabla E^{\sigma _{t}}\left( \mathbf{P}\right) ,t\right) }e^{-iE^{\sigma _{t}}t}\int _{\Gamma _{i}}G\left( \mathbf{P}\right) \psi _{\mathbf{P},\sigma _{t}}d^{3}P \)}\\
\marginpar{
{\large (33.3)}{\large \par}
}{\large }\\
{\large }\\
{\large I exploit lemma B1 for 33.1, lemma B2 for 33.2, and Huygens} {\large principle
in order to estimate \( sup_{\mathbf{x}}\left| \varphi \left( \mathbf{x},t\right) \chi ^{\left( t\right) }_{\mathbf{v}_{i}}\left( \frac{\mathbf{x}}{t},t\right) \right|  \)
in the expression (33.3).}{\large }\\
{\large }\\
{\large \par}
\end{itemize}
{\large The scattering subspaces}{\large ~H}{\large \( ^{out\left( in\right) }\equiv \overline{\bigvee \psi _{G,\varphi }^{out(in)}} \)
are invariant under space-time translations because the subspaces~}{\large  H}{\large \( ^{1out\left( in\right) } \)
are invariant.}{\large }\\
{\large }\\
{\large }\\
\textbf{\large Theorem 5.2}{\large }\\
{\large }\\
{\large The nucleon mean velocity functions \( f \), which are continuous and
of compact support, have asymptotically strong limits}{\large  ~~H}{\large \( ^{out} \);
in particular: }\\
{\large }\\
{\large \( s-lim_{t\rightarrow +\infty }e^{iHt}f\left( \frac{\mathbf{x}}{t}\right) e^{-iHt}\psi _{G,\varphi }^{out}=\psi _{G\cdot \widehat{f},\varphi }^{out} \)}
{\large ~~~~~~where~~\( \widehat{f}\left( \mathbf{P}\right) \equiv lim_{\sigma \rightarrow 0}f\left( \nabla E^{\sigma }\left( \mathbf{P}\right) \right)  \)~
}\\
{\large }\\
{\large Proof}\\
{\large }\\
{\large It is sufficient to prove it on the vectors \( \psi _{G,\varphi }^{out} \)
for functions \( f\in C_{0}^{\infty }\left( R^{3}\right)  \). Exploiting theorem
5.1 and the uniform boundedness in \( t \) of the operators \( f\left( \frac{\mathbf{x}}{t}\right)  \)
and \( e^{iHt}e^{i\left( a\left( \varphi _{t}\right) +a^{\dagger }\left( \varphi _{t}\right) \right) }e^{-iHt} \)
we obtain}\\
{\large }\\
{\large \( s-lim_{t\rightarrow +\infty }e^{iHt}f\left( \frac{\mathbf{x}}{t}\right) e^{-iHt}\psi _{G,\varphi }^{out}=s-lim_{t\rightarrow +\infty }e^{iHt}f\left( \frac{\mathbf{x}}{t}\right) e^{i\left( a\left( \varphi _{t}\right) +a^{\dagger }\left( \varphi _{t}\right) \right) }e^{-iHt}\psi ^{out}_{G}= \)}\\
{\large \( =s-lim_{t\rightarrow +\infty }e^{iHt}e^{i\left( a\left( \varphi _{t}\right) +a^{\dagger }\left( \varphi _{t}\right) \right) }f\left( \frac{\mathbf{x}}{t}\right) e^{-iHt}\psi _{G}^{out}= \)}\\
{\large }\\
{\large \( =s-lim_{t\rightarrow +\infty }e^{iHt}e^{i\left( a\left( \varphi _{t}\right) +a^{\dagger }\left( \varphi _{t}\right) \right) }e^{-iHt}\psi _{G\cdot \widehat{f}}^{out} \)
}\\
{\large (the last step is proved by the technique used in lemma B2)}\\
{\large }\\
{\large The extension to all of}{\large  ~H}{\large \( ^{out} \) is automatic
since \( f\left( \frac{\mathbf{x}}{t}\right)  \) is uniformly bounded in \( t \)
and moreover the set \( \bigvee \psi _{G,\varphi }^{out} \) is a dense set
in ~}{\large  H}{\large \( ^{out} \), by construction of~}{\large  H}{\large \( ^{out} \).}
 {\large }\textbf{\large }\\
\textbf{\large }\\
\textbf{\large Corollary 5.3}{\large }\\
{\large }\\
{\large In the space}{\large ~H}{\large \( ^{out} \), the asymptotic meson
algebra \( \mathcal{A}^{out} \)} {\large is defined as the norm closure of
the {*}algebra generated by the set of unitary operators \( \left\{ W^{out}\left( \mu \right) :\widetilde{\mu }\in C_{0}^{\infty }\left( R^{3}\setminus 0\right) \right\}  \)
constructed in}{\large  ~H}{\large \( ^{out} \) as follows: }\\
{\large 
\[
W^{out}\left( \mu \right) =s-\lim _{t\rightarrow +\infty }e^{iHt}e^{i\left( a\left( \mu _{t}\right) +a^{\dagger }\left( \mu _{t}\right) \right) }e^{-iHt}\]
}\marginpar{
{\large (34)}{\large \par}
} {\large }\\
{\large }\\
{\large The following properties hold:}\\
{\large }\\
{\large 1) the generators} {\large \( \left\{ W^{out}\left( \mu \right) :\widetilde{\mu }\in C_{0}^{\infty }\left( R^{3}\setminus 0\right) \right\}  \)}
{\large satisfy the Weyl commutation rules}\\
{\large }\\
{\large \( W^{out}\left( \mu \right) W^{out}\left( \eta \right) =W^{out}\left( \eta \right) W^{out}\left( \mu \right) e^{-h\left( \mu ,\eta \right) } \)
~~~where~~~\( h\left( \mu ,\eta \right) =2iIm\int \widetilde{\mu }\left( \mathbf{k}\right) \overline{\widetilde{\eta }}\left( \mathbf{k}\right) d^{3}k \);}\\
{\large }\\
{\large 2) for each fixed region \( R^{3}\setminus O_{\mathbf{a}} \) where
\( O_{\mathbf{a}} \) is a ball of radius \( a\neq 0 \) , centered in the origin
of \( R^{3} \), the group of the operators \( W^{out}\left( \mu \right)  \),
where \( \widetilde{\mu }\in C_{0}^{\infty }\left( R^{3}\setminus O_{\mathbf{a}}\right)  \),
is strongly continuous with respect to \( \widetilde{\mu } \) in the \( L^{2}\left( R^{3}\setminus O_{\mathbf{a}},d^{3}k\right)  \)
norm;}\\
{\large }\\
{\large 3) given the \( \tau  \)-evolved generators:} \\
\\
{\large \( W_{\tau }^{out}\left( \mu \right) =s-\lim _{t\rightarrow +\infty }e^{iH\left( t+\tau \right) }e^{i\left( a\left( \mu _{t}\right) +a^{\dagger }\left( \mu _{t}\right) \right) }e^{-iH\left( t+\tau \right) }=W^{out}\left( \mu _{-\tau }\right)  \)}\marginpar{
{\large (35)}{\large \par}
} {\large }\\
{\large }\\
{\large it is uniquely defined an automorphism \( \alpha _{\tau } \) of \( \mathcal{A}^{out} \)
starting from \( \alpha _{\tau }\left( W^{out}\left( \mu \right) \right) = \)}\\
{\large \( =W^{out}\left( \mu _{-\tau }\right)  \). Therefore, being \( \mu _{-\tau } \)
the test function \( \mu  \) free evolved in the time \( \tau  \), we can
conclude that \( \mathcal{A}^{out} \) is the Weyl algebra associated to the
scalar massless field.}\\
{\large }\\
{\large 4) the algebra \( \mathcal{A}^{out} \) commutes with the asymptotic
nucleon mean velocity defined through theorem 5.2.}\\
{\large 
\[
\Leftrightarrow \]
}\\
{\large }\\
{\large Proof}\\
{\large }\\
{\large The existence of \( s-\lim _{t\rightarrow +\infty }e^{iHt}e^{i\left( a\left( \mu _{t}\right) +a^{\dagger }\left( \mu _{t}\right) \right) }e^{-iHt}\psi _{G,\varphi }^{out} \)
is substantially the contents of theorem 5.1. The bounded operators \( W^{out}\left( \mu \right)  \),
defined on the dense set \( \left\{ \bigvee \psi _{G,\varphi }^{out}\right\}  \)
of}{\large  ~H}{\large \( ^{out} \), can be linearly extended to all of}{\large  ~H}{\large \( ^{out} \)
by continuity. They leave the space}{\large  ~H}{\large \( ^{out} \) invariant
and they are unitary in}{\large ~ H}{\large \( ^{out} \).}\\
{\large }\\
{\large 1) On the vectors in}{\large  ~H}{\large \( ^{out} \), \( W^{out}\left( \mu \right) W^{out}\left( \eta \right)  \)
is the limit of the product of the approximating vectors (34), at the same time
\( t \). The last ones satisfy the property by construction. Therefore, the
property is satisfied in the limit.}\\
{\large }\\
{\large 2) It is sufficient to prove that \( W^{out}\left( \mu \right)  \)
is strongly continuous with respect to \( \widetilde{\mu } \) if it is applied
to the total set \( \left\{ \psi _{G,\varphi }^{out}\right\}  \):} {\large }\\
{\large - \( W^{out}\left( \mu \right) \psi _{G,\varphi }^{out}=s-\lim _{t\rightarrow +\infty }e^{iHt}e^{i\left( a\left( \mu _{t}\right) +a^{\dagger }\left( \mu _{t}\right) \right) }e^{-iHt}\psi _{G,\varphi }\left( t\right)  \)
;}\\
{\large - at fixed \( t \), the vector \( e^{i\left( a\left( \mu _{t}\right) +a^{\dagger }\left( \mu _{t}\right) \right) }e^{-iHt}\psi _{G,\varphi }\left( t\right)  \)
is strongly continuous with respect to \( \widetilde{\mu }\in C_{0}^{\infty }\left( R^{3}\setminus O_{\mathbf{a}}\right)  \)
. This is due to the fact that}\\
{\large }\\
{\large \( e^{i\left( a\left( \mu _{t}\right) +a^{\dagger }\left( \mu _{t}\right) \right) }\sum _{i=1}^{N\left( t\right) }W_{\sigma _{t}}\left( \mathbf{v}_{i},t\right) e^{i\gamma _{\sigma _{t}}\left( \mathbf{v}_{i},\nabla E^{\sigma _{t}}\left( \mathbf{P}\right) ,t\right) }e^{-iE^{\sigma _{t}}t}\psi _{i,\sigma _{t}}^{\left( t\right) }= \)
}\\
{\large }\\
{\large \( =\sum _{i=1}^{N\left( t\right) }W_{\sigma _{t}}\left( \mathbf{v}_{i},t\right) e^{i\left( a\left( \mu _{t}\right) +a^{\dagger }\left( \mu _{t}\right) \right) }e^{-h\left( \mu _{t},\xi _{\mathbf{v}_{i}}\right) }e^{i\gamma _{\sigma _{t}}\left( \mathbf{v}_{i},\nabla E^{\sigma _{t}}\left( \mathbf{P}\right) ,t\right) }e^{-iE^{\sigma _{t}}t}\psi _{i,\sigma _{t}}^{\left( t\right) } \)}\\
{\large }\\
{\large where \( \xi _{\mathbf{v}_{i}}\equiv -ig\frac{\chi _{\sigma _{t}}^{\kappa _{1}}}{\sqrt{2}\left| \mathbf{k}\right| ^{\frac{3}{2}}\left( 1-\widehat{\mathbf{k}}\cdot \mathbf{v}_{i}\right) } \)
and because \( \psi _{i,\sigma _{t}}^{\left( t\right) } \) is in the domain
of the generator of \( e^{i\left( a\left( \mu _{t}\right) +a^{\dagger }\left( \mu _{t}\right) \right) } \).
Being \( \widetilde{\mu }_{t}\in C_{0}^{\infty }\left( R^{3}\setminus O_{\mathbf{a}}\right)  \)
and \( \left\Vert \widetilde{\mu }_{t}\right\Vert _{L^{2}\left( R^{3}\setminus O_{\mathbf{a}},d^{3}k\right) }=\left\Vert \widetilde{\mu }\right\Vert _{L^{2}\left( R^{3}\setminus O_{\mathbf{a}},d^{3}k\right) } \),
the vector is strongly continuous (with respect to \( \widetilde{\mu } \))
uniformly in \( t \). }\\
{\large Since \( \left\Vert W^{out}\left( \mu \right) \right\Vert =1 \) , the
property holds for each vector in~}{\large  H}{\large \( ^{out} \). }\\
{\large }\\
{\large 3) The \( \tau  \)-evolved generators \( e^{iH\tau }W^{out}\left( \mu \right) e^{-iH\tau } \)
are well defined because \( e^{-iH\tau } \)}{\large :H}{\large \( ^{out} \)\( \rightarrow  \)}{\large H}{\large \( ^{out} \).
By inserting the expression (34) for \( W^{out}\left( \mu \right)  \), we arrive
at the (35).} {\large The Weyl commutation rules are conserved by \( \alpha _{\tau } \)
since}\\
{\large }\\
{\large 
\[
h\left( \mu _{-\tau },\eta _{-\tau }\right) =2iIm\int \widetilde{\mu }\left( \mathbf{k}\right) \overline{\widetilde{\eta }}\left( \mathbf{k}\right) d^{3}k=h\left( \mu ,\eta \right) \]
}\\
{\large }\\
{\large 4) Such a property is implicit in the construction of the asymptotic
nucleon mean velocity (theorem 5.2).}\\
{\large 
\[
\Leftarrow \Rightarrow \]
}\\
{\large }\\
{\large }\\
{\large }\\
{\large }\\
{\large \par}

\textbf{\Large APPENDIX A  }\\
{\Large \par}

\emph{\large Preliminary remarks to the lemma A1}{\large }\\
{\large }\\
{\large Like in lemma 1.4, it will be proved that the operator \( \left( \Delta H_{\mathbf{P}}^{w}\right) ^{\epsilon ^{\frac{j+1}{2}}}_{\epsilon ^{\frac{j+2}{2}}}\equiv  \)
}\\
{\large \( \equiv \widehat{H}^{w}_{\mathbf{P},\epsilon ^{\frac{j+2}{2}}}+c_{\mathbf{P}}\left( j+1\right) -\widehat{c}_{\mathbf{P}}\left( j+2\right) -H^{w}_{\mathbf{P},\epsilon ^{\frac{j+1}{2}}} \)
\( \psi _{G}^{out\left( in\right) } \) is small of order \( 1 \) with respect
to \( H^{w}_{\mathbf{P},\epsilon ^{\frac{j+1}{2}}}\mid _{F_{\epsilon ^{\frac{j+2}{2}}}^{+}} \)
in a generalized sense for \( g \) sufficiently small. We aim at expanding
the resolvent}\\
{\large }\\
{\large 
\[
\frac{1}{\widehat{H}^{w}_{\mathbf{P},\epsilon ^{\frac{j+2}{2}}}-\left( E\left( j+1\right) +\widehat{c}_{\mathbf{P}}\left( j+2\right) -c_{\mathbf{P}}\left( j+1\right) \right) }\mid _{F_{\epsilon ^{\frac{j+2}{2}}}^{+}}\]
} ( \( E\left( j+1\right) \in \mathcal{C} \) s.t. \( \left| E\left( j+1\right) -E_{\mathbf{P}}^{\epsilon ^{\frac{j+1}{2}}}\right| =\frac{11\epsilon ^{\frac{j+2}{2}}}{20} \),
\( \widehat{c}_{\mathbf{P}}\left( j+2\right) -c_{\mathbf{P}}\left( j+1\right) =-g^{2}\int _{\epsilon ^{\frac{j+2}{2}}}^{\epsilon ^{\frac{j+1}{2}}}\frac{1}{2\left| \mathbf{k}\right| ^{2}\left( 1-\widehat{\mathbf{k}}\cdot \nabla E^{\epsilon ^{\frac{j+1}{2}}}\left( \mathbf{P}\right) \right) }d^{3}k \)){\large }\\
{\large }\\
{\large in terms of \( \frac{1}{H^{w}_{\epsilon ^{\frac{j+1}{2}}}-E\left( j+1\right) }\mid _{F_{\epsilon ^{\frac{j+2}{2}}}^{+}} \)and
\( \left( \Delta H_{\mathbf{P}}^{w}\right) ^{\epsilon ^{\frac{j+1}{2}}}_{\epsilon ^{\frac{j+2}{2}}} \).}\\
{\large }\\
{\large I will only treat the case \( \mathbf{P}=0 \), where the difference
operator is written~~~\( \left( \Delta H^{w}\right) ^{\epsilon ^{\frac{j+1}{2}}}_{\epsilon ^{\frac{j+2}{2}}}\equiv  \)}\\
{\large \( \equiv H^{w}_{\epsilon ^{\frac{j+2}{2}}}+c\left( j+1\right) -c\left( j+2\right) -H^{w}_{\epsilon ^{\frac{j+1}{2}}} \)
.}\\
{\large In the case \( \mathbf{P}\neq 0 \) the steps are analogous and well
defined since the module of \( \nabla E^{\sigma }\left( \mathbf{P}\right)  \)
is uniformly bounded in \( \sigma  \) by a positive constant less than 1, for
\( \mathbf{P}\in \Sigma  \) (see lemma A2). }\\
{\large }\\
{\large \par}

\textbf{\large Lemma A1}{\large }\\
{\large }\\
{\large Given the spectral properties in the beginning of paragraph 2.1, \( \left( \Delta H^{w}\right) ^{\epsilon ^{\frac{j+1}{2}}}_{\epsilon ^{\frac{j+2}{2}}}\mid _{F_{\epsilon ^{\frac{j+2}{2}}}^{+}} \)is
small with respect to \( H^{w}_{\epsilon ^{\frac{j+1}{2}}} \) for values of
\( g \) less than a proper \( \overline{g} \) and \( \mathbf{P}\in \Sigma  \),
in the following sense: }\\
{\large }\\
{\large }\\
{\large given \( E\left( j+1\right) \in C \) s.t.\( \left| E\left( j+1\right) -E^{\epsilon ^{\frac{j+1}{2}}}\right| =\frac{11\epsilon ^{\frac{j+2}{2}}}{20} \),}\\
{\large }\\
{\large }\\
{\large \( \frac{1}{H^{w}_{\epsilon ^{\frac{j+2}{2}}}-\left( E\left( j+1\right) +c\left( j+2\right) -c\left( j+1\right) \right) }\mid _{F_{\epsilon ^{\frac{j+2}{2}}}^{+}}= \)}{\small }\\
{\small }\\
{\large \( =\frac{1}{H^{w}_{\epsilon ^{\frac{j+1}{2}}}+\left( \Delta H^{w}\right) ^{\epsilon ^{\frac{j+1}{2}}}_{\epsilon ^{\frac{j+2}{2}}}-c\left( j+1\right) +c\left( j+2\right) -\left( E\left( j+1\right) -c\left( j+1\right) +c\left( j+2\right) \right) }\mid _{F_{\epsilon ^{\frac{j+2}{2}}}^{+}}= \)}{\small }\\
{\small }\\
{\large \( =\frac{1}{H^{w}_{\epsilon ^{\frac{j+1}{2}}}+\left( \Delta H^{w}\right) ^{\epsilon ^{\frac{j+1}{2}}}_{\epsilon ^{\frac{j+2}{2}}}-E\left( j+1\right) }\mid _{F_{\epsilon ^{\frac{j+2}{2}}}^{+}}= \)}{\small }\\
{\large }\\
{\large \( =\frac{1}{H^{w}_{\epsilon ^{\frac{j+1}{2}}}-E\left( j+1\right) }\mid _{F_{\epsilon ^{\frac{j+2}{2}}}^{+}}+\frac{1}{H^{w}_{\epsilon ^{\frac{j+1}{2}}}-E\left( j+1\right) }\sum ^{+\infty }_{n=1}\left( -\left( \Delta H^{w}\right) ^{\epsilon ^{\frac{j+1}{2}}}_{\epsilon ^{\frac{j+2}{2}}}\frac{1}{H^{w}_{\epsilon ^{\frac{j+1}{2}}}-E\left( j+1\right) }\right) ^{n}\mid _{F_{\epsilon ^{\frac{j+2}{2}}}^{+}} \)
}\\
{\large }\\
{\large where}\\
{\large \par}

\begin{itemize}
\item {\large \( \left\Vert \frac{1}{H^{w}_{\epsilon ^{\frac{j+1}{2}}}-E\left( j+1\right) }\left( -\left( \Delta H^{w}\right) ^{\epsilon ^{\frac{j+1}{2}}}_{\epsilon ^{\frac{j+2}{2}}}\frac{1}{H^{w}_{\epsilon ^{\frac{j+1}{2}}}-E\left( j+1\right) }\right) ^{n}\right\Vert _{F_{\epsilon ^{\frac{j+2}{2}}}^{+}}\leq \frac{20\left( C\left( \overline{g},\overline{m}\right) \right) ^{n}}{\epsilon ^{\frac{j+2}{2}}} \)
 }{\large \par}
\item {\large \( 0<C\left( \overline{g},m\right) <\frac{1}{12} \) .}\\
{\large }\\
 {\large }{\large \par}
\end{itemize}
{\large Proof}\\
{\large }\\
{\large Like in lemma 1.4 we can state}\\
{\large }\\
{\large \( \frac{1}{H^{w}_{\epsilon ^{\frac{j+1}{2}}}-E\left( j+1\right) }\left( -\left( \Delta H^{w}\right) ^{\epsilon ^{\frac{j+1}{2}}}_{\epsilon ^{\frac{j+2}{2}}}\frac{1}{H^{w}_{\epsilon ^{\frac{j+1}{2}}}-E\left( j+1\right) }\right) ^{n}= \)}\\
{\large }\\
{\large \( =\left( -1\right) ^{n}\frac{1}{H^{w}_{\epsilon ^{\frac{j+1}{2}}}-E\left( j+1\right) }\left( \Delta H^{w}\right) ^{\epsilon ^{\frac{j+1}{2}}}_{\epsilon ^{\frac{j+2}{2}}}\frac{1}{H^{w}_{\epsilon ^{\frac{j+1}{2}}}-E\left( j+1\right) }.......................\left( \Delta H^{w}\right) ^{\epsilon ^{\frac{j+1}{2}}}_{\epsilon ^{\frac{j+2}{2}}}\frac{1}{H^{w}_{\epsilon ^{\frac{j+1}{2}}}-E\left( j+1\right) }= \)}\\
{\large }\\
{\large \( =\left( -1\right) ^{n}\left( \frac{1}{H^{w}_{\epsilon ^{\frac{j+1}{2}}}-E\left( j+1\right) }\right) ^{\frac{1}{2}}....\left( \frac{1}{H^{w}_{\epsilon ^{\frac{j+1}{2}}}-E\left( j+1\right) }\right) ^{\frac{1}{2}}\left( \Delta H^{w}\right) ^{\epsilon ^{\frac{j+1}{2}}}_{\epsilon ^{\frac{j+2}{2}}}\left( \frac{1}{H^{w}_{\epsilon ^{\frac{j+1}{2}}}-E\left( j+1\right) }\right) ^{\frac{1}{2}}....\left( \frac{1}{H^{w}_{\epsilon ^{\frac{j+1}{2}}}-E\left( j+1\right) }\right) ^{\frac{1}{2}} \)}\textbf{\emph{\large }}\\
\textbf{\emph{\large }}\\
\textbf{\emph{\large }}\\
\textbf{\emph{\large Study of the norm of}} {\large ~~\( \left( \frac{1}{H^{w}_{\epsilon ^{\frac{j+1}{2}}}-E\left( j+1\right) }\right) ^{\frac{1}{2}}\left( \Delta H^{w}\right) ^{\epsilon ^{\frac{j+1}{2}}}_{\epsilon ^{\frac{j+2}{2}}}\left( \frac{1}{H^{w}_{\epsilon ^{\frac{j+1}{2}}}-E\left( j+1\right) }\right) ^{\frac{1}{2}}\mid _{F_{\epsilon ^{\frac{j+2}{2}}}^{+}} \).}\\
{\large }\\
{\large }\\
{\large \( \left( \Delta H^{w}\right) ^{\epsilon ^{\frac{j+1}{2}}}_{\epsilon ^{\frac{j+2}{2}}}=-\frac{g}{2m}\int ^{\epsilon ^{\frac{j+1}{2}}}_{\epsilon ^{\frac{j+2}{2}}}\frac{\mathbf{k}}{\sqrt{2}\left| \mathbf{k}\right| ^{\frac{3}{2}}}\left( b\left( \mathbf{k}\right) +b^{\dagger }\left( \mathbf{k}\right) \right) d^{3}k\cdot \Pi _{\epsilon ^{\frac{j+1}{2}}}+ \)}\\
{\large }\\
{\large \( -\Pi _{\epsilon ^{\frac{j+1}{2}}}\cdot \frac{g}{2m}\int ^{\epsilon ^{\frac{j+1}{2}}}_{\epsilon ^{\frac{j+2}{2}}}\frac{\mathbf{k}}{\sqrt{2}\left| \mathbf{k}\right| ^{\frac{3}{2}}}\left( b\left( \mathbf{k}\right) +b^{\dagger }\left( \mathbf{k}\right) \right) d^{3}k+\frac{g^{2}}{2m}\left( \int ^{\epsilon ^{\frac{j+1}{2}}}_{\epsilon ^{\frac{j+2}{2}}}\frac{\mathbf{k}}{\sqrt{2}\left| \mathbf{k}\right| ^{\frac{3}{2}}}\left( b\left( \mathbf{k}\right) +b^{\dagger }\left( \mathbf{k}\right) \right) d^{3}k\right) ^{2} \)}\\
{\large }\\
 {\large }{\large \par}

{\large Making the calculations we have:}\\
{\large }\\
{\large \( \frac{g^{2}}{2m}\left( \int ^{\epsilon ^{\frac{j+1}{2}}}_{\epsilon ^{\frac{j+2}{2}}}\frac{\mathbf{k}}{\sqrt{2}\left| \mathbf{k}\right| ^{\frac{3}{2}}}\left( b\left( \mathbf{k}\right) +b^{\dagger }\left( \mathbf{k}\right) \right) d^{3}k\right) ^{2}=\frac{g^{2}}{2m}\sum _{i}\int ^{\epsilon ^{\frac{j+1}{2}}}_{\epsilon ^{\frac{j+2}{2}}}k^{i}b\left( \mathbf{k}\right) \frac{d^{3}k}{\left| \mathbf{k}\right| \sqrt{2\left| \mathbf{k}\right| }}\int ^{\epsilon ^{\frac{j+1}{2}}}_{\epsilon ^{\frac{j+2}{2}}}k^{i}b\left( \mathbf{k}\right) \frac{d^{3}k}{\left| \mathbf{k}\right| \sqrt{2\left| \mathbf{k}\right| }}+ \)}\\
{\large }\\
{\large \( +\frac{g^{2}}{2m}\sum _{i}\int ^{\epsilon ^{\frac{j+1}{2}}}_{\epsilon ^{\frac{j+2}{2}}}k^{i}b^{\dagger }\left( \mathbf{k}\right) \frac{d^{3}\mathbf{k}}{\left| \mathbf{k}\right| \sqrt{2\left| \mathbf{k}\right| }}\int ^{\epsilon ^{\frac{j+1}{2}}}_{\epsilon ^{\frac{j+2}{2}}}k^{i}b^{\dagger }\left( \mathbf{k}\right) \frac{d^{3}\mathbf{k}}{\left| \mathbf{k}\right| \sqrt{2\left| \mathbf{k}\right| }}+ \)}\\
{\large }\\
{\large \( +\frac{g^{2}}{m}\sum _{i}\int ^{\epsilon ^{\frac{j+1}{2}}}_{\epsilon ^{\frac{j+2}{2}}}k^{i}b^{\dagger }\left( \mathbf{k}\right) \frac{d^{3}k}{\left| \mathbf{k}\right| \sqrt{2\left| \mathbf{k}\right| }}\int ^{\epsilon ^{\frac{j+1}{2}}}_{\epsilon ^{\frac{j+2}{2}}}k^{i}b\left( \mathbf{k}\right) \frac{d^{3}k}{\left| \mathbf{k}\right| \sqrt{2\left| \mathbf{k}\right| }}+\frac{g^{2}}{4m}\int ^{\epsilon ^{\frac{j+1}{2}}}_{\epsilon ^{\frac{j+2}{2}}}\frac{d^{3}k}{\left| \mathbf{k}\right| } \)}\\
{\large }\\
{\large }\\
 {\large }{\large \par}

{\large We have to examine:}\\
{\large }\\
\textbf{\large 1)} {\large \( \frac{g^{2}}{2m}\left\Vert \left( \frac{1}{H^{w}_{\epsilon ^{\frac{j+1}{2}}}-E\left( j+1\right) }\right) ^{\frac{1}{2}}\int ^{\epsilon ^{\frac{j+1}{2}}}_{\epsilon ^{\frac{j+2}{2}}}k^{i}b^{\dagger }\left( \mathbf{k}\right) \frac{d^{3}\mathbf{k}}{\left| \mathbf{k}\right| \sqrt{2\left| \mathbf{k}\right| }}\int ^{\epsilon ^{\frac{j+1}{2}}}_{\epsilon ^{\frac{j+2}{2}}}k^{i}b\left( \mathbf{k}\right) \frac{d^{3}\mathbf{k}}{\left| \mathbf{k}\right| \sqrt{2\left| \mathbf{k}\right| }}\left( \frac{1}{H^{w}_{\epsilon ^{\frac{j+1}{2}}}-E\left( j+1\right) }\right) ^{\frac{1}{2}}\right\Vert _{F_{\epsilon ^{\frac{j+2}{2}}}^{+}} \)}\\
{\large }\\
\textbf{\large 2)} {\large \( \frac{g^{2}}{2m}\left\Vert \left( \frac{1}{H^{w}_{\epsilon ^{\frac{j+1}{2}}}-E\left( j+1\right) }\right) ^{\frac{1}{2}}\int ^{\epsilon ^{\frac{j+1}{2}}}_{\epsilon ^{\frac{j+2}{2}}}k^{i}b\left( \mathbf{k}\right) \frac{d^{3}\mathbf{k}}{\left| \mathbf{k}\right| \sqrt{2\left| \mathbf{k}\right| }}\int ^{\epsilon ^{\frac{j+1}{2}}}_{\epsilon ^{\frac{j+2}{2}}}k^{i}b\left( \mathbf{k}\right) \frac{d^{3}\mathbf{k}}{\left| \mathbf{k}\right| \sqrt{2\left| \mathbf{k}\right| }}\left( \frac{1}{H^{w}_{\epsilon ^{\frac{j+1}{2}}}-E\left( j+1\right) }\right) ^{\frac{1}{2}}\right\Vert _{F_{\epsilon ^{\frac{j+2}{2}}}^{+}} \)}\\
{\large }\\
\textbf{\large 3)} {\large \( \frac{g^{2}}{m}\left\Vert \left( \frac{1}{H^{w}_{\epsilon ^{\frac{j+1}{2}}}-E\left( j+1\right) }\right) ^{\frac{1}{2}}k^{i}b^{\dagger }\left( \mathbf{k}\right) \frac{d^{3}\mathbf{k}}{\left| \mathbf{k}\right| \sqrt{2\left| \mathbf{k}\right| }}\int ^{\epsilon ^{\frac{j+1}{2}}}_{\epsilon ^{\frac{j+2}{2}}}k^{i}b^{\dagger }\left( \mathbf{k}\right) \frac{d^{3}\mathbf{k}}{\left| \mathbf{k}\right| \sqrt{2\left| \mathbf{k}\right| }}\left( \frac{1}{H^{w}_{\epsilon ^{\frac{j+1}{2}}}-E\left( j+1\right) }\right) ^{\frac{1}{2}}\right\Vert _{F_{\epsilon ^{\frac{j+2}{2}}}^{+}} \)}\\
{\large }\\
\textbf{\large 4)} {\large \( \frac{g^{2}}{4m}\left\Vert \left( \frac{1}{H^{w}_{\epsilon ^{\frac{j+1}{2}}}-E\left( j+1\right) }\right) ^{\frac{1}{2}}\int ^{\epsilon ^{\frac{j+1}{2}}}_{\epsilon ^{\frac{j+2}{2}}}\frac{d^{3}\mathbf{k}}{\left| \mathbf{k}\right| }\left( \frac{1}{H^{w}_{\epsilon ^{\frac{j+1}{2}}}-E\left( j+1\right) }\right) ^{\frac{1}{2}}\right\Vert _{F_{\epsilon ^{\frac{j+2}{2}}}^{+}} \)}\\
{\large }\\
\textbf{\large 5)} {\large \( \frac{g}{2m}\left\Vert \left( \frac{1}{H^{w}_{\epsilon ^{\frac{j+1}{2}}}-E\left( j+1\right) }\right) ^{\frac{1}{2}}\Pi ^{i}_{\epsilon ^{\frac{j+1}{2}}}\cdot \int ^{\epsilon ^{\frac{j+1}{2}}}_{\epsilon ^{\frac{j+2}{2}}}k^{i}\left( b\left( \mathbf{k}\right) +b^{\dagger }\left( \mathbf{k}\right) \right) \frac{d^{3}\mathbf{k}}{\left| \mathbf{k}\right| \sqrt{2\left| \mathbf{k}\right| }}\left( \frac{1}{H^{w}_{\epsilon ^{\frac{j+1}{2}}}-E\left( j+1\right) }\right) ^{\frac{1}{2}}\right\Vert _{F_{\epsilon ^{\frac{j+2}{2}}}^{+}} \)}\\
{\large }\\
\textbf{\large 6)} {\large \( \frac{g}{2m}\left\Vert \left( \frac{1}{H^{w}_{\epsilon ^{\frac{j+1}{2}}}-E\left( j+1\right) }\right) ^{\frac{1}{2}}\int ^{\epsilon ^{\frac{j+1}{2}}}_{\epsilon ^{\frac{j+2}{2}}}k^{i}\left( b\left( \mathbf{k}\right) +b^{\dagger }\left( \mathbf{k}\right) \right) \frac{d^{3}\mathbf{k}}{\left| \mathbf{k}\right| \sqrt{2\left| \mathbf{k}\right| }}\cdot \Pi ^{i}_{\epsilon ^{\frac{j+1}{2}}}\left( \frac{1}{H^{w}_{\epsilon ^{\frac{j+1}{2}}}-E\left( j+1\right) }\right) ^{\frac{1}{2}}\right\Vert _{F_{\epsilon ^{\frac{j+2}{2}}}^{+}} \)}\\
{\large }\\
{\large }\\
{\large In order to control the above quantities, I will use the following estimate
again and again }\\
{\large }\\
{\large \( \left\Vert \int ^{\epsilon ^{\frac{j+1}{2}}}_{\epsilon ^{\frac{j+2}{2}}}k^{i}b\left( \mathbf{k}\right) \frac{d^{3}\mathbf{k}}{\left| \mathbf{k}\right| \sqrt{2\left| \mathbf{k}\right| }}\left( \frac{1}{H^{w}_{\epsilon ^{\frac{j+1}{2}}}-E\left( j+1\right) }\right) ^{\frac{1}{2}}\right\Vert _{F_{\epsilon ^{\frac{j+2}{2}}}^{+}}\leq \sqrt{40\pi }\cdot \epsilon ^{\frac{j+1}{4}} \)}\marginpar{
{\large (a1)}{\large \par}
}{\large }\\
{\large }\\
{\large which is proved like the estimate (4) of lemma 1.4., by performing an
unitary transformation.}\\
{\large }\\
{\large }\\
{\large }\\
{\large I study the quadratic quantities (1,2,3,4) and then the mixed terms
(5,6) which contain the \( \Pi _{\epsilon ^{\frac{j+1}{2}}} \). The following
estimates are worked out:}\\
{\large }\\
{\large }\\
\textbf{\large 1)}{\large }\\
{\large }\\
{\large \( \left\Vert \left( \frac{1}{H^{w}_{\epsilon ^{\frac{j+1}{2}}}-E\left( j+1\right) }\right) ^{\frac{1}{2}}\int ^{\epsilon ^{\frac{j+1}{2}}}_{\epsilon ^{\frac{j+2}{2}}}k^{i}b^{\dagger }\left( \mathbf{k}\right) \frac{d^{3}k}{\left| \mathbf{k}\right| \sqrt{2\left| \mathbf{k}\right| }}\int ^{\epsilon ^{\frac{j+1}{2}}}_{\epsilon ^{\frac{j+2}{2}}}k^{i}b\left( \mathbf{k}\right) \frac{d^{3}k}{\left| \mathbf{k}\right| \sqrt{2\left| \mathbf{k}\right| }}\left( \frac{1}{H^{w}_{\epsilon ^{\frac{j+1}{2}}}-E\left( j+1\right) }\right) ^{\frac{1}{2}}\right\Vert _{F_{\epsilon ^{\frac{j+2}{2}}}^{+}}\leq  \)}
\\
\\
{\large \( \leq \left\Vert \left( \frac{1}{H^{w}_{\epsilon ^{\frac{j+1}{2}}}-E\left( j+1\right) }\right) ^{\frac{1}{2}}\int ^{\epsilon ^{\frac{j+1}{2}}}_{\epsilon ^{\frac{j+2}{2}}}k^{i}b^{\dagger }\left( \mathbf{k}\right) \frac{d^{3}k}{\left| \mathbf{k}\right| \sqrt{2\left| \mathbf{k}\right| }}\right\Vert _{F_{\epsilon ^{\frac{j+2}{2}}}^{+}}\cdot \left\Vert \int ^{\epsilon ^{\frac{j+1}{2}}}_{\epsilon ^{\frac{j+2}{2}}}k^{i}b\left( \mathbf{k}\right) \frac{d^{3}k}{\left| \mathbf{k}\right| \sqrt{2\left| \mathbf{k}\right| }}\left( \frac{1}{H^{w}_{\epsilon ^{\frac{j+1}{2}}}-E\left( j+1\right) }\right) ^{\frac{1}{2}}\right\Vert _{F_{\epsilon ^{\frac{j+2}{2}}}^{+}} \)}\\
{\large }\\
{\large if the norms on the right hand side exist.}\\
{\large }\\
{\large Note that}\\
{\large }\\
{\large \( \left\Vert \left( \frac{1}{H^{w}_{\epsilon ^{\frac{j+1}{2}}}-E\left( j+1\right) }\right) ^{\frac{1}{2}}\int ^{\epsilon ^{\frac{j+1}{2}}}_{\epsilon ^{\frac{j+2}{2}}}k^{i}b^{\dagger }\left( \mathbf{k}\right) \frac{d^{3}k}{\left| \mathbf{k}\right| \sqrt{2\left| \mathbf{k}\right| }}\right\Vert _{F_{\epsilon ^{\frac{j+2}{2}}}^{+}}=\left\Vert \int ^{\epsilon ^{\frac{j+1}{2}}}_{\epsilon ^{\frac{j+2}{2}}}k^{i}b\left( \mathbf{k}\right) \frac{d^{3}k}{\left| \mathbf{k}\right| \sqrt{2\left| \mathbf{k}\right| }}\left[ \left( \frac{1}{H^{w}_{\epsilon ^{\frac{j+1}{2}}}-E\left( j+1\right) }\right) ^{\frac{1}{2}}\right] ^{\dagger }\right\Vert _{F_{\epsilon ^{\frac{j+2}{2}}}^{+}} \)}\\
{\large }\\
{\large the norm on the right hand side is controlled like}\\
{\large }\\
{\large \( \left\Vert \int ^{\epsilon ^{\frac{j+1}{2}}}_{\epsilon ^{\frac{j+2}{2}}}k^{i}b\left( \mathbf{k}\right) \frac{d^{3}k}{\left| \mathbf{k}\right| \sqrt{2\left| \mathbf{k}\right| }}\left( \frac{1}{H^{w}_{\epsilon ^{\frac{j+1}{2}}}-E\left( j+1\right) }\right) ^{\frac{1}{2}}\right\Vert _{F_{\epsilon ^{\frac{j+2}{2}}}^{+}} \).}\\
{\large }\\
{\large }\\
{\large In conclusion}\\
{\large }\\
{\large \( \frac{g^{2}}{2m}\left\Vert \left( \frac{1}{H^{w}_{\epsilon ^{\frac{j+1}{2}}}-E\left( j+1\right) }\right) ^{\frac{1}{2}}\int ^{\epsilon ^{\frac{j+1}{2}}}_{\epsilon ^{\frac{j+2}{2}}}k^{i}b^{\dagger }\left( \mathbf{k}\right) \frac{d^{3}k}{\left| \mathbf{k}\right| \sqrt{2\left| \mathbf{k}\right| }}\int ^{\epsilon ^{\frac{j+1}{2}}}_{\epsilon ^{\frac{j+2}{2}}}k^{i}b\left( \mathbf{k}\right) \frac{d^{3}k}{\left| \mathbf{k}\right| \sqrt{2\left| \mathbf{k}\right| }}\left( \frac{1}{H^{w}_{\epsilon ^{\frac{j+1}{2}}}-E\left( j+1\right) }\right) ^{\frac{1}{2}}\right\Vert _{F_{\epsilon ^{\frac{j+2}{2}}}^{+}}\leq  \)}\\
{\large }\\
{\large \( \leq C^{i}_{1}\left( g,m\right) \cdot \epsilon ^{\frac{j+1}{2}} \)}\\
{\large }\\
{\large }\\
\textbf{\large 2)}{\large }\\
{\large }\\
{\large \( \left\Vert \left( \frac{1}{H^{w}_{\epsilon ^{\frac{j+1}{2}}}-E\left( j+1\right) }\right) ^{\frac{1}{2}}\int ^{\epsilon ^{\frac{j+1}{2}}}_{\epsilon ^{\frac{j+2}{2}}}k^{i}b\left( \mathbf{k}\right) \frac{d^{3}k}{\left| \mathbf{k}\right| \sqrt{2\left| \mathbf{k}\right| }}\int ^{\epsilon ^{\frac{j+1}{2}}}_{\epsilon ^{\frac{j+2}{2}}}k^{i}b\left( \mathbf{k}\right) \frac{d^{3}k}{\left| \mathbf{k}\right| \sqrt{2\left| \mathbf{k}\right| }}\left( \frac{1}{H^{w}_{\epsilon ^{\frac{j+1}{2}}}-E\left( j+1\right) }\right) ^{\frac{1}{2}}\right\Vert _{F_{\epsilon ^{\frac{j+2}{2}}}^{+}}\leq  \)}\\
{\large }\\
{\large \( \leq \left\Vert \left( \frac{1}{H^{w}_{\epsilon ^{\frac{j+1}{2}}}-E\left( j+1\right) }\right) ^{\frac{1}{2}}\int ^{\epsilon ^{\frac{j+1}{2}}}_{\epsilon ^{\frac{j+2}{2}}}k^{i}b\left( \mathbf{k}\right) \frac{d^{3}k}{\left| \mathbf{k}\right| \sqrt{2\left| \mathbf{k}\right| }}\right\Vert _{F_{\epsilon ^{\frac{j+2}{2}}}^{+}}\cdot \left\Vert \int ^{\epsilon ^{\frac{j+1}{2}}}_{\epsilon ^{\frac{j+2}{2}}}k^{i}b\left( \mathbf{k}\right) \frac{d^{3}k}{\left| \mathbf{k}\right| \sqrt{2\left| \mathbf{k}\right| }}\left( \frac{1}{H^{w}_{\epsilon ^{\frac{j+1}{2}}}-E\left( j+1\right) }\right) ^{\frac{1}{2}}\right\Vert _{F_{\epsilon ^{\frac{j+2}{2}}}^{+}} \)}\\
{\large }\\
{\large if the norms on the right hand side exist.} {\large }\\
{\large I evaluate the norm of \( \int ^{\epsilon ^{\frac{j+1}{2}}}_{\epsilon ^{\frac{j+2}{2}}}k^{i}b^{\dagger }\left( \mathbf{k}\right) \frac{d^{3}k}{\left| \mathbf{k}\right| \sqrt{2\left| \mathbf{k}\right| }}\left[ \left( \frac{1}{H^{w}_{\epsilon ^{\frac{j+1}{2}}}-E\left( j+1\right) }\right) ^{\frac{1}{2}}\right] ^{\dagger } \)
restricted to \( F_{\epsilon ^{\frac{j+2}{2}}}^{+} \).}\\
{\large }\\
{\large Given \( \varphi \in D^{b}\bigcap F_{\epsilon ^{\frac{j+2}{2}}}^{+} \):}
{\large }\\
{\large }\\
{\large \( \left\Vert \int ^{\epsilon ^{\frac{j+1}{2}}}_{\epsilon ^{\frac{j+2}{2}}}k^{i}b^{\dagger }\left( \mathbf{k}\right) \frac{d^{3}k}{\left| \mathbf{k}\right| \sqrt{2\left| \mathbf{k}\right| }}\left[ \left( \frac{1}{H^{w}_{\epsilon ^{\frac{j+1}{2}}}-E\left( j+1\right) }\right) ^{\frac{1}{2}}\right] ^{\dagger }\varphi \right\Vert ^{2}= \)
}\\
{\large }\\
{\large \( =\left( \int ^{\epsilon ^{\frac{j+1}{2}}}_{\epsilon ^{\frac{j+2}{2}}}k^{i}b^{\dagger }\left( \mathbf{k}\right) \frac{d^{3}k}{\left| \mathbf{k}\right| \sqrt{2\left| \mathbf{k}\right| }}\left[ \left( \frac{1}{H^{w}_{\epsilon ^{\frac{j+1}{2}}}-E\left( j+1\right) }\right) ^{\frac{1}{2}}\right] ^{\dagger }\varphi \: ,\: \int ^{\epsilon ^{\frac{j+1}{2}}}_{\epsilon ^{\frac{j+2}{2}}}k^{i}b^{\dagger }\left( \mathbf{k}\right) \frac{d^{3}k}{\left| \mathbf{k}\right| \sqrt{2\left| \mathbf{k}\right| }}\left[ \left( \frac{1}{H^{w}_{\epsilon ^{\frac{j+1}{2}}}-E\left( j+1\right) }\right) ^{\frac{1}{2}}\right] ^{\dagger }\varphi \right) = \)}\\
{\large }\\
{\large \( \leq \int ^{\epsilon ^{\frac{j+1}{2}}}_{\epsilon ^{\frac{j+2}{2}}}\frac{\left( k^{i}\right) ^{2}d^{3}k}{2\left| \mathbf{k}\right| ^{3}}\left( \left[ \left( \frac{1}{H^{w}_{\epsilon ^{\frac{j+1}{2}}}-E\left( j+1\right) }\right) ^{\frac{1}{2}}\right] ^{\dagger }\varphi \: ,\: \left[ \left( \frac{1}{H^{w}_{\epsilon ^{\frac{j+1}{2}}}-E\left( j+1\right) }\right) ^{\frac{1}{2}}\right] ^{\dagger }\varphi \right) + \)}\\
{\large }\\
{\large \( +\left( \int ^{\epsilon ^{\frac{j+1}{2}}}_{\epsilon ^{\frac{j+2}{2}}}k^{i}b\left( \mathbf{k}\right) \frac{d^{3}k}{\left| \mathbf{k}\right| \sqrt{2\left| \mathbf{k}\right| }}\left[ \left( \frac{1}{H^{w}_{\epsilon ^{\frac{j+1}{2}}}-E\left( j+1\right) }\right) ^{\frac{1}{2}}\right] ^{\dagger }\varphi \: ,\: \int ^{\epsilon ^{\frac{j+1}{2}}}_{\epsilon ^{\frac{j+2}{2}}}k^{i}b\left( \mathbf{k}\right) \frac{d^{3}k}{\left| \mathbf{k}\right| \sqrt{2\left| \mathbf{k}\right| }}\left[ \left( \frac{1}{H^{w}_{\epsilon ^{\frac{j+1}{2}}}-E\left( j+1\right) }\right) ^{\frac{1}{2}}\right] ^{\dagger }\varphi \right) \leq  \)}\\
{\large }\\
{\large \( \leq \pi \cdot \epsilon ^{j+1}\cdot \frac{20}{\epsilon ^{\frac{j+2}{2}}}\cdot \left\Vert \varphi \right\Vert ^{2}+40\pi \epsilon ^{\frac{j+1}{2}}\cdot \left\Vert \varphi \right\Vert ^{2} \)}\\
{\large }\\
{\large Then}\\
{\large }\\
{\large \( \frac{g^{2}}{2m}\left\Vert \left( \frac{1}{H^{w}_{\epsilon ^{\frac{j+1}{2}}}-E\left( j+1\right) }\right) ^{\frac{1}{2}}\int ^{\epsilon ^{\frac{j+1}{2}}}_{\epsilon ^{\frac{j+2}{2}}}k^{i}b\left( \mathbf{k}\right) \frac{d^{3}\mathbf{k}}{\left| \mathbf{k}\right| \sqrt{2\left| \mathbf{k}\right| }}\int ^{\epsilon ^{\frac{j+1}{2}}}_{\epsilon ^{\frac{j+2}{2}}}k^{i}b\left( \mathbf{k}\right) \frac{d^{3}\mathbf{k}}{\left| \mathbf{k}\right| \sqrt{2\left| \mathbf{k}\right| }}\left( \frac{1}{H^{w}_{\epsilon ^{\frac{j+1}{2}}}-E\left( j+1\right) }\right) ^{\frac{1}{2}}\right\Vert _{F_{\epsilon ^{\frac{j+2}{2}}}^{+}}\leq  \)}\\
{\large }\\
{\large \( \leq C^{i}_{2}\left( g,m\right) \cdot \epsilon ^{\frac{j}{2}} \)}\\
{\large }\\
\textbf{\large 3)} {\large }\\
{\large }\\
{\large \( \left\Vert \left( \frac{1}{H^{w}_{\epsilon ^{\frac{j+1}{2}}}-E\left( j+1\right) }\right) ^{\frac{1}{2}}\int ^{\epsilon ^{\frac{j+1}{2}}}_{\epsilon ^{\frac{j+2}{2}}}k^{i}b^{\dagger }\left( \mathbf{k}\right) \frac{d^{3}\mathbf{k}}{\left| \mathbf{k}\right| \sqrt{2\left| \mathbf{k}\right| }}\int ^{\epsilon ^{\frac{j+1}{2}}}_{\epsilon ^{\frac{j+2}{2}}}k^{i}b^{\dagger }\left( \mathbf{k}\right) \frac{d^{3}\mathbf{k}}{\left| \mathbf{k}\right| \sqrt{2\left| \mathbf{k}\right| }}\left( \frac{1}{H^{w}_{\epsilon ^{\frac{j+1}{2}}}-E\left( j+1\right) }\right) ^{\frac{1}{2}}\right\Vert _{F_{\epsilon ^{\frac{j+2}{2}}}^{+}}\leq  \)}\emph{}\\
\emph{}\\
\emph{\large \( \leq \left\Vert \left( \frac{1}{H^{w}_{\epsilon ^{\frac{j+1}{2}}}-E\left( j+1\right) }\right) ^{\frac{1}{2}}\int ^{\epsilon ^{\frac{j+1}{2}}}_{\epsilon ^{\frac{j+2}{2}}}k^{i}b^{\dagger }\left( \mathbf{k}\right) \frac{d^{3}\mathbf{k}}{\left| \mathbf{k}\right| \sqrt{2\left| \mathbf{k}\right| }}\right\Vert _{F_{\epsilon ^{\frac{j+2}{2}}}^{+}}\cdot \left\Vert \int ^{\epsilon ^{\frac{j+1}{2}}}_{\epsilon ^{\frac{j+2}{2}}}k^{i}b^{\dagger }\left( \mathbf{k}\right) \frac{d^{3}\mathbf{k}}{\left| \mathbf{k}\right| \sqrt{2\left| \mathbf{k}\right| }}\left( \frac{1}{H^{w}_{\epsilon ^{\frac{j+1}{2}}}-E\left( j+1\right) }\right) ^{\frac{1}{2}}\right\Vert _{F_{\epsilon ^{\frac{j+2}{2}}}^{+}} \)}\emph{}\\
 \emph{\large }\\
{\large \par}

\begin{itemize}
\item {\large the norm \( \left\Vert \left( \frac{1}{H^{w}_{\epsilon ^{\frac{j+1}{2}}}-E\left( j+1\right) }\right) ^{\frac{1}{2}}\int ^{\epsilon ^{\frac{j+1}{2}}}_{\epsilon ^{\frac{j+2}{2}}}k^{i}b^{\dagger }\left( \mathbf{k}\right) \frac{d^{3}k}{\left| \mathbf{k}\right| \sqrt{2\left| \mathbf{k}\right| }}\right\Vert _{F_{\epsilon ^{\frac{j+2}{2}}}^{+}} \)was
treated at point} \textbf{\large 1}{\large )}{\large ; }{\large \par}
\item {\large the norm \( \left\Vert \int ^{\epsilon ^{\frac{j+1}{2}}}_{\epsilon ^{\frac{j+2}{2}}}k^{i}b^{\dagger }\left( \mathbf{k}\right) \frac{d^{3}k}{\left| \mathbf{k}\right| \sqrt{2\left| \mathbf{k}\right| }}\left( \frac{1}{H^{w}_{\epsilon ^{\frac{j+1}{2}}}-E\left( j+1\right) }\right) ^{\frac{1}{2}}\right\Vert _{F_{\epsilon ^{\frac{j+2}{2}}}^{+}} \)can
be controlled like the norm \( \left\Vert \int ^{\epsilon ^{\frac{j+1}{2}}}_{\epsilon ^{\frac{j+2}{2}}}k^{i}b^{\dagger }\left( \mathbf{k}\right) \frac{d^{3}k}{\left| \mathbf{k}\right| \sqrt{2\left| \mathbf{k}\right| }}\left[ \left( \frac{1}{H^{w}_{\epsilon ^{\frac{j+1}{2}}}-E\left( j+1\right) }\right) ^{\frac{1}{2}}\right] ^{\dagger }\right\Vert _{F_{\epsilon ^{\frac{j+2}{2}}}^{+}} \)
 studied at point} \textbf{\large 2)}{\large .}\\
{\large \par}
\end{itemize}
{\large In conclusion:}\\
{\large }\\
{\large \( \frac{g}{m}\left\Vert \left( \frac{1}{H^{w}_{\epsilon ^{\frac{j+1}{2}}}-E\left( j+1\right) }\right) ^{\frac{1}{2}}\int ^{\epsilon ^{\frac{j+1}{2}}}_{\epsilon ^{\frac{j+2}{2}}}k^{i}b^{\dagger }\left( \mathbf{k}\right) \frac{d^{3}k}{\left| \mathbf{k}\right| \sqrt{2\left| \mathbf{k}\right| }}\int ^{\epsilon ^{\frac{j+1}{2}}}_{\epsilon ^{\frac{j+2}{2}}}k^{i}b^{\dagger }\left( \mathbf{k}\right) \frac{d^{3}k}{\left| \mathbf{k}\right| \sqrt{2\left| \mathbf{k}\right| }}\left( \frac{1}{H^{w}_{\epsilon ^{\frac{j+1}{2}}}-E\left( j+1\right) }\right) ^{\frac{1}{2}}\right\Vert _{F_{\epsilon ^{\frac{j+2}{2}}}^{+}}\leq  \)}\\
{\large }\\
\emph{\large \( \leq C^{i}_{3}\left( g,m\right) \cdot \epsilon ^{\frac{j}{2}} \)}\\
\emph{\large }\\
\emph{\large }\\
\textbf{\large 4)}{\large }\\
{\large }\\
{\large \( \frac{g^{2}}{4m}\left\Vert \left( \frac{1}{H^{w}_{\epsilon ^{\frac{j+1}{2}}}-E\left( j+1\right) }\right) ^{\frac{1}{2}}\int ^{\epsilon ^{\frac{j+1}{2}}}_{\epsilon ^{\frac{j+2}{2}}}\frac{d^{3}k}{\left| \mathbf{k}\right| }\left( \frac{1}{H^{w}_{\epsilon ^{\frac{j+1}{2}}}-E\left( j+1\right) }\right) ^{\frac{1}{2}}\right\Vert _{F_{\epsilon ^{\frac{j+2}{2}}}^{+}}= \)}\\
{\large }\\
{\large \( =\frac{g^{2}}{4m}\int ^{\epsilon ^{\frac{j+1}{2}}}_{\epsilon ^{\frac{j+2}{2}}}\frac{d^{3}k}{\left| \mathbf{k}\right| }\left\Vert \frac{1}{H^{w}_{\epsilon ^{\frac{j+1}{2}}}-E\left( j+1\right) }\right\Vert _{F_{\epsilon ^{\frac{j+2}{2}}}^{+}}\leq C^{i}_{4}\left( g,m\right) \cdot \epsilon ^{\frac{j}{2}} \)}
\emph{\large }\\
\emph{\large }\\
\textbf{\large 5)}{\large }\\
{\large }\\
\emph{\large \( \frac{g^{2}}{2m}\left\Vert \left( \frac{1}{H^{w}_{\epsilon ^{\frac{j+1}{2}}}-E\left( j+1\right) }\right) ^{\frac{1}{2}}\Pi ^{i}_{\epsilon ^{\frac{j+1}{2}}}\cdot \int ^{\epsilon ^{\frac{j+1}{2}}}_{\epsilon ^{\frac{j+2}{2}}}k^{i}\left( b\left( \mathbf{k}\right) +b^{\dagger }\left( \mathbf{k}\right) \right) \frac{d^{3}k}{\left| \mathbf{k}\right| \sqrt{2\left| \mathbf{k}\right| }}\left( \frac{1}{H^{w}_{\epsilon ^{\frac{j+1}{2}}}-E\left( j+1\right) }\right) ^{\frac{1}{2}}\right\Vert _{F_{\epsilon ^{\frac{j+2}{2}}}^{+}} \)}\\
\emph{\large }\\
{\large 5A)}\emph{\large }\\
\emph{\large }\\
\emph{\large \( \frac{1}{\sqrt{2m}}\cdot \left\Vert \left( \frac{1}{H^{w}_{\epsilon ^{\frac{j+1}{2}}}-E\left( j+1\right) }\right) ^{\frac{1}{2}}\Pi ^{i}_{\epsilon ^{\frac{j+1}{2}}}\cdot \int ^{\epsilon ^{\frac{j+1}{2}}}_{\epsilon ^{\frac{j+2}{2}}}k^{i}b\left( \mathbf{k}\right) \frac{d^{3}\mathbf{k}}{\left| \mathbf{k}\right| \sqrt{2\left| \mathbf{k}\right| }}\left( \frac{1}{H^{w}_{\epsilon ^{\frac{j+1}{2}}}-E\left( j+1\right) }\right) ^{\frac{1}{2}}\right\Vert _{F_{\epsilon ^{\frac{j+2}{2}}}^{+}}\leq  \)}\emph{}\\
\emph{}\\
{\large \( \leq \frac{1}{\sqrt{2m}}\cdot \left\Vert \left( \frac{1}{H^{w}_{\epsilon ^{\frac{j+1}{2}}}-E\left( j+1\right) }\right) ^{\frac{1}{2}}\Pi ^{i}_{\epsilon ^{\frac{j+1}{2}}}\right\Vert _{F_{\epsilon ^{\frac{j+2}{2}}}^{+}}\cdot \left\Vert \int ^{\epsilon ^{\frac{j+1}{2}}}_{\epsilon ^{\frac{j+2}{2}}}k^{i}b\left( \mathbf{k}\right) \frac{d^{3}\mathbf{k}}{\left| \mathbf{k}\right| \sqrt{2\left| \mathbf{k}\right| }}\left( \frac{1}{H^{w}_{\epsilon ^{\frac{j+1}{2}}}-E\left( j+1\right) }\right) ^{\frac{1}{2}}\right\Vert _{F_{\epsilon ^{\frac{j+2}{2}}}^{+}} \)}\emph{}\\
\emph{\large }\\
{\large if the two norms on the right hand side exist. }\\
{\large }\\
{\large In order to prove the existence and the bound of the second one see}
\textbf{\large 1)}{\large . }\\
{\large As regards the bound of the first one, we can start from the equality
(verified ``a posteriori'' thanks to the existence of the norm on the right
hand side)}\\
{\large }\\
{\large 
\[
\left\Vert \left( \frac{1}{H^{w}_{\epsilon ^{\frac{j+1}{2}}}-E\left( j+1\right) }\right) ^{\frac{1}{2}}\Pi ^{i}_{\epsilon ^{\frac{j+1}{2}}}\right\Vert _{F_{\epsilon ^{\frac{j+2}{2}}}^{+}}=\left\Vert \Pi ^{i}_{\epsilon ^{\frac{j+1}{2}}}\left[ \left( \frac{1}{H^{w}_{\epsilon ^{\frac{j+1}{2}}}-E\left( j+1\right) }\right) ^{\frac{1}{2}}\right] ^{\dagger }\right\Vert _{F_{\epsilon ^{\frac{j+2}{2}}}^{+}}\]
\(  \)}\\
{\large and from the estimate (}{\large \( \varphi \in D^{b}\bigcap F_{\epsilon ^{\frac{j+2}{2}}}^{+} \)}{\large )}\\
{\large }\\
{\large \( \frac{1}{2m}\left\Vert \Pi ^{i}_{\epsilon ^{\frac{j+1}{2}}}\left[ \left( \frac{1}{H^{w}_{\epsilon ^{\frac{j+1}{2}}}-E\left( j+1\right) }\right) ^{\frac{1}{2}}\right] ^{\dagger }\varphi \right\Vert ^{2}= \)}\\
{\large }\\
{\large \( =\frac{1}{2m}\left( \Pi ^{i}_{\epsilon ^{\frac{j+1}{2}}}\left[ \left( \frac{1}{H^{w}_{\epsilon ^{\frac{j+1}{2}}}-E\left( j+1\right) }\right) ^{\frac{1}{2}}\right] ^{\dagger }\varphi \: ,\: \Pi ^{i}_{\epsilon ^{\frac{j+1}{2}}}\left[ \left( \frac{1}{H^{w}_{\epsilon ^{\frac{j+1}{2}}}-E\left( j+1\right) }\right) ^{\frac{1}{2}}\right] ^{\dagger }\varphi \right) \leq  \)
}\\
{\large \( \leq \frac{1}{2m}\sum _{i}\left( \Pi ^{i}_{\epsilon ^{\frac{j+1}{2}}}\left[ \left( \frac{1}{H^{w}_{\epsilon ^{\frac{j+1}{2}}}-E\left( j+1\right) }\right) ^{\frac{1}{2}}\right] ^{\dagger }\varphi \: ,\: \Pi ^{i}_{\epsilon ^{\frac{j+1}{2}}}\left[ \left( \frac{1}{H^{w}_{\epsilon ^{\frac{j+1}{2}}}-E\left( j+1\right) }\right) ^{\frac{1}{2}}\right] ^{\dagger }\varphi \right) \leq  \)}\\
{\large }\\
{\large \( \leq \left( \left[ \left( \frac{1}{H^{w}_{\epsilon ^{\frac{j+1}{2}}}-E\left( j+1\right) }\right) ^{\frac{1}{2}}\right] ^{\dagger }\varphi \: ,\: \left( \frac{1}{2m}\cdot \Pi ^{2}_{\epsilon ^{\frac{j+1}{2}}}+H^{mes}\right) \left[ \left( \frac{1}{H^{w}_{\epsilon ^{\frac{j+1}{2}}}-E\left( j+1\right) }\right) ^{\frac{1}{2}}\right] \varphi \right) \leq  \)}\\
{\large }\\
{\large \( =\left( \varphi \: ,\: \left( H_{\epsilon ^{\frac{j+1}{2}}}^{w}-c\left( j+1\right) \right) \left( \frac{1}{H^{w}_{\epsilon ^{\frac{j+1}{2}}}-E\left( j+1\right) }\right) ^{\frac{1}{2}}\left[ \left( \frac{1}{H^{w}_{\epsilon ^{\frac{j+1}{2}}}-E\left( j+1\right) }\right) ^{\frac{1}{2}}\right] ^{\dagger }\varphi \right) \leq \frac{40}{\epsilon ^{\frac{j+2}{2}}}\left\Vert \varphi \right\Vert ^{2} \)}\\
{\large }\\
{\large ( I used the fact that \( \left| c\left( j+1\right) \right| \leq 1 \)and
\( \left| E^{\epsilon ^{\frac{j+1}{2}}}\right| \leq 1 \))}\\

{\large 5B)}\\
{\large }\\
{\large Commuting and exploiting the properties of the norm we have:} {\large }\\
{\large }\\
{\large \( \left\Vert \left( \frac{1}{H^{w}_{\epsilon ^{\frac{j+1}{2}}}-E\left( j+1\right) }\right) ^{\frac{1}{2}}\Pi ^{i}_{\epsilon ^{\frac{j+1}{2}}}\cdot \int ^{\epsilon ^{\frac{j+1}{2}}}_{\epsilon ^{\frac{j+2}{2}}}k^{i}b^{\dagger }\left( \mathbf{k}\right) \frac{d^{3}k}{\left| \mathbf{k}\right| \sqrt{2\left| \mathbf{k}\right| }}\left( \frac{1}{H^{w}_{\epsilon ^{\frac{j+1}{2}}}-E\left( j+1\right) }\right) ^{\frac{1}{2}}\right\Vert _{F_{\epsilon ^{\frac{j+2}{2}}}^{+}}\leq  \)}\\
{\large }\\
{\large \( \leq \left\Vert \left( \frac{1}{H^{w}_{\epsilon ^{\frac{j+1}{2}}}-E\left( j+1\right) }\right) ^{\frac{1}{2}}\int ^{\epsilon ^{\frac{j+1}{2}}}_{\epsilon ^{\frac{j+2}{2}}}k^{i}b^{\dagger }\left( \mathbf{k}\right) \frac{d^{3}k}{\left| \mathbf{k}\right| \sqrt{2\left| \mathbf{k}\right| }}\Pi ^{i}_{\epsilon ^{\frac{j+1}{2}}}\left( \frac{1}{H^{w}_{\epsilon ^{\frac{j+1}{2}}}-E\left( j+1\right) }\right) ^{\frac{1}{2}}\right\Vert _{F_{\epsilon ^{\frac{j+2}{2}}}^{+}}+ \)}\\
{\large }\\
{\large \( +\left\Vert \left( \frac{1}{H^{w}_{\epsilon ^{\frac{j+1}{2}}}-E\left( j+1\right) }\right) ^{\frac{1}{2}}\int ^{\epsilon ^{\frac{j+1}{2}}}_{\epsilon ^{\frac{j+2}{2}}}\left( k^{i}\right) ^{2}b^{\dagger }\left( \mathbf{k}\right) \frac{d^{3}k}{\left| \mathbf{k}\right| \sqrt{2\left| \mathbf{k}\right| }}\left( \frac{1}{H^{w}_{\epsilon ^{\frac{j+1}{2}}}-E\left( j+1\right) }\right) ^{\frac{1}{2}}\right\Vert _{F_{\epsilon ^{\frac{j+2}{2}}}^{+}} \)}\\
\\
{\large i) the term\( \left\Vert \left( \frac{1}{H^{w}_{\epsilon ^{\frac{j+1}{2}}}-E\left( j+1\right) }\right) ^{\frac{1}{2}}\int ^{\epsilon ^{\frac{j+1}{2}}}_{\epsilon ^{\frac{j+2}{2}}}k^{i}b^{\dagger }\left( \mathbf{k}\right) \frac{d^{3}k}{\left| \mathbf{k}\right| \sqrt{2\left| \mathbf{k}\right| }}\Pi ^{i}_{\epsilon ^{\frac{j+1}{2}}}\left( \frac{1}{H^{w}_{\epsilon ^{\frac{j+1}{2}}}-E\left( j+1\right) }\right) ^{\frac{1}{2}}\right\Vert _{F_{\epsilon ^{\frac{j+2}{2}}}^{+}} \)}
{\large }\\
{\large }\\
{\large can be controlled by \( \left\Vert \left( \frac{1}{H^{w}_{\epsilon ^{\frac{j+1}{2}}}-E\left( j+1\right) }\right) ^{\frac{1}{2}}\Pi ^{i}_{\epsilon ^{\frac{j+1}{2}}}\cdot \int ^{\epsilon ^{\frac{j+1}{2}}}_{\epsilon ^{\frac{j+2}{2}}}k^{i}b\left( \mathbf{k}\right) \frac{d^{3}k}{\left| \mathbf{k}\right| \sqrt{2\left| \mathbf{k}\right| }}\left( \frac{1}{H^{w}_{\epsilon ^{\frac{j+1}{2}}}-E\left( j+1\right) }\right) ^{\frac{1}{2}}\right\Vert _{F_{\epsilon ^{\frac{j+2}{2}}}^{+}} \);
}\\
{\large }\\
{\large ii) the term \( \left\Vert \left( \frac{1}{H^{w}_{\epsilon ^{\frac{j+1}{2}}}-E\left( j+1\right) }\right) ^{\frac{1}{2}}\int ^{\epsilon ^{\frac{j+1}{2}}}_{\epsilon ^{\frac{j+2}{2}}}\left( k^{i}\right) ^{2}b^{\dagger }\left( \mathbf{k}\right) \frac{d^{3}k}{\left| \mathbf{k}\right| \sqrt{2\left| \mathbf{k}\right| }}\left( \frac{1}{H^{w}_{\epsilon ^{\frac{j+1}{2}}}-E\left( j+1\right) }\right) ^{\frac{1}{2}}\right\Vert _{F_{\epsilon ^{\frac{j+2}{2}}}^{+}} \)
}\\
{\large is bounded by}\\
{\large }\\
{\large \( \left\Vert \left( \frac{1}{H^{w}_{\epsilon ^{\frac{j+1}{2}}}-E\left( j+1\right) }\right) ^{\frac{1}{2}}\int ^{\epsilon ^{\frac{j+1}{2}}}_{\epsilon ^{\frac{j+2}{2}}}\left( k^{i}\right) ^{2}b^{\dagger }\left( \mathbf{k}\right) \frac{d^{3}k}{\left| \mathbf{k}\right| \sqrt{2\left| \mathbf{k}\right| }}\right\Vert _{_{F_{\epsilon ^{\frac{j+2}{2}}}^{+}}}\cdot \left\Vert \left( \frac{1}{H^{w}_{\epsilon ^{\frac{j+1}{2}}}-E\left( j+1\right) }\right) ^{\frac{1}{2}}\right\Vert _{_{F_{\epsilon ^{\frac{j+2}{2}}}^{+}}} \)}
{\small }\\
{\large that is of order \( \frac{\epsilon ^{\frac{j+1}{2}}}{\epsilon ^{\frac{j+2}{4}}} \).}\\
{\large }\\
{\large }\\
{\large Summarizing:}\\
{\large }\\
{\large \( \left\Vert \left( \frac{1}{H^{w}_{\epsilon ^{\frac{j+1}{2}}}-E\left( j+1\right) }\right) ^{\frac{1}{2}}\Pi ^{i}_{\epsilon ^{\frac{j+1}{2}}}\cdot \int ^{\epsilon ^{\frac{j+1}{2}}}_{\epsilon ^{\frac{j+2}{2}}}k^{i}\left( b\left( \mathbf{k}\right) +b^{\dagger }\left( \mathbf{k}\right) \right) \frac{d^{3}k}{\left| \mathbf{k}\right| \sqrt{2\left| \mathbf{k}\right| }}\left( \frac{1}{H^{w}_{\epsilon ^{\frac{j+1}{2}}}-E\left( j+1\right) }\right) ^{\frac{1}{2}}\right\Vert _{_{F_{\epsilon ^{\frac{j+2}{2}}}^{+}}}\leq  \)}\\
{\large }\\
{\large \( \leq C^{i}_{5}\left( g,m\right) \cdot \frac{\epsilon ^{\frac{j+1}{4}}}{\epsilon ^{\frac{j+2}{4}}}=C^{i}_{5}\left( g,m\right) \cdot \epsilon ^{-\frac{1}{4}} \)}\\
{\large }\\
\textbf{\large 6)}{\large }\\
{\large }\\
{\large It is controlled like the expression} \textbf{\large 5)}{\large }\\
{\large }\\
{\large }\\
\textbf{\emph{\large Conclusion}}{\large }\\
{\large }\\
{\large If \( g \) is less than a limit value \( \overline{g} \) , the \( \epsilon ^{\frac{j+1}{2}}- \)}
{\large independent constants \( C^{i}_{1}\left( g,m\right) ,.....,C^{i}_{6}\left( g,m\right)  \)
can be tuned in order to arrive at the thesis.}\\
{\large }\\
{\large }\\
{\large }\\
\textbf{\large Lemma A2}{\large }\\
{\large }\\
{\large Results about \( \nabla E^{\sigma }\left( \mathbf{P}\right)  \) where
\( \mathbf{P}\in \Sigma  \):}\\
{\large }\\
\textbf{\large 1)} {\large \( \left| \nabla E^{\sigma }\left( \mathbf{P}\right) \right| <v^{max}<1 \)\( \, \, \, \, \, \, \, \forall \sigma  \)
;}\\
{\large }\\
\textbf{\large 2)} {\large \( \left| \nabla E^{\epsilon ^{\frac{j+1}{2}}}\left( \mathbf{P}\right) -\nabla E^{\epsilon ^{\frac{j+2}{2}}}\left( \mathbf{P}\right) \right|  \)\( <C\left( g\right) \cdot \epsilon ^{\frac{j+1}{8}} \)
for \( g \) sufficiently small and uniform in \( j \);}\\
{\large }\\
\textbf{\large 3)} {\large \( \nabla E^{\sigma }\left( \mathbf{P}\right)  \)
and \( \frac{1}{\left\Vert \widetilde{\phi }_{\mathbf{P}}^{\sigma }\right\Vert ^{2}}\left( \widetilde{\phi }_{\mathbf{P}}^{\sigma },\! \Pi _{\mathbf{P},\sigma }\! \widetilde{\phi }_{\mathbf{P}}^{\sigma }\right)  \)
converge for \( \sigma \rightarrow 0 \)~(for the definition of \( \widetilde{\phi }^{\sigma }_{\mathbf{P}} \)
see chapter 3} \textbf{\large ~~``Spectral regularity''}{\large ).}\\
{\large }\\
{\large 
\[
\Leftrightarrow \]
}\\
{\large Proof}\\
{\large }\\
\textbf{\large 1)} {\large \( \left| \nabla E^{\sigma }\left( \mathbf{P}\right) \right| =\frac{\left| \left( \psi ^{\sigma }_{\mathbf{P}},\, \mathbf{P}-\mathbf{P}^{mes}\psi ^{\sigma }_{\mathbf{P}}\right) \right| }{m\left\Vert \psi _{\mathbf{P}}^{\sigma }\right\Vert ^{2}}\leq \sqrt{\frac{2}{m}}\cdot \frac{1}{\left\Vert \psi _{\mathbf{P}}^{\sigma }\right\Vert }\cdot \left| \left( \psi ^{\sigma }_{\mathbf{P}},\! H_{\mathbf{P},\sigma }+2\pi g^{2}\kappa \, \psi ^{\sigma }_{\mathbf{P}}\right) \right| ^{\frac{1}{2}}<1 \)\(  \)}\\
{\large }\\
{\large since, according to the initial hypotheses, we have}\\
{\large }\\
{\large \( \frac{\left| \left( \psi ^{\sigma }_{\mathbf{P}},\! H_{\mathbf{P},\sigma }\! \psi ^{\sigma }_{\mathbf{P}}\right) \right| }{\left\Vert \psi _{\mathbf{P}}^{\sigma }\right\Vert ^{2}}\leq \left| \left( \psi _{0},\! H_{\mathbf{P},\sigma }\! \psi _{0}\right) \right| =\frac{\mathbf{P}^{2}}{2m}\leq \frac{1}{2} \)}\\
{\large }\\
{\large \( \frac{\left| \left( \psi ^{\sigma }_{\mathbf{P}},\! 2\pi g^{2}\kappa \, \psi ^{\sigma }_{\mathbf{P}}\right) \right| }{\left\Vert \psi _{\mathbf{P}}^{\sigma }\right\Vert ^{2}}\leq \frac{1}{4} \)}\\
{\large }\\
\textbf{\large 2)} {\large }\\
{\large The property to be proved is involved in the proof of theorem 2.3bis.
We assume an implication of the inductive hypothesis}\\
{\large }\\
\( \left\Vert \frac{\widehat{\phi }_{\mathbf{P}}^{\epsilon ^{\frac{j+2}{2}}}}{\left\Vert \widehat{\phi }_{\mathbf{P}}^{\epsilon ^{\frac{j+2}{2}}}\right\Vert }-\frac{\phi _{\mathbf{P}}^{\epsilon ^{\frac{j+1}{2}}}}{\left\Vert \phi _{\mathbf{P}}^{\epsilon ^{\frac{j+1}{2}}}\right\Vert }\right\Vert =\left\Vert \frac{\widehat{\phi }_{\mathbf{P}}^{\epsilon ^{\frac{j+2}{2}}}\left\Vert \phi ^{\epsilon ^{\frac{j+1}{2}}}_{\mathbf{P}}\right\Vert -\phi _{\mathbf{P}}^{\epsilon ^{\frac{j+1}{2}}}\left\Vert \widehat{\phi }^{\epsilon ^{\frac{j+2}{2}}}_{\mathbf{P}}\right\Vert }{\left\Vert \widehat{\phi }_{\mathbf{P}}^{\epsilon ^{\frac{j+2}{2}}}\right\Vert \left\Vert \phi ^{\epsilon ^{\frac{j+1}{2}}}_{\mathbf{P}}\right\Vert }\right\Vert \leq  \)\\
\\
\( \leq \frac{\left\Vert \widehat{\phi }_{\mathbf{P}}^{\epsilon ^{\frac{j+2}{2}}}-\phi _{\mathbf{P}}^{\epsilon ^{\frac{j+1}{2}}}\right\Vert }{\left\Vert \widehat{\phi }_{\mathbf{P}}^{\epsilon ^{\frac{j+2}{2}}}\right\Vert }+\frac{\left| \left\Vert \widehat{\phi }_{\mathbf{P}}^{\epsilon ^{\frac{j+2}{2}}}\right\Vert -\left\Vert \phi _{\mathbf{P}}^{\epsilon ^{\frac{j+1}{2}}}\right\Vert \right| }{\left\Vert \widehat{\phi }_{\mathbf{P}}^{\epsilon ^{\frac{j+2}{2}}}\right\Vert }\leq  \)\\
\\
\( \leq \frac{2\left\Vert \widehat{\phi }_{\mathbf{P}}^{\epsilon ^{\frac{j+2}{2}}}-\phi _{\mathbf{P}}^{\epsilon ^{\frac{j+1}{2}}}\right\Vert }{\left\Vert \widehat{\phi }_{\mathbf{P}}^{\epsilon ^{\frac{j+2}{2}}}\right\Vert }\leq 3\left\Vert -\widehat{\phi }_{\mathbf{P}}^{\epsilon ^{\frac{j+2}{2}}}+\phi _{\mathbf{P}}^{\epsilon ^{\frac{j+1}{2}}}\right\Vert  \)\\
{\large }\\
{\large ( by induction, from theorem 2.3bis ( case \( \mathbf{P}\neq 0 \))
we have:}\\
{\large }\\
{\large \( \left\Vert \phi _{\mathbf{P}}^{\epsilon }\right\Vert \leq \left\Vert \widehat{\phi }_{\mathbf{P}}^{\epsilon ^{\frac{j+2}{2}}}\right\Vert +\left\Vert -\widehat{\phi }_{\mathbf{P}}^{\epsilon ^{\frac{j+2}{2}}}+\phi _{\mathbf{P}}^{\epsilon ^{\frac{j+1}{2}}}\right\Vert +\left\Vert -\widehat{\phi }_{\mathbf{P}}^{\epsilon ^{\frac{j+1}{2}}}+\phi _{\mathbf{P}}^{\epsilon ^{\frac{j+1}{2}}}\right\Vert +.......+\left\Vert -\widehat{\phi }^{\epsilon }_{\mathbf{P}}+\phi _{\mathbf{P}}^{\epsilon }\right\Vert  \)}\\
{\large }\\
\textbf{\large \( \left\Vert \widehat{\phi }_{\mathbf{P}}^{\epsilon ^{\frac{j+2}{2}}}\right\Vert \geq \left\Vert \phi _{\mathbf{P}}^{\epsilon }\right\Vert -\left( \epsilon ^{\frac{1}{16}}+\epsilon ^{\frac{2}{16}}+....+\epsilon ^{\frac{j+1}{16}}\right) \geq 1-\left( \frac{\epsilon ^{\frac{1}{16}}}{1-\epsilon ^{\frac{1}{16}}}\right) \geq \frac{1-2\epsilon ^{\frac{1}{16}}}{1-\epsilon ^{\frac{1}{16}}}>\frac{2}{3} \)}{\large )}\\
{\large }\\
{\large }\\
{\large }\\
{\large Let us analyze the difference between the gradients of energy.}\\
{\large }\\
\( m\nabla E_{\mathbf{P}}^{\epsilon ^{\frac{j+1}{2}}}=\mathbf{P}-\frac{\left( \phi ^{\epsilon ^{\frac{j+1}{2}}}_{\mathbf{P}},\mathbf{P}^{mes}\phi ^{\epsilon ^{\frac{j+1}{2}}}_{\mathbf{P}}\right) }{\left\Vert \phi _{\mathbf{P}}^{\epsilon ^{\frac{j+1}{2}}}\right\Vert ^{2}}+\frac{1}{\left\Vert \phi _{\mathbf{P}}^{\epsilon ^{\frac{j+1}{2}}}\right\Vert ^{2}}\left( \phi ^{\epsilon ^{\frac{j+1}{2}}}_{\mathbf{P}},g\int _{\epsilon ^{\frac{j+1}{2}}}^{\kappa }\frac{\mathbf{k}}{\sqrt{2}\left| \mathbf{k}\right| ^{\frac{3}{2}}\left( 1-\widehat{\mathbf{k}}\cdot \nabla E_{\mathbf{P}}^{\sigma }\right) }\left( b\left( \mathbf{k}\right) +b^{\dagger }\left( \mathbf{k}\right) \right) d^{3}k\phi ^{\epsilon ^{\frac{j+1}{2}}}_{\mathbf{P}}\right) + \){\small }\\
{\small }\\
\( -g^{2}\int _{\epsilon ^{\frac{j+1}{2}}}^{\kappa }\frac{\mathbf{k}}{2\left| \mathbf{k}\right| ^{3}\left( 1-\widehat{\mathbf{k}}\cdot \nabla E_{\mathbf{P}}^{\epsilon ^{\frac{j+1}{2}}}\right) ^{2}}d^{3}k= \)\\
\\
\( =\mathbf{P}-\frac{\left( \phi ^{\epsilon ^{\frac{j+1}{2}}}_{\mathbf{P}},\, \Pi _{\mathbf{P},\epsilon ^{\frac{j+1}{2}}}\phi ^{\epsilon ^{\frac{j+1}{2}}}_{\mathbf{P}}\right) }{\left\Vert \phi _{\mathbf{P}}^{\epsilon ^{\frac{j+1}{2}}}\right\Vert ^{2}}-g^{2}\int _{\epsilon ^{\frac{j+1}{2}}}^{\kappa }\frac{\mathbf{k}}{2\left| \mathbf{k}\right| ^{3}\left( 1-\widehat{\mathbf{k}}\cdot \nabla E_{\mathbf{P}}^{\epsilon ^{\frac{j+1}{2}}}\right) ^{2}}d^{3}k \){\large }\\
{\large }\\
{\large while}\\
{\large }\\
\( m\nabla E_{\mathbf{P}}^{\epsilon ^{\frac{j+2}{2}}}=\mathbf{P}-\frac{\left( \widehat{\phi }^{\epsilon ^{\frac{j+2}{2}}}_{\mathbf{P}},\, \widehat{\Pi }_{\mathbf{P},\epsilon ^{\frac{j+2}{2}}}\widehat{\phi }^{\epsilon ^{\frac{j+2}{2}}}_{\mathbf{P}}\right) }{\left\Vert \widehat{\phi }_{\mathbf{P}}^{\epsilon ^{\frac{j+2}{2}}}\right\Vert ^{2}}-g^{2}\int _{\epsilon ^{\frac{j+2}{2}}}^{\kappa }\frac{\mathbf{k}}{2\left| \mathbf{k}\right| ^{3}\left( 1-\widehat{\mathbf{k}}\cdot \nabla E_{\mathbf{P}}^{\epsilon ^{\frac{j+2}{2}}}\right) ^{2}}d^{3}k \){\large }\\
{\large }\\
{\large then}\\
{\large }\\
\( m\nabla E_{\mathbf{P}}^{\epsilon ^{\frac{j+1}{2}}}-m\nabla E_{\mathbf{P}}^{\epsilon ^{\frac{j+2}{2}}}-g^{2}\int _{\epsilon ^{\frac{j+2}{2}}}^{\kappa }\frac{\mathbf{k}}{2\left| \mathbf{k}\right| ^{3}\left( 1-\widehat{\mathbf{k}}\cdot \nabla E_{\mathbf{P}}^{\epsilon ^{\frac{j+2}{2}}}\right) ^{2}}d^{3}k+g^{2}\int _{\epsilon ^{\frac{j+1}{2}}}^{\kappa }\frac{\mathbf{k}}{2\left| \mathbf{k}\right| ^{3}\left( 1-\widehat{\mathbf{k}}\cdot \nabla E^{\epsilon ^{\frac{j+1}{2}}}\right) ^{2}}d^{3}k= \)\\
\marginpar{
{\large (a2)}{\large \par}
}\\
\( =\frac{1}{\left\Vert \widehat{\phi }_{\mathbf{P}}^{\epsilon ^{\frac{j+2}{2}}}\right\Vert ^{2}}\left( \widehat{\phi }_{\mathbf{P}}^{\epsilon ^{\frac{j+2}{2}}},\! \widehat{\Pi }_{\mathbf{P},\epsilon ^{\frac{j+2}{2}}}\! \widehat{\phi }_{\mathbf{P}}^{\epsilon ^{\frac{j+2}{2}}}\right) -\frac{1}{\left\Vert \phi _{\mathbf{P}}^{\epsilon ^{\frac{j+1}{2}}}\right\Vert ^{2}}\left( \phi _{\mathbf{P}}^{\epsilon ^{\frac{j+1}{2}}},\! \Pi _{\mathbf{P},\epsilon ^{\frac{j+1}{2}}}\! \phi _{\mathbf{P}}^{\epsilon ^{\frac{j+1}{2}}}\right)  \)\\
\\
\\
{\small \( m\nabla E_{\mathbf{P}}^{\epsilon ^{\frac{j+1}{2}}}-m\nabla E_{\mathbf{P}}^{\epsilon ^{\frac{j+2}{2}}}+g^{2}\int _{\epsilon ^{\frac{j+1}{2}}}^{\kappa }\frac{\mathbf{k}\left( -\widehat{\mathbf{k}}\cdot \nabla E_{\mathbf{P}}^{\epsilon ^{\frac{j+2}{2}}}+\widehat{\mathbf{k}}\cdot \nabla E_{\mathbf{P}}^{\epsilon ^{\frac{j+1}{2}}}\right) \left( 2-\widehat{\mathbf{k}}\cdot \nabla E_{\mathbf{P}}^{\epsilon ^{\frac{j+1}{2}}}-\widehat{\mathbf{k}}\cdot \nabla E_{\mathbf{P}}^{\epsilon ^{\frac{j+2}{2}}}\right) }{2\left| \mathbf{k}\right| ^{3}\left( 1-\widehat{\mathbf{k}}\cdot \nabla E_{\mathbf{P}}^{\epsilon ^{\frac{j+2}{2}}}\right) ^{2}\left( 1-\widehat{\mathbf{k}}\cdot \nabla E_{\mathbf{P}}^{\epsilon ^{\frac{j+1}{2}}}\right) ^{2}}d^{3}k-g^{2}\int _{\epsilon ^{\frac{j+2}{2}}}^{\epsilon ^{\frac{j+1}{2}}}\frac{\mathbf{k}}{2\left| \mathbf{k}\right| ^{3}\left( 1-\widehat{\mathbf{k}}\cdot \nabla E_{\mathbf{P}}^{\epsilon ^{\frac{j+2}{2}}}\right) ^{2}}d^{3}k= \)}{\large }\\
{\large }\\
{\small \( =\frac{1}{\left\Vert \widehat{\phi }_{\mathbf{P}}^{\epsilon ^{\frac{j+2}{2}}}\right\Vert }\left( \widehat{\phi }_{\mathbf{P}}^{\epsilon ^{\frac{j+2}{2}}},\! \widehat{\Pi }_{\mathbf{P},\epsilon ^{\frac{j+2}{2}}}\! \left( \frac{\widehat{\phi }_{\mathbf{P}}^{\epsilon ^{\frac{j+2}{2}}}}{\left\Vert \widehat{\phi }_{\mathbf{P}}^{\epsilon ^{\frac{j+2}{2}}}\right\Vert }-\frac{\phi _{\mathbf{P}}^{\epsilon ^{\frac{j+1}{2}}}}{\left\Vert \phi _{\mathbf{P}}^{\epsilon ^{\frac{j+1}{2}}}\right\Vert }\right) \right) + \)}\\
{\small }\\
{\small \( +\frac{1}{\left\Vert \widehat{\phi }_{\mathbf{P}}^{\epsilon ^{\frac{j+2}{2}}}\right\Vert \left\Vert \phi _{\mathbf{P}}^{\epsilon ^{\frac{j+1}{2}}}\right\Vert }\left( \widehat{\phi }_{\mathbf{P}}^{\epsilon ^{\frac{j+2}{2}}},\! \widehat{\Pi }_{\mathbf{P},\epsilon ^{\frac{j+2}{2}}}\! \phi _{\mathbf{P}}^{\epsilon ^{\frac{j+1}{2}}}\right) -\frac{1}{\left\Vert \widehat{\phi }_{\mathbf{P}}^{\epsilon ^{\frac{j+2}{2}}}\right\Vert \left\Vert \phi _{\mathbf{P}}^{\epsilon ^{\frac{j+1}{2}}}\right\Vert }\left( \widehat{\phi }_{\mathbf{P}}^{\epsilon ^{\frac{j+2}{2}}},\! \Pi _{\mathbf{P},\epsilon ^{\frac{j+1}{2}}}\! \phi _{\mathbf{P}}^{\epsilon ^{\frac{j+1}{2}}}\right) + \)}\\
{\small }\\
{\small \( +\frac{1}{\left\Vert \widehat{\phi }_{\mathbf{P}}^{\epsilon ^{\frac{j+2}{2}}}\right\Vert \left\Vert \phi _{\mathbf{P}}^{\epsilon ^{\frac{j+1}{2}}}\right\Vert }\left( \widehat{\phi }_{\mathbf{P}}^{\epsilon ^{\frac{j+2}{2}}},\! \Pi _{\mathbf{P},\epsilon ^{\frac{j+1}{2}}}\! \phi _{\mathbf{P}}^{\epsilon ^{\frac{j+1}{2}}}\right) -\frac{1}{\left\Vert \phi _{\mathbf{P}}^{\epsilon ^{\frac{j+1}{2}}}\right\Vert ^{2}}\left( \phi _{\mathbf{P}}^{\epsilon ^{\frac{j+1}{2}}},\! \Pi _{\mathbf{P},\epsilon ^{\frac{j+1}{2}}}\! \phi _{\mathbf{P}}^{\epsilon ^{\frac{j+1}{2}}}\right)  \)}\\
{\small }\\
{\small }\\
{\large Considering that}{\small }\\
{\small }\\
{\small \( \widehat{\Pi }^{i}_{\mathbf{P},\epsilon ^{\frac{j+2}{2}}}-\Pi ^{i}_{\mathbf{P},\epsilon ^{\frac{j+1}{2}}}= \)}\\
{\small }\\
{\small \( =-g\int ^{\epsilon ^{\frac{j+1}{2}}}_{\epsilon ^{\frac{j+2}{2}}}\frac{k^{i}\left( b\left( \mathbf{k}\right) +b^{\dagger }\left( \mathbf{k}\right) \right) }{\sqrt{2}\left| \mathbf{k}\right| ^{\frac{3}{2}}\left( 1-\widehat{\mathbf{k}}\nabla E^{\epsilon ^{\frac{j+1}{2}}}_{\mathbf{P}}\right) }d^{3}k+\frac{g^{2}}{2}\int ^{\kappa }_{\epsilon ^{\frac{j+2}{2}}}\frac{k^{i}\left( -\widehat{\mathbf{k}}\cdot \nabla E_{\mathbf{P}}^{\epsilon ^{\frac{j+2}{2}}}+\widehat{\mathbf{k}}\cdot \nabla E_{\mathbf{P}}^{\epsilon ^{\frac{j+1}{2}}}\right) \left( 2-\widehat{\mathbf{k}}\cdot \nabla E_{\mathbf{P}}^{\epsilon ^{\frac{j+1}{2}}}-\widehat{\mathbf{k}}\cdot \nabla E_{\mathbf{P}}^{\epsilon ^{\frac{j+2}{2}}}\right) }{2\left| \mathbf{k}\right| ^{3}\left( 1-\widehat{\mathbf{k}}\cdot \nabla E_{\mathbf{P}}^{\epsilon ^{\frac{j+2}{2}}}\right) ^{2}\left( 1-\widehat{\mathbf{k}}\cdot \nabla E_{\mathbf{P}}^{\epsilon ^{\frac{j+1}{2}}}\right) ^{2}}d^{3}k \)}{\small }\\
{\small }\\
{\large the equation (a2) can be written in the following way}\\
{\large }\\
{\small \( m\nabla E_{\mathbf{P}}^{\epsilon ^{\frac{j+1}{2}}}-m\nabla E_{\mathbf{P}}^{\epsilon ^{\frac{j+2}{2}}}+g^{2}\int _{\epsilon ^{\frac{j+1}{2}}}^{\kappa }\frac{\mathbf{k}\left( \widehat{\mathbf{k}}\cdot \nabla E_{\mathbf{P}}^{\epsilon ^{\frac{j+2}{2}}}-\widehat{\mathbf{k}}\cdot \nabla E_{\mathbf{P}}^{\epsilon ^{\frac{j+1}{2}}}\right) \left( 2-\widehat{\mathbf{k}}\cdot \nabla E_{\mathbf{P}}^{\epsilon ^{\frac{j+1}{2}}}-\widehat{\mathbf{k}}\cdot \nabla E_{\mathbf{P}}^{\epsilon ^{\frac{j+2}{2}}}\right) }{2\left| \mathbf{k}\right| ^{3}\left( 1-\widehat{\mathbf{k}}\cdot \nabla E_{\mathbf{P}}^{\epsilon ^{\frac{j+2}{2}}}\right) ^{2}\left( 1-\widehat{\mathbf{k}}\cdot \nabla E_{\mathbf{P}}^{\epsilon ^{\frac{j+1}{2}}}\right) ^{2}}d^{3}k+ \)}{\large }\\
{\large }\\
{\large }\\
{\small \( -\frac{1}{\left\Vert \widehat{\phi }_{\mathbf{P}}^{\epsilon ^{\frac{j+2}{2}}}\right\Vert \left\Vert \phi _{\mathbf{P}}^{\epsilon ^{\frac{j+1}{2}}}\right\Vert }\left( \widehat{\phi }_{\mathbf{P}}^{\epsilon ^{\frac{j+2}{2}}},\phi _{\mathbf{P}}^{\epsilon ^{\frac{j+1}{2}}}\right) \frac{g^{2}}{2}\int _{\epsilon ^{\frac{j+1}{2}}}^{\kappa }\frac{\mathbf{k}\left( -\widehat{\mathbf{k}}\cdot \nabla E_{\mathbf{P}}^{\epsilon ^{\frac{j+2}{2}}}+\widehat{\mathbf{k}}\cdot \nabla E_{\mathbf{P}}^{\epsilon ^{\frac{j+1}{2}}}\right) \left( 2-\widehat{\mathbf{k}}\cdot \nabla E_{\mathbf{P}}^{\epsilon ^{\frac{j+1}{2}}}-\widehat{\mathbf{k}}\cdot \nabla E_{\mathbf{P}}^{\epsilon ^{\frac{j+2}{2}}}\right) }{2\left| \mathbf{k}\right| ^{3}\left( 1-\widehat{\mathbf{k}}\cdot \nabla E_{\mathbf{P}}^{\epsilon ^{\frac{j+2}{2}}}\right) ^{2}\left( 1-\widehat{\mathbf{k}}\cdot \nabla E_{\mathbf{P}}^{\epsilon ^{\frac{j+1}{2}}}\right) ^{2}}d^{3}k= \)}{\large }\\
{\large }\\
{\large }\\
{\small \( =\frac{1}{\left\Vert \widehat{\phi }_{\mathbf{P}}^{\epsilon ^{\frac{j+2}{2}}}\right\Vert }\left( \widehat{\phi }_{\mathbf{P}}^{\epsilon ^{\frac{j+2}{2}}},\! \widehat{\Pi }_{\mathbf{P},\epsilon ^{\frac{j+2}{2}}}\! \left( \frac{\widehat{\phi }_{\mathbf{P}}^{\epsilon ^{\frac{j+2}{2}}}}{\left\Vert \widehat{\phi }_{\mathbf{P}}^{\epsilon ^{\frac{j+2}{2}}}\right\Vert }-\frac{\phi _{\mathbf{P}}^{\epsilon ^{\frac{j+1}{2}}}}{\left\Vert \phi _{\mathbf{P}}^{\epsilon ^{\frac{j+1}{2}}}\right\Vert }\right) \right) +\frac{1}{\left\Vert \phi _{\mathbf{P}}^{\epsilon ^{\frac{j+1}{2}}}\right\Vert }\left( \left( \frac{\widehat{\phi }_{\mathbf{P}}^{\epsilon ^{\frac{j+2}{2}}}}{\left\Vert \widehat{\phi }_{\mathbf{P}}^{\epsilon ^{\frac{j+2}{2}}}\right\Vert }-\frac{\phi _{\mathbf{P}}^{\epsilon ^{\frac{j+1}{2}}}}{\left\Vert \phi _{\mathbf{P}}^{\epsilon ^{\frac{j+1}{2}}}\right\Vert }\right) \! \Pi _{\mathbf{P},\epsilon ^{\frac{j+1}{2}}}\! \phi _{\mathbf{P}}^{\epsilon ^{\frac{j+1}{2}}}\right) + \)
}\\
{\small }\\
{\small \( +g^{2}\int _{\epsilon ^{\frac{j+2}{2}}}^{\epsilon ^{\frac{j+1}{2}}}\frac{\mathbf{k}}{2\left| \mathbf{k}\right| ^{3}\left( 1-\widehat{\mathbf{k}}\cdot \nabla E_{\mathbf{P}}^{\epsilon ^{\frac{j+2}{2}}}\right) ^{2}}d^{3}k-\frac{1}{\left\Vert \widehat{\phi }_{\mathbf{P}}^{\epsilon ^{\frac{j+2}{2}}}\right\Vert \left\Vert \phi _{\mathbf{P}}^{\epsilon ^{\frac{j+1}{2}}}\right\Vert }\left( \widehat{\phi }_{\mathbf{P}}^{\epsilon ^{\frac{j+2}{2}}},\! g\int ^{\epsilon ^{\frac{j+1}{2}}}_{\epsilon ^{\frac{j+2}{2}}}\frac{\mathbf{k}\left( b\left( \mathbf{k}\right) +b^{\dagger }\left( \mathbf{k}\right) \right) }{\sqrt{2}\left| \mathbf{k}\right| ^{\frac{3}{2}}\left( 1-\widehat{\mathbf{k}}\cdot \nabla E^{\epsilon ^{\frac{j+1}{2}}}_{\mathbf{P}}\right) }d^{3}k\! \phi _{\mathbf{P}}^{\epsilon ^{\frac{j+1}{2}}}\right)  \)}{\large }\\
{\large }\\
{\large }\\
{\large On the left hand side of the equation, there is a quantity whose module
is bigger than \( C\cdot \left| \nabla E_{\mathbf{P}}^{\epsilon ^{\frac{j+1}{2}}}-\nabla E_{\mathbf{P}}^{\epsilon ^{\frac{j+2}{2}}}\right|  \)
for \( g\rightarrow 0 \), where \( C \) is a positive constant that is uniform
in \( j \) and converges to \( m \) for \( g\rightarrow 0 \). It is due to
the result in point 1). On the right hand side, there is a quantity whose module
is bounded by a \( g- \)dependent constant times \( \epsilon ^{\frac{j+1}{8}} \).
Looking at the proof of theorem 2.3bis, the norm \( \left\Vert \widehat{\phi }_{\mathbf{P}}^{\epsilon ^{\frac{j+2}{2}}}-\phi _{\mathbf{P}}^{\epsilon ^{\frac{j+1}{2}}}\right\Vert  \)
is of order \( \epsilon ^{\frac{j+1}{8}} \)(the multiplicative constant, which
is uniform in j, gets smaller by reducing \( g \)) . Moreover the following
bounds hold:}\\
{\large }\\
{\large }\\
~\( \left( \widehat{\phi }_{\mathbf{P}}^{\epsilon ^{\frac{j+2}{2}}},\! g\int ^{\epsilon ^{\frac{j+1}{2}}}_{\epsilon ^{\frac{j+2}{2}}}\frac{k^{i}\left( b\left( \mathbf{k}\right) +b^{\dagger }\left( \mathbf{k}\right) \right) }{\sqrt{2}\left| \mathbf{k}\right| ^{\frac{3}{2}}\left( 1-\widehat{\mathbf{k}}\cdot \nabla E^{\epsilon ^{\frac{j+1}{2}}}_{\mathbf{P}}\right) }d^{3}k\! \phi _{\mathbf{P}}^{\epsilon ^{\frac{j+1}{2}}}\right) =\left( \widehat{\phi }_{\mathbf{P}}^{\epsilon ^{\frac{j+2}{2}}},\! g\int ^{\epsilon ^{\frac{j+1}{2}}}_{\epsilon ^{\frac{j+2}{2}}}\frac{k^{i}b^{\dagger }\left( \mathbf{k}\right) }{\sqrt{2}\left| \mathbf{k}\right| ^{\frac{3}{2}}\left( 1-\widehat{\mathbf{k}}\cdot \nabla E^{\epsilon ^{\frac{j+1}{2}}}_{\mathbf{P}}\right) }d^{3}k\! \phi _{\mathbf{P}}^{\epsilon ^{\frac{j+1}{2}}}\right) = \)\\
\\
\( =g\int ^{\epsilon ^{\frac{j+1}{2}}}_{\epsilon ^{\frac{j+2}{2}}}\frac{k^{i}}{\sqrt{2}\left| \mathbf{k}\right| ^{2}\left( 1-\widehat{\mathbf{k}}\cdot \nabla E^{\epsilon ^{\frac{j+1}{2}}}_{\mathbf{P}}\right) }\left( \widehat{\phi }_{\mathbf{P}}^{\epsilon ^{\frac{j+2}{2}}},\! \left| \mathbf{k}\right| ^{\frac{1}{2}}b^{\dagger }\left( \mathbf{k}\right) \! \phi _{\mathbf{P}}^{\epsilon ^{\frac{j+1}{2}}}\right) d^{3}k\leq  \)\\
\\
\( \leq g\left( \int ^{\epsilon ^{\frac{j+1}{2}}}_{\epsilon ^{\frac{j+2}{2}}}\frac{\left( k^{i}\right) ^{2}}{2\left| \mathbf{k}\right| ^{4}\left( 1-\widehat{\mathbf{k}}\cdot \nabla E^{\epsilon ^{\frac{j+1}{2}}}_{\mathbf{P}}\right) ^{2}}d^{3}k\right) ^{\frac{1}{2}}\left( \widehat{\phi }_{\mathbf{P}}^{\epsilon ^{\frac{j+2}{2}}},\! H^{mes}\! \widehat{\phi }_{\mathbf{P}}^{\epsilon ^{\frac{j+2}{2}}}\right) ^{\frac{1}{2}}\leq  \)\\
\\
\( \leq g\left( \int ^{\epsilon ^{\frac{j+1}{2}}}_{\epsilon ^{\frac{j+2}{2}}}\frac{\left( k^{i}\right) ^{2}}{2\left| \mathbf{k}\right| ^{4}\left( 1-\widehat{\mathbf{k}}\cdot \nabla E^{\epsilon ^{\frac{j+1}{2}}}_{\mathbf{P}}\right) ^{2}}d^{3}k\right) ^{\frac{1}{2}}\left( \widehat{\phi }_{\mathbf{P}}^{\epsilon ^{\frac{j+2}{2}}},\! H^{mes}\! \widehat{\phi }_{\mathbf{P}}^{\epsilon ^{\frac{j+2}{2}}}\right) ^{\frac{1}{2}}\leq  \)\\
\\
\( \leq g\left( \int ^{\epsilon ^{\frac{j+1}{2}}}_{\epsilon ^{\frac{j+2}{2}}}\frac{\left( k^{i}\right) ^{2}}{2\left| \mathbf{k}\right| ^{4}\left( 1-\widehat{\mathbf{k}}\cdot \nabla E^{\epsilon ^{\frac{j+1}{2}}}_{\mathbf{P}}\right) ^{2}}d^{3}k\right) ^{\frac{1}{2}}\left( E^{\epsilon ^{\frac{j+2}{2}}}_{\mathbf{P}}-\widehat{c}_{\mathbf{P}}\left( j+2\right) \right) ^{\frac{1}{2}}\left( \widehat{\phi }_{\mathbf{P}}^{\epsilon ^{\frac{j+2}{2}}},\! \widehat{\phi }_{\mathbf{P}}^{\epsilon ^{\frac{j+2}{2}}}\right) ^{\frac{1}{2}} \)\\
\\
\\
\( \frac{\left\Vert \widehat{\Pi }^{i}_{\mathbf{P},\epsilon ^{\frac{j+2}{2}}}\widehat{\phi }^{\epsilon ^{\frac{j+2}{2}}}_{\mathbf{P}}\right\Vert }{\left\Vert \widehat{\phi }_{\mathbf{P}}^{\epsilon ^{\frac{j+2}{2}}}\right\Vert }\leq \sqrt{2m}\cdot \left( E^{\epsilon ^{\frac{j+2}{2}}}_{\mathbf{P}}-c_{\mathbf{P}}\left( j+2\right) \right) ^{\frac{1}{2}} \)
~~,~~\( \frac{\left\Vert \Pi ^{i}_{\mathbf{P},\epsilon ^{\frac{j+1}{2}}}\phi ^{\epsilon ^{\frac{j+1}{2}}}_{\mathbf{P}}\right\Vert }{\left\Vert \phi _{\mathbf{P}}^{\epsilon ^{\frac{j+1}{2}}}\right\Vert }\leq \sqrt{2m}\cdot \left( E^{\epsilon ^{\frac{j+1}{2}}}_{\mathbf{P}}-c_{\mathbf{P}}\left( j+1\right) \right) ^{\frac{1}{2}} \)
\\
{\large }\\
{\large }\\
{\large Having established the bound \( \left| \nabla E^{\epsilon ^{\frac{j+1}{2}}}\left( \mathbf{P}\right) -\nabla E^{\epsilon ^{\frac{j+2}{2}}}\left( \mathbf{P}\right) \right|  \)\( <C\left( g\right) \cdot \epsilon ^{\frac{j+1}{8}} \)
for \( g \) sufficiently small, the \( j \) uniformity of \( C\left( g\right)  \)
follows from the inductive procedure and the related tuning of the constant
\( g \) . About the last point I do not give the details but a substantially
analogous procedure is used in theorem 2.3. In conclusion the result is that
for \( g \) less than a proper \( \overline{g} \), the thesis of the point
2) of the lemma is proved jointly with theorem 2.3bis.}\\
{\large }\\
{\large }\\
\textbf{\large 3)}{\large }\\
{\large The convergence of \( \nabla E^{\sigma }\left( \mathbf{P}\right)  \)
and of \( \frac{1}{\left\Vert \widetilde{\phi }_{\mathbf{P}}^{\sigma }\right\Vert ^{2}}\left( \widetilde{\phi }_{\mathbf{P}}^{\sigma },\! \Pi _{\mathbf{P},\sigma }\! \widetilde{\phi }_{\mathbf{P}}^{\sigma }\right)  \)
follows from the convergence proved in theorem 2.3bis. I will prove the second
limit only. }\\
{\large }\\
{\large I estimate the difference}\\
{\large }\\
{\large 
\[
\frac{1}{\left\Vert \widetilde{\phi }^{\sigma _{2}}_{\mathbf{P}}\right\Vert ^{2}}\left( \widetilde{\phi }^{\sigma _{2}}_{\mathbf{P}},\! \Pi _{\mathbf{P},\sigma _{2}}\! \widetilde{\phi }^{\sigma _{2}}_{\mathbf{P}}\right) -\frac{1}{\left\Vert \widetilde{\phi }^{\sigma _{1}}_{\mathbf{P}}\right\Vert ^{2}}\left( \widetilde{\phi }^{\sigma _{1}}_{\mathbf{P}},\! \Pi _{\mathbf{P},\sigma _{1}}\! \widetilde{\phi }^{\sigma _{1}}_{\mathbf{P}}\right) \]
}\\
{\large }\\
{\large I recall the definition of \( \widetilde{\phi }^{\sigma }_{\mathbf{P}} \)
:}\\
{\large given \( \sigma  \) between \( \epsilon ^{\frac{j+2}{2}} \)and \( \epsilon ^{\frac{j+3}{2}} \)
, it can be written as \( \epsilon '^{\frac{j+2}{2}} \) where \( \epsilon '=\sigma ^{\frac{2}{j+2}} \)
\( \left\{ \epsilon ':\; \epsilon \geq \epsilon '\geq \epsilon \sqrt{\epsilon }\right\}  \)
and it can be defined \( \widetilde{\phi }^{\sigma }_{\mathbf{P}}\equiv \phi _{\mathbf{P}}^{\epsilon '\left( \sigma \right) ^{\frac{j+2}{2}}} \),
where we obtain \( \phi _{\mathbf{P}}^{\epsilon '\left( \sigma \right) ^{\frac{j+2}{2}}} \)
by iteration starting from the cut-off~\( \epsilon ' \). }\\
{\large }\\
{\large I rewrite \( \widetilde{\phi }^{\sigma _{2}}_{\mathbf{P}}-\widetilde{\phi }^{\sigma _{1}}_{\mathbf{P}} \)
in the following way}\\
{\large }\\
{\large \( \widetilde{\phi }^{\sigma _{2}}_{\mathbf{P}}-\widetilde{\phi }^{\sigma _{1}}_{\mathbf{P}}=\widetilde{\phi }^{\sigma _{2}}_{\mathbf{P}}-\phi _{\mathbf{P}}^{\epsilon _{2}\left( \sigma _{2}\right) ^{\frac{l+2}{2}}}+\phi _{\mathbf{P}}^{\epsilon _{2}\left( \sigma _{2}\right) ^{\frac{l+2}{2}}}-\phi _{\mathbf{P}}^{\epsilon _{1}\left( \sigma _{1}\right) ^{\frac{m+2}{2}}}+\phi _{\mathbf{P}}^{\epsilon _{1}\left( \sigma _{1}\right) ^{\frac{m+2}{2}}}-\widetilde{\phi }^{\sigma _{1}}_{\mathbf{P}} \)
}\\
{\large }\\
{\large Now, fixed an arbitrarily small \( \delta  \), there exist \( l\left( \delta \right) ,m\left( \delta \right)  \)
sufficiently large and a phase \( e^{i\eta \left( \delta \right) } \) such
that}\\
{\large }\\
{\large \( \left\Vert \frac{\phi _{\mathbf{P}}^{\epsilon _{2}\left( \sigma _{2}\right) ^{\frac{l+2}{2}}}}{\left\Vert \phi _{\mathbf{P}}^{\epsilon _{2}\left( \sigma _{2}\right) ^{\frac{l+2}{2}}}\right\Vert }-e^{i\eta \left( \delta \right) }\frac{\phi _{\mathbf{P}}^{\epsilon _{1}\left( \sigma _{1}\right) ^{\frac{m+2}{2}}}}{\left\Vert \phi _{\mathbf{P}}^{\epsilon _{1}\left( \sigma _{1}\right) ^{\frac{m+2}{2}}}\right\Vert }\right\Vert \leq \delta  \)
}\\
{\large }\\
{\large This is substantially due to the convergence stated in the theorem 2.3bis
and to the unicity of the ground state till there is a cut-off, by construction.}\\
{\large }\\
{\large Then \( \left\Vert \frac{\widetilde{\phi }^{\sigma _{2}}_{\mathbf{P}}}{\left\Vert \widetilde{\phi }^{\sigma _{2}}_{\mathbf{P}}\right\Vert }-e^{i\eta \left( \delta \right) }\frac{\widetilde{\phi }^{\sigma _{1}}_{\mathbf{P}}}{\left\Vert \widetilde{\phi }^{\sigma _{1}}_{\mathbf{P}}\right\Vert }\right\Vert  \)
is bounded by a quantity of order \( \sigma ^{\frac{1}{8}}_{2}+\sigma ^{\frac{1}{8}}_{1}+\delta  \).}\\
{\large }\\
{\large Since}\\
{\large }\\
{\large \( \frac{1}{\left\Vert \widetilde{\phi }^{\sigma _{2}}_{\mathbf{P}}\right\Vert ^{2}}\left( \widetilde{\phi }^{\sigma _{2}}_{\mathbf{P}},\! \Pi _{\mathbf{P},\sigma _{2}}\! \widetilde{\phi }^{\sigma _{2}}_{\mathbf{P}}\right) -\frac{1}{\left\Vert \widetilde{\phi }^{\sigma _{1}}_{\mathbf{P}}\right\Vert ^{2}}\left( \widetilde{\phi }^{\sigma _{1}}_{\mathbf{P}},\! \Pi _{\mathbf{P},\sigma _{1}}\! \widetilde{\phi }^{\sigma _{1}}_{\mathbf{P}}\right) =\frac{1}{\left\Vert \widetilde{\phi }^{\sigma _{2}}_{\mathbf{P}}\right\Vert ^{2}}\left( \widetilde{\phi }^{\sigma _{2}}_{\mathbf{P}},\! \Pi _{\mathbf{P},\sigma _{2}}\! \widetilde{\phi }^{\sigma _{2}}_{\mathbf{P}}\right) -\frac{1}{\left\Vert e^{i\eta }\widetilde{\phi }^{\sigma _{1}}_{\mathbf{P}}\right\Vert ^{2}}\left( e^{i\eta }\widetilde{\phi }^{\sigma _{1}}_{\mathbf{P}},\! \Pi _{\mathbf{P},\sigma _{1}}\! e^{i\eta }\widetilde{\phi }^{\sigma _{1}}_{\mathbf{P}}\right)  \)}\\
{\large }\\
{\large }\\
{\large we can conclude that the limit exists by exploiting the bounds (from
above) showed in point} \textbf{\large 2)}{\large .}\\
{\large }\\
{\large }\\
{\large }\\
\emph{\large Calculation of} {\large \( \widehat{H}_{\mathbf{P},\epsilon ^{\frac{j+2}{2}}} \).}\\
{\large }\\
{\small \( \widehat{H}_{\mathbf{P},\epsilon ^{\frac{j+2}{2}}}=\frac{\left( \mathbf{P}_{1}+\mathbf{P}_{2}\right) ^{2}}{2m}+\frac{1}{2m}\left( \mathbf{P}^{mes}-g\int _{\epsilon ^{\frac{j+2}{2}}}^{\kappa }\frac{\mathbf{k}\left( b\left( \mathbf{k}\right) +b^{\dagger }\left( \mathbf{k}\right) \right) }{\sqrt{2}\left| \mathbf{k}\right| ^{\frac{3}{2}}\left( 1-\widehat{\mathbf{k}}\cdot \nabla E^{\epsilon ^{\frac{j+1}{2}}}\left( \mathbf{P}\right) \right) }d^{3}k+g^{2}\int _{\epsilon ^{\frac{j+2}{2}}}^{\kappa }\frac{\mathbf{k}}{2\left| \mathbf{k}\right| ^{3}\left( 1-\widehat{\mathbf{k}}\cdot \nabla E^{\epsilon ^{\frac{j+1}{2}}}\left( \mathbf{P}\right) \right) ^{2}}d^{3}k\right) ^{2}+ \)}\\
{\small }\\
{\small \( -\frac{\mathbf{P}_{2}}{m}\cdot \left( \mathbf{P}^{mes}-g\int _{\epsilon ^{\frac{j+2}{2}}}^{\kappa }\frac{\mathbf{k}\left( b\left( \mathbf{k}\right) +b^{\dagger }\left( \mathbf{k}\right) \right) }{\sqrt{2}\left| \mathbf{k}\right| ^{\frac{3}{2}}\left( 1-\widehat{\mathbf{k}}\cdot \nabla E^{\epsilon ^{\frac{j+1}{2}}}\left( \mathbf{P}\right) \right) }d^{3}k+g^{2}\int _{\epsilon ^{\frac{j+2}{2}}}^{\kappa }\frac{\mathbf{k}}{2\left| \mathbf{k}\right| ^{3}\left( 1-\widehat{\mathbf{k}}\cdot \nabla E^{\epsilon ^{\frac{j+1}{2}}}\left( \mathbf{P}\right) \right) ^{2}}d^{3}k\right) + \)}\\
{\small }\\
{\small \( +\int ^{\infty }_{0}\left( \left| \mathbf{k}\right| -\mathbf{k}\cdot \nabla E^{\epsilon ^{\frac{j+1}{2}}}\left( \mathbf{P}\right) \right) b^{\dagger }\left( \mathbf{k}\right) b\left( \mathbf{k}\right) d^{3}k-g^{2}\int _{\epsilon ^{\frac{j+2}{2}}}^{\kappa }\frac{1}{2\left| \mathbf{k}\right| ^{2}\left( 1-\widehat{\mathbf{k}}\cdot \nabla E^{\epsilon ^{\frac{j+1}{2}}}\left( \mathbf{P}\right) \right) }d^{3}k= \)}\\
{\small }\\
{\small \( =\frac{1}{2m}\left( \Pi _{\mathbf{P},\epsilon ^{\frac{j+1}{2}}}-g\int _{\epsilon ^{\frac{j+2}{2}}}^{\epsilon ^{\frac{j+1}{2}}}\frac{\mathbf{k}}{\sqrt{2}\left| \mathbf{k}\right| ^{\frac{3}{2}}\left( 1-\widehat{\mathbf{k}}\cdot \nabla E^{\epsilon ^{\frac{j+1}{2}}}\left( \mathbf{P}\right) \right) }\left( b\left( \mathbf{k}\right) +b^{\dagger }\left( \mathbf{k}\right) \right) d^{3}k+g^{2}\int _{\epsilon ^{\frac{j+2}{2}}}^{\kappa }\frac{\mathbf{k}}{2\left| \mathbf{k}\right| ^{3}\left( 1-\widehat{\mathbf{k}}\cdot \nabla E^{\epsilon ^{\frac{j+1}{2}}}\left( \mathbf{P}\right) \right) ^{2}}d^{3}k\right) ^{2}+ \)}{\large }\\
{\large }\\
{\small \( -\frac{1}{m}\left( \frac{1}{\left\Vert \phi _{\mathbf{P}}^{\epsilon ^{\frac{j+1}{2}}}\right\Vert ^{2}}\left( \phi ^{\epsilon ^{\frac{j+1}{2}}}_{\mathbf{P}},\Pi _{\mathbf{P},\epsilon ^{\frac{j+1}{2}}}\phi ^{\epsilon ^{\frac{j+1}{2}}}_{\mathbf{P}}\right) +g^{2}\int _{\epsilon ^{\frac{j+1}{2}}}^{\kappa }\frac{\mathbf{k}}{2\left| \mathbf{k}\right| ^{3}\left( 1-\widehat{\mathbf{k}}\cdot \nabla E^{\epsilon ^{\frac{j+1}{2}}}\left( \mathbf{P}\right) \right) ^{2}}d^{3}k\right) \cdot  \)}{\large }\\
{\large }\\
{\small \( \cdot \left( \Pi _{\mathbf{P},\epsilon ^{\frac{j+1}{2}}}-g\int _{\epsilon ^{\frac{j+2}{2}}}^{\epsilon ^{\frac{j+1}{2}}}\frac{\mathbf{k}}{\sqrt{2}\left| \mathbf{k}\right| ^{\frac{3}{2}}\left( 1-\widehat{\mathbf{k}}\cdot \nabla E^{\epsilon ^{\frac{j+1}{2}}}\left( \mathbf{P}\right) \right) }\left( b\left( \mathbf{k}\right) +b^{\dagger }\left( \mathbf{k}\right) \right) d^{3}k+g^{2}\int _{\epsilon ^{\frac{j+2}{2}}}^{\kappa }\frac{\mathbf{k}}{2\left| \mathbf{k}\right| ^{3}\left( 1-\widehat{\mathbf{k}}\cdot \nabla E^{\epsilon ^{\frac{j+1}{2}}}\left( \mathbf{P}\right) \right) ^{2}}d^{3}k\right)  \)}{\large }\\
{\large }\\
{\small \( +\int ^{\infty }_{0}\left( \left| \mathbf{k}\right| -\mathbf{k}\cdot \nabla E^{\epsilon ^{\frac{j+1}{2}}}\left( \mathbf{P}\right) \right) b^{\dagger }\left( \mathbf{k}\right) b\left( \mathbf{k}\right) d^{3}k-g^{2}\int _{\epsilon ^{\frac{j+2}{2}}}^{\kappa }\frac{1}{2\left| \mathbf{k}\right| ^{2}\left( 1-\widehat{\mathbf{k}}\cdot \nabla E^{\epsilon ^{\frac{j+1}{2}}}\left( \mathbf{P}\right) \right) }d^{3}k= \)}{\large }\\
{\small }\\
{\small \( =\frac{1}{2m}\left( \Pi _{\mathbf{P},\epsilon ^{\frac{j+1}{2}}}-\frac{\left( \phi _{\mathbf{P}}^{\epsilon ^{\frac{j+1}{2}}},\! \Pi _{\mathbf{P},\epsilon ^{\frac{j+1}{2}}}\! \phi _{\mathbf{P}}^{\epsilon ^{\frac{j+1}{2}}}\right) }{\left\Vert \phi _{\mathbf{P}}^{\epsilon ^{\frac{j+1}{2}}}\right\Vert ^{2}}-g\int _{\epsilon ^{\frac{j+2}{2}}}^{\epsilon ^{\frac{j+1}{2}}}\frac{\mathbf{k}\left( b\left( \mathbf{k}\right) +b^{\dagger }\left( \mathbf{k}\right) \right) }{\sqrt{2}\left| \mathbf{k}\right| ^{\frac{3}{2}}\left( 1-\widehat{\mathbf{k}}\cdot \nabla E^{\epsilon ^{\frac{j+1}{2}}}\left( \mathbf{P}\right) \right) }d^{3}k+g^{2}\int _{\epsilon ^{\frac{j+2}{2}}}^{\epsilon ^{\frac{j+1}{2}}}\frac{\mathbf{k}}{2\left| \mathbf{k}\right| ^{3}\left( 1-\widehat{\mathbf{k}}\cdot \nabla E^{\epsilon ^{\frac{j+1}{2}}}\left( \mathbf{P}\right) \right) ^{2}}d^{3}k\right) ^{2}+ \)}\\
{\small }\\
{\small \( +\int ^{\infty }_{0}\left( \left| \mathbf{k}\right| -\mathbf{k}\cdot \nabla E^{\epsilon ^{\frac{j+1}{2}}}\left( \mathbf{P}\right) \right) b^{\dagger }\left( \mathbf{k}\right) b\left( \mathbf{k}\right) d^{3}k-g^{2}\int _{\epsilon ^{\frac{j+2}{2}}}^{\kappa }\frac{1}{2\left| \mathbf{k}\right| ^{2}\left( 1-\widehat{\mathbf{k}}\cdot \nabla E^{\epsilon ^{\frac{j+1}{2}}}\left( \mathbf{P}\right) \right) }d^{3}k+ \)}\\
{\small }\\
{\small \( -\frac{1}{2m}\left( \frac{1}{\left\Vert \phi _{\mathbf{P}}^{\epsilon ^{\frac{j+1}{2}}}\right\Vert ^{2}}\left( \phi ^{\epsilon ^{\frac{j+1}{2}}}_{\mathbf{P}},\Pi _{\mathbf{P},\epsilon ^{\frac{j+1}{2}}}\phi ^{\epsilon ^{\frac{j+1}{2}}}_{\mathbf{P}}\right) +g^{2}\int _{\epsilon ^{\frac{j+1}{2}}}^{\kappa }\frac{\mathbf{k}}{2\left| \mathbf{k}\right| ^{3}\left( 1-\widehat{\mathbf{k}}\cdot \nabla E^{\epsilon ^{\frac{j+1}{2}}}\left( \mathbf{P}\right) \right) ^{2}}d^{3}k\right) ^{2}+\frac{\mathbf{P}^{2}}{2m} \)}\\
{\small }\\
{\large }\\
{\large }\\
\textbf{\Large APPENDIX B}\\
{\Large \par}

\emph{\large Preliminary remarks}{\large }\\
{\large }\\
{\large In the next lemmas I will consider an implicit hypothesis which is not
proved in the spectral analysis but it is physically reasonable: }\\
{\large for \( \mathbf{P}\in \Sigma  \) , there exists a constant \( \frac{1}{m_{r}} \)
(\( m_{r} \) means renormalized mass) such that the following inequalities
hold, uniformly in \( \sigma >0 \):}\\
\textbf{\large }\\
\textbf{\large hypothesis B1~~~~~~~~~~~~~~~~~~~~~~~\( \frac{\partial E^{\sigma }\left( \mathbf{P}\right) }{\partial \left| \mathbf{P}\right| }\geq \frac{\left| \mathbf{P}\right| }{m_{r}} \)~~~~~}{\large and}
\textbf{\large ~~~~\( \frac{\partial ^{2}E^{\sigma }\left( \mathbf{P}\right) }{\partial ^{2}\left| \mathbf{P}\right| }\geq \frac{1}{m_{r}} \)}{\large }\\
{\large }\\
{\large Starting from this hypothesis, we obtain that the application \( \mathbf{J}_{\sigma }:\mathbf{P}\rightarrow \nabla E^{\sigma }\left( \mathbf{P}\right)  \)
is one to one and that the determinant of jacobian satisfies the inequality:}\textbf{\large }\\
\textbf{\large }\\
\textbf{\large 
\[
det\, \mathbf{dJ}_{\sigma }=\frac{1}{\left| \mathbf{P}\right| ^{2}}\left( \frac{\partial E^{\sigma }\left( \mathbf{P}\right) }{\partial \left| \mathbf{P}\right| }\right) ^{2}\cdot \frac{\partial ^{2}E^{\sigma }\left( \mathbf{P}\right) }{\partial ^{2}\left| \mathbf{P}\right| }\geq \frac{1}{m^{3}_{r}}\]
}{\large }\\
{\large ( I recall that the function \( E^{\sigma }\left( \mathbf{P}\right)  \)
is invariant under rotations and that it belongs to \( C^{\infty }\left( R^{3}\right)  \)
, see {[}2{]}) .}\textbf{\large }\\
{\large Then, given the region \( O_{\nabla E^{\sigma }} \) and the corresponding
\( O_{\mathbf{P}}=\mathbf{J}_{\sigma }^{-1}\left( O_{\nabla E^{\sigma }}\right)  \)
~\( O_{\mathbf{P}}\subset \Sigma  \), we have the following relation between
their volumes:}\\
\textbf{\large }\\
\textbf{\large 
\[
V_{O_{\mathbf{P}}}\leq m^{3}_{r}V_{O_{\nabla E^{\sigma }}}\]
}\\
\textbf{\large }\\
\emph{\large Remark on the notations}\\
{\large }\\
{\large As in the previous chapters, I use the convention to generically call
\( C \) the constants which are uniform in the variables we are treating. The
bounds are intended from above, up to a different explicit warning.}\\
\textbf{\large }\\
\textbf{\large }\\
\textbf{\large Definition}\\
\textbf{\large }\\
{\large As anticipated in the paragraph 4.1, the function \( \chi ^{\left( t\right) }_{\mathbf{v}_{i}}\left( \nabla E^{\sigma _{t}},s\right)  \)
has to approximate the characteristic function of \( \mathbf{J}_{\sigma _{t}}\left( \Gamma _{i}\right)  \)
for \( s\rightarrow +\infty  \) (where \( s \) is bigger or equal to the \( t \)
(\( \gg 1 \)) of the partition, the most general expression is \( \chi ^{\left( t_{1}\right) }_{\mathbf{v}_{i}}\left( \nabla E^{\sigma _{t_{2}}},s\right)  \)
where the constraint is \( s\geq t_{1} \)). In particular, in order to approximate
the region \( \mathbf{J}_{\sigma _{t}}\left( \Gamma _{i}\right)  \) from inside,
I define \( \chi ^{\left( t\right) }_{\mathbf{v}_{i}}\left( \nabla E^{\sigma _{t}},s\right) =\sum _{l}\chi ^{l\left( t\right) }_{\mathbf{v}_{l\left( i\right) }}\left( \nabla E^{\sigma _{t}},s\right)  \)
where \( supp_{\nabla E^{\sigma _{t}}}\chi ^{\left( t\right) }_{\mathbf{v}_{i}}\left( \nabla E^{\sigma _{t}},s\right) \subset supp\mathbf{J}_{\sigma _{t}}\left( \Gamma _{i}\right)  \)
and where the \( \chi ^{l\left( t\right) }_{\mathbf{v}_{l\left( i\right) }}\left( \nabla E^{\sigma _{t}},s\right)  \)
are constructed starting from the ``model'' function}\\
{\large }\\
{\large \( \chi ^{l\left( t\right) }\left( \mathbf{z},s\right) =\chi ^{l\left( t\right) }_{1}\left( z_{1},s\right) \cdot \chi ^{l\left( t\right) }_{2}\left( z_{2},s\right) \cdot \chi ^{l\left( t\right) }_{3}\left( z_{3},s\right)  \)}\\
{\large }\\
{\large \( \chi ^{l\left( t\right) }_{k}\left( z_{k},s\right)  \)}  \( \begin{array}{ccc}
 & -s^{\frac{\delta }{2}}z_{k}+s^{\frac{\delta }{2}}\cdot \frac{1}{2s^{\frac{\delta }{6}}} & for\: -\frac{1}{s^{\frac{\delta }{2}}}+\frac{1}{2s^{\frac{\delta }{6}}}<z_{k}\leq \frac{1}{2s^{\frac{\delta }{6}}}\\
\equiv \left\{ \right.  & 1 & for\: -\frac{1}{2s^{\frac{\delta }{6}}}+\frac{1}{s^{\frac{\delta }{2}}}<z_{k}\leq \frac{1}{2s^{\frac{\delta }{6}}}-\frac{1}{s^{\frac{\delta }{2}}}\\
 & s^{\frac{\delta }{2}}z_{k}+s^{\frac{\delta }{2}}\cdot \frac{1}{2s^{\frac{\delta }{6}}} & for\: -\frac{1}{2s^{\frac{\delta }{6}}}<z_{k}\leq -\frac{1}{2s^{\frac{\delta }{6}}}+\frac{1}{s^{\frac{\delta }{2}}}
\end{array} \){\large }\\
{\large }\\
{\large }\\
{\large through a translation which sends the origin of the coordinates \( \nabla E^{\sigma _{t}} \)
space in \( \mathbf{v}_{l\left( i\right) }\in \mathbf{J}_{\sigma _{t}}\left( \Gamma _{i}\right)  \).
Note that the support of \( \mathbf{J}_{\sigma _{t}}\left( \Gamma _{i}\right)  \)
has a volume of order \( \frac{1}{t^{3\epsilon }} \). In order to have a well
defined \( \chi ^{\left( t\right) }_{\mathbf{v}_{i}}\left( \nabla E^{\sigma _{t}},s\right)  \)
two requirements are necessary:}\\
{\large - the inequality \( \delta >6\epsilon  \) }\\
{\large - a finite scale factor (related to \( m_{r} \)) for the variable \( z_{k} \)
in the function \( \chi ^{l\left( t\right) }_{k}\left( z_{k},s\right)  \).
To simplify the notations I will assume this factor equal to \( 1 \). }\\
{\large It comes out that the \( \widetilde{\chi }^{l\left( t\right) }_{\mathbf{v}_{l\left( i\right) }}\left( \mathbf{q},s\right)  \)
have a behavior similar to}\\
{\large }\\
{\large \( \widetilde{\chi }^{l\left( t\right) }\left( \mathbf{q},s\right) =\widetilde{\chi }^{l\left( t\right) }_{1}\left( q_{1},s\right) \cdot \widetilde{\chi }^{l\left( t\right) }_{2}\left( q_{2},s\right) \cdot \widetilde{\chi }^{l\left( t\right) }_{3}\left( q_{3},s\right)  \)
~~~}\\
{\large where \( \widetilde{\chi }^{l\left( t\right) }_{k}\left( q_{k},s\right) \propto s^{\frac{\delta }{2}}\cdot \frac{\cos \left( q_{k}\left( \frac{1}{2s^{\frac{\delta }{6}}}-\frac{1}{s^{\frac{\delta }{2}}}\right) \right) -\cos \left( q_{k}\cdot \frac{1}{2s^{\frac{\delta }{6}}}\right) }{q^{2}_{k}} \)}\\
{\large }\\
{\large and then}\\
{\large }\\
{\large \( \int \left| \widetilde{\chi }^{l\left( t\right) }\left( \mathbf{q},s\right) \right| d^{3}q<C\cdot s^{3\frac{\delta }{2}} \)~~
e~~ \( \int ^{\infty }_{a}\left| \widetilde{\chi }^{l\left( t\right) }\left( \mathbf{q},s\right) \right| d^{3}q<C\cdot \frac{1}{a}\cdot s^{\frac{\delta }{2}}\cdot s^{\delta } \)
~}\\
{\large }\\
{\large from which}\\
{\large }\\
{\large \( \int \left| \widetilde{\chi }^{\left( t\right) }_{\mathbf{v}_{i}}\left( \mathbf{q},s\right) \right| d^{3}q\leq \sum _{l}\int \left| \widetilde{\chi }^{l\left( t\right) }_{\mathbf{v}_{i}}\left( \mathbf{q},s\right) \right| d^{3}q\leq C\cdot \frac{L^{3}}{t^{3\epsilon }}\cdot s^{3\cdot \frac{\delta }{6}}\cdot s^{3\frac{\delta }{2}}\leq C\cdot s^{2\delta } \)
}\\
{\large }\\
{\large \( \int ^{+\infty }_{a}\left| \widetilde{\chi }^{\left( t\right) }_{\mathbf{v}_{i}}\left( \mathbf{q},s\right) \right| d^{3}q\leq \sum _{l}\int ^{\infty }_{a}\left| \widetilde{\chi }^{l\left( t\right) }_{\mathbf{v}_{l\left( i\right) }}\left( q,s\right) \right| d^{3}q\leq C\cdot \frac{L^{3}}{t^{3\epsilon }}\cdot s^{3\frac{\delta }{6}}\frac{1}{a}\cdot s^{\frac{3\delta }{2}}<C\cdot \frac{1}{a}\cdot s^{2\delta } \)
}\\
\textbf{\large }\\
\textbf{\large }\\
\textbf{\large }\\
\textbf{\large Lemma B1}{\large }\\
{\large }\\
{\large The norm \( \left\Vert \left( 1_{\Gamma _{i}}\left( \mathbf{P}\right) -\chi ^{\left( t\right) }_{\mathbf{v}_{i}}\left( \nabla E^{\sigma _{t}}\left( \mathbf{P}\right) ,s\right) \right) \psi _{i,\sigma _{t}}^{\left( t\right) }\right\Vert  \)
is bounded by a quantity of order \( \frac{1}{s^{\frac{\delta }{12}}}\cdot t^{-\epsilon } \).}\\
{\large }\\
{\large Proof}\\
{\large }\\
{\large I define \( \mathbf{J}^{-1}_{\sigma _{t}}\left( \widehat{O}^{i}_{\nabla E^{\sigma _{t}}}\right) \equiv \left\{ \mathbf{P}\in \Sigma :\nabla E^{\sigma _{t}}\left( \mathbf{P}\right) \in supp\chi ^{\left( t\right) }_{\mathbf{v}_{i}}\left( \nabla E^{\sigma _{t}},s\right) \, and\, \chi ^{\left( t\right) }_{\mathbf{v}_{i}}\left( \nabla E^{\sigma _{t}},s\right) \neq 1\right\}  \).
Taking into account the definition of \( \chi ^{\left( t\right) }_{\mathbf{v}_{i}}\left( \nabla E^{\sigma _{t}},s\right)  \),
the definition of the application \( \mathbf{J}_{\sigma _{t}} \) (\( \mathbf{J}_{\sigma _{t}}\left( \mathbf{P}\right) \propto \mathbf{P} \))
and the hypothesis B1, the volume \( \mathbf{J}^{-1}_{\sigma _{t}}\left( \widehat{O}^{i}_{\nabla E^{\sigma _{t}}}\right)  \)
is bounded by a quantity of order \( \frac{1}{s^{\frac{\delta }{3}}}\cdot t^{-3\epsilon } \).
On the other hand, the volume of the region \( supp\mathbf{J}_{\sigma _{t}}\left( \Gamma _{i}\right) \setminus supp_{\nabla E^{\sigma _{t}}}\chi ^{\left( t\right) }_{\mathbf{v}_{i}}\left( \nabla E^{\sigma _{t}},s\right)  \)
is bounded by a quantity of order \( \frac{1}{s^{\frac{\delta }{6}}}\cdot t^{-2\epsilon } \).
}\\
{\large Therefore:}\\
{\large }\\
{\large \( \left\Vert \left( 1_{\Gamma _{i}}\left( \mathbf{P}\right) -\chi ^{\left( t\right) }_{\mathbf{v}_{i}}\left( \nabla E^{\sigma _{t}}\left( \mathbf{P}\right) ,s\right) \right) \psi _{i,\sigma _{t}}^{\left( t\right) }\right\Vert ^{2}=\int _{\Gamma _{i}}\left| 1_{\Gamma _{i}}\left( \mathbf{P}\right) -\chi ^{\left( t\right) }_{\mathbf{v}_{i}}\left( \nabla E^{\sigma _{t}},s\right) \right| ^{2}\left| G\left( \mathbf{P}\right) \right| ^{2}\left\Vert \psi _{\mathbf{P},\sigma _{t}}\right\Vert ^{2}d^{3}P\leq  \)}\\
{\large }\\
{\large \( \leq C\cdot \frac{1}{s^{\frac{\delta }{6}}}\cdot t^{-2\epsilon } \)}\\
{\large }\\
{\large from which the thesis follows.}\\
{\large }\\
\textbf{\large Lemma B2}{\large }\\
{\large }\\
{\large In the constructive hypothesis fixed at the beginning of chapter 4,
we have}\\
{\large }\\
{\large \( \left\Vert \int \widetilde{\chi }^{\left( t\right) }_{\mathbf{v}_{i}}\left( \mathbf{q},s\right) \left( e^{-i\mathbf{q}\cdot \nabla E^{\sigma _{t}}}-e^{-i\mathbf{q}\cdot \frac{\mathbf{x}}{s}}\right) d^{3}qe^{-iE^{\sigma _{t}}\left( \mathbf{P}\right) s}e^{i\gamma _{\sigma _{t}}\left( \mathbf{v}_{i},\nabla E^{\sigma _{t}},s\right) }\psi _{i,\sigma _{t}}^{\left( t\right) }\right\Vert \leq  \)}\\
{\large }\\
{\large \( \leq C\cdot s^{-\frac{1}{112}}\cdot s^{2\delta -\frac{3\epsilon }{2}}\cdot \left( \ln \left( \sigma _{t}\right) \right) ^{2} \)}\\
{\large }\\
{\large (\( C \) is the same constant for all the cells and it is uniform in
the partitions).}\\
{\large }\\
{\large Proof}\\
{\large }\\
{\large \( \left\Vert \int \widetilde{\chi }^{\left( t\right) }_{\mathbf{v}_{i}}\left( \mathbf{q},s\right) \left( e^{-i\mathbf{q}\cdot \nabla E^{\sigma _{t}}}-e^{-i\mathbf{q}\cdot \frac{\mathbf{x}}{s}}\right) d^{3}qe^{-iE^{\sigma _{t}}\left( \mathbf{P}\right) s}e^{i\gamma _{\sigma _{t}}\left( \mathbf{v}_{i},\nabla E^{\sigma _{t}},s\right) }\psi _{i,\sigma _{t}}^{\left( t\right) }\right\Vert = \)}\\
{\large }\\
{\large \( =\left\Vert e^{iE^{\sigma _{t}}\left( \mathbf{P}\right) s}\int \widetilde{\chi }^{\left( t\right) }_{\mathbf{v}_{i}}\left( \mathbf{q},s\right) \left( e^{-i\mathbf{q}\cdot \nabla E^{\sigma _{t}}}-e^{-i\mathbf{q}\cdot \frac{\mathbf{x}}{s}}\right) d^{3}qe^{-iE^{\sigma _{t}}\left( \mathbf{P}\right) s}e^{i\gamma _{\sigma _{t}}\left( \mathbf{v}_{i},\nabla E^{\sigma _{t}},s\right) }\psi _{i,\sigma _{t}}^{\left( t\right) }\right\Vert = \)}\\
{\large }\\
{\small \( =\left\Vert \int \widetilde{\chi }^{\left( t\right) }_{\mathbf{v}_{i}}\left( \mathbf{q},s\right) \left[ \left( e^{-i\mathbf{q}\cdot \nabla E^{\sigma _{t}}}-e^{i\left( E^{\sigma _{t}}\left( \mathbf{P}\right) -E^{\sigma _{t}}\left( \mathbf{P}+\frac{\mathbf{q}}{s}\right) \right) s}-e^{i\left( E^{\sigma _{t}}\left( \mathbf{P}\right) -E^{\sigma _{t}}\left( \mathbf{P}+\frac{\mathbf{q}}{s}\right) \right) s}\left( e^{-i\mathbf{q}\cdot \frac{\mathbf{x}}{s}}-1\right) \right) \right] d^{3}qe^{i\gamma _{\sigma _{t}}\left( \mathbf{v}_{i},\nabla E^{\sigma _{t}},s\right) }\psi _{i,\sigma _{t}}^{\left( t\right) }\right\Vert \leq  \)}{\large }\\
{\large }\\
{\large \( \leq \left\Vert \int \widetilde{\chi }^{\left( t\right) }_{\mathbf{v}_{i}}\left( \mathbf{q},s\right) \left( e^{-i\mathbf{q}\cdot \nabla E^{\sigma _{t}}}-e^{i\left( E^{\sigma _{t}}\left( \mathbf{P}\right) -E^{\sigma _{t}}\left( \mathbf{P}+\frac{\mathbf{q}}{s}\right) \right) s}\right) d^{3}qe^{i\gamma _{\sigma _{t}}\left( \mathbf{v}_{i},\nabla E^{\sigma _{t}},s\right) }\psi _{i,\sigma _{t}}^{\left( t\right) }\right\Vert + \)}\marginpar{
\textbf{\large i)}{\large \par}
}{\large }\\
{\large }\\
{\large \( +\left\Vert \int \widetilde{\chi }^{\left( t\right) }_{\mathbf{v}_{i}}\left( \mathbf{q},s\right) e^{i\left( E^{\sigma _{t}}\left( \mathbf{P}\right) -E^{\sigma _{t}}\left( \mathbf{P}+\frac{\mathbf{q}}{s}\right) \right) s}\left( e^{-i\mathbf{q}\cdot \frac{\mathbf{x}}{s}}-1\right) d^{3}qe^{i\gamma _{\sigma _{t}}\left( \mathbf{v}_{i},\nabla E^{\sigma _{t}},s\right) }\psi _{i,\sigma _{t}}^{\left( t\right) }\right\Vert  \)}\marginpar{
\textbf{\large ii)}{\large \par}
}\textbf{\large }\\
\textbf{\large }\\
\textbf{\large }\\
\textbf{\large }\\
\textbf{\large i)} {\large \( \left\Vert \int \widetilde{\chi }^{\left( t\right) }_{\mathbf{v}_{i}}\left( \mathbf{q},s\right) \left( e^{-i\mathbf{q}\cdot \nabla E^{\sigma _{t}}}-e^{i\left( E^{\sigma _{t}}\left( \mathbf{P}\right) -E^{\sigma _{t}}\left( \mathbf{P}+\frac{\mathbf{q}}{s}\right) \right) s}\right) d^{3}qe^{i\gamma _{\sigma _{t}}\left( \mathbf{v}_{i},\nabla E^{\sigma _{t}},s\right) }\psi _{i,\sigma _{t}}^{\left( t\right) }\right\Vert \leq  \)}\\
{\large \( \leq C\cdot \left\Vert \psi _{i,\sigma }^{\left( t\right) }\right\Vert \cdot \int ^{+\infty }_{s^{\frac{1}{34}}}\left| \widetilde{\chi }^{\left( t\right) }_{\mathbf{v}_{i}}\left( \mathbf{q},s\right) \right| d^{3}q+C\cdot \left\Vert \psi _{i,\sigma }^{\left( t\right) }\right\Vert \cdot \frac{\int ^{s^{\frac{1}{34}}}_{0}\left| \widetilde{\chi }^{\left( t\right) }_{\mathbf{v}_{i}}\left( \mathbf{q},s\right) \right| d^{3}q}{s^{\frac{1}{32}}} \)}\\
{\large }\\
 {\large It follows for these reasons:}\\
{\large \par}

\begin{itemize}
\item {\large being the energy differentiable: }\\
{\large \( sE^{\sigma _{t}}\left( \mathbf{P}\right) -sE^{\sigma _{t}}\left( \mathbf{P}+\frac{\mathbf{q}}{s}\right) =-\mathbf{q}\cdot \nabla E^{\sigma _{t}}\left( \mathbf{P}'\right)  \)
where \( \mathbf{P}' \) is such that \( \left| \mathbf{P}-\mathbf{P}'\right| \leq \left| \frac{\mathbf{q}}{s}\right|  \)}{\large \par}
\item {\large for the lemma 3.3: }\\
{\large \( \left| \nabla E^{\sigma _{t}}\left( \mathbf{P}\right) -\nabla E^{\sigma _{t}}\left( \mathbf{P}'\right) \right| \leq C\cdot \left| \mathbf{P}-\mathbf{P}'\right| ^{\frac{1}{16}}\leq C\cdot \left| \frac{\mathbf{q}}{s}\right| ^{\frac{1}{16}} \)}\\
{\large \par}
\item {\large \( \int \left| \widetilde{\chi }^{\left( t\right) }_{\mathbf{v}_{i}}\left( \mathbf{q},s\right) \right| d^{3}q\leq C\cdot s^{2\delta } \)
}\\
{\large }\\
{\large \( \int ^{+\infty }_{a}\left| \widetilde{\chi }^{\left( t\right) }_{\mathbf{v}_{i}}\left( \mathbf{q},s\right) \right| d^{3}q<C\cdot \frac{1}{a}\cdot s^{2\delta } \)
}\\
{\large \par}
\end{itemize}
{\large Then the term} \textbf{\large i)} {\large is surely bounded by a quantity
of order \( \frac{s^{2\delta }}{s^{\frac{1}{34}}} \).} \textbf{\large }\\
\textbf{\large }\\
\textbf{\large ii)} {\large }\\
{\large }\\
{\large \( \left\Vert \int \widetilde{\chi }^{\left( t\right) }_{\mathbf{v}_{i}}\left( \mathbf{q},s\right) e^{i\left( E^{\sigma _{t}}\left( \mathbf{P}\right) -E^{\sigma _{t}}\left( \mathbf{P}+\frac{\mathbf{q}}{s}\right) \right) s}\cdot \left( e^{-i\mathbf{q}\cdot \frac{\mathbf{x}}{s}}-1\right) d^{3}qe^{i\gamma _{\sigma _{t}}\left( \mathbf{v}_{i},\nabla E^{\sigma _{t}},s\right) }\psi _{i,\sigma _{t}}^{\left( t\right) }\right\Vert \leq  \)}\\
{\large }\\
{\large \( \leq \left\Vert \int ^{s^{\frac{1}{20}}}_{0}\widetilde{\chi }^{\left( t\right) }_{\mathbf{v}_{i}}\left( \mathbf{q},s\right) e^{i\left( E^{\sigma _{t}}\left( \mathbf{P}\right) -E^{\sigma _{t}}\left( \mathbf{P}+\frac{\mathbf{q}}{s}\right) \right) s}\cdot \left( e^{-i\mathbf{q}\cdot \frac{\mathbf{x}}{s}}-1\right) d^{3}qe^{i\gamma _{\sigma _{t}}\left( \mathbf{v}_{i},\nabla E^{\sigma _{t}},s\right) }\psi _{i,\sigma _{t}}^{\left( t\right) }\right\Vert + \)}\\
\marginpar{
{\large (b2)}{\large \par}
}{\large }\\
\emph{\large \( +\left\Vert \int ^{+\infty }_{s^{\frac{1}{20}}}\widetilde{\chi }^{\left( t\right) }_{\mathbf{v}_{i}}\left( \mathbf{q},s\right) e^{i\left( E^{\sigma _{t}}\left( \mathbf{P}\right) -E^{\sigma _{t}}\left( \mathbf{P}+\frac{\mathbf{q}}{s}\right) \right) s}\cdot \left( e^{-i\mathbf{q}\cdot \frac{\mathbf{x}}{s}}-1\right) d^{3}qe^{i\gamma _{\sigma _{t}}\left( \mathbf{v}_{i},\nabla E^{\sigma _{t}},s\right) }\psi _{i,\sigma _{t}}^{\left( t\right) }\right\Vert  \)}\\
\emph{\large }\\
\emph{\large first term of (b2)} {\large }\\
{\large }\\
{\large \( \left\Vert \int ^{s^{\frac{1}{20}}}_{0}\widetilde{\chi }^{\left( t\right) }_{\mathbf{v}_{i}}\left( \mathbf{q},s\right) e^{i\left( E^{\sigma _{t}}\left( \mathbf{P}\right) -E^{\sigma _{t}}\left( \mathbf{P}+\frac{\mathbf{q}}{s}\right) \right) s}\cdot \left( e^{-i\mathbf{q}\cdot \frac{\mathbf{x}}{s}}-1\right) d^{3}qe^{i\gamma _{\sigma _{t}}\left( \mathbf{v}_{i},\nabla E_{\mathbf{P}}^{\sigma _{t}},s\right) }\psi _{i,\sigma _{t}}^{\left( t\right) }\right\Vert = \)}\\
{\large }\\
{\large \( =\left\Vert \int ^{s^{\frac{1}{20}}}_{0}\widetilde{\chi }^{\left( t\right) }_{\mathbf{v}_{i}}\left( \mathbf{q},s\right) \int _{\Gamma _{i}}e^{i\left( E^{\sigma _{t}}\left( \mathbf{P}\right) -E^{\sigma _{t}}\left( \mathbf{P}+\frac{\mathbf{q}}{s}\right) \right) s}\cdot \left( e^{-i\mathbf{q}\cdot \frac{\mathbf{x}}{s}}-1\right) G\left( \mathbf{P}\right) e^{i\gamma _{\sigma _{t}}\left( \mathbf{v}_{i},\nabla E_{\mathbf{P}}^{\sigma _{t}},s\right) }\psi _{\mathbf{P},\sigma _{t}}d^{3}Pd^{3}q\right\Vert = \)}\\
{\large }\\
{\large \( =\left\Vert \int ^{s^{\frac{1}{20}}}_{0}\widetilde{\chi }^{\left( t\right) }_{\mathbf{v}_{i}}\left( \mathbf{q},s\right) \int _{\Gamma _{i}}e^{i\left( E^{\sigma _{t}}\left( \mathbf{P}\right) -E^{\sigma _{t}}\left( \mathbf{P}+\frac{\mathbf{q}}{s}\right) \right) s}e^{-i\mathbf{q}\cdot \frac{\mathbf{x}}{s}}G\left( \mathbf{P}\right) e^{i\gamma _{\sigma _{t}}\left( \mathbf{v}_{i},\nabla E_{\mathbf{P}}^{\sigma _{t}},s\right) }\psi _{\mathbf{P},\sigma _{t}}d^{3}Pd^{3}q+\right.  \)}\\
{\large }\\
{\large \( \left. -\int ^{s^{\frac{1}{20}}}_{0}\widetilde{\chi }^{\left( t\right) }_{\mathbf{v}_{i}}\left( \mathbf{q},s\right) \int _{\Gamma _{i}}e^{i\left( E^{\sigma _{t}}\left( \mathbf{P}\right) -E^{\sigma _{t}}\left( \mathbf{P}+\frac{\mathbf{q}}{s}\right) \right) s}G\left( \mathbf{P}\right) e^{i\gamma _{\sigma _{t}}\left( \mathbf{v}_{i},\nabla E_{\mathbf{P}}^{\sigma _{t}},s\right) }\psi _{\mathbf{P},\sigma _{t}}d^{3}Pd^{3}q\right\Vert = \)}
{\large }\\
{\large }\\
{\large \( =\left\Vert \int ^{s^{\frac{1}{20}}}_{0}\widetilde{\chi }^{\left( t\right) }_{\mathbf{v}_{i}}\left( \mathbf{q},s\right) \int _{\Gamma _{i}}e^{i\left( E^{\sigma _{t}}\left( \mathbf{P}\right) -E^{\sigma _{t}}\left( \mathbf{P}+\frac{\mathbf{q}}{s}\right) \right) s}e^{-i\mathbf{q}\cdot \frac{\mathbf{x}}{s}}G\left( \mathbf{P}\right) e^{i\gamma _{\sigma _{t}}\left( \mathbf{v}_{i},\nabla E_{\mathbf{P}}^{\sigma _{t}},s\right) }\psi _{\mathbf{P},\sigma _{t}}d^{3}Pd^{3}q+\right.  \)}\\
{\large }\\
{\large \( -\int ^{s^{\frac{1}{20}}}_{0}\widetilde{\chi }^{\left( t\right) }_{\mathbf{v}_{i}}\left( \mathbf{q},s\right) \int _{\Gamma _{i}}e^{i\left( E^{\sigma _{t}}\left( \mathbf{P}\right) -E^{\sigma _{t}}\left( \mathbf{P}+\frac{\mathbf{q}}{s}\right) \right) s}G\left( \mathbf{P}\right) e^{i\gamma _{\sigma _{t}}\left( \mathbf{v}_{i},\nabla E_{\mathbf{P}-\frac{\mathbf{q}}{s}}^{\sigma _{t}},s\right) }\psi _{\mathbf{P}-\frac{\mathbf{q}}{s},\sigma _{t}}d^{3}Pd^{3}q+ \)}\\
{\large }\\
{\large \( +\int ^{s^{\frac{1}{20}}}_{0}\widetilde{\chi }^{\left( t\right) }_{\mathbf{v}_{i}}\left( \mathbf{q},s\right) \int _{\Gamma _{i}}e^{i\left( E^{\sigma _{t}}\left( \mathbf{P}\right) -E^{\sigma _{t}}\left( \mathbf{P}+\frac{\mathbf{q}}{s}\right) \right) s}G\left( \mathbf{P}\right) e^{i\gamma _{\sigma _{t}}\left( \mathbf{v}_{i},\nabla E_{\mathbf{P}-\frac{\mathbf{q}}{s}}^{\sigma _{t}},s\right) }\psi _{\mathbf{P}-\frac{\mathbf{q}}{s},\sigma _{t}}d^{3}Pd^{3}q+ \)}\\
{\large }\\
{\large \( -\int ^{s^{\frac{1}{20}}}_{0}\widetilde{\chi }^{\left( t\right) }_{\mathbf{v}_{i}}\left( \mathbf{q},s\right) \int _{\Gamma _{i}}e^{i\left( E^{\sigma _{t}}\left( \mathbf{P}-\frac{\mathbf{q}}{s}\right) -E^{\sigma _{t}}\left( \mathbf{P}\right) \right) s}G\left( \mathbf{P}-\frac{\mathbf{q}}{s}\right) e^{i\gamma _{\sigma _{t}}\left( \mathbf{v}_{i},\nabla E_{\mathbf{P}-\frac{\mathbf{q}}{s}}^{\sigma _{t}},s\right) }\psi _{\mathbf{P}-\frac{\mathbf{q}}{s},\sigma _{t}}d^{3}Pd^{3}q+ \)}\\
{\large }\\
{\large \( +\int ^{s^{\frac{1}{20}}}_{0}\widetilde{\chi }^{\left( t\right) }_{\mathbf{v}_{i}}\left( \mathbf{q},s\right) \int _{\Gamma _{i}}e^{i\left( E^{\sigma _{t}}\left( \mathbf{P}-\frac{\mathbf{q}}{s}\right) -E^{\sigma _{t}}\left( \mathbf{P}\right) \right) s}G\left( \mathbf{P}-\frac{\mathbf{q}}{s}\right) e^{i\gamma _{\sigma _{t}}\left( \mathbf{v}_{i},\nabla E_{\mathbf{P}-\frac{\mathbf{q}}{s}}^{\sigma _{t}},s\right) }\psi _{\mathbf{P}-\frac{\mathbf{q}}{s},\sigma _{t}}d^{3}Pd^{3}q+ \)}\\
{\large }\\
{\large \( \left. -\int ^{s^{\frac{1}{20}}}_{0}\widetilde{\chi }^{\left( t\right) }_{\mathbf{v}_{i}}\left( \mathbf{q},s\right) \int _{\Gamma _{i}}e^{i\left( E^{\sigma _{t}}\left( \mathbf{P}\right) -E^{\sigma _{t}}\left( \mathbf{P}+\frac{\mathbf{q}}{s}\right) \right) s}G\left( \mathbf{P}\right) e^{i\gamma _{\sigma _{t}}\left( \mathbf{v}_{i},\nabla E_{\mathbf{P}}^{\sigma _{t}},s\right) }\psi _{\mathbf{P},\sigma _{t}}d^{3}Pd^{3}q\right\Vert \leq  \)}\\
{\large }\\
\( \leq \int ^{s^{\frac{1}{20}}}_{0}\left| \widetilde{\chi }^{\left( t\right) }_{\mathbf{v}_{i}}\left( \mathbf{q},s\right) \right| \left\{ \int _{\Gamma _{i}}\left| G\left( \mathbf{P}\right) \right| ^{2}\left\Vert e^{i\gamma _{\sigma _{t}}\left( \mathbf{v}_{i},\nabla E_{\mathbf{P}}^{\sigma _{t}},s\right) }I_{\mathbf{P}}\left( \psi _{\mathbf{P},\sigma _{t}}\right) -e^{i\gamma _{\sigma _{t}}\left( \mathbf{v}_{i},\nabla E_{\mathbf{P}-\frac{\mathbf{q}}{s}}^{\sigma _{t}},s\right) }I_{\mathbf{P}-\frac{\mathbf{q}}{s}}\left( \psi _{\mathbf{P}-\frac{\mathbf{q}}{s},\sigma _{t}}\right) \right\Vert ^{2}_{F}d^{3}P\right\} ^{\frac{1}{2}}d^{3}q+ \){\small }\\
{\small }\\
\marginpar{
{\large (b3.1)}{\large \par}
} {\large }\\
{\small \( +\int ^{s^{\frac{1}{20}}}_{0}\left| \widetilde{\chi }^{\left( t\right) }_{\mathbf{v}_{i}}\left( \mathbf{q},s\right) \right| \left\{ \int _{\Gamma _{i}}\left| G\left( \mathbf{P}-\frac{\mathbf{q}}{s}\right) e^{i\left( E^{\sigma _{t}}\left( \mathbf{P}-\frac{\mathbf{q}}{s}\right) -E^{\sigma _{t}}\left( \mathbf{P}\right) \right) s}-G\left( \mathbf{P}\right) e^{i\left( E^{\sigma _{t}}\left( \mathbf{P}\right) -E^{\sigma _{t}}\left( \mathbf{P}+\frac{\mathbf{q}}{s}\right) \right) s}\right| ^{2}\left\Vert I_{\mathbf{P}-\frac{\mathbf{q}}{s}}\left( \psi _{\mathbf{P}-\frac{\mathbf{q}}{s},\sigma _{t}}\right) \right\Vert _{F}^{2}d^{3}P\right\} ^{\frac{1}{2}}d^{3}q+ \)}\\
\\
\marginpar{
{\large (b3.2)}{\large \par}
}{\large }\\
\( +\int ^{s^{\frac{1}{20}}}_{0}\left| \widetilde{\chi }^{\left( t\right) }_{\mathbf{v}_{i}}\left( \mathbf{q},s\right) \right| \left\{ \int _{O_{\frac{\mathbf{q}}{s}}}\left| G\left( \mathbf{P}\right) \right| ^{2}\left\Vert I_{\mathbf{P}}\left( \psi _{\mathbf{P},\sigma _{t}}\right) \right\Vert _{F}^{2}d^{3}P\right\} ^{\frac{1}{2}}d^{3}q \)\\
\marginpar{
{\large (b3.3)}{\large \par}
}{\large }\\
{\large }\\
\emph{\large bound of the term (b3.1)} {\large }\\
{\large }\\
{\large \( \left\Vert e^{i\gamma _{\sigma _{t}}\left( \mathbf{v}_{i},\nabla E_{\mathbf{P}}^{\sigma _{t}},s\right) }I_{\mathbf{P}}\left( \psi _{\mathbf{P},\sigma _{t}}\right) -e^{i\gamma _{\sigma _{t}}\left( \mathbf{v}_{i},\nabla E_{\mathbf{P}-\frac{\mathbf{q}}{s}}^{\sigma _{t}},s\right) }I_{\mathbf{P}-\frac{\mathbf{q}}{s}}\left( \psi _{\mathbf{P}-\frac{\mathbf{q}}{s},\sigma _{t}}\right) \right\Vert _{F}\leq  \)}\\
{\large }\\
{\large \( \leq \left\Vert e^{i\gamma _{\sigma _{t}}\left( \mathbf{v}_{i},\nabla E_{\mathbf{P}}^{\sigma _{t}},s\right) }I_{\mathbf{P}}\left( \psi _{\mathbf{P},\sigma _{t}}\right) -e^{i\gamma _{\sigma _{t}}\left( \mathbf{v}_{i},\nabla E_{\mathbf{P}}^{\sigma _{t}},s\right) }I_{\mathbf{P}-\frac{\mathbf{q}}{s}}\left( \psi _{\mathbf{P}-\frac{\mathbf{q}}{s},\sigma _{t}}\right) \right\Vert _{F}+ \)}\\
{\large }\\
{\large \( +\left\Vert e^{i\gamma _{\sigma _{t}}\left( \mathbf{v}_{i},\nabla E_{\mathbf{P}}^{\sigma _{t}},s\right) }I_{\mathbf{P}-\frac{\mathbf{q}}{s}}\left( \psi _{\mathbf{P}-\frac{\mathbf{q}}{s},\sigma _{t}}\right) -e^{i\gamma _{\sigma _{t}}\left( \mathbf{v}_{i},\nabla E_{\mathbf{P}-\frac{\mathbf{q}}{s}}^{\sigma _{t}},s\right) }I_{\mathbf{P}-\frac{\mathbf{q}}{s}}\left( \psi _{\mathbf{P}-\frac{\mathbf{q}}{s},\sigma _{t}}\right) \right\Vert _{F}\leq  \)
\(  \)}\\
{\large }\\
{\large \( \leq \left\Vert I_{\mathbf{P}}\left( W_{\sigma _{t}}^{b^{\dagger }}\left( \nabla E_{\mathbf{P}}^{\sigma _{t}}\right) W_{\sigma _{t}}^{b}\left( \nabla E_{\mathbf{P}}^{\sigma _{t}}\right) \psi _{\mathbf{P},\sigma _{t}}\right) -I_{\mathbf{P}-\frac{\mathbf{q}}{s}}\left( W_{\sigma _{t}}^{b^{\dagger }}\left( \nabla E_{\mathbf{P}-\frac{\mathbf{q}}{s}}^{\sigma _{t}}\right) W_{\sigma _{t}}^{b}\left( \nabla E_{\mathbf{P}-\frac{\mathbf{q}}{s}}^{\sigma _{t}}\right) \psi _{\mathbf{P}-\frac{\mathbf{q}}{s},\sigma _{t}}\right) \right\Vert _{F}+ \)}\\
{\large }\\
{\large \( +\left| e^{i\gamma _{\sigma _{t}}\left( \mathbf{v}_{i},\nabla E_{\mathbf{P}}^{\sigma _{t}},s\right) }-e^{i\gamma _{\sigma _{t}}\left( \mathbf{v}_{i},\nabla E_{\mathbf{P}-\frac{\mathbf{q}}{s}}^{\sigma _{t}},s\right) }\right| \left\Vert I_{\mathbf{P}-\frac{\mathbf{q}}{s}}\left( \psi _{\mathbf{P}-\frac{\mathbf{q}}{s},\sigma _{t}}\right) \right\Vert _{F}\leq  \)}\\
{\large }\\
{\large \( \leq \left\Vert I_{\mathbf{P}}\left( W_{\sigma _{t}}^{b}\left( \nabla E_{\mathbf{P}}^{\sigma _{t}}\right) \psi _{\mathbf{P},\sigma _{t}}\right) -I_{\mathbf{P}-\frac{\mathbf{q}}{s}}\left( W_{\sigma _{t}}^{b}\left( \nabla E_{\mathbf{P}-\frac{\mathbf{q}}{s}}^{\sigma _{t}}\right) \psi _{\mathbf{P}-\frac{\mathbf{q}}{s},\sigma _{t}}\right) \right\Vert _{F}+ \)}\\
{\large }\\
{\large \( +\left\Vert I_{\mathbf{P}-\frac{\mathbf{q}}{s}}\left( \left( W_{\sigma _{t}}^{b^{\dagger }}\left( \nabla E_{\mathbf{P}}^{\sigma _{t}}\right) -W_{\sigma _{t}}^{b^{\dagger }}\left( \nabla E_{\mathbf{P}-\frac{\mathbf{q}}{s}}^{\sigma _{t}}\right) \right) W_{\sigma _{t}}^{b}\left( \nabla E_{\mathbf{P}-\frac{\mathbf{q}}{s}}^{\sigma _{t}}\right) \psi _{\mathbf{P}-\frac{\mathbf{q}}{s},\sigma _{t}}\right) \right\Vert _{F}+ \)}\\
{\large }\\
{\large \( +\left| e^{i\gamma _{\sigma _{t}}\left( \mathbf{v}_{i},\nabla E_{\mathbf{P}}^{\sigma _{t}},s\right) }-e^{i\gamma _{\sigma _{t}}\left( \mathbf{v}_{i},\nabla E_{\mathbf{P}-\frac{\mathbf{q}}{s}}^{\sigma _{t}},s\right) }\right| \left\Vert I_{\mathbf{P}-\frac{\mathbf{q}}{s}}\left( \psi _{\mathbf{P}-\frac{\mathbf{q}}{s},\sigma _{t}}\right) \right\Vert _{F} \)}\\
{\large }\\
{\large }\\
{\large The norm \( \left\Vert I_{\mathbf{P}}\left( W_{\sigma _{t}}^{b}\left( \nabla E_{\mathbf{P}}^{\sigma _{t}}\right) \psi _{\mathbf{P},\sigma _{t}}\right) -I_{\mathbf{P}-\frac{\mathbf{q}}{s}}\left( W_{\sigma _{t}}^{b}\left( \nabla E_{\mathbf{P}-\frac{\mathbf{q}}{s}}^{\sigma _{t}}\right) \psi _{\mathbf{P}-\frac{\mathbf{q}}{s},\sigma _{t}}\right) \right\Vert _{F} \)is
bounded by a quantity of order \( \left( \frac{\left| \mathbf{q}\right| }{s}\right) ^{\frac{1}{32}} \)
(see theorem 3.4).}\\
{\large }\\
{\large }\\
{\large }\\
{\large The norm of}\\
{\large }\\
{\large \( \left\Vert I_{\mathbf{P}-\frac{\mathbf{q}}{s}}\left( \left( W_{\sigma _{t}}^{b^{\dagger }}\left( \nabla E_{\mathbf{P}}^{\sigma _{t}}\right) -W_{\sigma _{t}}^{b^{\dagger }}\left( \nabla E_{\mathbf{P}-\frac{\mathbf{q}}{s}}^{\sigma _{t}}\right) \right) W_{\sigma _{t}}^{b}\left( \nabla E_{\mathbf{P}-\frac{\mathbf{q}}{s}}^{\sigma _{t}}\right) \psi _{\mathbf{P}-\frac{\mathbf{q}}{s},\sigma _{t}}\right) \right\Vert _{F} \)}\marginpar{
{\large (b4)}{\large \par}
}{\large }\\
{\large }\\
{\large can be estimated by the norm of}{\small }\\
{\small }\\
{\small \( g\int ^{\kappa }_{\sigma _{t}}\frac{\widehat{\mathbf{k}}\cdot \left( \nabla E^{\sigma _{t}}\left( \mathbf{P}\right) -\nabla E^{\sigma _{t}}\left( \mathbf{P}-\frac{\mathbf{q}}{s}\right) \right) }{\left| \mathbf{k}\right| \left( 1-\widehat{\mathbf{k}}\cdot \nabla E^{\sigma _{t}}\left( \mathbf{P}\right) \right) \left( 1-\widehat{\mathbf{k}}\cdot \nabla E^{\sigma _{t}}\left( \mathbf{P}-\frac{\mathbf{q}}{s}\right) \right) }\left( b\left( \mathbf{k}\right) -b^{\dagger }\left( \mathbf{k}\right) \right) \frac{d^{3}k}{\sqrt{2\left| \mathbf{k}\right| }}I_{\mathbf{P}-\frac{\mathbf{q}}{s}}\left( W^{b}_{\sigma _{t}}\left( \nabla E^{\sigma _{t}}\left( \mathbf{P}-\frac{\mathbf{q}}{s}\right) \right) \psi _{\mathbf{P}-\frac{\mathbf{q}}{s},\sigma _{t}}\right)  \)}
{\large }\\
{\large then it is substantially the product of the following quantities:}{\large \par}

\begin{itemize}
\item {\large \( \left( \int ^{\kappa }_{\sigma _{t}}\left( \frac{g\widehat{\mathbf{k}}\cdot \left( \nabla E^{\sigma _{t}}\left( \mathbf{P}\right) -\nabla E^{\sigma _{t}}\left( \mathbf{P}-\frac{\mathbf{q}}{s}\right) \right) }{\sqrt{2}\left| \mathbf{k}\right| ^{\frac{3}{2}}\left( 1-\widehat{\mathbf{k}}\cdot \nabla E^{\sigma _{t}}\left( \mathbf{P}\right) \right) \left( 1-\widehat{\mathbf{k}}\cdot \nabla E^{\sigma _{t}}\left( \mathbf{P}-\frac{\mathbf{q}}{s}\right) \right) }\right) ^{2}d^{3}k\right) ^{\frac{1}{2}} \)}\\
{\large }\\
{\large it is bounded by \( C\cdot \left| \nabla E^{\sigma _{t}}\left( \mathbf{P}-\frac{\mathbf{q}}{s}\right) -\nabla E^{\sigma _{t}}\left( \mathbf{P}\right) \right| \cdot \left( \ln \left( \sigma _{t}\right) \right) ^{\frac{1}{2}}\leq  \)}\\
{\large \( \leq C\cdot \left| s^{-\frac{19}{20}}\right| ^{\frac{1}{16}}\cdot \left( \ln \left( \sigma _{t}\right) \right) ^{\frac{1}{2}} \)
( see lemma 3.3);}\\
{\large \par}
\item {\large \( \left( \int ^{\kappa }_{\sigma _{t}}\left\Vert b\left( \mathbf{k}\right) I_{\mathbf{P}-\frac{\mathbf{q}}{s}}\left( W^{b}_{\sigma _{t}}\left( \nabla E^{\sigma _{t}}\left( \mathbf{P}-\frac{\mathbf{q}}{s}\right) \right) \psi _{\mathbf{P}-\frac{\mathbf{q}}{s},\sigma _{t}}\right) \right\Vert _{F}^{2}d^{3}k\right) ^{\frac{1}{2}}= \)}\\
{\large }\\
\( =\left( \int ^{\kappa }_{\sigma _{t}}\left\Vert I_{\mathbf{P}-\frac{\mathbf{q}}{s}}\left( W^{b}_{\sigma _{t}}\left( \nabla E^{\sigma _{t}}\left( \mathbf{P}-\frac{\mathbf{q}}{s}\right) \right) \left( b\left( \mathbf{k}\right) +\frac{g\chi ^{\kappa }_{\sigma _{t}}\left( \mathbf{k}\right) }{\sqrt{2}\left| \mathbf{k}\right| ^{\frac{3}{2}}\left( 1-\widehat{\mathbf{k}}\cdot \nabla E_{\mathbf{P}-\frac{\mathbf{q}}{s}}^{\sigma _{t}}\right) }\right) \psi _{\mathbf{P}+\frac{\mathbf{q}}{s},\sigma _{t}}\right) \right\Vert _{F}^{2}d^{3}k\right) ^{\frac{1}{2}} \){\large }\\
{\large }\\
{\large by using techniques like in {[}2{]}, it is possible to give a bound
with a uniform constant in \( t \) and in \( s \). For our purposes, a bound
of order \( \left( \ln \left( \sigma _{t}\right) \right) ^{\frac{1}{2}} \)
(uniform in \( s \)) is sufficient. For \( \mathbf{P}\in \Sigma  \), it comes
from the following formula obtained in {[}2{]}}\\
{\large }\\
{\large 
\[
b\left( \mathbf{k}\right) \psi _{\mathbf{P},\sigma _{t}}=\frac{g}{\sqrt{2\left| \mathbf{k}\right| }}\left( \frac{1}{E^{\sigma _{t}}\left( \mathbf{P}\right) -\left| \mathbf{k}\right| -H_{\mathbf{P}-\mathbf{k},\sigma _{t}}}\right) \psi _{\mathbf{P},\sigma _{t}}\]
}\\
{\large }\\
{\large \par}
\end{itemize}
{\large For \( s\leq \sigma _{t}^{-\frac{40}{39}} \), the module of \( \left| e^{i\gamma _{\sigma _{t}}\left( \mathbf{v}_{i},\nabla E_{\mathbf{P}}^{\sigma _{t}},s\right) }-e^{i\gamma _{\sigma _{t}}\left( \mathbf{v}_{i},\nabla E_{\mathbf{P}+\frac{\mathbf{q}}{s}}^{\sigma _{t}},s\right) }\right|  \)
can be estimated by the module of the difference of the exponents:}\\
\\
{\large \( \left| \int ^{s}_{1}\left\{ g^{2}\int ^{\sigma ^{L}_{\tau }}_{\sigma _{t}}\frac{\cos \left( \mathbf{k}\cdot \nabla E_{\mathbf{P}}^{\sigma _{t}}\tau -\left| \mathbf{k}\right| \tau \right) -\cos \left( \mathbf{k}\cdot \nabla E_{\mathbf{P}-\frac{\mathbf{q}}{s}}^{\sigma _{t}}\cdot \tau -\left| \mathbf{k}\right| \tau \right) }{\left( 1-\widehat{\mathbf{k}}\cdot \mathbf{v}_{i}\right) }d\Omega d\left| \mathbf{k}\right| \right\} d\tau \right| = \)}
{\large }\\
{\large }\\
{\large \( =\left| 2\int ^{s}_{1}\left\{ g^{2}\int ^{\sigma ^{L}_{\tau }}_{\sigma _{t}}\frac{\sin \left( \frac{\mathbf{k}\cdot \left( \nabla E_{\mathbf{P}}^{\sigma _{t}}-\nabla E_{\mathbf{P}-\frac{\mathbf{q}}{s}}^{\sigma _{t}}\right) \tau }{2}\right) \sin \left( \frac{\mathbf{k}\cdot \left( \nabla E_{\mathbf{P}}^{\sigma _{t}}+\nabla E_{\mathbf{P}-\frac{\mathbf{q}}{s}}^{\sigma _{t}}\right) \tau -2\left| \mathbf{k}\right| \tau }{2}\right) }{\left( 1-\widehat{\mathbf{k}}\cdot \mathbf{v}_{i}\right) }d\Omega d\left| \mathbf{k}\right| \right\} d\tau \right| \leq  \)}\\
{\large }\\
{\large \( \leq C\cdot \left| \nabla E_{\mathbf{P}}^{\sigma _{t}}-\nabla E_{\mathbf{P}-\frac{\mathbf{q}}{s}}^{\sigma _{t}}\right| \cdot \int ^{s}_{1}\left( \sigma ^{L}_{\tau }\right) ^{2}\cdot \tau d\tau \leq C\cdot \left| \frac{\mathbf{q}}{s}\right| ^{\frac{1}{16}}\cdot \int ^{s}_{1}\tau ^{-\frac{19}{20}}d\tau \leq  \)}\\
{\large }\\
{\large \( \leq C\cdot \left| \frac{s^{\frac{1}{20}}}{s}\right| ^{\frac{1}{16}}\cdot s^{\frac{1}{20}}<C\cdot s^{-\frac{1}{112}} \)}\\
\\
{\large An analogous estimate holds for \( s>\sigma _{t}^{-\frac{40}{39}} \).}\\
{\large }\\
{\large }\\
{\large Summarizing, the term (b3.1) is bounded by}\\
{\large }\\
{\large \( C\cdot \left( s^{-\frac{19}{20}}\right) ^{\frac{1}{32}}\cdot s^{2\delta }\cdot t^{-\frac{3\epsilon }{2}}+C\cdot \left( s^{-\frac{19}{20}}\right) ^{\frac{1}{16}}\cdot s^{2\delta }\cdot t^{-\frac{3\epsilon }{2}}\cdot \left| \ln \sigma _{t}\right| +C\cdot s^{-\frac{1}{112}}\cdot s^{2\delta }\cdot t^{-\frac{3\epsilon }{2}}\leq  \)}\\
{\large }\\
{\large \( \leq C\cdot s^{-\frac{1}{112}}\cdot s^{2\delta }\cdot t^{-\frac{3\epsilon }{2}}\cdot \left| \ln \sigma _{t}\right|  \)}
{\large }\\
{\large }\\
{\large }\\
\emph{\large bound of the term (b3.2)}{\large }\\
{\large }\\
{\large Being \( G\in C_{0}^{1}\left( R^{3}\setminus 0\right)  \) and for the
lemma 3.3 applied to }\\
{\large \( e^{i\left( E^{\sigma _{t}}\left( \mathbf{P}-\frac{\mathbf{q}}{s}\right) -E^{\sigma _{t}}\left( \mathbf{P}\right) \right) s}-e^{i\left( E^{\sigma _{t}}\left( \mathbf{P}\right) -E^{\sigma _{t}}\left( \mathbf{P}+\frac{\mathbf{q}}{s}\right) \right) s} \)}
{\large there is a bounding with the quantity}\\
{\large 
\[
C\cdot \left| \mathbf{q}\right| \cdot \left( \frac{\left| \mathbf{q}\right| }{s}\right) ^{\frac{1}{16}}\cdot s^{2\delta }\cdot t^{-\frac{3\epsilon }{2}}\leq C\cdot s^{-\frac{1}{20}}\cdot \left( s^{-\frac{19}{20}}\right) ^{\frac{1}{16}}\cdot s^{2\delta }\cdot t^{-\frac{3\epsilon }{2}}\leq C\cdot s^{2\delta }\cdot s^{-\frac{1}{112}}\cdot t^{-\frac{3\epsilon }{2}}\]
}\\
{\large }\\
{\large }\\
\emph{\large bound of the term (b3.3)}{\large }\\
{\large }\\
{\large Starting from a difference between volumes, the expression (b3.3) is
bounded by a quantity of order~ \( \left( \frac{\left| \mathbf{q}\right| }{s}\right) ^{\frac{1}{2}}\cdot s^{2\delta }\cdot t^{-\epsilon }\leq s^{-\frac{19}{40}}\cdot s^{2\delta }\cdot t^{-\epsilon } \)}
{\large }\\
{\large }\\
{\large }\\
\emph{\large second term of the (b2)}{\large }\\
{\large }\\
{\large It is bounded by \( \left\Vert \psi _{i,\sigma _{t}}^{\left( t\right) }\right\Vert \cdot \int ^{+\infty }_{s^{\frac{1}{20}}}\left| \widetilde{\chi }^{\left( t\right) }_{\mathbf{v}_{i}}\left( \mathbf{q},s\right) \right| d^{3}q \)
then by
\[
C\cdot s^{-\frac{1}{20}}\cdot s^{2\delta }\cdot t^{-\frac{3\epsilon }{2}}\]
}\\
{\large }\\
{\large }\\
{\large In conclusion the sum of the terms} \textbf{\large i)} {\large and}
\textbf{\large ii)} {\large is bounded by}  {\large }\\
{\large 
\[
C\cdot s^{-\frac{1}{112}}\cdot s^{2\delta }\cdot \left| \ln \sigma _{t}\right| \cdot t^{-\frac{3\epsilon }{2}}\]
}\\
{\large having assumed \( \frac{\epsilon }{2}\leq \frac{19}{40}-\frac{1}{112} \).}\\
{\large }\\
\textbf{\emph{\large Corollary B2}}{\large }\\
{\large }\\
{\large From the previous proof it follows that for \( s>t\gg 1 \) the norm
of }\\
{\large }\\
\textbf{\( \left( g^{2}\int ^{s\cdot s^{-\frac{39}{40}}}_{\sigma _{t}\cdot s}\int \frac{\cos \left( \mathbf{q}\cdot \frac{\mathbf{x}}{s}-\left| \mathbf{q}\right| \right) }{\left( 1-\widehat{\mathbf{q}}\cdot \mathbf{v}_{i}\right) }d\Omega \frac{d\left| \mathbf{q}\right| }{s}-g^{2}\int ^{s\cdot s^{-\frac{39}{40}}}_{\sigma _{t}\cdot s}\int \frac{\cos \left( \mathbf{q}\cdot \nabla E^{\sigma _{t}}-\left| \mathbf{q}\right| \right) }{\left( 1-\widehat{\mathbf{q}}\cdot \mathbf{v}_{i}\right) }d\Omega \frac{d\left| \mathbf{q}\right| }{s}\right) e^{-iE^{\sigma _{t}}s}e^{i\gamma _{\sigma _{t}}\left( \mathbf{v}_{i},\nabla E_{\mathbf{P}}^{\sigma _{t}},s\right) }\psi _{i,\sigma _{t}}^{\left( t\right) } \)}\\
\\
\\
{\large is surely bounded by a quantity of order \( s^{-1}\cdot s^{-\frac{1}{112}}\cdot s^{2\delta }\cdot \left| \ln \sigma _{t}\right| \cdot t^{-\frac{3\epsilon }{2}} \).}\\
{\large }\\
\textbf{\large }\\
\textbf{\large Lemma B3} {\large }\\
{\large }\\
{\large I study the function \( \varphi _{\sigma _{t},\mathbf{v}_{i}}\left( \mathbf{x},t\right)  \)}
{\large ~~\( (t\gg 1) \) ~~~where \( \sigma _{t}=\frac{1}{t^{\alpha }} \)
and \( \alpha >0 \)}\\
{\large }\\
{\large \( \varphi _{\sigma _{t},\mathbf{v}_{i}}\left( \mathbf{x},t\right) =g^{2}\int ^{\kappa _{1}}_{\sigma _{t}}\int \frac{\cos \left( \mathbf{k}\cdot \mathbf{x}-\left| \mathbf{k}\right| t\right) }{\left( 1-\widehat{\mathbf{k}}\cdot \mathbf{v}_{i}\right) }d\Omega d\left| \mathbf{k}\right| =g^{2}\int \frac{\sin \left( \kappa _{1}\widehat{\mathbf{k}}\cdot \mathbf{x}-\kappa _{1}t\right) -\sin \left( \sigma _{t}\widehat{\mathbf{k}}\cdot \mathbf{x}-\sigma _{t}\cdot t\right) }{\left( 1-\widehat{\mathbf{k}}\cdot \mathbf{v}_{i}\right) \cdot \left( \widehat{\mathbf{k}}\cdot \mathbf{x}-t\right) }d\Omega  \)}\\
{\large }\\
\textbf{\emph{\large Observation}} \textbf{\large 1} {\large }\\
{\large }\\
{\large I analyze the behavior of \( \varphi _{\sigma _{t},\mathbf{v}_{i}}\left( \mathbf{x},t\right)  \)
for \( \mathbf{x}\in R^{3} \)}\emph{\large .}\textbf{\emph{\large }}\\
\textbf{\emph{\large }}\\
\textbf{\emph{\large region}} \textbf{\large \( \mathbf{x}:\; \left| \mathbf{x}\right| <\left( 1-\eta \right) t \)
, \( 0<\eta <1 \)}:{\large }\\
{\large }\\
{\large \( \left| g^{2}\int \frac{\sin \left( \kappa \widehat{_{1}\mathbf{k}}\cdot \mathbf{x}-\kappa _{1}t\right) -\sin \left( \sigma _{t}\widehat{\mathbf{k}}\cdot \mathbf{x}-\sigma _{t}\cdot t\right) }{\left( 1-\widehat{\mathbf{k}}\cdot \mathbf{v}_{i}\right) \cdot \left( \widehat{\mathbf{k}}\cdot \mathbf{x}-t\right) }\cdot d\Omega \right| \leq \int \left| g^{2}\frac{2}{\left( 1-\widehat{\mathbf{k}}\cdot \mathbf{v}_{i}\right) \cdot \left( \widehat{\mathbf{k}}\cdot \mathbf{x}-t\right) }\right| d\Omega \leq  \)}\\
{\large \( \leq \frac{1}{\eta t}\int \left| g^{2}\frac{2}{\left( 1-\widehat{\mathbf{k}}\cdot \mathbf{v}_{i}\right) }\right| d\Omega  \)}\\
\textbf{\Large }\\
\textbf{\Large }\\
\textbf{\Large }\\
\textbf{\emph{\large region}} \textbf{\large \( \mathbf{x}:\; \left( 1-\eta \right) t<\left| \mathbf{x}\right|  \)
, \( 0<\eta <1 \)}\textbf{:} {\large }\\
{\large }\\
{\large \( \varphi _{\sigma _{t},\mathbf{v}_{i}}\left( \mathbf{x},t\right) =g^{2}\int ^{\kappa _{1}}_{\sigma _{t}}\int \frac{\cos \left( \mathbf{k}\cdot \mathbf{x}-\left| \mathbf{k}\right| t\right) }{\left( 1-\widehat{\mathbf{k}}\cdot \mathbf{v}_{i}\right) }d\Omega d\left| \mathbf{k}\right| =g^{2}\int ^{\kappa _{1}}_{\sigma _{t}}\int \cos \left( \mathbf{k}\cdot \mathbf{x}-\left| \mathbf{k}\right| t\right) \cdot \xi \left( \widehat{\mathbf{k}},\mathbf{v}_{i}\right) d\Omega d\left| \mathbf{k}\right|  \)}\\
{\large }\\
{\large where \( \xi \left( \widehat{\mathbf{k}},\mathbf{v}_{i}\right) =\xi \left( \theta ,\varphi ,\mathbf{v}_{i}\right) \equiv \frac{1}{\left( 1-\widehat{\mathbf{k}}\cdot \mathbf{v}_{i}\right) } \).
Note that being \( \left| \mathbf{v}_{i}\right| \leq v^{max}<1 \) }\\
{\large \( \exists \quad M\geq 0,M'\geq 0 \) \( \! \Longrightarrow \qquad \left| \xi \left( \widehat{\mathbf{k}},\mathbf{v}_{i}\right) \right| <M \)
and \( \left| \frac{d}{d\cos \theta }\left[ \xi \left( \theta ,\varphi ,\mathbf{v}_{i}\right) \right] \right| <M' \));}
{\large }\\
{\large }\\
{\large I integrate by parts with respect to \( d\cos \theta  \)}\\
{\large }\\
{\large \( g^{2}\int ^{\kappa _{1}}_{\sigma _{t}}\int \frac{e^{i\left( \mathbf{k}\cdot \mathbf{x}-\left| \mathbf{k}\right| t\right) }}{\left( 1-\widehat{\mathbf{k}}\cdot \mathbf{v}_{i}\right) }d\Omega d\left| \mathbf{k}\right| = \)
}\\
{\large }\\
{\large \( =\int ^{\kappa _{1}}_{\sigma _{t}}\int \frac{e^{-i\left( \left| \mathbf{k}\right| \cdot \left| \mathbf{x}\right| +\left| \mathbf{k}\right| t\right) }}{\left| \mathbf{x}\right| }\cdot \frac{\xi \left( \pi ,\varphi ,\mathbf{v}_{i}\right) }{i\left| \mathbf{k}\right| }d\left| \mathbf{k}\right| d\varphi -\int ^{\kappa _{1}}_{\sigma _{t}}\int \frac{e^{i\left( \left| \mathbf{k}\right| \cdot \left| \mathbf{x}\right| -\left| \mathbf{k}\right| t\right) }}{\left| \mathbf{x}\right| }\cdot \frac{\xi \left( 0,\varphi ,\mathbf{v}_{i}\right) }{i\left| \mathbf{k}\right| }d\left| \mathbf{k}\right| d\varphi + \)}\\
{\large }\\
{\large \( -\int ^{\kappa _{1}}_{\sigma _{t}}\int \frac{e^{i\left( \mathbf{k}\cdot \mathbf{x}-\left| \mathbf{k}\right| t\right) }}{\left| \mathbf{x}\right| }\cdot \frac{1}{i\left| \mathbf{k}\right| }\frac{d}{d\cos \theta }\left[ \xi \left( \theta ,\varphi ,\mathbf{v}_{i}\right) \right] d\left| \mathbf{k}\right| d\Omega  \)} \marginpar{
{\large (b5)}{\large \par}
} {\large }\\
{\large }\\
{\large this is bounded by a quantity of order \( \frac{\left| \ln \sigma _{t}\right| }{t-\eta t} \).
}\\
{\large }\\
{\large In conclusion there exists a constant \( C \), that is uniform in \( \mathbf{v}_{i} \)
belonging to the region \( \left| \mathbf{v}_{i}\right| <v^{max}<1 \), such
that \( \forall \mathbf{x}\in R^{3} \)~~ \( \left| \varphi _{\sigma _{t},\mathbf{v}_{i}}\left( \mathbf{x},t\right) \right| \leq C\cdot \frac{\left| \ln \sigma _{t}\right| }{t-\eta t} \).}\marginpar{
{\large (b6)}{\large \par}
}\textbf{\large }\\
\textbf{\large }\\
{\large }\\
\textbf{\emph{\large Observation 2}}{\large }\\
{\large }\\
{\large }\\
{\large I analyze the behavior of \( \varphi _{\sigma _{t},\mathbf{v}_{i}}\left( \mathbf{x},t\right)  \)
for \( \left( 1-\eta '\right) t<\left| \mathbf{x}\right| <\left( 1-\eta \right) t \),
where}\\
{\large \( 0<\eta <\eta '<1 \). I study the term (b5) (the other term has an
analogous behavior)}\\
{\large }\\
{\large \( \int ^{\kappa _{1}}_{\frac{1}{t^{\alpha }}}\int \frac{e^{-i\left( \left| \mathbf{k}\right| \cdot \left| \mathbf{x}\right| +\left| \mathbf{k}\right| t\right) }}{\left| \mathbf{x}\right| }\cdot \frac{\xi \left( \pi ,\varphi ,\mathbf{v}_{i}\right) }{i\left| \mathbf{k}\right| }d\left| \mathbf{k}\right| d\varphi -\int ^{\kappa _{1}}_{\frac{1}{t^{\alpha }}}\int \frac{e^{i\left( \left| \mathbf{k}\right| \cdot \left| \mathbf{x}\right| -\left| \mathbf{k}\right| t\right) }}{\left| \mathbf{x}\right| }\cdot \frac{\xi \left( 0,\varphi ,\mathbf{v}_{i}\right) }{i\left| \mathbf{k}\right| }d\left| \mathbf{k}\right| d\varphi + \)}
{\large }\\
{\large \( -\int ^{\kappa _{1}}_{\frac{1}{t^{\alpha }}}\int \frac{e^{i\left( \mathbf{k}\cdot \mathbf{x}-\left| \mathbf{k}\right| t\right) }}{\left| \mathbf{x}\right| }\cdot \frac{1}{i\left| \mathbf{k}\right| }\frac{d}{d\cos \theta }\left[ \xi \left( \theta ,\varphi ,\mathbf{v}_{i}\right) \right] d\left| \mathbf{k}\right| d\Omega  \)}\marginpar{
{\large (b7)}{\large \par}
}{\large }\\
{\large }\\
{\large By hypothesis, the following inequalities hold:}\\
{\large }\\
{\large \( \left| \mathbf{x}\right| +t>t \) }\\
{\large \( \left| \mathbf{x}\right| -t<\left( 1-\eta \right) t-t \) \( \Rightarrow  \)
\( \left| \mathbf{x}\right| -t<-\eta t \)\( \Rightarrow  \)\( \left| \left| \mathbf{x}\right| -t\right| >\eta t \)}
{\large }\\
{\large \( \left| \mathbf{x}\right| >\left( 1-\eta '\right) t \)}\\
{\large }\\
{\large I consider, for example, the first term of the (b7) (the other ones
have an analogous behavior) :}\\
{\large }\\
{\large \( -i\left( \left| \mathbf{x}\right| +t\right) \cdot \int ^{\kappa _{1}}_{\frac{1}{t^{\alpha }}}\int \frac{e^{-i\left| \mathbf{k}\right| \cdot \left( \left| \mathbf{x}\right| +t\right) }}{\left| \mathbf{x}\right| }\cdot \frac{\xi \left( \pi ,\varphi ,\mathbf{v}_{i}\right) }{i\left| \mathbf{k}\right| }d\left| \mathbf{k}\right| d\varphi =\int ^{\kappa _{1}}_{\frac{1}{t^{\alpha }}}\int \frac{d}{d\left| \mathbf{k}\right| }\left( \frac{e^{-i\left| \mathbf{k}\right| \cdot \left( \left| \mathbf{x}\right| +t\right) }}{\left| \mathbf{x}\right| }\right) \cdot \frac{\xi \left( \pi ,\varphi ,\mathbf{v}_{i}\right) }{i\left| \mathbf{k}\right| }d\left| \mathbf{k}\right| d\varphi = \)}\\
{\large }\\
{\large (now I integrate by part with respect to \( d\left| \mathbf{k}\right|  \))}\\
{\large }\\
{\large \( =\int \frac{e^{-i\kappa _{1}\cdot \left( \left| \mathbf{x}\right| +t\right) }}{\left| \mathbf{x}\right| }\cdot \frac{\xi \left( \pi ,\varphi ,\mathbf{v}_{i}\right) }{i\kappa _{1}}d\varphi -\int \frac{e^{-it^{-\alpha }\cdot \left( \left| \mathbf{x}\right| +t\right) }}{\left| \mathbf{x}\right| }\cdot \frac{\xi \left( \pi ,\varphi ,\mathbf{v}_{i}\right) }{it^{-\alpha }}d\varphi + \)}\\
{\large }\\
{\large \( +\int ^{\kappa _{1}}_{\frac{1}{t^{\alpha }}}\int \frac{e^{-i\left| \mathbf{k}\right| \cdot \left( \left| \mathbf{x}\right| +t\right) }}{\left| \mathbf{x}\right| }\cdot \xi \left( \pi ,\varphi ,\mathbf{v}_{i}\right) \cdot \frac{1}{i\left| \mathbf{k}\right| ^{2}}\cdot d\left| \mathbf{k}\right| d\varphi  \)}\\
{\large }\\
{\large from which}\\
{\large }\\
{\large \( \int ^{\kappa _{1}}_{\frac{1}{t^{\alpha }}}\int \frac{e^{-i\left| \mathbf{k}\right| \cdot \left( \left| \mathbf{x}\right| +t\right) }}{\left| \mathbf{x}\right| }\cdot \frac{\xi \left( \pi ,\varphi ,\mathbf{v}_{i}\right) }{i\left| \mathbf{k}\right| }d\left| \mathbf{k}\right| d\varphi =\int \frac{e^{-i\kappa _{1}\cdot \left( \left| \mathbf{x}\right| +t\right) }}{\left| \mathbf{x}\right| \left( \left| \mathbf{x}\right| +t\right) }\cdot \frac{\xi \left( \pi ,\varphi ,\mathbf{v}_{i}\right) }{\left| \kappa _{1}\right| }d\varphi -\int \frac{e^{-it^{-\alpha }\cdot \left( \left| \mathbf{x}\right| +t\right) }}{\left| \mathbf{x}\right| \left( \left| \mathbf{x}\right| +t\right) }\cdot \frac{\xi \left( \pi ,\varphi ,\mathbf{v}_{i}\right) }{t^{-\alpha }}d\varphi + \)}\\
{\large }\\
{\large \( +\int ^{\kappa _{1}}_{\frac{1}{t^{\alpha }}}\int \frac{e^{-i\left| \mathbf{k}\right| \cdot \left( \left| \mathbf{x}\right| +t\right) }}{\left| \mathbf{x}\right| \left( \left| \mathbf{x}\right| +t\right) }\cdot \frac{\xi \left( \pi ,\varphi ,\mathbf{v}_{i}\right) }{\left| \mathbf{k}\right| ^{2}}d\left| \mathbf{k}\right| d\varphi  \)}\\
{\large }\\
{\large where each term on the right hand side is bounded by a quantity of order
\( \frac{t^{\alpha }}{t^{2}} \).}\\
{\large }\\
{\large Conclusion: }\\
{\large in the region} {\large \( \left( 1-\eta '\right) t<\left| \mathbf{x}\right| <\left( 1-\eta \right) t \)}
{\large ~~we have that} {\large \( \left| \varphi _{\sigma _{t},\mathbf{v}_{i}}\left( \mathbf{x},t\right) \right| \leq C_{\eta ,\eta '}\cdot \frac{t^{\alpha }}{t^{2}} \)}{\large ~~(\( \left| \mathbf{v}_{i}\right| <v^{max}<1 \)).}
\textbf{\large }{\large }\\
{\large }\\
{\large }\\
\textbf{\large Theorem B4}{\large }\\
{\large }\\
{\large Taking into account lemma B2}, {\large one can prove the existence of}
\\
\\
{\large \( s-\lim _{s\rightarrow +\infty }e^{iH_{\sigma _{t}}s}\int ^{\kappa _{1}}_{\sigma _{t}}\frac{a\left( \mathbf{k}\right) e^{i\left| \mathbf{k}\right| s}}{\left| \mathbf{k}\right| }\cdot h_{i,j}\left( \widehat{\mathbf{k}}\right) \frac{d^{3}k}{\sqrt{2\left| \mathbf{k}\right| }}e^{-iH_{\sigma _{t}}s}\psi _{i,\sigma _{t}}^{\left( t\right) }\equiv a_{\sigma _{t}}^{out}\left( h\right) \psi _{i,\sigma _{t}}^{\left( t\right) } \)}\\
\\
{\large The vectors \( a_{\sigma _{t}}^{out\left( in\right) }\left( h\right) \psi _{i,\sigma _{t}}^{\left( t\right) } \)are
in \( D\left( H_{\sigma _{t}}\right)  \).}\\
\\
{\large Proof}\\
\\
{\large In order to prove the strong convergence, we check that the following
quantity is integrable with respect to \( s \): }\\
{\large }\\
{\large 
\[
\left\Vert \frac{d\left\{ e^{iH_{\sigma _{t}}s}\int ^{\kappa _{1}}_{\sigma _{t}}\frac{a\left( \mathbf{k}\right) e^{i\left| \mathbf{k}\right| s}}{\left| \mathbf{k}\right| }\cdot h_{i,j}\left( \widehat{\mathbf{k}}\right) \frac{d^{3}k}{\sqrt{2\left| \mathbf{k}\right| }}e^{-iH_{\sigma _{t}}s}\psi _{i,\sigma _{t}}^{\left( t\right) }\right\} }{ds}\right\Vert \]
}\\
{\large }\\
{\large }\\
{\large Formally :}\\
{\large }\\
{\large \( \frac{d\left( e^{iH_{\sigma _{t}}s}a\left( h_{s}\right) e^{-iH_{\sigma _{t}}s}\right) }{dt}=\frac{d\left( e^{iH_{\sigma _{t}}s}e^{-iH^{mes}s}a\left( h\right) e^{iH^{mes}s}e^{-iH_{\sigma _{t}}s}\right) }{dt}= \)}\\
{\large }\\
{\large \( =e^{iH_{\sigma _{t}}s}\left\{ i\int ^{\kappa _{1}}_{\sigma _{t}}\widetilde{h}_{i,j}\left( \mathbf{k}\right) e^{i\left( \left| \mathbf{k}\right| s-\mathbf{k}\cdot \mathbf{x}\right) }\frac{1}{2\left| \mathbf{k}\right| ^{2}}d^{3}k\right\} e^{-iH_{\sigma _{t}}s} \)}\\
{\large }\\
{\large The formal expressions are well defined from an operatorial point of
view in \( D\left( H_{\sigma _{t}}\right)  \). }\\
{\large }\\
{\large Having defined \( \widehat{h}\left( \mathbf{x},s\right) \equiv \left\{ i\int ^{\kappa _{1}}_{\sigma _{t}}\widetilde{h}_{i,j}\left( \mathbf{k}\right) e^{i\left( \left| \mathbf{k}\right| s-\mathbf{k}\cdot \mathbf{x}\right) }\frac{1}{2\left| \mathbf{k}\right| ^{2}}d^{3}k\right\}  \),
I consider the Hilbert inequality: }\\
{\large }\\
\( \left\Vert \frac{d\left( e^{iH_{\sigma _{t}}s}a\left( h\right) e^{-iH_{\sigma _{t}}s}\right) }{dt}\psi _{i,\sigma _{t}}^{\left( t\right) }\right\Vert \leq  \)\\
\\
\( \leq \left\Vert \widehat{h}\left( \mathbf{x},s\right) \left( 1_{\Gamma _{i}}\left( \mathbf{P}\right) -\chi ^{\left( t\right) }_{\mathbf{v}_{i}}\left( \nabla E^{\sigma _{t}},s\right) \right) e^{-iH_{\sigma _{t}}s}\psi _{i,\sigma _{t}}^{\left( t\right) }\right\Vert +\left\Vert \widehat{h}\left( \mathbf{x},s\right) \left( \chi ^{\left( t\right) }_{\mathbf{v}_{i}}\left( \nabla E^{\sigma _{t}},s\right) -\chi ^{\left( t\right) }_{\mathbf{v}_{i}}\left( \frac{\mathbf{x}}{s},s\right) \right) e^{-iH_{\sigma _{t}}s}\psi _{i,\sigma _{t}}^{\left( t\right) }\right\Vert + \)\\
\\
\( +\left\Vert \widehat{h}\left( \mathbf{x},s\right) \chi ^{\left( t\right) }_{\mathbf{v}_{i}}\left( \frac{\mathbf{x}}{s},s\right) e^{-iH_{\sigma _{t}}s}\psi _{i,\sigma _{t}}^{\left( t\right) }\right\Vert \leq  \)\\
\\
\( \leq \sup _{\mathbf{x}}\left| \widehat{h}\left( \mathbf{x},s\right) \right| \cdot \left\Vert \left( 1_{\Gamma _{i}}\left( \mathbf{P}\right) -\chi ^{\left( t\right) }_{\mathbf{v}_{i}}\left( \nabla E^{\sigma _{t}},s\right) \right) e^{-iH_{\sigma _{t}}s}\psi _{i,\sigma _{t}}^{\left( t\right) }\right\Vert +\sup _{\mathbf{x}}\left| \widehat{h}\left( \mathbf{x},s\right) \right| \cdot \left\Vert \left( \chi ^{\left( t\right) }_{\mathbf{v}_{i}}\left( \nabla E^{\sigma _{t}},s\right) -\chi ^{\left( t\right) }_{\mathbf{v}_{i}}\left( \frac{\mathbf{x}}{s},s\right) \right) e^{-iH_{\sigma _{t}}s}\psi _{i,\sigma _{t}}^{\left( t\right) }\right\Vert + \)\\
\\
\( +\sup _{\frac{\mathbf{x}}{s}\in \mathbf{J}_{\sigma _{t}}\left( \Gamma _{i}\right) }\left| \widehat{h}\left( \mathbf{x},s\right) \cdot \chi _{\mathbf{v}_{i}}\left( \frac{\mathbf{x}}{s},s\right) \right| \cdot \left\Vert \psi _{i,\sigma _{t}}^{\left( t\right) }\right\Vert  \){\large }\\
{\large }\\
{\large Because of the results of lemmas B1, B2 and of the estimate in lemma
B3, the first two terms on the right hand side are respectively bounded by \( C\cdot s^{-\frac{\delta }{12}}\cdot \frac{\left| \ln \sigma _{t}\right| }{s}\cdot t^{-\epsilon } \)
(\( \delta >24\epsilon  \)) and by \( C\cdot s^{-1}\cdot s^{-\frac{1}{112}}\cdot s^{2\delta -\frac{3\epsilon }{2}}\cdot \left( \ln \sigma _{t}\right) ^{2} \)
(we assume the constraint \( 2\delta +3\epsilon <\frac{1}{112} \)). As regards
the third term, the hypotheses on \( G\left( \mathbf{P}\right)  \), \( \mathbf{v}_{i} \)
(see the note at pag.69) and} {\large the} \emph{\large observation 2} {\large of
lemma B3} {\large ensure a vanishing of order \( \frac{\sigma _{t}}{s^{2}}\cdot t^{-\frac{3\epsilon }{2}} \),
for \( s\rightarrow +\infty  \).}\\
{\large }\\
\textbf{\large }\\
\emph{\large The vectors \( a_{\sigma _{t}}^{out\left( in\right) }\left( h\right) \psi _{i,\sigma _{t}}^{\left( t\right) } \)belong
to \( D\left( H\right)  \)}{\large .}\\
{\large }\\
{\large For each \( s \), \( H_{\sigma _{t}}e^{iH_{\sigma _{t}}s}a\left( h_{s}\right) e^{-iH_{\sigma _{t}}s}\psi _{i,\sigma _{t}}^{\left( t\right) } \)
is well defined because \( a\left( h_{s}\right) \psi _{i,\sigma _{t}}^{\left( t\right) }\subset D\left( H_{\sigma _{t}}\right)  \).
\( H_{\sigma _{t}} \) is a closed operator. Therefore it is sufficient to prove
the convergence, for \( s\rightarrow +\infty  \), of:}\\
{\large }\\
{\large \( H_{\sigma _{t}}e^{iH_{\sigma _{t}}s}a\left( h_{s}\right) e^{-iH_{\sigma _{t}}s}\psi _{i,\sigma _{t}}^{\left( t\right) }=e^{iH_{\sigma _{t}}s}H_{\sigma _{t}}a\left( h_{s}\right) e^{-iH_{\sigma _{t}}s}\psi _{i,\sigma _{t}}^{\left( t\right) }= \)}\\
{\large }\\
\( =\left\{ e^{iH_{\sigma _{t}}s}a\left( h_{s}\right) e^{-iH_{\sigma _{t}}s}E^{\sigma _{t}}\left( \mathbf{P}\right) \psi _{i,\sigma _{t}}^{\left( t\right) }+e^{iH_{\sigma _{t}}s}\left[ H_{\sigma _{t}}-H^{mes},a\left( h_{s}\right) \right] e^{-iH_{\sigma _{t}}s}\psi _{i,\sigma _{t}}^{\left( t\right) }+e^{iH_{\sigma _{t}}s}\left[ H^{e.m.},a\left( h_{s}\right) \right] e^{-iH_{\sigma _{t}}s}\psi _{i,\sigma _{t}}^{\left( t\right) }\right\}  \)\marginpar{

}{\large }\\
{\large }\\
{\large Looking at the first part of the theorem and being} \emph{\large \( \left[ H^{e.m.},a\left( h_{s}\right) \right] =-\int ^{\kappa _{1}}_{\sigma _{t}}a\left( \mathbf{k}\right) e^{i\left| \mathbf{k}\right| s}\frac{\widetilde{h}_{i,j}\left( \mathbf{k}\right) }{\sqrt{2\left| \mathbf{k}\right| }}d^{3}k \)}{\large ,
each term in the above expression has limit.}\\
{\large }\\
{\large }\\
{\large }\\
 {\large }\textbf{\large Theorem B5}{\large }\\
{\large }\\
{\large If for each \( \mathbf{P} \) in \( \Gamma _{i} \) and for each \( \mathbf{k} \)
in \( supp\widetilde{h}\left( \mathbf{k}\right)  \), \( \widetilde{h}\left( \mathbf{k}\right) =\frac{h_{i,j}\left( \widehat{\mathbf{k}}\right) }{\left| \mathbf{k}\right| \sqrt{2\left| \mathbf{k}\right| }}\chi ^{\kappa _{1}}_{\sigma _{t}}\left( \mathbf{k}\right)  \),
it happens that \( \mathbf{P}+\mathbf{k}\in \Sigma  \), then }\\
{\large 
\[
a_{\sigma _{t}}^{out\left( in\right) }\left( h\right) \psi _{i,\sigma _{t}}^{\left( t\right) }=0\]
} {\large }\\
{\large Proof }\\
{\large }\\
{\large Starting from the spectral decomposition with respect to \( \mathbf{P} \)
operators, we obtain that \( \int a_{\sigma _{t}}^{out}\left( \mathbf{k}\right) \widetilde{h}\left( \mathbf{k}\right) \left( \psi _{i,\sigma _{t}}^{\left( t\right) }\right) _{\mathbf{P}+\mathbf{k}}d^{3}k \)
~is a vector in}{\large  ~H}{\large \( _{\mathbf{P}} \) and that it belongs
to the domain of \( H_{\mathbf{P},\sigma _{t}} \).   Then the procedure consists
in studying the mean value of the positive operator \( H_{\mathbf{P},\sigma _{t}}-E^{\sigma _{t}}\left( \mathbf{P}\right)  \)
on it and in taking into account the condition \( \left| \nabla E^{\sigma _{t}}\mid _{\mathbf{P}+\mathbf{q}}\right| <1 \)
\( \forall \mathbf{q}\in supp\widetilde{h} \) (if \( \mathbf{P}\in \Sigma  \))
to estimate \( E^{\sigma _{t}}\left( \mathbf{P}+\mathbf{k}\right) -\left| \mathbf{k}\right| -E^{\sigma _{t}}\left( \mathbf{P}\right) <0 \)
and to conclude that the vector is zero.}\\
{\large }\\
{\large }\\
{\large }\\
 {\large }

\end{document}